\documentclass[phd,bottom,nosig]{usbthesis}
\usepackage{graphicx}
\usepackage{color}
\usepackage{bm}
\usepackage{amsmath}
\usepackage{amssymb}
\usepackage{setspace}
\usepackage{array}
\usepackage{float}
\usepackage[percent]{overpic}
\usepackage{tabularx}
\usepackage{multirow}
\usepackage{afterpage}
\usepackage{caption}
\usepackage{subcaption}
\usepackage{paralist} 
\usepackage{hhline}
\usepackage{url}
\usepackage{hyperref} 
\usepackage{footmisc} 
\usepackage{xspace}
\newcommand{\tb}{\ensuremath{\tau_1^b}\xspace}
\newcommand{\GIM}{\ensuremath{M_{Gr.}^{2}/Q^2}\xspace}
\usepackage[numbers,sort&compress]{natbib}
\usepackage{pdfpages}

\def\simge{%
    \mathrel{\rlap{\raise 0.511ex
    \hbox{$>$}}{\lower 0.511ex \hbox{$\sim$}}}}
\def\simle{%
    \mathrel{\rlap{\raise 0.511ex
    \hbox{$<$}}{\lower 0.511ex \hbox{$\sim$}}}}

\DeclareMathAccent{\ring}{\mathalpha}{operators}{"17}

\providecommand*{\degree}{\ensuremath{^\circ}}

\providecommand{\renewoperator}[3]{\renewcommand*{#1}{\mathop{#2}#3}}
\renewoperator{\Re}{\mathrm{Re}}{\nolimits}
\renewoperator{\Im}{\mathrm{Im}}{\nolimits}

\makeatletter
\providecommand*{\diff}{\@ifnextchar^{\DIfF}{\DIfF^{}}}
\def\DIfF^#1{\mathop{\mathrm{\mathstrut d}}\nolimits^{#1}\gobblespace}
\def\gobblespace{\futurelet\diffarg\opspace}
\def\opspace{%
    \let\DiffSpace\!%
    \ifx\diffarg(%
        \let\DiffSpace\relax
    \else
        \ifx\diffarg[%
            \let\DiffSpace\relax
        \else
            \ifx\diffarg\{%
                \let\DiffSpace\relax
            \fi\fi\fi\DiffSpace}

\author{Henry Klest}%
\title{Groomed Event Shapes at HERA and the sPHENIX TPC}%

\month{\textbf{August}}
\year{\textbf{2023}}%
\program{Physics}%
\director{Thomas K. Hemmick}{Distinguished Teaching Professor, Department of Physics and Astronomy\\Stony Brook University}%
\chairman{Axel Drees}{Distinguished Professor,  Department of Physics and Astronomy\\Stony Brook University}%
\fstmember{Abhay Deshpande}{Distinguished Professor,  Department of Physics and Astronomy\\Stony Brook University}%
\sndmember{Rouven Essig}{Professor, C. N. Yang Institute for Theoretical Physics\\Stony Brook University}%
\outmember{Berndt O. M{\"u}ller}{James B. Duke Distinguished Professor of Physics\\Trinity College of Arts $\&$ Sciences, Duke University}

\begin{document}

\singlespacing %
\pagenumbering{roman} %
\maketitle %
\makeapproval %

\begin{abstract}
This thesis summarizes the work of the author in two directions, both aimed at the study of quantum chromodynamics (QCD). 

The first topic presented is a measurement of groomed event shapes using archived data collected by the H1 experiment at HERA. The data analysis methods and physics implications of the results are discussed, with the goal of improving the theoretical description of the hadronic final state in electron-hadron collisions before the construction of the Electron-Ion Collider (EIC).

The second topic concerns the sPHENIX experiment, which is now installed at the Relativistic Heavy Ion Collider (RHIC). The sPHENIX physics program and apparatus will be discussed, along with a description of each of the subdetectors. Special attention will be dedicated to the operating principles, design, and construction of the time projection chamber (TPC).
\end{abstract}
\tableofcontents %
\listoffigures %
%
%
\pagestyle{thesis}
\newpage
\pagenumbering{arabic}
\chapter{Deep Inelastic Scattering}
The vast majority of the discoveries within the realm of particle physics in the 20th and 21st centuries have been made by scattering experiments. Since the discovery and characterization of atomic nuclei by Rutherford, the technique of scattering has become ubiquitous in the study of particles and their interactions. The ability of mankind to accelerate particles to high energies for the purposes of scattering experiments has facilitated the construction and elucidation of the immensely successful Standard Model of particle physics. \par
\begin{figure}
    \centering
    \includegraphics[width=13cm]{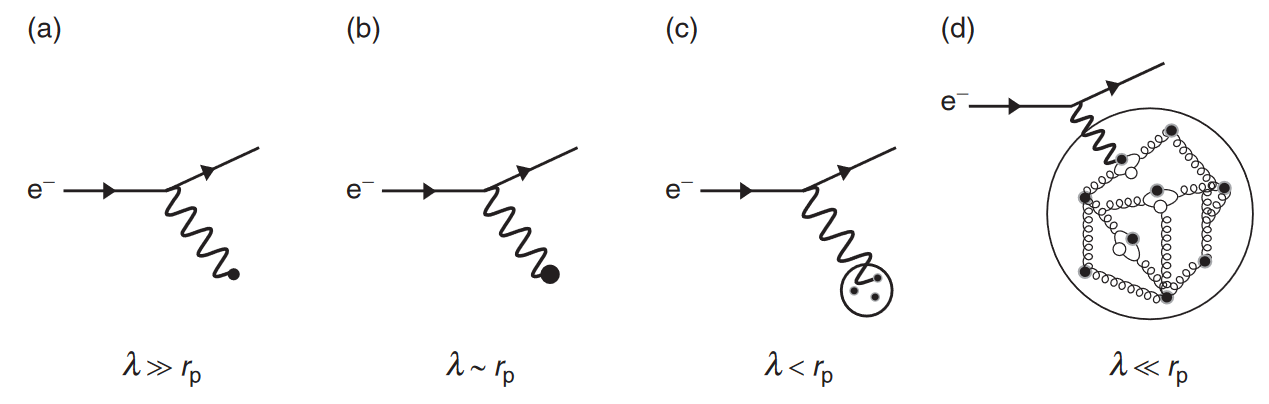}
    \caption{Diagram of electron scattering as a function of energy. $\lambda$ represents the wavelength of the exchanged boson. As the energy of the probing electron increases, smaller boson wavelengths, and therefore smaller length scales, are accessible. Case (a) describes low energy scattering, where the probe's wavelength is significantly larger than the scale of the target, and is therefore unable to resolve any structure. Case (b) represents what was seen at the first Stanford electron scattering experiments, where the extension of the proton was observed. Case (c) shows the results of the SLAC DIS experiments, where partons within the proton could be resolved. Case (d) is the case in high energy DIS, where partons deep inside the proton with small momenta can be resolved. Figure from~\cite{Thomson:2013zua}.}
    \label{fig:ResolutionScale}
\end{figure}

Particularly fruitful have been the experiments employing deep inelastic scattering of electrons on protons. The electron, at the energies accessible to physics thus far, has been determined to be a ``fundamental" particle, in the sense that it lacks substructure. In scattering experiments, the electron interacts primarily via the well-understood electromagnetic force, and the detection of the electron after the scattering allows for determination of the kinematics. Both of these aspects make the electron a useful projectile with which to study the structure of a target, from tabletop electron microscopes to high energy colliders. Higher energy projectiles can exchange more momentum with the target, and thus probe smaller scales, as pictorially represented by Fig.~\ref{fig:ResolutionScale}.  \par
\par
\section{History of DIS}
By the early 1960s, evidence had been mounting that protons and neutrons were not point-like~\cite{Mcallister:1956ng}. Elastic scattering results at Stanford with a 1 GeV electron beam showed a distinct reduction in the cross section at large scattering angles compared to what would be expected from a point charge.  
\begin{figure}[htbp]
    \centering
    \includegraphics[width=13.9cm]{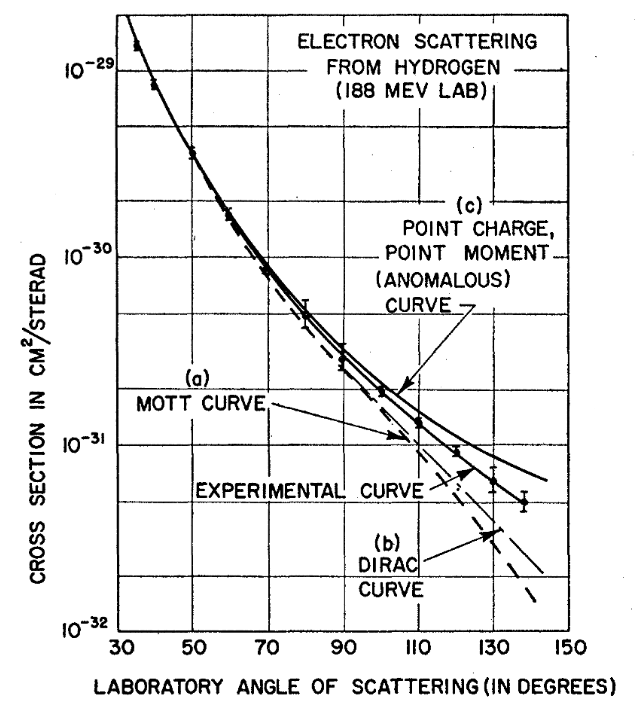}
    \caption{Results of the elastic scattering experiments performed at Stanford in the 1950s compared to models of the proton as a point charge. Curve (a) is Mott scattering, where the proton is point-like and spinless, and the electron is relativistic but the proton recoil can be neglected, i.e. when $m_e \ll E \ll m_p$. Curve (b) describes scattering on a point-like proton with a Dirac magnetic moment. Curve (c) describes scattering on a point-like proton with spin and a magnetic moment, which was the state-of-the-art understanding at the time. The data lying below curve (c) demonstrates that the proton must be an extended object.~\cite{Mcallister:1956ng}}
    \label{fig:Hofstadter}
\end{figure}
After the Stanford results, the prevailing models viewed the proton as an extended charge cloud. The prediction of these models was that as projectile energies increased, the cross section for large-angle scattering would drop significantly as the ability for the probe to ``penetrate" the cloud increased. Similar experiments were thereafter performed at DESY and CEA, observing similar results that confirmed the extension of the proton, but these experiments were unable to make a strong claim about substructure~\cite{Riordan:1992hr}.\par
Additional evidence for hadrons, including protons and neutrons, being composite particles came from the rapidly expanding taxonomy of hadronic states being discovered at accelerators and in cosmic ray experiments. The plethora of hadronic states being discovered provided theorists with the insight that these hadrons may not be fundamental, but instead could be different arrangements of more fundamental constituents. It was proposed in 1961 that the spectrum of particles being discovered could be the manifestation of an SU(3) flavor symmetry, leading Gell-Mann and others to propose the existence of three light, spin 1/2, fractionally charged, fundamental constituents of hadrons called quarks~\cite{Glashow:1961ep}\cite{Gell-Mann:1964ewy}. It was recognized that, like the nucleons in a nucleus, the quarks must have an additional strong interaction between each other to maintain stability against coulomb repulsion. \par 
These considerations and more motivated the series of electron-proton scattering experiments at SLAC with a 20 GeV electron beam~\cite{Bloom:1969kc}~\cite{Breidenbach:1969kd}~\cite{Riordan:1992hr}. The initial SLAC experiments studying elastic scattering showed no signs of substructure, as the elastic cross-section continued to drop precipitously at higher momentum transfer. In Autumn of 1967, the SLAC-MIT collaboration performed the first set of inelastic scattering experiments at 20 GeV. It was observed that the cross section for inelastic scattering decreased with momentum transfer much more slowly than expected, suggesting the electrons were scattering off a hard ``core" inside the proton that could only be accessed in inelastic collisions. If the virtual photons mediating inelastic scattering were interacting coherently with the electromagnetic structure of the proton, a cross section that falls as steeply as elastic scattering would be expected.
\begin{figure}[htbp]
    \centering
    \includegraphics[width=11cm]{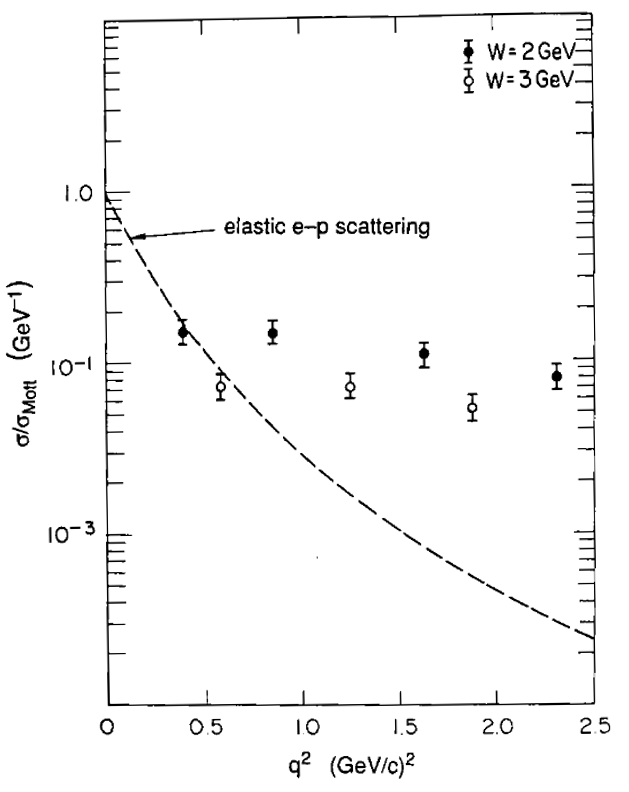}
    \caption{Data of the SLAC-MIT experiment taken at two values of the hadronic invariant mass $W$ demonstrating the unexpectedly slow fall of the inelastic scattering cross section as a function of momentum transfer $q^{2}$.~\cite{Bloom:1969kc}}
    \label{fig:SLAC}
\end{figure}
The initial findings were confirmed through subsequent experiments, including those performed in neutrino DIS at the Gargamelle bubble chamber at CERN~\cite{GargamelleNeutrino:1974exc}, and the data as a whole was in good agreement with predictions in which the electron is scattering on point-like constituents inside the proton.
\section{DIS Kinematics}
\label{DISKinematics}
\begin{figure}[htbp]
    \centering
    \includegraphics[width=12cm]{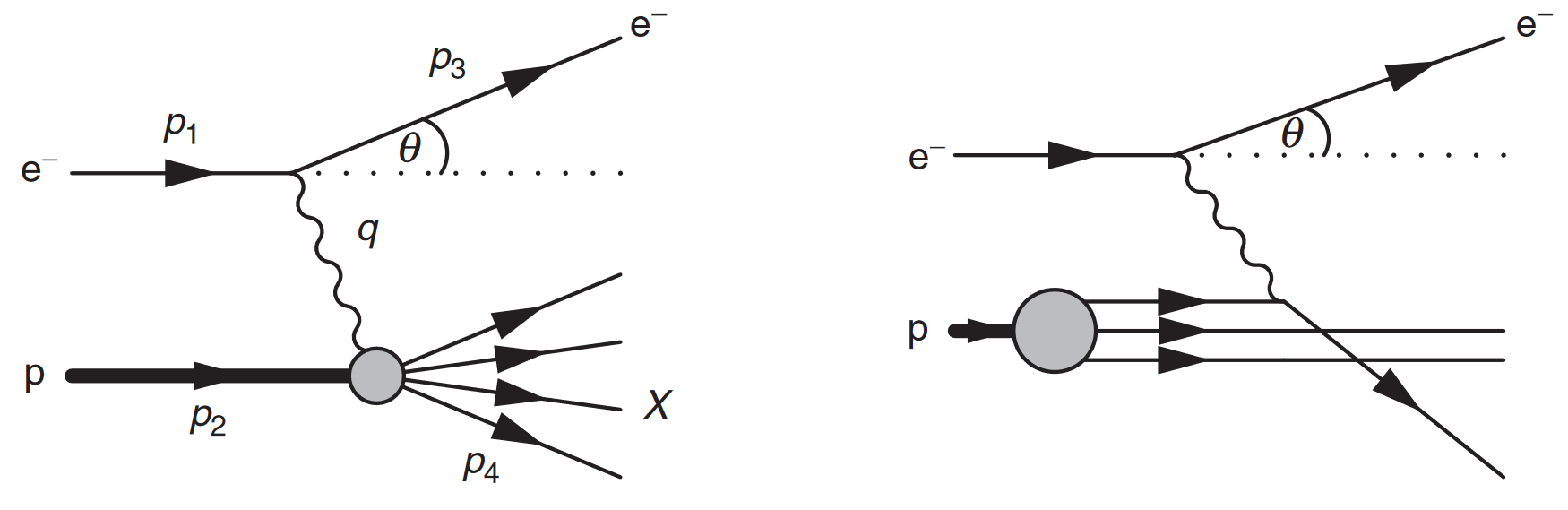}
    \caption{Schematic diagram of neutral current deep inelastic scattering of an electron on a proton. The hadronic final state is represented by $X$. Figure from~\cite{Thomson:2013zua}.}
    \label{fig:DISdiagram}
\end{figure}
The three variables that uniquely identify the kinematics of a standard DIS event are $Q^2, x,$ and $ y$. $Q^2$ is the negative four-momentum squared of the virtual boson exchanged in the collision. In high energy experiments where the electron mass can be neglected, $Q^2$ can be determined by the scattering angle and energy of the scattered electron via Eq. \ref{Q2}

\begin{equation}
\label{Q2}
Q^2 = 2EE'(1-\cos\theta)
\end{equation}

where $E$ is the incoming electron energy, $E'$ is the outgoing electron energy, and $\theta$ is the outgoing electron's scattering angle. 
$y$ is the so-called ``inelasticity" of the collision, defined as:

\begin{equation}
\label{y}
y = \frac{P \cdot q}{P \cdot k}
\end{equation}

Where $P$ is the incoming proton four momentum, $q$ is the exchanged boson four momentum, and $k$ is the incoming electron four momentum. $y$ is constrained to be between 0 and 1, and can be interpreted as the fractional energy transfer of the electron to the hadronic final state in the proton rest frame. 
The variable $x$, often denoted as Bjorken $x$, after James Bjorken who first introduced it~\cite{Bjorken:1968dy}, is defined as:

\begin{equation}
\label{x}
x = \frac{Q^2}{2P \cdot q}
\end{equation}

and is similarly bounded by 0 and 1 for single nucleon targets. $x$ can be naively interpreted as the fraction of the proton's longitudinal momentum that participated in the collision in the frame where the proton has very high energy.
A useful relationship between the above variables in the high energy DIS limit is 

\begin{equation}
\label{sxy}
Q^2 = sxy
\end{equation}

where $s$ is the squared center of mass energy of the collision.\par

The experimental results of deep inelastic scattering experiments can be interpreted in terms of the quark-parton model, wherein the scattering is fundamentally that of an electron scattering elastically on a charged fermion with spin $\frac{1}{2}$ within the proton. The leading-order differential cross section for the elastic scattering of an electron on a fermion with charge $q_q$ whose mass can be neglected is:

\begin{equation}
\label{eqscattering}
\frac{d^{2}\sigma}{dQ^{2}} = \frac{4\pi\alpha^{2}q_q^2}{Q^4}[(1-y)+\frac{y^2}{2}]
\end{equation}

If one assumes that the proton is composed of an unknown variety of these charged fermions, the scattering cross section of an electron and a proton must be parameterized in a way that accounts for the lack of knowledge about the constituents. The cross section for the reaction $e+p \rightarrow e+X$ mediated by the exchange of a single virtual photon can be represented as:

\begin{equation}
\label{F2}
\frac{d^{2}\sigma}{dxdQ^{2}} = \frac{4\pi\alpha^{2}}{Q^4}[(1-y-\frac{m^2_py^2}{Q^2})\frac{F_2(x,Q^2)}{x}+y^2F_1(x,Q^2)]
\end{equation}

Where $F_2$ and $F_1$ are known as the proton structure functions. The observations at SLAC showed that the structure functions at a fixed value of $x$ were roughly independent of $Q^2$ over the accessible kinematics, exhibiting what became known as ``Bjorken scaling". The phenomenon of Bjorken scaling results from the scale invariance of the scattering of two structure-less free particles, in this case the electron and valence quarks in the proton. While the quarks in the proton are clearly not free particles when probed at large length scales, as $Q^2$ increases the quarks begin to appear as free particles within the proton.
\par
\begin{figure}[htbp]
    \centering
    \includegraphics[width=13.9cm]{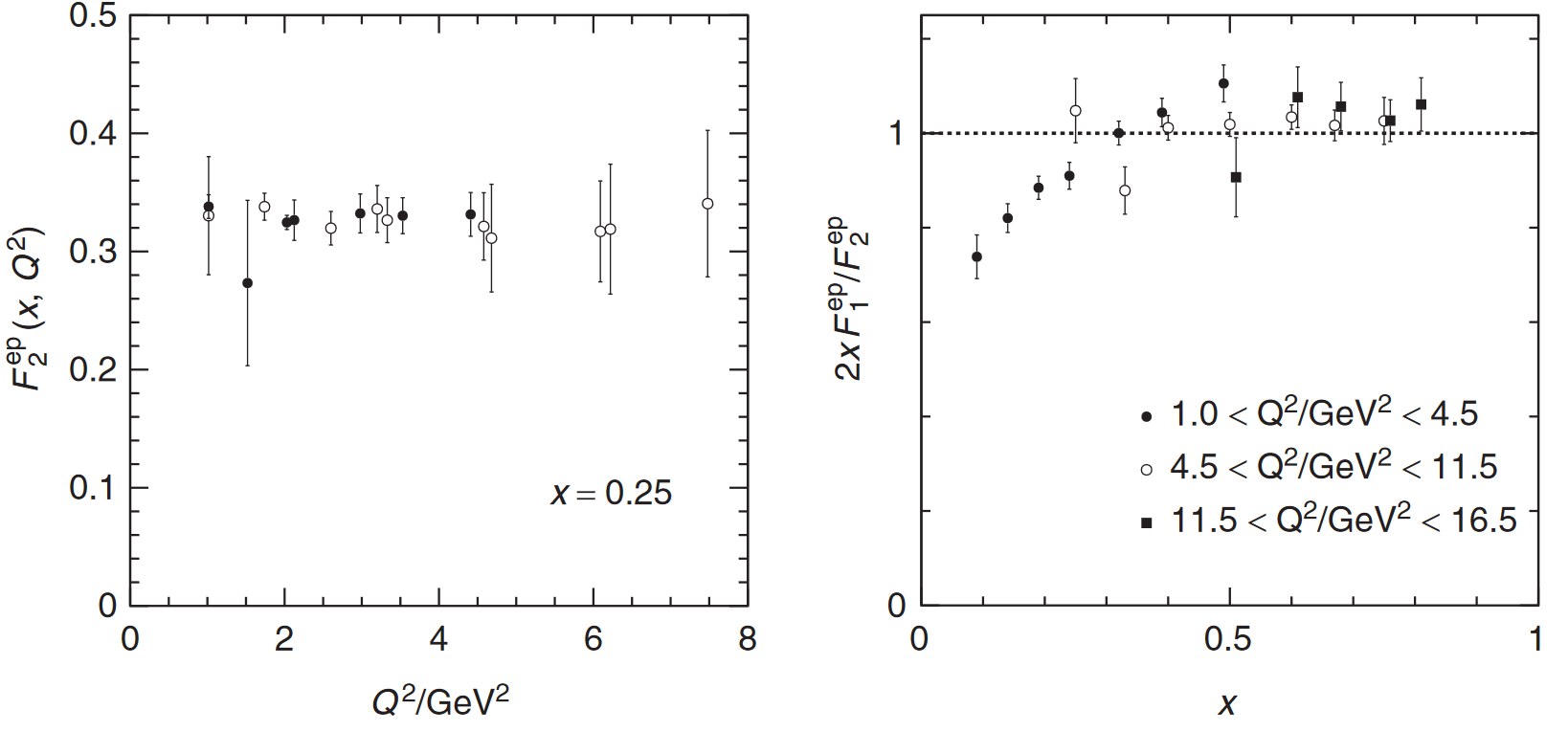}
    \caption{Experimental data demonstrating Bjorken scaling (left panel), and the Callan-Gross relation (right panel). Figure from~\cite{Thomson:2013zua}, adapted from~\cite{Friedman:1972sy} and~\cite{Bodek:1979rx}.}
    \label{fig:BJCGR}
\end{figure}

In addition to Bjorken scaling, it was observed that $F_2$ and $F_1$ satisfied the Callan-Gross relation, $F_2(x) = 2xF_1(x)$. This relation results from the spin properties of the quarks. $F_1$ corresponds to the absorption of transversely polarized bosons, while the absorption of longitudinally polarized bosons is represented by:
\begin{equation}
\label{FL}
F_L(x,q^2) = (1+\frac{4M^2x^2}{Q^2})F_2(x,Q^2) - 2xF_1(x,Q^2)
\end{equation}
which in the Bjorken limit of $Q^2 \rightarrow \infty$ reduces to $F_2-2xF_1$. A spin 0 quark cannot absorb a transversely polarized boson, while a spin 1/2 quark cannot absorb a longitudinally polarized boson. The result is that if quarks were spin 0, $F_1$ would be zero, and $F_L$ would be equal to $F_2$, while if quarks were spin 1/2, $F_L$ would be zero, meaning $F_2(x)=2xF_1(x)$. The experimental data are shown in Fig. \ref{fig:BJCGR}, clearly validating both Bjorken scaling and the Callan-Gross relation in the valence-quark dominated kinematic regions.
 These results solidified the success of the naive quark-parton model of the proton, which states that partons within the proton carry a fraction $x$ of the proton's total longitudinal momentum, and in DIS the exchanged virtual photon elastically scatters incoherently off the quarks within the proton. 
\section{Quantum Chromodynamics}
As early as 1964 it had been suggested that quarks possessed another quantum number that allowed three identical strange quarks to arrange themselves into the $\Omega^-$ hyperon and three identical up quarks to be a $\Delta^{++}$ baryon~\cite{Tkachov:2009na}. This quantum number became known as color, and the theory describing it quantum chromodynamics (QCD). QCD is a non-abelian gauge theory based on an SU(3) color symmetry. Quarks come in three colors, labelled red, blue, and green in analogy with visible light. The quarks themselves carry color charge, and they arrange themselves in such a way that all hadrons are color neutral. The colorful bosons that mediate the interactions between quarks are the gluons, which are massless. There exist eight different gluons which carry a combination of color charge and anti-charge, e.g. red-antiblue, blue-antired, etc. that allow the gluons to change the color of quarks when both emitted and absorbed, thus conserving total color neutrality when exchanged within hadrons. \par
In addition to the valence quarks, protons contain a significant fraction of quarks at low-x known as sea quarks that arise from the splitting of gluons~\cite{Dokshitzer:1977sg}. The introduction of QCD into the quark-parton model allows for violation of the previously mentioned scaling relations via the fluctuations of gluons emitted by valence quarks into quark-antiquark pairs~\cite{Manohar:1992tz}.  The result of these gluon emissions and subsequent pair-conversions is a relative depletion of the number of partons in the valence region of $x \approx 0.3$ and a relative enhancement of partons at low x, i.e $x < 0.01$. Gluons carry no electric charge, so electromagnetic scattering of charged leptons on protons cannot probe the gluon density directly. The $g\rightarrow q\Bar{q}$ process allows the gluon density to be inferred from the various quark densities.
\begin{figure}[htbp]
    \centering
    \includegraphics[width=13cm]{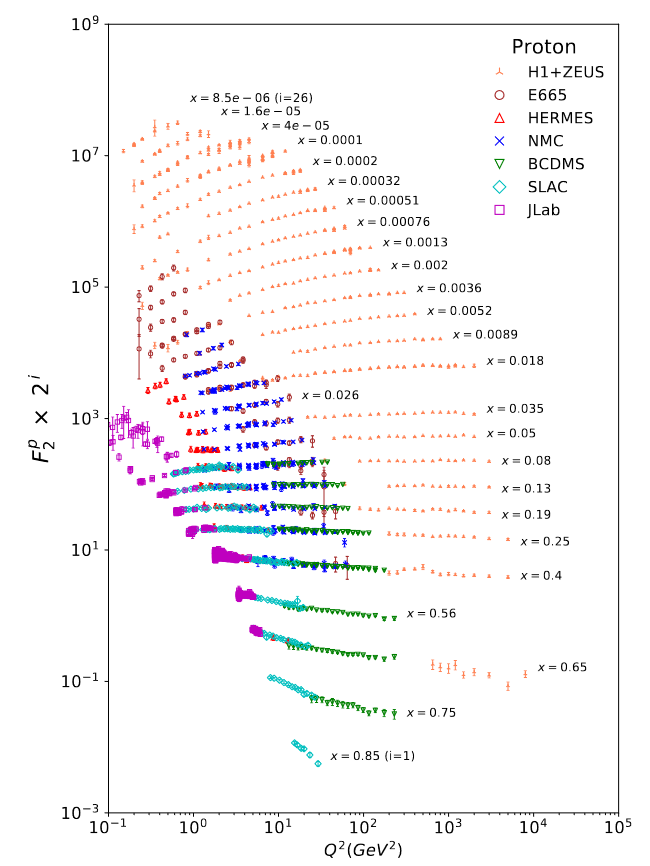}
    \caption{A selection of the world data on the $F_2$ structure function. The curves at different values of $x$ have different slopes, demonstrating scaling violation. Negative slope indicates a depletion of charged partons, while positive slope indicates an enhancement. The depletion at high-x and the corresponding enhancement at low-x can be seen as the redistribution of longitudinal momentum from the valence quarks via gluon emission and splitting~\cite{ParticleDataGroup:2018ovx}.}
    \label{fig:PDGF2}
\end{figure}
\subsection{Confinement and Asymptotic Freedom}
 In quantum electrodynamics (QED), a charged particle is constantly surrounded by a cloud of virtual particle-antiparticle pairs that polarize and serve to ``screen" the effective charge of the primary particle when probed at large scales. The effective electromagnetic coupling constant $\alpha_{EM}$ thus is small ($\approx 1/137$) at low energies, and increases as the energy of the probe increases. When probed at smaller length scales, the vacuum polarization contribution decreases and the probe ``sees" more of the bare charge of the primary particle.
\par
\begin{figure}[ht!]
    \centering
    \includegraphics[width=14cm]{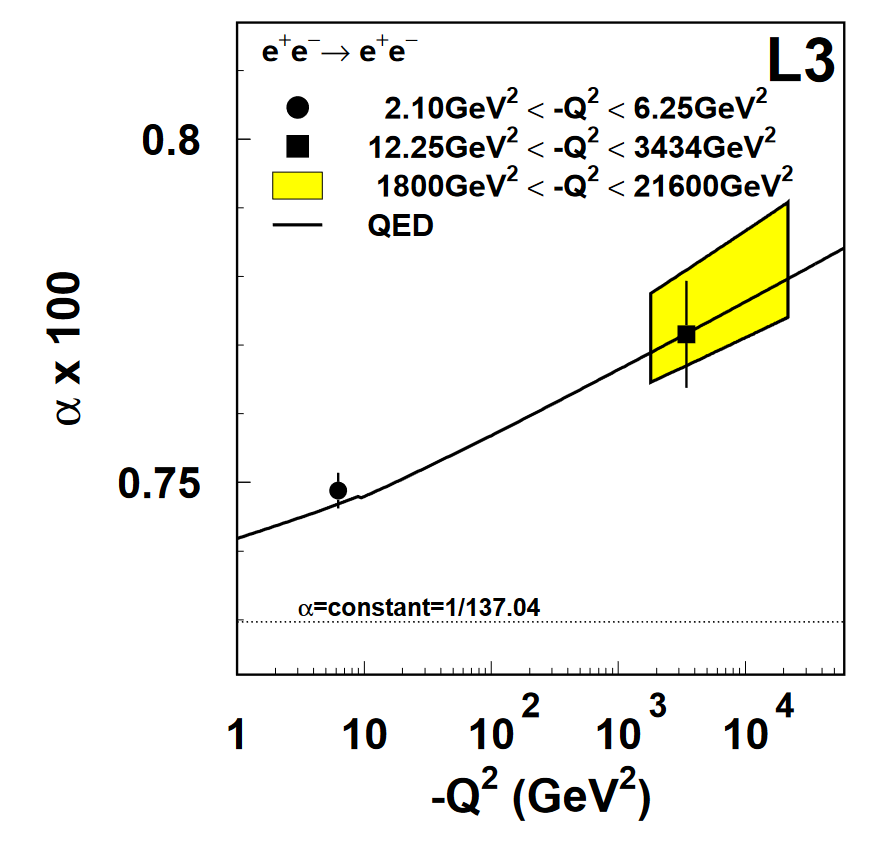}
    \caption{Running of the electromagnetic coupling constant with energy as measured by the L3 collaboration at LEP. It can be seen that the coupling strength increases as a function of probe energy scale~\cite{L3:2005tsb}.}
    \label{fig:AlphaEM}
\end{figure}
\begin{figure}[htbp]
    \centering
    \includegraphics[width=10cm]{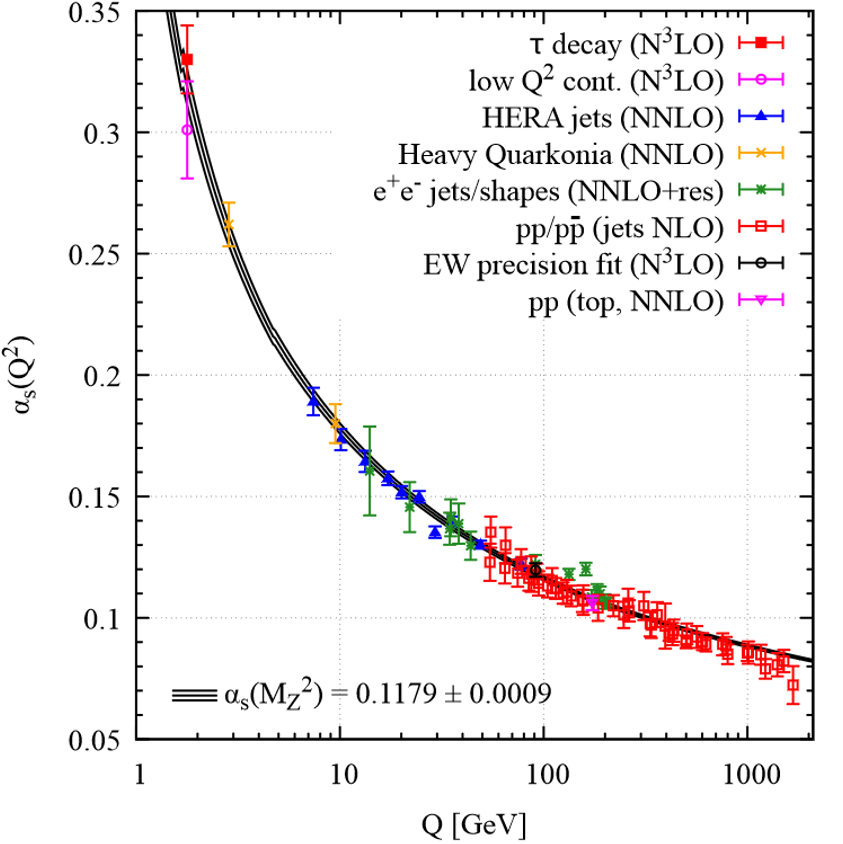}
    \caption{A selection of world experimental data on the value of the strong coupling constant $\alpha_s$ as a function of hard scale $Q$, from Ref.~\cite{ParticleDataGroup:2018ovx}.}
    \label{fig:AlphaS}
\end{figure}

One of the key aspects of QCD is the gluon self-interaction. It is this self-coupling that produces many of the complex emergent phenomena that characterize the strong interaction. Opposite to QED, where the photons do not directly self-couple, gluons couple strongly to one another, and thus manifest an anti-screening effect around color charges. The QCD vacuum is filled not only with quarks and antiquarks, but also with these self-interacting virtual gluon pairs. The net result is that the amount of color charge increases with distance from the primary color charge. Therefore, at large distances the coupling between color charges increases indefinitely, meaning the energy of a lone quark is infinite. This apparently catastrophic growth in the strength of the color field is tamed by the observation that as these color fields become very strong, the amount of energy required to sustain them becomes greater than the threshold to produce new particles that can partner with the lone quark and reduce the overall energy of the system. This lower energy configuration is simply a hadron, and all quarks are therefore ``confined" within hadrons. This phenomenon is known as confinement. On the other end of the spectrum, at high energies and thus short distances, the primary color charge re-emerges at its naturally small coupling strength. This effect is summarized in Fig. \ref{fig:AlphaS}, which shows the world data on the strong coupling constant $\alpha_s$ as a function of energy scale. Color charges thus regain their weak coupling at asymptotically high energies, leading to so-called ``asymptotic freedom". As elegantly quipped by Wilczek, ``quarks are born free, but everywhere they are in chains."~\cite{Wilczek:2005az}. 

\section{Jet Physics} \label{sec:JetPhysics}
 In high energy particle collisions, it is often the case that a parton can participate in the scattering and be ejected into the final-state with high momentum. As these partons pass through the QCD vacuum, they can radiate additional softer partons before the partonic color charge is confined into one or more hadrons. The result is a ``spray" of hadrons, each carrying some fraction of the momentum of the initially struck parton. Since these hadrons all find their origin from the same initiating parton, they are typically collimated around the direction of that parton. These collimated ejections of particles were first noted in cosmic ray experiments as early as the 1950s. In the 1960s the early accelerator experiments at CERN and BNL observed similar event topologies, and theory work began on explaining the phenomenon~\cite{Brandt:1964sa}. As the parton picture emerged and was confirmed by DIS experiments, the natural next step was to search for partons in the final states of particle collisions~\cite{Cabibbo:1970mh}. It was not until the 1975 that unambiguous evidence of jets was observed at SPEAR in electron-positron annihilation at $\sqrt s=7.4$ GeV~\cite{Hanson:1975fe}. The SPEAR results demonstrated not only that jets are produced in electron-positron annihilation, but also that the initiators of these jets were most likely spin $1/2$ partons. In 1977, Sterman and Weinberg calculated that the properties of these jets matched roughly what could be expected from using QCD perturbation theory~\cite{Sterman:1977wj}. This proved to be a major stepping stone in the development of jet physics as a field, as it was not clear a priori that perturbation theory could be useful in describing the final-state manifestations of the strong interaction.
 \par
\begin{figure}[htbp]
    \centering
    \includegraphics[width=10cm]{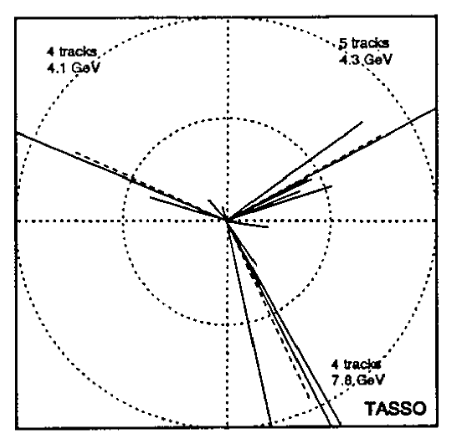}
    \caption{First three jet event observed by the TASSO experiment at PETRA~\cite{Haatuft:1979ne}.}
    \label{fig:ThreeJet}
\end{figure}

 In the late 1970s, the $e^+e^-$ collider PETRA began to operate at $\sqrt{s} \approx 30$ GeV. At that point, there was strong theoretical evidence for QCD, but the vital gauge boson had not yet been convincingly discovered. At high enough energies, it was expected that the gluon bremsstrahlung process $e^+e^- \rightarrow q\Bar{q}g$ could manifest as three spatially distinguishable jets of hadrons. This phenomenon was observed by the PETRA collaborations in 1979, an example of which is shown in Fig. \ref{fig:ThreeJet}.
The discovery of the gluon in the three-jet channel solidified jets as useful tools for gaining information about partons. In the coming decades, jets would be used in the discovery of the top quark, the study of the W, Z, and Higgs bosons, and as probes to study the hot, dense quark-gluon plasma produced in ultrarelativistic heavy ion collisions.
\subsubsection{Perturbative QCD}
\label{subsec:pQCD}
In general, for the successful use of perturbation theory, the parameter in which the perturbative expansion is performed must be small. If the parameter is too large, the perturbative expansion fails to converge, as terms of higher order in the parameter contribute significantly to the overall result. For making predictions in perturbative quantum field theory, the parameter of interest is typically the coupling constant of the theory. In the case of QED, due to the smallness of $\alpha_{EM}$ perturbation theory works at all but the very smallest of scales, which are regardless experimentally inaccessible. For QCD the case is opposite, at low energies the coupling constant is large and perturbation theory cannot make reliable predictions. However, at energy scales on the order of several GeV, $\alpha_s$ begins to become small enough to perform a perturbative expansion. The general form of these expansions for an observable $X(Q)$ are 
\begin{equation}
\label{PertExp}
X(Q) = c_0 + c_1\cdot\alpha_s(Q) + c_2\cdot\alpha_s^2(Q) + c_3\cdot\alpha_s^3(Q) + ...
\end{equation}
Where $c_n$ is a coefficient that must be calculated by taking into account all the Feynman diagrams that contribute at that order. However, given the sharply increasing difficulty of calculating the higher order coefficients in the expansion, at the present time only a handful of calculations have been done to next-to-next-to-next-to-leading order, or N$^3$LO. There exists a ``theoretical uncertainty" associated with perturbative calculations that accounts for missing higher order terms in the expansion. This uncertainty clearly increases with the value of the expansion parameter, thus making perturbative calculations less precise at lower values of $Q$.\par
In addition to so-called ``fixed order" expansions directly in the coupling constant, there exist a class of perturbative expansions in terms of $\alpha_sL$, where $L$ is the logarithm of a large number, e.g $ln(1/X)$, where $X$ is small. By reshuffling the perturbative expansion, these calculations can extend the validity of perturbative predictions to regions where fixed-order perturbation theory typically breaks down, such as the 2-jet limit in $e^+e^-$ annihilation. These calculations are known as ``resummed", by virtue of the fact that they resum the large logarithms to all orders. At the present time there exist several N$^3$LL QCD calculations.\par

\subsection{Soft-Collinear Effective Theory}
\label{subsec:SCET}
An additional tool for making predictions for QCD final states is Soft-Collinear Effective Theory, or SCET. SCET is an ``effective" field theory, in the sense that scales far different from the relevant scale of the process being studied can be treated as small or ignored. The goal of effective theories is to define appropriately the relevant degrees of freedom for a problem, and utilize them to make predictions where ``full" theories cannot. In the case of SCET, the goal is to describe soft and/or collinear degrees of freedom in the presence of a hard scale, for example soft gluon emission in a jet. SCET has been successful in describing various QCD observables, ranging from decays of $B$ mesons to event shapes in $e^+e^-$ annihilation. The SCET formalism has allowed some of the most precise extractions of the strong coupling constant~\cite{Abbate:2010xh}.\par

\subsubsection{Jet Algorithms}
\label{subsec:JA}
Prior to 1990, jets were defined differently by different experiments and theorists, such that quantitative comparison between experiment and theory, or even between different experiments, was challenging or impossible. Jet definitions were often chosen by experimental collaborations to minimize the systematic uncertainty of their detector, and by theorists in ways that were infeasible for experiments to measure. The community came together to settle on a set of standards that could be used to facilitate quantitative comparisons between experiments and theory~\cite{Huth:1990mi}. The resulting agreement became known as the Snowmass Accord, and it laid the groundwork for high-precision jet phenomenology. \par
The essential requirement of the Snowmass Accord is that any jet definition which purports to be for the purposes of QCD phenomenology must be so-called infrared and collinear safe. Infrared and collinear safety mean that the resulting jets produced by the definition should not be sensitive to soft or collinear emissions, respectively. Since soft and/or collinear emissions typically produce divergences in theoretical calculations, if a jet definition is sensitive to these emissions it cannot be compared to theoretical calculations.\par
By far the most used algorithms at present are IRC-safe algorithms based on sequential recombination. The $k_T$ family of algorithms are most common, with the anti-$k_T$ algorithm being the de facto standard at the LHC. These algorithms take particles from an event and iteratively cluster them first into groups of particles sometimes known as pseudojets, then finally into jets. The procedure for producing a jet out of a distribution of particles is as follows:
\begin{enumerate}
  \item For each pair of objects in the event, compute a distance measure $d_{ij}$ between object $i$ and object $j$ using some combination of momentum and position information
  \item Do the same for each particle with respect to the beam, producing a distance to the beam $d_{iB}$
  \item If the smallest distance is a $d_{ij}$, combine particles $i$ and $j$ into one object and repeat the first step
  \item If the smallest distance is a $d_{iB}$, call $i$ a jet and remove it from the list of objects.
  \item Repeat the procedure until no objects remain
\end{enumerate}
The form for $d_{ij}$ and $d_{iB}$ in $k_T$-style algorithms is:
\begin{equation}
\label{recomb}
\begin{split}
&d_{ij} = min(k_{T,i}^p, k_{T,j}^p)\frac{\Delta_{ij}}{R^2}\\
&d_{iB} = k^p_{T,i}
\end{split}
\end{equation}
Where $\Delta_{ij}$ is the geometric distance between objects in rapidity and azimuthal angle, and $R$ is the jet radius parameter. At hadron colliders, the typical form of $\Delta_{ij}$ is $(y_i-y_j)^2+(\phi_i-\phi_j)^2$, due to the fact that radiation associated with the breakup of the hadron beam is ``undesirable", in the sense that it is typically unassociated with the relevant hard process being studied. Since rapidity $y$ rapidly grows near to the direction of the beam, beam remnant particles are typically single particle jets at low $p_T$, and thus easily rejected with a cut on $p_T$. In $e^+e^-$ collisions, since no preferred directions exist, jets can be clustered directly in $E,\theta,\phi$ space, as opposed to $p_T, y, \phi$ space. The $e^+e^-$ collider mode of these algorithms is typically denoted as spherically invariant (SI), whereas the hadron collider mode is longitudinally invarint (LI).\par
The scheme by which one generates a pseudojet out of the pair of particles or pseudojets combined into it is known as the combination scheme, the most common of which is the E-scheme, where the four-momenta of the objects are simply added. Similarly, there is the $p_T$ scheme, where the objects are treated as massless and thus the 3-momenta of the objects are added. Another scheme which has received considerable interest recently is the winner-take-all scheme, where the energies of the two objects are summed, but the direction vector of the combined object is instead taken to be that of the more energetic of the two objects. This scheme reduces the sensitivity of the jet direction to soft radiation. \par
\begin{figure}[htbp]
    \centering
    \includegraphics[width=14cm]{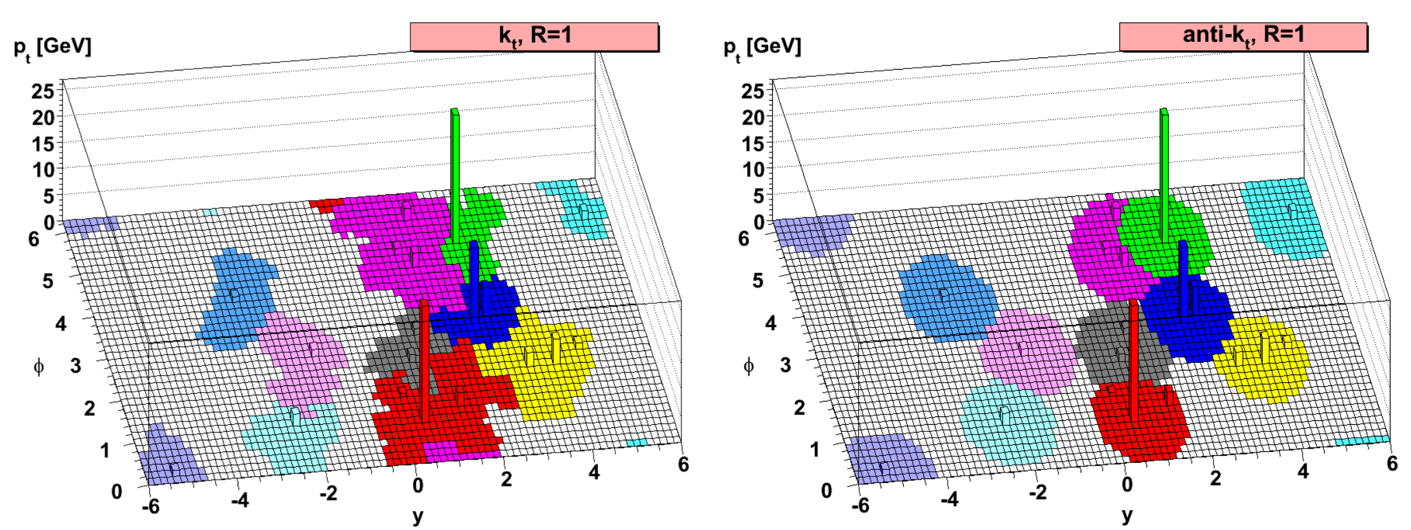}
    \caption{A comparison of jets produced with the $k_T$ and anti-$k_T$ jet clustering algorithms in the presence of a uniform soft background. The $k_T$ algorithm produces jets with irregular shapes and larger ambiguity in the presence of a background of soft particles. The anti-$k_T$ algorithm produces jets that are cone-like, centered on hard particles. Figure adapted from Ref.~\cite{Cacciari:2008gp}.}
    \label{fig:JetAlgo}
\end{figure}

The case $p=2$ in eq. \ref{recomb} corresponds to the $k_T$ algorithm~\cite{Ellis:1993tq}, in which $d_{ij}$ is smallest for soft particles and thus they are clustered first. The $k_T$ algorithm reflects the structure of QCD dynamics in the soft-collinear limit, and thus reflects sequential parton branchings. However, the $k_T$ algorithm has significant disadvantages in environments such at the LHC, where there are many soft particles that are uncorrelated with the hard scattering. These soft particles can easily be lost by the detector, while the signals from hard particles are typically better reconstructed. The fact that $k_T$ clusters soft particles first means the evolution of the clustering tree is soft-sensitive, and the resulting jets have irregular shapes. This effect can be seen in Fig. \ref{fig:JetAlgo}. Irregularly shaped jets are more challenging to calibrate in the presence of detector noise, and thus are experimentally disfavored.  \par
The widely used alternative is the so-called anti-$k_T$ algorithm, where the $p$ of Eq. \ref{recomb} is taken to be -2. Anti-$k_T$ produces jets that are effectively cones in $\eta,\phi$ space, and have very regular areas jet-by-jet. The jet boundary is sensitive to additional hard particles, while gaining insensitivity to soft particles. While it does not necessarily follow the tree of successive branchings expected from a QCD jet, it offers the experimental and theoretical advantage of soft-resiliency, and thus has become the standard jet algorithm used in hadron collisions. 

\subsection{Monte Carlo Event Generators}
The theoretical formalisms described previously are all purely analytic techniques that can be used to make predictions for QCD jet observables. Comparison between analytic theory and data typically occurs at the level of published distributions of data. A complementary approach is the use of so-called Monte Carlo event generators (MCEGs), which can simulate individual events. The outcome of a single event in an experiment depends on a variety of essentially random and statistically independent processes that follow some underlying distribution. Since the events themselves contain these random components, the use of Monte Carlo methods in particle physics is especially well-justified. MCEGs generally use a theoretical model, combined with some experimental inputs, to produce individual events that reproduce that underlying physics distribution as accurately as possible. The underlying theoretical model can be falsified if the experimental data and the events generated by the MCEG are sufficiently incompatible.\par
The most relevant components of MCEGs for jet production at colliders are the initial-state model, the hard scattering matrix elements, the parton shower, and the hadronization model. The initial state of particle collisions is most complex in the case of hadron collisions, where the parton distribution functions play an essential role. There exist parton distribution functions for other collision systems as well, but they are generally less important for producing accurate results. The MCEG selects one or more partons from the initial-state parton distributions that will participate in the hard process. The hard process describes the scattering of the selected partons themselves, and can be calculated either analytically in perturbative QCD or numerically. Once a parton has been scattered into the final-state, it enters the so-called parton shower stage, where it iteratively radiates additional partons, which can then themselves radiate. Each radiation step depletes the momentum of the initial parton, until it reaches a scale at which the parton shower stops and hadronization occurs. To produce reliable results, each of these components requires theoretical and experimental input, and they can in general be degenerate with one another for a single experimental observable. For this reason, in order to rigorously test any individual component of the models, either a single targeted measurement or multiple complementary measurements must be made.\par
An ideal environment to test the MCEG components mentioned above are high-energy electron-proton collisions. The initial-state proton parton distribution functions can be measured most cleanly in $e+p$ collisions. The hard process scale is set by the momentum transfer $Q$ which can be measured directly via the scattered electron. The typical hadronic final state in high $Q^2$ events is a single jet, whose structure is not experimentally complicated by underlying event as in proton-proton collisions. 


\chapter{HERA and the H1 Experiment}
The following section describes the salient features of an analysis performed on archived HERA data. The preservation of the H1 data and software was performed in the context of the DPHEP collaboration~\cite{South:2012vh,DPHEP:2023blx}, which seeks to maintain the wealth of data collected by high-energy physics experiments. A debt of gratitude is thereby owed to those performing the responsible stewardship of the unique HERA data, without whom this analysis would have been impossible.
\section{HERA}
The HERA collider was first proposed in 1981, shortly after the discovery of the gluon at DESY. The concept was for a high energy electron-proton collider, using PETRA as an injector. HERA began operation in 1991, and physics running in 1992. HERA initially ran with a 820 GeV proton beam, and a 26.7 GeV positron beam\footnote{Throughout this section, ``electron" will refer interchangeably to electrons and positrons} beam. After 1994, the electron beam energy was stable around 27.6 GeV. In 1998, the proton beam energy was increased to 920 GeV. 
\begin{figure}[htbp]
    \centering
    \includegraphics[width=10cm]{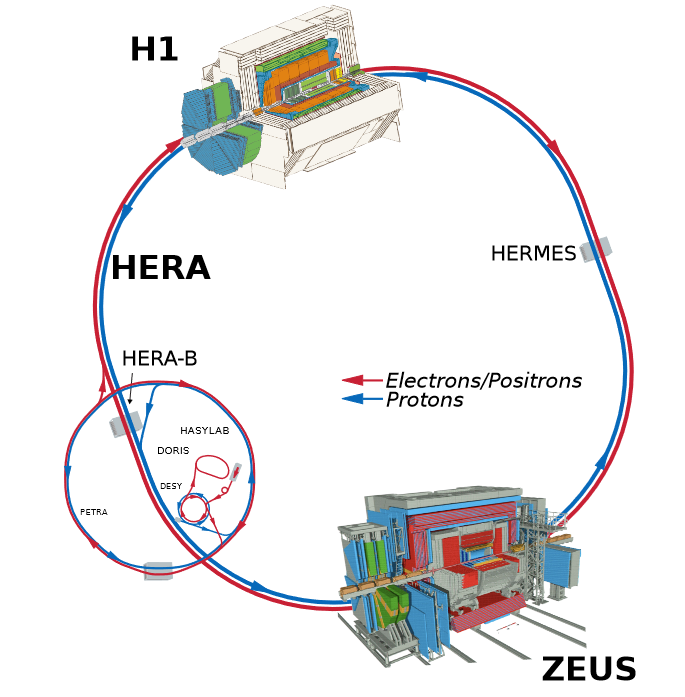}
    \caption{The HERA accelerator complex, with injection scheme. Figure from~\cite{Bolz:2019cvq}}
    \label{fig:HERA}
\end{figure}
HERA supported four experiments, H1, ZEUS, HERMES, and HERA-B. H1 and ZEUS were general-purpose, hermetic detectors, designed primarily to seek new phenomena at the energy frontier. HERMES and HERA-B were fixed target experiments, HERMES utilized the electron beam of HERA to study the structure of protons and nuclei, while HERA-B utilized the proton beam to study $B$-mesons.\par
\begin{figure}[htbp]
    \centering
    \includegraphics[width=13.9cm]{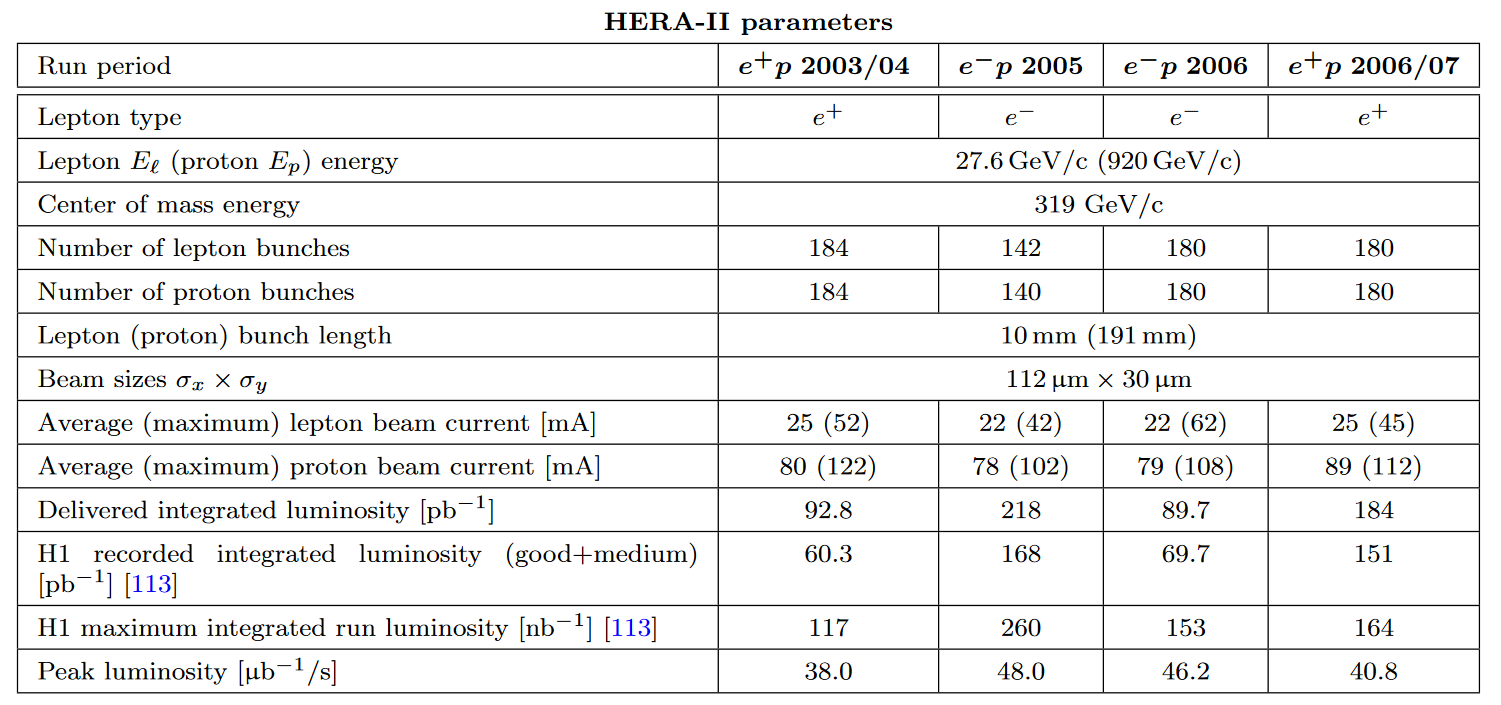}
    \caption{HERA parameters during the HERA-II run period.}
    \label{fig:HERAParameters}
\end{figure}

The HERA-I run period ended in 2000, having delivered around 100 pb$^{-1}$ to the two general purpose experiments. After HERA-I, there was a shutdown of HERA to perform upgrades with the goal of increasing the luminosity by a factor of $\sim$5. When the beams first restarted for commissioning in late 2001, there were unexpectedly large beam backgrounds registered in the detectors, that precluded physics running until 2003. In the period 2003-2007, around 400 pb$^{-1}$ were delivered to H1 and ZEUS each. The analysis described in Chapter \ref{Chap:GES} utilizes 352 pb$^{-1}$ of data collected between 2003 and 2007, split roughly equally between electron and positron running at $\sqrt{s} \cong \sqrt{4\cdot E_p \cdot E_e} = \sqrt{4\cdot 920 \cdot 27.6} \cong 319$ GeV. A summary of the HERA running periods and the H1 integrated luminosity is shown in Fig.~\ref{fig:HERAParameters}.

\section{The H1 Detector}
Since the second half of this thesis is dedicated in large part to detector physics, a didactic overview of the operating principles of particle detectors is left to sections~\ref{Chap:TPC} and~\ref{Chap:sPHENIX}. In this section, the most relevant properties of the H1 detector are described, assuming some pre-existing knowledge of detector physics. \par
\begin{figure}[htbp]
    \centering
    \includegraphics[width=13.9cm]{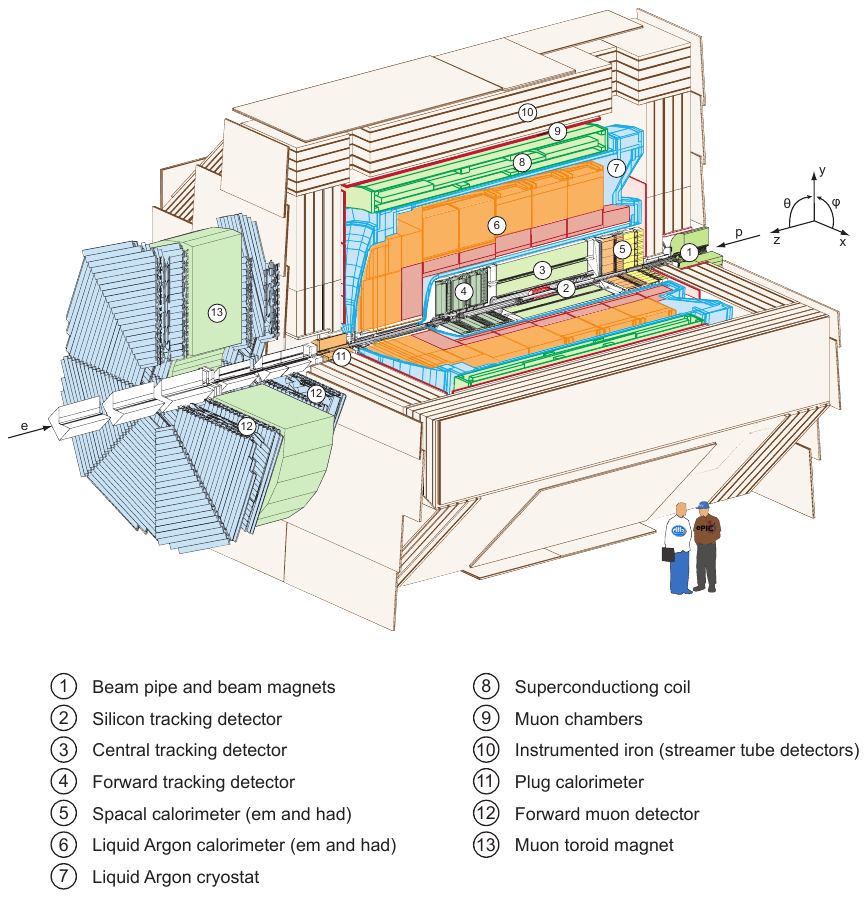}
    \caption{The H1 central detector.}
    \label{fig:H1Detector}
\end{figure}

The H1 coordinate system has the +z direction pointing along the proton beam, the +x direction pointing toward the center of HERA, and the +y direction pointing upwards. A diagram of the H1 central detector is shown in fig.~\ref{fig:H1Detector}.
\begin{figure}[htbp]
    \centering
    \includegraphics[width=10cm, angle=270]{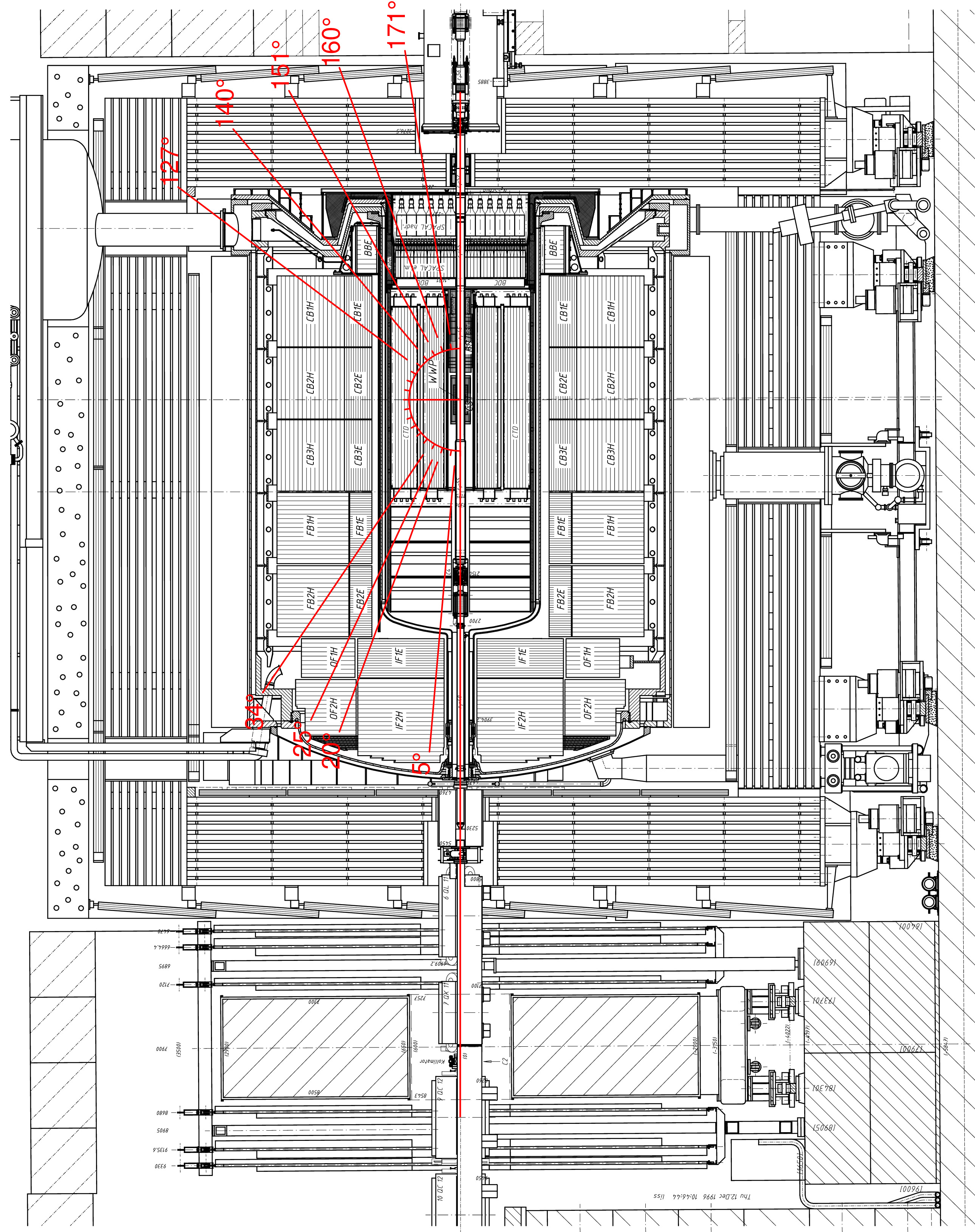}
    \caption{Technical drawing of the H1 detector, with polar angles of detector boundaries overlaid.}
    \label{fig:H1TechDetector}
\end{figure}
The central detector is hermetic and has acceptance for particles originating at the nominal interaction vertex at polar angles between $4\degree < \theta < 178\degree$, which corresponds to pseudorapidities of $3.35 > \eta > -3.8$. In order of radial distance from the beampipe, the detector consists of tracking, calorimetry, the superconducting solenoid which provides a magnetic field of $\sim$1.2T in the z-direction, and the instrumented magnet return yoke. The detector is around 15 m long, and 12 m in diameter. A more detailed description of the central detector than the one provided here is given in Ref.~\cite{H1:1996jzy}. In addition to the central detector shown in Fig.~\ref{fig:H1Detector}, several ancillary systems instrument regions of the beamline outside the H1 hall. These include the detectors which measure the luminosity delivered to H1, consisting of two electron taggers and a photon calorimeter in the backward direction, as well as two forward proton spectrometers and a forward neutron detector which provide acceptance for elastic and diffractive events. 
\subsection{Calorimeter Systems}
Both general purpose experiments at HERA put significant emphasis on the design and performance of their respective calorimeter systems. Since $e+p$ DIS events typically contain an energetic scattered electron, and one or more energetic jets, the calorimeters were designed to have good resolution for both hadronic and electromagnetic energy deposits. The H1 solution to the problem of calorimetry at HERA was a sampling calorimeter utilizing liquid argon readout covering the forward\footnote{"Forward" here and henceforth refers to the proton-going direction, and ``backward" refers to the electron-going direction} and central portions of the detector. A ``spaghetti" calorimeter (SpaCal) consisting of scintillating fibers embedded in lead furnished the backward region, where the requirements on the position resolution and electromagnetic energy resolution were stricter.
\subsubsection{Liquid Argon Calorimeter}
The H1 liquid argon  (LAr) calorimeter~\cite{H1CalorimeterGroup:1993boq} is comprised of two longitudinal sections, an electromagnetic section and a hadronic section. For events with $Q^2 >$ 150 GeV$^2$, as used in the analysis of Chapter~\ref{Chap:GES}, the DIS scattered electron is deflected at large enough angles that it falls within the acceptance of the LAr, thus making the LAr vital to inclusive analyses in this kinematic region. Both the hadronic and electromagnetic sections of the calorimeter are housed in a single cryostat filled with liquid argon at 90 K, which is inside the large-bore superconducting magnet. The electromagnetic section of the LAr uses lead as the absorber, while the hadronic section utilizes steel. Between the absorbers is liquid argon, which produces ionization when traversed by charged particles. The ionization is then collected on readout pads. The electromagnetic portion has a depth of around 20-30 radiation lengths ($X_0$), while the hadronic part has around 5-7 hadronic interaction lengths ($\lambda_0$). The LAr is grouped into 256 so-called ``Big Towers", which are capable of providing a trigger on an energetic cluster of electromagnetic or hadronic energy. The total number of channels in the electromagnetic portion amounts to 30784, and the hadronic section has 13568. 
\begin{figure}[htbp]
    \centering
    \includegraphics[width=13.9cm]{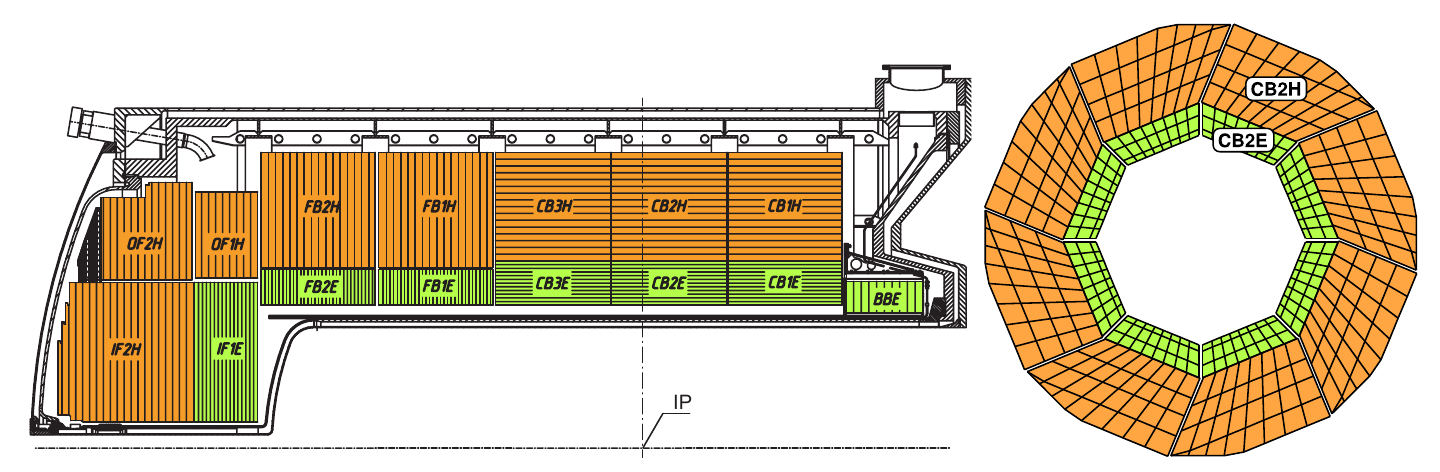}
    \caption{Side cross section (left) and radial cross section (right) of the LAr, displaying the segmentation scheme. The central barrel (CB) section has absorbers oriented longitudinally and an eight-fold azimuthal segmentation. The cracks between the hadronic sections are non-projective to the interaction point, to prevent hadrons channeling through the cracks. The forward sections (FB, IF, OF) of the LAr have absorbers oriented transversely, to make the angle of incident particles more perpendicular and thus reduce energy loss into the absorber.}
    \label{fig:H1LAr}
\end{figure}
In the electromagnetic portion, the lead absorbers are 2.4 mm thick, with a 2.35 mm gap of liquid argon between them. In the hadronic section, the steel absorbers are 16 mm thick and the liquid argon gap is $\sim5$ mm. The longitudinal and transverse segmentation of the calorimeter allows for shower shape analyses to distinguish electromagnetic energy deposits from hadronic ones. The calorimeter is non-compensating, meaning the response is different for hadronic and electromagnetic showers. In general, without any additional measures, the response ratio $\frac{e}{h}$ is around $\sim$ 1.3, meaning hadronic showers of equal energy will be reconstructed as having $\sim$ 30\% less energy than electromagnetic ones. Thus, proper reconstruction of hadronic energy deposits required a software reweighting of clusters based on their estimated fraction of hadronic and electromagnetic energy. The noise in the calorimeter cells was on the order of 10-30 MeV per cell.\par 
The purity of the liquid argon is important to reduce absorption of ionization electrons and to maintain stability of the signal over time. Certain contaminants at concentrations as small as parts per million can significantly decrease the transparency of the LAr to the drifting electrons. The purity of the liquid argon in the calorimeter was constantly monitored with radioactive sources, and fluctuations in the amount of charge collected were on the order of $0.1\%$. The degradation of the signal due to impurities was less than $0.2\%$ per year. Additional calibrations were performed to correct for any ionization loss due to impurities. \par
The overall energy resolution of the LAr was studied extensively at test beam. For incident electrons, the energy resolution was determined to be $\frac{\sigma(E)}{E} = \frac{11\%}{\sqrt{E}} \oplus \frac{0.15\text{ GeV}}{E} \oplus 0.6\%$. The energy resolution for incident pions was determined to be $\frac{\sigma(E)}{E} = \frac{55\%}{\sqrt{E}}\oplus 1.6\%$.
\subsubsection{SpaCal}
The SpaCal~\cite{H1SPACALGroup:1996ziw} utilizes lead absorber with longitudinally running scintillating fibers. For events with $Q^2 <$ 100 GeV$^2$, the scattered electron typically falls in the SpaCal acceptance. The most plentiful DIS events occur at low $Q^2$, where the scattered electron energy is typically close to the nominal electron beam energy of 27.6 GeV. Reconstruction of the DIS kinematics is very sensitive to the scattered electron energy in this region, so the SpaCal must have good resolution for electrons of $\sim$ 30 GeV. The position resolution of the SpaCal is also important for scattering angle reconstruction and track/cluster matching. The SpaCal consists of a 25 cm long electromagnetic section of 28 $X_0$ and $\sim 1 \lambda_0$, as well as a less granular hadronic section consisting of an additional $\sim 1 \lambda_0$. The hadronic section is important for identifying the scattered electron at low-x, where both the hadronic final state and the scattered electron tend to fall within the SpaCal acceptance. Electrons can be identified by the absence of energy being deposited in the hadronic section. The electromagnetic portion has a lead/fiber ratio of 2.3:1, and the hadronic portion has a lead/fiber ratio of 3.4:1. The electromagnetic portion is segmented into ~ 4x4 cm blocks which are readout by a single photomultiplier tube. The position resolution achieved is around 3.5 mm. The electromagnetic energy resolution was found to be $\frac{\sigma(E)}{E} = \frac{7.5\%}{\sqrt{E}}\oplus 2\%$~\cite{Glazov:2010zza}.
\subsection{Tracking Systems}
\label{sec:H1tracking}
Tracking is achieved in H1 using a variety of wire-based gaseous detectors and solid-state detectors. The tracking system is subdivided into forward, central, and backward regions, denoted as FTD, CTD, and BPC respectively. 
\begin{figure}[htbp]
    \centering
    \includegraphics[width=13.9cm]{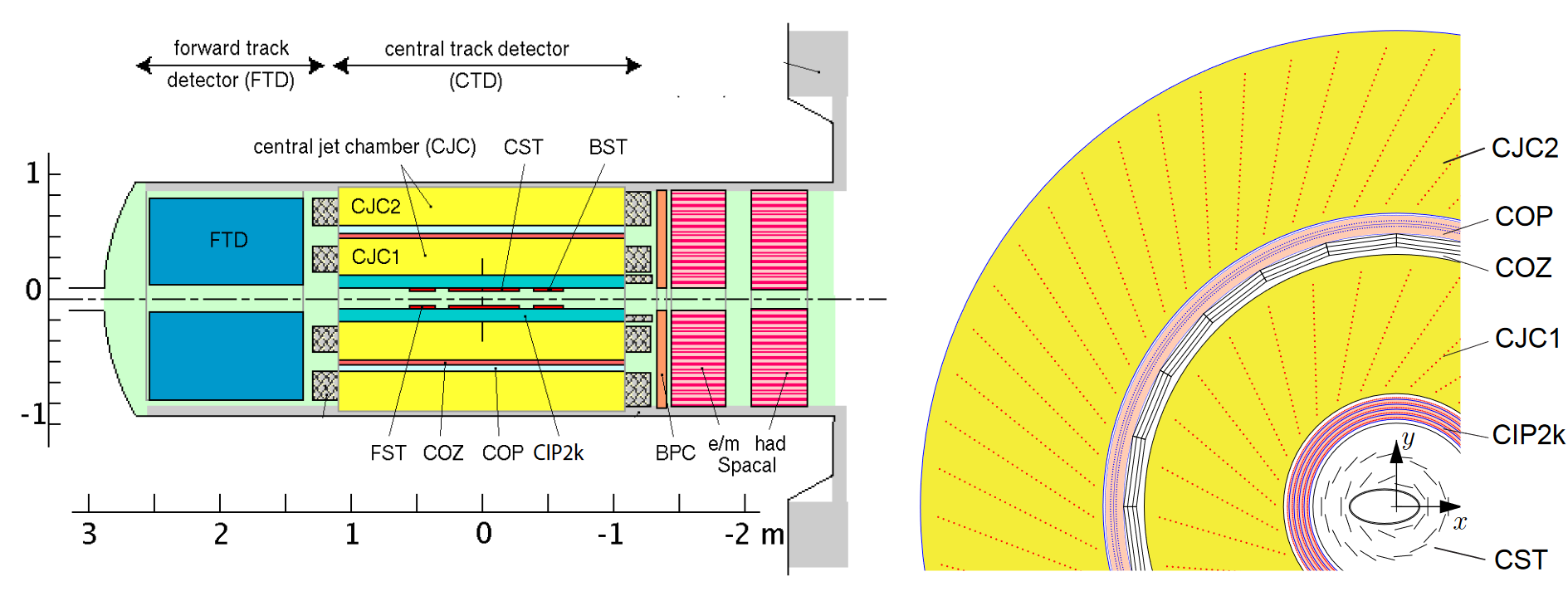}
    \caption{Side cross section (left) and radial cross section (right) of the H1 tracking system in the HERA-II configuration. In yellow are the two concentric sections of the CJC, denoted as CJC1 and CJC2. The CJC is complemented with the CST, COZ, COP, and CST trackers. The FTD is located at forward rapidities and is shown in blue. }
    \label{fig:H1Tracker}
\end{figure}
The primary components of the CTD are the two concentric drift chambers known as the central jet chambers (CJC). The CJC consists of 2640 total sense wires strung roughly parallel to the beamline, which each provide measurements of position, pulse height, and pulse time-of-arrival. The pulse height can be used for particle identification via specific energy loss (dE/dx). The signals from the CJC are sampled at 104 MHz, and 256 samples of each signal are stored. Since the CJC is designed to precisely the $r$ and $\phi$ components of a particle's location, the corresponding measurement of the z-location comes from the pulse time-of-arrival, but the resolution is fairly poor. The position resolution in the bending plane of the CJC is around $\sigma_{r\phi} = 170$ $\mu$m, and the position resolution in z is around $\sigma_{z} = 2.2$ cm. To supplement the CJC, two sets of MWPCs (COZ, CIP2K), provide precise measurements of z by having the wires oriented differently than the CJC. These proportional chambers additionally provided good time resolution for rejection of non-$ep$ backgrounds produced by the HERA machine, which was generally out of time from the bunch crossings which occured every 96 ns.\par
The silicon tracking system of H1 consists of the central-, backwards-, and forward silicon detectors, denoted as CST, BST, and FST respectively. The detectors are constructed of silicon strips with resolutions of $\sigma = 22 $ $\mu$m along the strips and $\sigma = 12 $ $\mu$m perpendicular to the strips. The CST is two layers, arranged roughly elliptically around the elliptical beampipe as shown in Fig. \ref{fig:H1Tracker}. The FST is comprised of five planes arranged in disks perpendicular to the beamline, designed to measure particles at very forward angles. The BST is comprised of four planes arranged similarly to the FST.\par
The forward tracking detector is comprised of 14 layers of drift chambers with wires running radially. The outer radius of the FTD was around 80 cm. The drift chambers are grouped into three supermodules, each of which contains 4 or 5 modules with the wires of each module rotated to allow for unambiguous measurement of the locations of incoming tracks. The efficiency for track finding in the FTD was only around 70\%, in part due to hardware failures of some modules and the substantial multiple scattering of tracks passing through the endcap of the CJC. The backward proportional chamber consists of six wire layers, and provides a spatial resolution of $\sigma_x = 1 mm$ in front of the SpaCal to aid in reconstruction of the scattered electron, and to discriminate electrons from photons.\par
Combined, the tracking detector provides a resolution on the transverse momentum of particles that can be parameterized as $\frac{\sigma_{p_T}}{p_T} \approx 0.2\% \cdot p_T \oplus 1.5\%$. The resolution on dE/dx is around 6.5\% for ``long" tracks which reach the CJC outer radius. The tracking efficiency is around $90\%$ or greater for central tracks with $p_T > 1$ GeV.
\subsection{Luminosity Measurement}
The analysis presented in Chapter~\ref{Chap:GES} reports absolute cross sections, which necessitates measurement of the integrated luminosity of the event sample. The relation between the measured number of events in the detector and the absolute cross section is:
\begin{equation}
\label{lumi}
N = \mathcal{L}\sigma
\end{equation}
Where $\mathcal{L}$ is the integrated luminosity recorded by the detector. The luminosity delivered to the experiment needs to be measured constantly during operation, since the complex dynamics of the beams preclude a priori calculations of the luminosity based on the accelerator parameters. The luminosity is time-dependent even during stable operation, due to the depletion and de-focusing of the beams as they collide at the interaction regions. The luminosity measurement is typically done by measuring the number of events of a theoretically well-understood process, where the uncertainty on the absolute cross section $\sigma$ is small. In H1, the primary technique for measuring the luminosity relied on the Bethe-Heitler process, $ep\rightarrow ep\gamma$. The Bethe-Heitler process typically produces photons at very small angles with respect to the electron beam, and a photon detector (PD) located at $z\approx -130$ m measured the yield of these photons. The signal in the PD could be compared to the precise theoretical expectation for the number of photons produced within its acceptance to determine the luminosity. Since the Bethe-Heitler process has a large cross section, the luminosity could be actively monitored during HERA running.\par
A second method of luminosity measurement was used to determine the absolute normalization of the entire HERA-II data set. This method employed large-angle QED Compton scattering, which produces rare but unique event signatures in the H1 central detector with a precisely known cross section. In these events, the electron is scattered at large angles and is azimuthally back-to-back with a photon. This analysis reached an uncertainty of around 2\% on the integrated luminosity collected during HERA-II.
\chapter{Groomed Event Shapes}
\label{Chap:GES}
This chapter reports the first measurement of groomed event shapes\footnote{All of the results presented here are not yet official H1 results. An H1 preliminary result containing a subset of the results of the analysis (H1prelim-22-033) is publicly available. At the time of writing of this thesis, a paper draft for the analysis has been written and is undergoing review by the collaboration.}. The aspects of the analysis described in Secs.~\ref{Sec:IK},~\ref{Sec:Unf},~\ref{Sec:RadCor},~\ref{Sec:SysUnc}, and \ref{Sec:Results} were performed by the author, based on the predetermined energy flow objects described in Sec.~\ref{Sec:HFS} and code inherited from the H1 inclusive jet measurement performed in Ref.~\cite{H1:2014cbm}. \par
Event shapes are inclusive observables that quantify the distribution of energy within an event. They are some of the earliest QCD observables, dating back to the late 1970s, and have been used extensively in $e^+e^-$ and $e+p$ collisions to measure the strong coupling constant $\alpha_s$, test QCD matrix elements and resummed calculations, and study non-perturbative effects such as hadronization. We report here on new observables, namely \emph{groomed} event shapes, which utilize the highly successful technique of jet grooming to differentially study the dynamics of QCD jets produced in DIS events. The cross sections are reported in the kinematic phase space of $Q^2 > 150$ GeV$^2$, and $0.2<y<0.7$. The data is unfolded to the non-radiative particle-level using TUnfold~\cite{Schmitt:2012kp} to correct for detector effects and bin-by-bin correction to account for QED effects. The reported cross section for a given bin of event shape observable $e$ is defined as 
\begin{equation}
  \frac{d\sigma}{de} = \frac{A^+(\vec{n}_\text{data} - \vec{n}_\text{Bkg})} {\mathcal{L} \cdot
    \Delta}\cdot c_{\text{QED}}\,,
\end{equation}
Where $\Delta$ is the bin width, $A^+$ is the inverse of the detector response matrix, $\mathcal{L}$ is the measured integrated luminosity, and $c_{QED}$ is the bin QED correction factor. No hadronization corrections are applied to the data.\par
The important components of this analysis are described below, including the boost to the Breit frame in Sec.~\ref{Sec:BF}, the definition of the measured observables in Sec.~\ref{Sec:Obs}, reconstruction of particles in Sec.~\ref{Sec:HFS} and the event kinematics in Sec.~\ref{Sec:IK}, unfolding of the detector response in Sec.~\ref{Sec:Unf}, radiative corrections in Sec.~\ref{Sec:RadCor}, systematic uncertainties in Sec.~\ref{Sec:SysUnc}, and finally the results and comparisons to various theoretical predictions in Sec.~\ref{Sec:Results}.
\section{The Breit Frame}
\label{Sec:BF}
With knowledge of the kinematics of a DIS event, a Lorentz transformation can be performed to a more natural frame for describing an interaction. The physics of DIS is, of course, frame independent. However, certain reference frames make results more qualitatively interpretable, or even enable theoretical calculations that are excessively complicated or impossible in the lab frame. In $e^+e^-$ collisions, where the center-of-mass of the incoming particles is usually at rest in the lab frame, and the energy of an outgoing parton is close to the energy of a jet measured in the detector. However, in high energy $e+p$ collisions, where a parton with momentum fraction $x$ is struck by an exchanged boson $\vec{q}$, the center-of-mass frame of the partonic collision can be highly boosted in the lab frame. Moreover, the difference in the resulting partonic center-of-mass frame from collision-to-collision can vary dramatically. \par
\begin{figure}[ht!]
    \centering
    \includegraphics[width=13.9cm]{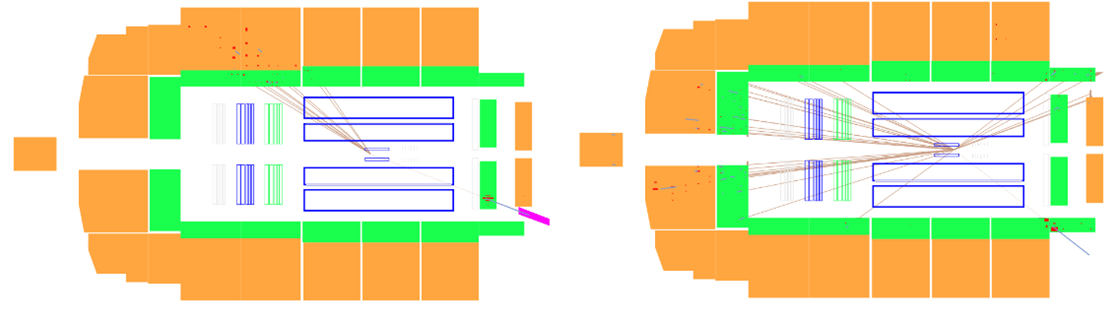}
    \caption{Two event displays of DIS events in the H1 detector demonstrating the difficulties of working in the lab frame. On the left is a Born-level DIS event where the current jet is easily identifiable. The jet (top) and the scattered electron (bottom) are back-to-back in azimuth and balanced in transverse momentum. On the right is an event where the interpretation is more difficult. In the lab frame, this event may be clustered as a three-jet event, yet without more information it is unclear whether one or more of these jets are arising from fragmentation of the proton beam remnant. The extra information provided by the scattered electron allows separation of the contributions to an event from the struck parton and those from the beam remnant. A typical way of doing this is by Lorentz boosting to a frame where there exists a large geometrical separation between the struck parton and the proton remnant.}
    \label{fig:H1EventDisplays}
\end{figure}

In low-$x$ events, which at HERA reached values of $x$ $\sim 10^{-5}$, the portion of the proton which participates in the collision can be very nearly at rest in the lab frame. In the case of an $x$ $\sim 10^{-5}$ event for a proton beam of 920 GeV, the longitudinal momentum of the object struck by the electron is only 92 MeV! This low momentum object is colliding with an electron beam of 27.6 GeV, producing an almost fixed-target collision which is highly directed in the electron-going direction. Conversely, in events at high-x, a substantial fraction of the 920 GeV available to a single constituent can participate in the interaction. A recent ZEUS result~\cite{ZEUS:2020ddd} measured $x$ values all the way up to 1, with a bin centered on 0.8. In such collisions, the electron energy is almost 30x smaller than the parton energy, and the resulting collision is highly boosted in the proton-going direction.  Performing measurements in the lab frame integrates over the possible range of event kinematics, which often makes formulating theoretical predictions challenging. In addition, the situation is further complicated by the fact that the most energetic jets measured in the detector are typically those arising from the fragmentation of the proton beam remnant or QCD ISR, which are not directly associated with the actual struck parton. The large boost of the proton beam means that the jets arising from these processes can have hundreds of GeV of energy in the lab frame. This effect is demonstrated in Fig.~\ref{fig:H1EventDisplays}. For measurements that utilize the kinematics of partons, it is generally preferred to perform a Lorentz boost to a frame in which the background of target fragmentation and QCD ISR can be separated, and calculations can be performed more easily. The ability to measure the partonic kinematics precisely and perform these Lorentz boosts is one of the key strengths of $e+p$ DIS as a tool for jet physics.\par

Two such reference frames were utilized extensively at HERA, the hadronic center-of-mass (HCM) frame and the Breit frame. The HCM frame is defined as the frame in which the virtual photon and the proton collide head-on, defined as: 
\begin{align}
   \vec{P} + \vec{q} = 0
\end{align}
where $\vec{P}$ is the proton beam four vector and $\vec{q}$ is the four vector of the exchanged boson~\cite{Grebenyuk:2012jwa}. The HCM frame effectively removes the need for integration over $Q^2$, since each event is re-defined in terms of the virtual photon kinematics. In this frame, particles which fall in the boson-going direction, i.e. $p_z < 0$, where negative z is defined as the direction of the virtual boson, can be interpreted naively as originating from the scattering of the photon on a constituent of the proton. Particles falling in the proton-going direction $p_z > 0$, can be interpreted as being associated with the target fragmentation of the proton remnant. One deficit of the HCM frame is that it does not account directly for variations in $x$, which can also fluctuate greatly from event-to-event, and thus interpreting results in the HCM frame still requires some integration over the event kinematics.\par
An almost\footnote{The Breit frame utilizes only collinear information, the intrinsic $k_T$ of partons inside the proton is not considered. Additionally, quark masses are neglected. 
} fully ``kinematics aware" reference frame is provided in the Breit frame. The Breit frame is defined as:
\begin{align}
   2x\vec{P} + \vec{q} = 0
\end{align}
the naive interpretation of which is that the incoming parton with longitudinal momentum $xP$ is struck by the photon $\vec{q}$ and leaves with longitudinal momentum $xP$ in the opposite direction. The Breit frame is often referred to as the ``brick wall frame", since it is the frame where the incoming parton has its momentum reversed, as if it were a ball reflecting off a brick wall. \par
\begin{figure}[ht!]
    \centering
    \includegraphics[width=13.9cm]{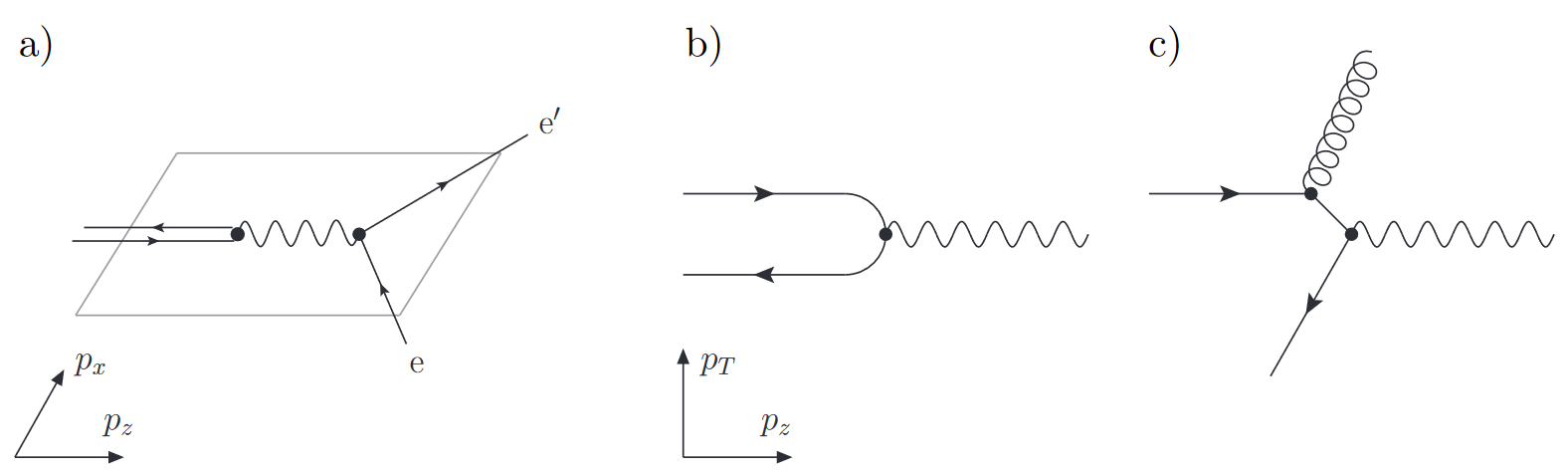}
    \caption{Features of the Breit frame. Panel a) demonstrates the kinematics of the struck parton with respect to the scattered electron. The incoming and out-going parton momenta define the z-axis. Panel b) shows the Born-level process, wherein the exchanged boson strikes the parton and it leaves in the -z direction with $p_T = 0$ GeV. Panel c) visualizes the QCD compton process, which generates large values of $p_T$ in the Breit frame. Figure from~\cite{Kogler:2011zz}}
    \label{fig:BreitJets}
\end{figure}
 In the Breit frame, similarly to the HCM frame, there can be defined two hemispheres, one of which is associated with the struck parton (henceforth the current hemisphere) and one associated with the target fragmentation of the proton remnant (the target hemisphere). By convention, the current hemisphere will be defined to be the hemisphere which contains particles with $p_z < 0$, and the target hemisphere that which contains $p_z > 0$ particles. At Born-level, the current hemisphere contains only the struck parton, carrying $p_z = Q/2$. The current hemisphere of the Breit frame is essentially one hemisphere of an $e^+e^-$ event.\par
 The Breit frame is particularly well suited for studies of jets and their substructure~\cite{Streng:1979pv}. $e+p$ scattering events are clean compared to $p+p$ collisions, and the direction of the current jet is, to first order, known from the scattered electron. This fact is an advantage over even $e^+e^-$ events, where the jet direction must be determined post hoc. The sequential recombination jet algorithms described in section~\ref{subsec:JA} can also be applied to particles in the Breit frame. Experimentally, performing the Lorentz boost to the Breit frame comes at the cost of additional uncertainty on the four momenta of the particles entering the clustering procedure, which arises from the uncertainty on the kinematic variables which constructed the boost. The $k_T$-style algorithms, which cluster based on the transverse momenta of particles, excel at finding di-jet events such as those from the QCD compton process shown schematically in Fig.~\ref{fig:BreitJets}. This technique is by far the standard one that was used for finding jets at HERA. Jets in the Breit frame have been used by both H1 and ZEUS to measure $\alpha_s$. It is worth noting that the application of the $k_T$ style algorithms will not cluster the Born-level configuration into a jet. This is a result of the Born-level parton having momentum only in the $z$ direction, and thus lying at pseudorapidity of $\eta = -\infty$. As a result, the leading order DIS jet topology was historically never studied using jet algorithms at HERA. Only recently did H1 perform a measurement of this configuration~\cite{H1:2021wkz}, with the goal of studying predictions from transverse momentum dependent (TMD) factorization.
\par
A note of caution should be mentioned with regard to the use of the Breit frame. While it is conceptually straight-forward and easily interpretable, the definition is rooted deeply in the naive quark-parton model. QCD effects such as partonic intrinsic $k_T$ inside the proton, higher twist configurations, and parton masses, among others, are not taken into account. At the high values of $Q^2$ considered in this analysis, the magnitudes of the effects which can sully the interpretability of the Breit frame are fairly small. However, when used at lower scales, the Breit frame should be interpreted as giving only a qualitative picture of the more complex reality. 
\section{Observables}
\label{Sec:Obs}
The observables which will be studied here are event shape variables, which will be measured on groomed events. Event shapes are inclusive observables to which every particle in an event nominally contributes~\cite{Fox:1978vu}. They have been measured previously in $e+p$ DIS by both H1 and ZEUS~\cite{Aktas:2005tz,Adloff:1999gn,Adloff:1997gq,ZEUS:2006vwm,ZEUS:2002tyf}. A large number of event shape variables exist, and the generic form of an event shape is simply a weighted sum over all the four vectors of the particles in an event~\cite{Korchemsky:1999kt}. Each event thus has a single value for a given event shape. The choice of weight given to a particle is essentially the only the difference between different event shapes. For all non-pathological choices of the weight factor, event shapes are by construction infrared and collinear safe~\cite{Dasgupta:2003iq}. Event shapes are therefore calculable in perturbation theory, and historically they have provided some of the best tests of QCD.\par
\begin{figure}[ht!]
    \centering
    \includegraphics[width=13.9cm]{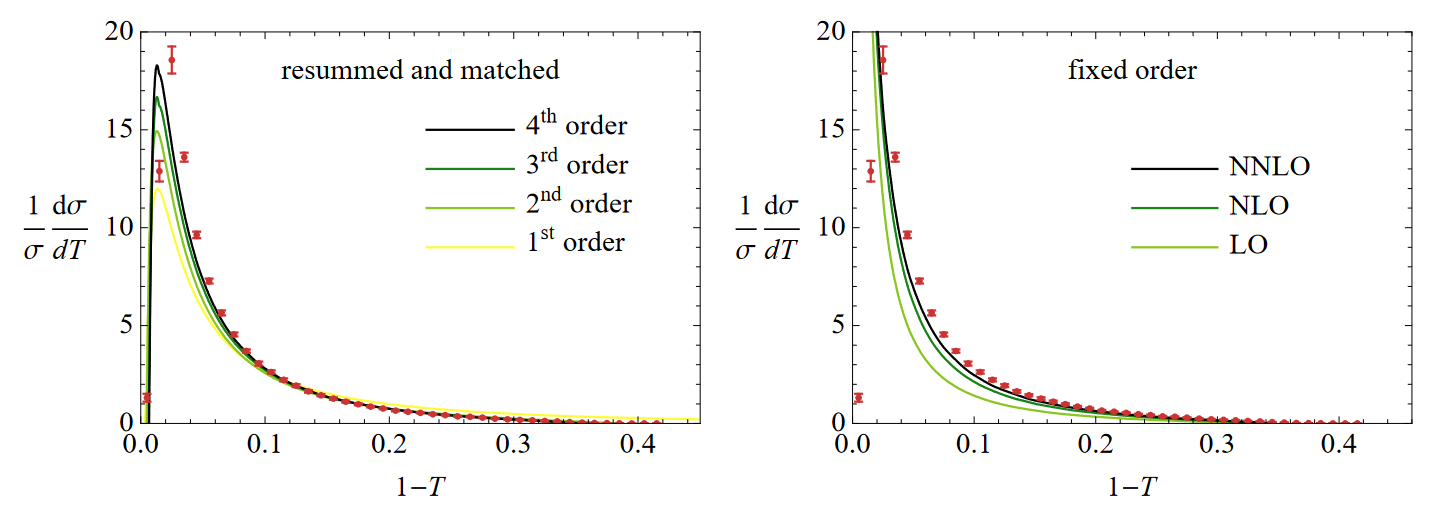}
    \caption{Comparison of resummed (left) and fixed order (right) calculations for the thrust, compared to $e^+e^-$ annihilation data from ALEPH at $\sqrt{s}=91.2$ GeV (red points). The resummed calculations are matched onto the fixed order results by requiring that they reproduce the fixed order calculations at large values of $1-T$. The fixed order divergence as $1-T\rightarrow0$ appears at all orders in the fixed order calculation. It is only through the addition of the resummed terms that the analytic prediction reproduces the smeared peak region of the data at low values of $1-T$. Figure adapted from Ref.~\cite{Becher:2008cf}}
    \label{fig:ResummedThrust}
\end{figure}
\begin{figure}[ht!]
    \centering
    \includegraphics[width=13.9cm]{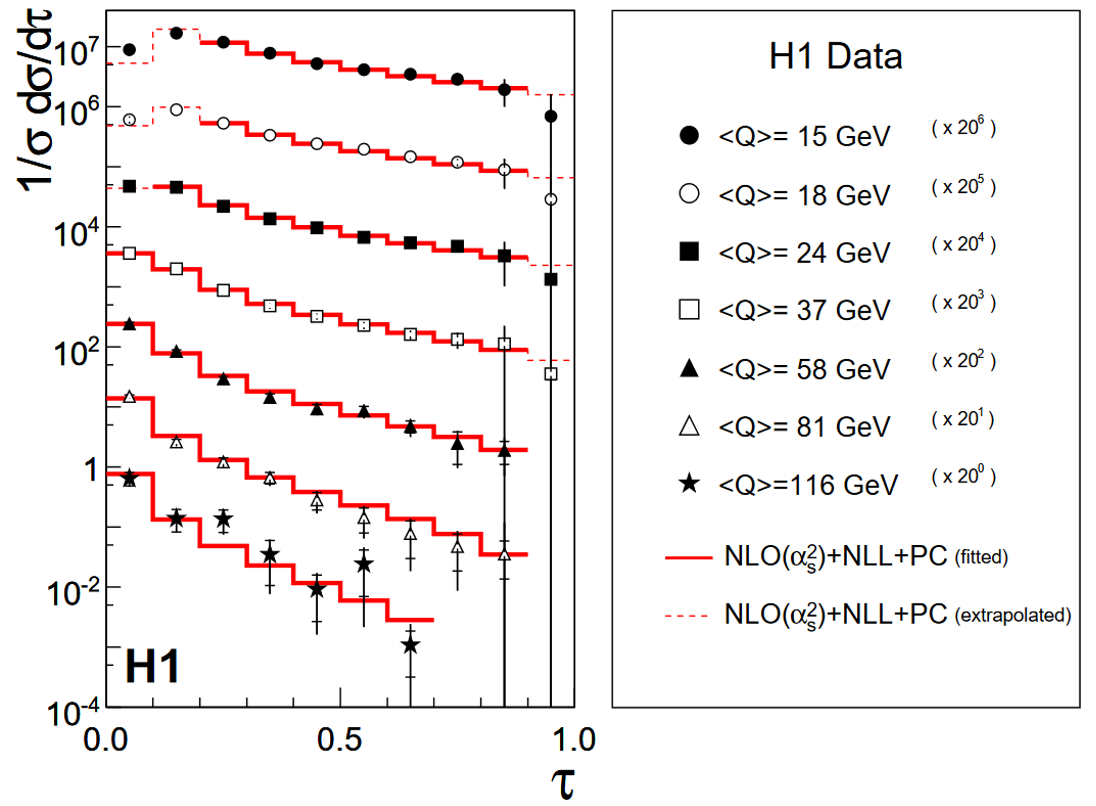}
    \caption{Thrust distribution $\tau = 1-T$, measured differentially in bins of $Q^2$. $\tau = 0$ represents the Born-level configuration. The data are compared to a calculation at next-to-leading order in $\alpha_s$ with resummed terms at next-to-leading log. The data confirm the expectations of QCD, that events become more collimated and ``1-jet-like" on average at higher $Q^2$. Figure adapted from Ref.~\cite{H1:2005zsk}}
    \label{fig:H1Thrust}
\end{figure}

The most commonly discussed event shape in the context of both $e^+e^-$ and Breit frame $e+p$ collisions is the thrust, often denoted as $T$. Although there exist a variety of definitions for the thrust, varying slightly in the axis used and the normalization factor, we will work briefly with the definition previously used by H1 in Ref.~\cite{H1:2005zsk}, reproduced below:
\begin{align}
   T = \frac{\sum_h |\vec{p_{z,h}}|}{\sum_h |\vec{p_{h}}|}
\end{align}
Where the sum runs over all particles in the current hemisphere, $p_{z,h}$ is the longitudinal momentum of a particle in the Breit frame, and $p_{h}$ is the total momentum of a particle in the current hemisphere. $p_{z,h}$ can equivalently be seen as the projection of the particle's momentum on the axis defined by the momentum of the virtual boson. In this definition, the limit of $T\rightarrow1$ is the Born-level configuation, and $T\rightarrow0$ is the case where the current hemisphere is empty, which can be produced by hard gluon radiation kicking both jets into the remnant hemisphere. At lowest order then, the normalized thrust cross section is simply: 
\begin{align}
  \frac{1}{\sigma}\frac{d\sigma}{dT} = \delta(1-T)
\end{align}
To the first non-trivial order in fixed-order perturbation theory, $O(\alpha_s)$, the thrust distribution is:
\begin{align}
  \frac{1}{\sigma}\frac{d\sigma}{dT} = C_F\frac{\alpha_s}{2\pi}[\frac{2(3T^2-3T+2)}{T(1-T)}\text{ln}(\frac{2T-1}{1-T})-\frac{3T(3T-2)(2-T)}{(1-T)}]
\end{align}
which diverges as $T\rightarrow1$! This demonstrates the soft and collinear singularities of QCD. Arbitrarily soft gluon emissions, and those which are collinear to the direction of the emitter, occur with probability 1. In reality, this apparent singularity at $T\rightarrow1$ is an artifact of fixed-order perturbation theory; the peak can be tamed at intermediate scales by introducing resummed calculations which handle these soft and collinear gluon emissions, and at an even lower scale by hadronization. The fixed order results have been extended to higher orders in $\alpha_s$ numerically, and the expansion takes the familiar form shown in Sec.~\ref{subsec:pQCD}.
\begin{align}
  \frac{1}{\sigma}\frac{d\sigma}{d\tau} = \delta(\tau)+\frac{\alpha_s}{2\pi}A(\tau)+(\frac{\alpha_s}{2\pi})^2B(\tau)+(\frac{\alpha_s}{2\pi})^3C(\tau) + ...
\end{align}
Where the substitution $\tau = 1 - T$ has been made. The effects of adding resummed calculations for the thrust distribution in $e^+e^-$ collisions is shown in Fig.~\ref{fig:ResummedThrust}.
The H1 results on the thrust (defined as $\tau = 1 - T$) are shown in Fig.~\ref{fig:H1Thrust}.
\par
Another event shape, the broadening $B$ can be defined as: 
\begin{align}
   B = \frac{\sum_h |\vec{p_{t,h}}|}{\sum_h |\vec{p_{h}}|}
\end{align}
The broadening and the thrust both can be measured on the same event sample, and they provide complementary information to one another. The final of the ``classical" event shapes that will be discussed here is the jet mass $\rho$:
\begin{align}
   \rho = \frac{(\sum_h E_h)^2 - (\sum_h \vec{p}_h)^2}{(\sum_h \vec{|p|}_h)^2)}
\end{align}
Which is essentially the invariant mass of the final state in the current hemisphere. In the definition from Ref.~\cite{H1:2005zsk}, the final state particles are treated as massless.
\par
For all the above mentioned event shapes, the sum runs only over the current hemisphere to suppress the poorly measured target hemisphere, and to facilitate a simpler comparison with $e^+e^-$ collision results. The considered normalization factors are typically sums over the particles in the current hemisphere, which can cancel some experimental uncertainties. A challenge arises in events with small energies in the current hemisphere, in such configurations these observables can become infrared and collinear unsafe. Therefore a requirement was imposed that an energy greater than $Q/10$ had to be reconstructed in the current hemisphere. This requirement imposes direct limits on the phase space for hard gluon emissions, which needs to be considered in the theoretical calculations. That only the current hemisphere contributes has experimental benefits, however it has disadvantages in terms of the formulation of theoretical predictions. Choosing only one hemisphere introduces a discrete boundary; an infinitesimal change in the four momentum of a particle very near to the boundary can cause it to cross the boundary and begin contribute to the sum. This discrete boundary and the chosen normalization factors render this definition \emph{non-global} in the language of QCD phenomenology~\cite{Kang:2013nha}. Radiation produced outside the boundary can fall inside of it, and radiation produced inside can fall outside. Non-global observables can still be predicted, but at the cost of introducing so-called non-global logarithms, which must additionally be calculated. These logarithms can spoil the precision of otherwise extremely precise calculations. It is worth noting that the standard definition of a jet using a jet algorithm is similarly non-global.\par
\begin{figure}[ht!]
    \centering
    \includegraphics[width=13.9cm]{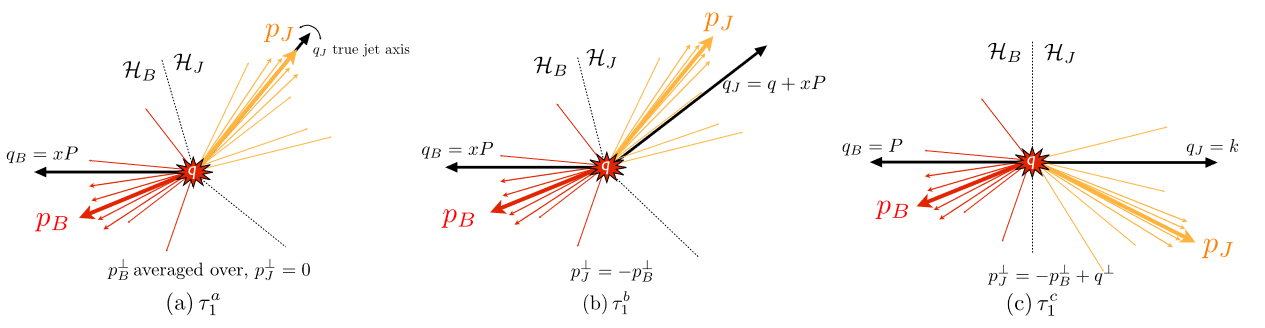}
    \caption{Pictorial representation of the three 1-jettiness definitions. Figure from Ref.~\cite{Kang:2013nha}}
    \label{fig:1jettiness}
\end{figure}

Since one of the original motivations for event shapes was their inherent globalness, it would be preferable to utilize formulations of these observables which maintain that feature. An example of a global observable very similar to the thrust is the so-called ``1-jettiness"~\cite{Kang:2013nha}\footnote{A clarifying comment should be made with regard to the number of jets in an event. In the parlance of QCD phenomenology, the beam remnant is also often referred to as a jet. Some would therefore classify the DIS Born-level configuration as a two jet event, one current jet and one beam jet. In the case of the 1-jettiness, the ``1" refers to the single jet in the current hemisphere, and neglects the beam jet. One could call this configuration a ``1+1" jet configuration, since the beam jet is always present. Therefore the general form for the $N$-jettiness is $N = $ the number of jets produced in the hard scattering minus the number of incoming hadronic beams.}. Three different definitions of the 1-jettiness have been formulated, and they are denoted as $\tau_1^a$, $\tau_1^b$, and $\tau_1^c$. All the definitions have the same basic form of:
\begin{align}
   \tau_1 = \sum_i\text{min}({\vec{q}_B\cdot \vec{p}_i,\vec{q}_J\cdot \vec{p}_i})
\end{align}
Where the sum over $i$ indicates a sum over all particles in the hadronic final state, $\vec{q}_J$ represents a vector somehow associated with the current jet, and $\vec{q}_B$ is a vector associated with the proton beam. The difference between $\tau_1^a$, $\tau_1^b$, and $\tau_1^c$ lies in the different choices for the vectors $\vec{q}_J$ and $\vec{q}_B$. The choices are:
\begin{align}
   \vec{q}_{B,a} &= x\vec{P} & 
   \vec{q}_{B,b} &= x\vec{P} &
   \vec{q}_{B,c} &= \vec{P}  & \\ 
   \vec{q}_{J,a} &= \vec{p}_J &
   \vec{q}_{J,b} &= x\vec{P} + \vec{q} &
   \vec{q}_{J,c} &= \vec{k} & 
\end{align}
Where $\vec{P}$ is the proton beam four momentum, $\vec{p}_J$ is the four momentum of a jet found using a jet algorithm, and $\vec{k}$ is the momentum of the electron beam. A sketch of the three observables is offered in Fig.~\ref{fig:1jettiness}\par
The analysis presented here will focus on $\tau_1^b$, where the relevant axes to which radiation shall be compared are the boson axis and proton remnant axis of the Breit frame. Of the formulations of the 1-jettiness, $\tau_1^b$ offers the highest degree of theoretical predictability. Since comparisons of QCD measurements to theory are often dominated by the uncertainty on the theory, it is prudent to measure the observables which can be predicted the most precisely.\par
One of the experimental challenges in measuring these global observables is the inevitable gap in the detector acceptance in the proton-going direction. This produces a new artificial boundary, similar to the one imposed by the current hemisphere constraint. To complicate matters further, the location of this boundary in the Breit frame is not fixed, and changes from event-to-event. Therefore, any measurement of a global observable in a realistic detector will have some element of non-globalness that must be corrected for. Two possible approaches can be imagined. The acceptance gap can be treated as a detector effect and corrected for. The challenge with this approach is that it requires extrapolating beyond the acceptance of the detector. This is generally avoided, since it relies on the assumption that the MC generators give an accurate description of the event structure outside the acceptance. Since this region generally lacks measurements with which to tune a MC, the quality of the description is effectively unknown, and the uncertainty on the result is difficult to quantify. Another method of circumventing non-global logarithms is to utilize a grooming procedure~\cite{Larkoski:2015zka} which removes the sensitivity to out-of-boundary radiation. In DIS events, the standard out-of-boundary emitter is the proton target fragment and the color string connecting it to the current parton. Utilizing the geometric separation between this radiation and the current jet in the Breit frame, grooming procedures can be defined that operate on the entire DIS event and remove these contributions. This procedure was first suggested in Ref.~\cite{Makris:2021drz}, and the measurement presented in the following analysis will utilize the specific grooming procedure described in that reference. The grooming procedure is based on the Centauro jet algorithm~\cite{MakrisLBL}, which is described in the following section.\par
\subsection{Centauro}
\label{subsec:Centauro}
\begin{figure}[ht!]
    \centering
    \includegraphics[width=10cm]{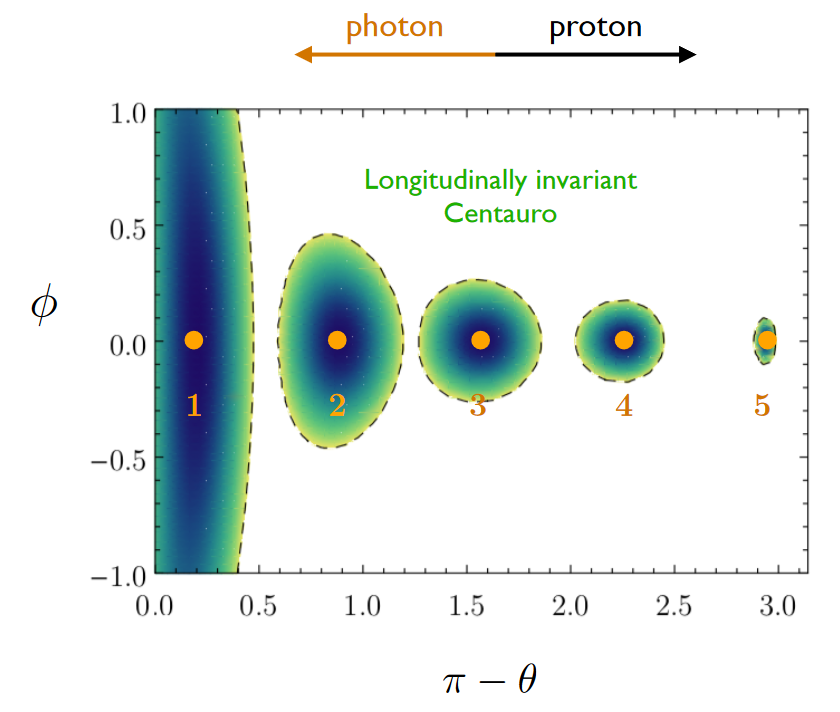}
    \caption{Isolines of constant $d_{ij}$ in the Centauro algorithm. objects geometrically close to the axis of the exchanged boson have smaller $d_{ij}$ with respect to each other than objects close to the proton remnant. The clustering procedure will tend to cluster particles in the boson-going direction with each other first. Figure from Ref.~\cite{MakrisLBL}}
    \label{fig:CentDist}
\end{figure}

The Centauro jet algorithm~\cite{Arratia:2020ssx} was introduced recently to provide a solution to the issue of clustering the Born-level configuration in the Breit frame. For the physics of the EIC, especially for measurements pertaining to TMDs, reliably measuring the DIS single-jet configuration is crucial. Centauro is a sequential combination algorithm like Anti-$k_T$, however it utilizes a distance measure which is asymmetric in the photon- and proton-going directions. The distance $d_{ij}$ between two objects in Centauro is given as:
\begin{equation}
d_{ij} = (\Delta\Bar{\eta}_{ij})^2+2\Bar{\eta}_i\Bar{\eta}_j(1-\cos{\Delta\phi_{ij}})
\end{equation}
where 
\begin{equation}
\Bar{\eta} = 2\sqrt{1+\frac{q\cdot p_i}{x_BP\cdot p_i}} \quad \underrightarrow{\text{Breit}} \quad \frac{2p_i^\perp}{p_i^+}
\end{equation}
Which effectively reproduces the spherically invariant mode of $k_T$-style algorithms in the current hemisphere and the longitudinally invariant mode in the target hemisphere. The behavior of this distance measure as a function of $\theta$ and $\phi$ is shown in Fig.~\ref{fig:CentDist}.
It can be seen that objects near to the axis of the virtual boson will have small values of $d_{ij}$ between them, whereas objects in the target hemisphere will have diverging values of $d_{ij}$ the farther in the proton-going direction they lie.
\begin{figure}[ht!]
    \centering
    \includegraphics[width=13.9cm]{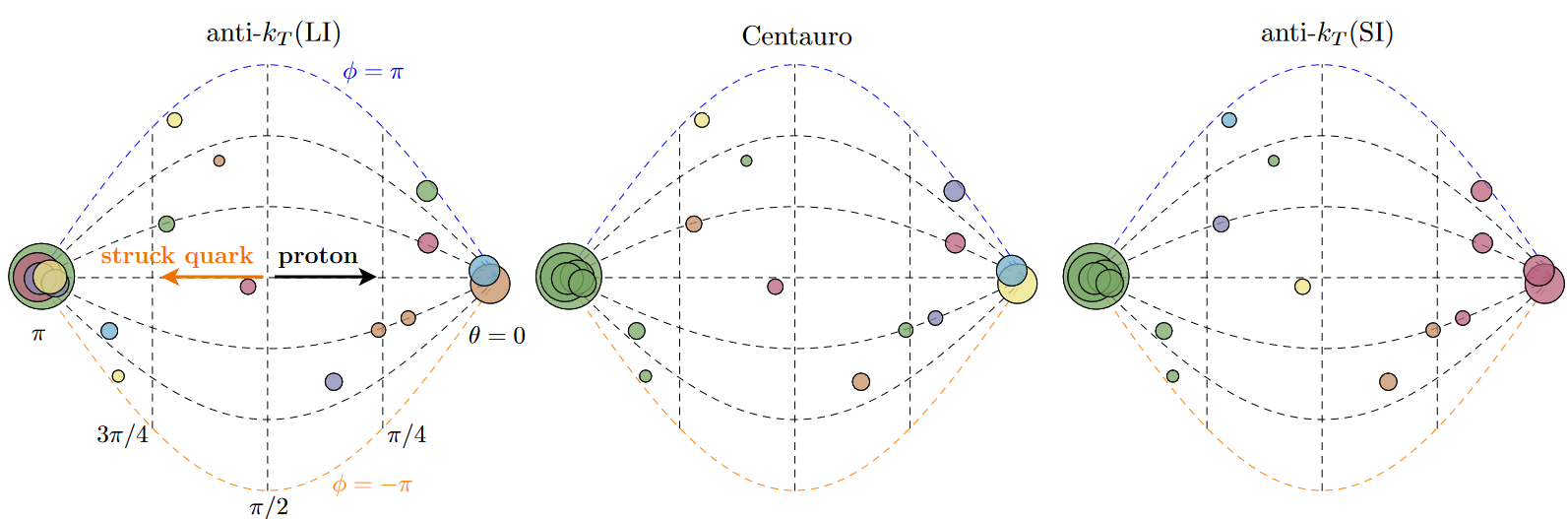}
    \caption{Illustration of the same single-jet event in the Breit frame clustered using different jet algorithms. The size of the circle is proportional to the energy of the particles, and the color corresponds to the jet it was clustered into. Figure from Ref.~\cite{Arratia:2020ssx}.}
    \label{fig:CentES}
\end{figure}
Sequential jet clustering algorithms almost always return a large number of jets. It is up to the user to introduce a cut that removes jets that are not likely to originate from a hard process. In the longitudinally invariant $k_T$-style algorithms, this cut is most often a cut on the $p_T$ of the jets. Requiring jets to have large transverse momentum effectively eliminates the single-particle jets at high $\eta$ that are produced in the target fragmentation process. In the spherically invariant case, the jet energy typically plays this role. In the case of Breit frame DIS, the relevant quantity for comparing jets is the fraction of the exchanged boson momentum carried by the jet, $z_i$ defined as:
\begin{equation}
z_i = \frac{P\cdot p_i}{P\cdot q} \quad \underrightarrow{\text{Breit}} \quad z_i = n\cdot p_i/Q = p^+_i/Q
\end{equation}
Where $P$ is the proton beam four momentum, $p_i$ is the four momentum of the relevant object, $q$ is the virtual boson four momentum, $n$ is the boson direction, and $p^+_i$ is the component of the objects momentum along the boson axis, i.e. $-p_z$ in the Breit frame. When using Centauro as a jet algorithm, the user must specify that only an object with $z_i$ greater than some value should be classified as a jet. The behavior of Centauro to cluster single-jet events is well understood, but it remains to be quantified if Centauro as an algorithm can reliably handle multi-jet configurations. 
\subsection{Grooming Procedure}
\begin{figure}[ht!]
    \centering
    \includegraphics[width=13.9cm]{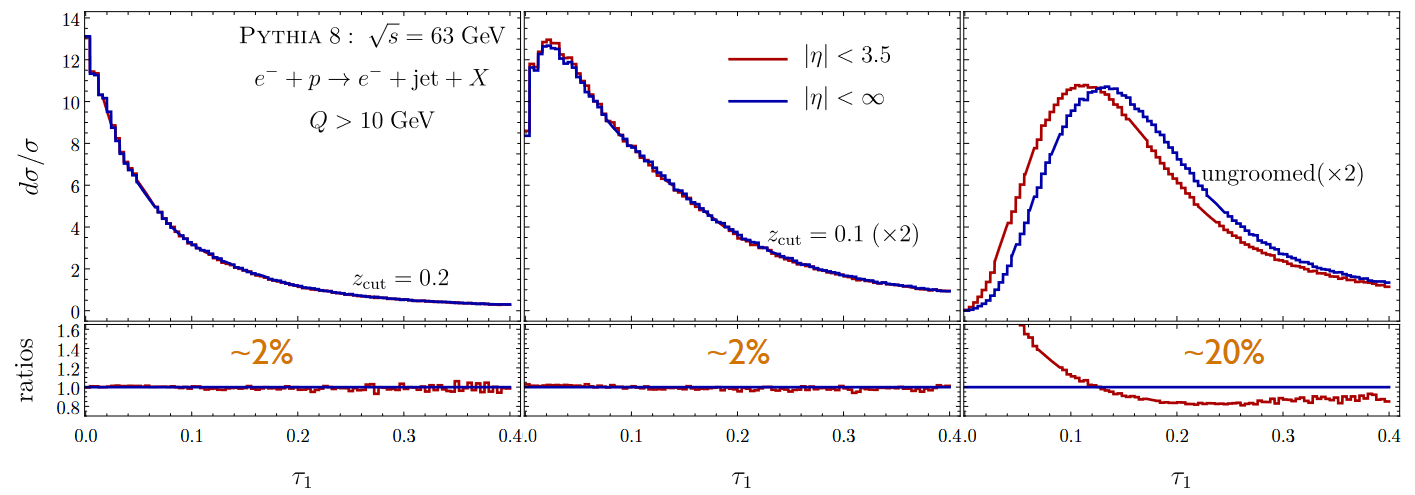}
    \caption{1-jettiness spectra in DIS at various strengths of grooming condition $z_{cut}$. The ungroomed case sees significant deviations produced by particles leaving the acceptance of a detector constrained to $|\eta| < 3.5$, while in the groomed case for both $z_{cut} = 0.2$ and $z_{cut} = 0.1$ the particles which escape the detector acceptance are groomed away, resulting in spectra that are nearly identical when the detector acceptance is imposed. Figure from Ref.~\cite{Makris:2021drz}.}
    \label{fig:Groomed1J}
\end{figure}

\begin{figure}[ht!]
    \centering
    \includegraphics[width=10cm]{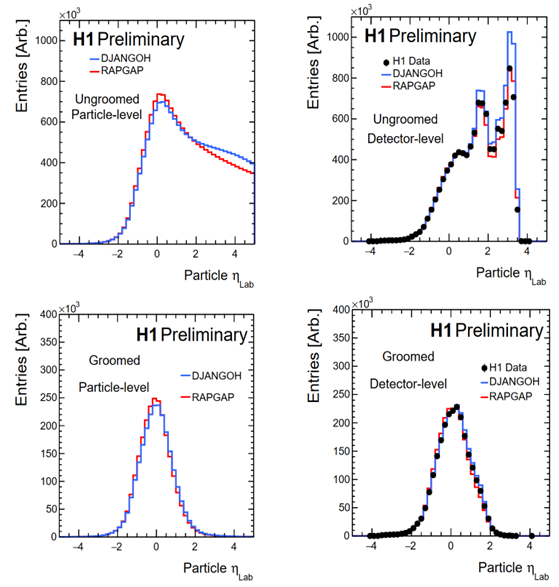}
    \caption{The particle- and detector-level pseudorapidity distributions of groomed and ungroomed particles. The ungroomed particle-level (top left) shows a long tail of particles out to high pseudorapidity, beyond the acceptance of the detector. The ungroomed detector-level (top right) shows the limits of the detector acceptance. The two bottom panels show the particle-level (left) and detector-level (right) groomed particle distributions, which are nearly identical. This demonstrates the ability of the grooming to restrict the contributing portion of the hadronic final state to that which can be measured precisely.}
    \label{fig:DetLevelGrooming}
\end{figure}

\begin{figure}[ht!]
    \centering
    \includegraphics[width=13.9cm]{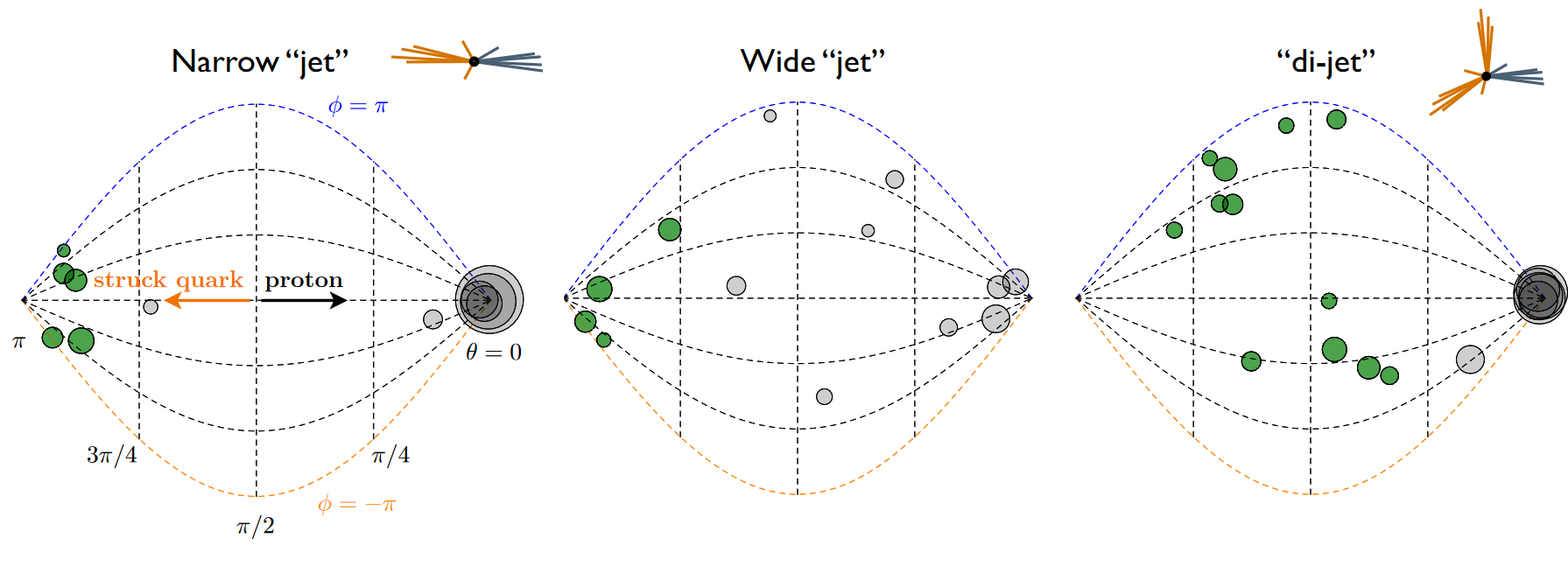}
    \caption{Three event displays demonstrating the effects of the grooming on different DIS event topologies. The particles which pass the grooming are denoted by the green circles, while those removed by the grooming procedure are grey. Figure from Ref.~\cite{MakrisLBL}.}
    \label{fig:GroomedED}
\end{figure}

With some understanding of Centauro, we can proceed to describe the grooming procedure used in this analysis. The goal is to determine which subset of the particles in a DIS event should be used in the determination of event shapes, without violating the globalness of the observables. The basic steps involved in the grooming process are summarized below:
\begin{enumerate}
  \item Compute the $d_{ij}$ between all particles in the event using the Centauro measure.
  \item Find the minimum $d_{ij}$ and merge those particles together into a new object.
  \item Repeat until all particles in the event are clustered into a single object. This is the same as clustering jets with the jet radius parameter $R=\infty$.
  \item Iteratively decluster the tree in the order in which it was clustered. At each step, compare the two branches using the grooming condition:
  \begin{equation}
  \frac{\text{min}(z_i,z_j)}{z_i+z_j} > z_{cut}
  \end{equation}
  If the two branches satisfy the grooming condition, terminate the grooming and return the remaining particles as the groomed final state. If they do not satisfy the condition, discard the branch with the smaller $z$ and continue with the declustering on the remaining branch.
  \item If the grooming condition is never passed, discard the event.
\end{enumerate}
The grooming condition imposes the requirement that the two branches being compared both share a substantial portion of the boson momentum. Particles in the proton-going direction and soft particles will typically carry either none or small amounts of the boson momentum, and thus will not satisfy the grooming condition by virtue of the fact that they have small $z_i$. The typical case is that the first comparison is a single particle very far in the proton-going direction with the entire rest of the final state. This results in a typical grooming condition of: 
\begin{equation}
  \frac{\text{min}(z_i,z_j)}{z_i+z_j} \sim \frac{\text{min}(0,1)}{1} = 0 < z_{cut}
\end{equation}
As the particles in the proton-going direction are removed, typically one at a time, the value of $\text{min}(z_i,z_j)/(z_i+z_j)$ will slowly grow. $(z_i+z_j)$ will slowly shrink, as some of the event's total longitudinal momentum will typically be removed. The values of $z_{cut}$ chosen need to be large enough to remove radiation from the beam and soft radiation, but values of $z_{cut}$ which are too large will fail often, and cause a bias in the fragmentation pattern of the surviving events. To underline the benefits of the grooming procedure, Fig.~\ref{fig:Groomed1J} shows the groomed and ungroomed 1-jettiness spectra for a realistic detector acceptance, and Fig.~\ref{fig:DetLevelGrooming} shows the distributions of groomed and ungroomed particles in the lab frame at particle- and detector-level. Around 30-40\% of particles produced in the high $Q^2$ $e+p$ DIS events considered here fall in the forward region of $\eta>3.5$.
 The considered values in the following analysis are $z_{cut} = 0.05, 0.1,$ and $0.2$. Some event displays showing what the groomed final state looks like are shown in Fig.~\ref{fig:GroomedED}. \par
Ref.~\cite{Makris:2021drz} formulates various predictions for the invariant mass of the groomed final state, henceforth referred to as the groomed invariant mass (GIM), defined as:
\begin{equation}
m_{gr.}^{2} = (\sum_{i}p_{i})^{2}
\end{equation}
It is worth noting that most event shape observables, including the GIM, are purely additive, i.e. the removal of particles via the grooming procedure serves only to decrease their total values. Thus one expects that at higher values of $z_{cut}$, the event shape distributions will shift to lower values. To build some intuition for the invariant masses of groups of particles, it is instructive to consider a few specific cases. The simplest is the case of a single particle carrying all the virtual boson's longitudinal momentum, i.e. $z_i = 1$. The invariant mass of this system is simply the mass of the particle. A cluster of particles all sitting at similar momenta close to the boson direction will have a small invariant mass, by virtue of the fact that a Lorentz boost can be performed to the relative rest frame of the cluster and the resulting momenta of the particles will be small. Events like di-jets, or events which include the proton beam remnant, will generally have very large invariant masses by virtue of the fact that no boost exists than can remove the relative momentum of the jets. These predictions are able to re-use aspects of the modified-massdrop tagger~\cite{Dasgupta:2013ihk} (mMDT) groomed jet mass observable for $e^+e^-$ collisions, for which calculations exist at NNLO in fixed-order QCD and NNNLL in resummation. The groomed invariant mass will typically appear normalized by $Q^2$, of the form $m^2_{Gr.}/Q^2$. Two regions of the GIM distribution are studied analytically.
\begin{align}
\begin{split}
  &\text{Region 1: } 1\gg z_{cut} \gg \text{m}^2_{Gr.} \\
  &\text{Region 2: } 1\gg z_{cut} \sim \text{m}^2_{Gr.}   
  \end{split}
  \label{eq:Factorization}
\end{align}
In Region 1, the GIM is very small, and thus the groomed final state is constrained to be very collimated along the boson axis. Since this region is important for the conclusions of the analysis, the salient features of the SCET calculation will be reproduced here. In this region, there exist four ``modes" of radiation which contribute, denoted as $n$-collinear, $\Bar{n}$-collinear, soft, and collinear soft. An important variable for determining the scalings of these various modes is $\lambda = m^2_{gr.}/Q^2$. The momentum scaling of these contributions is provided below, in the notation $p=(p^+,p^-,p^{\perp})$, where $p^+$ is the boson-going direction and $p^-$ is the proton-going direction.
\begin{align}
\begin{split}
  &\text{n-collinear}: p_n \sim Q(z_{cut},1,\sqrt{z_{cut}}) \\
  &\text{soft}: p_s \sim Qz_{cut}(1,1,1) \\
  &\text{collinear-soft}: p_{cs} \sim Q(\lambda,z_{cut},\sqrt{z_{cut}\lambda}) \\
  &\Bar{n}\text{-collinear}: p_{\Bar{n}} \sim Q(1,\lambda,\sqrt{\lambda})
  \end{split}
\end{align}
With these modes contributing, the cross section in this region can be considered as a convolution of a variety of functions, each with an association to a particular mode. The functions relevant here are the hard function $H(Q,y,\mu)$, the soft function $S(Qz_{cut},\mu)$, the quark flavor-dependent groomed beam function $\mathcal{B}_{f/P}(x,Q^2z_{cut},\mu$, the quark thrust jet function $J(e_{\Bar{n}},\mu^2)$, and finally the  SoftDrop collinear-soft function for use with jet-mass type observables $\mathcal{C}(e_{cs},z_{cut},\mu^2)$. Each of these functions has been compiled or calculated directly in Refs.~\cite{Kang:2013nha,Kardos:2020gty,Bauer:2011uc,Stewart:2009yx}. The cross section for region 1, to first order in terms $m^2_{gr.}/(z_{cut}Q^2)$, and as a function of $x$ and $Q^2$ is given as:
\begin{equation}
 \frac{d\sigma}{dxdQ^2dm^2_{gr.}} = H\cdot S \cdot \sum_f \mathcal{B}_{f/P}\int de_{\Bar{n}} de_{cs} \delta(m^2_{gr.} - e_{cs} - e_{\Bar{n}}) \cdot J \cdot \mathcal{C}
 \label{eq:GIMRegion1}
\end{equation}
where the aforementioned dependencies of the functions on other variables has been omitted for brevity. One can see that the only functions that depend on $m^2_{gr.}$ are the jet function $J$ and the collinear-soft function $\mathcal{C}$. The functions out front contribute only to the normalization of the cross section within the measured kinematic range. Thus, if the \emph{normalized} cross section is studied, the shape of the resulting distribution depends on only the jet and collinear-soft functions. These two functions have the unique property that they have no dependence on the event kinematics, i.e. $Q^2$ or $x$. Thus the prediction is that when normalized to this region, the shapes of the GIM spectra measured at different values of $Q^2$ should fall on top of each other. Qualitatively, this is due to the fact that the internal dynamics of jet evolution at soft scales don't ``see" the hard scattering.\par
In addition to the prediction of the $Q^2$ independence of region 1, there exist also analytic predictions for the cross section at various values of $z_{cut}$. The SCET calculation effectively produces a prediction for the locations and momenta of a collection of soft partons. This parton-level calculations needs to be corrected for hadronization in order to compare directly to measurements. In principle the data can also be corrected to the parton-level and compared to parton-level theory, but the pitfalls of doing this for groomed events will be discussed later. The SCET prediction includes hadronization corrections via convolution with a shape function~\cite{Hoang:2019ceu,Bosch:2004th}. The hadronization corrected cross section is given as:
\begin{equation}
 \frac{d\sigma_{Had.}}{dxdQ^2dm^2_{gr.}} =\int d\epsilon \frac{d\sigma}{dxdQ^2dm^2_{gr.}} (m^2_{gr.} - \frac{\epsilon^2}{z_{cut}}) \mathit{f}_{mod.}(\epsilon)
\end{equation}
Where $\mathit{f}_{mod.}(\epsilon)$ is the shape function, here taken to be of the functional form:
\begin{equation}
\mathit{f}_{mod.}(\epsilon) = N_{mod.}\frac{4\epsilon}{\Omega_{NP}^2}\text{exp}(\frac{2\epsilon}{\Omega_{NP}})
\end{equation}
where $N_{mod.}$ is the normalization and $\Omega_{NP}$ is the shape function mean. Hadronization serves to ``smear" the parton-level result to higher masses, and the parameter $\Omega_{NP}$ determines the magnitude of that smearing. The impact of the shape function will be studied in the analysis in the following sections.\par
An important component of the predictions of Ref.~\cite{Makris:2021drz} in region 1 is that the jet function $J$ is the quark jet thrust function. There is an implicit assumption that in the region of small GIM, only quark jets contribute to the cross section. The ingredients for formulating a fixed-order prediction for the GIM are included in~\cite{Makris:2021drz}, but at present the matching and full incorporation of these ingredients has not been done. This means that the predictions of Ref.~\cite{Makris:2021drz} are only valid the relatively low-mass region. \par
Two groomed event shapes are considered in the following section, the GIM and the groomed 1-jettiness. While the majority of the theoretical discussion above pertains to the GIM, the groomed 1-jettiness represents a complementary channel through which to study the structure of DIS events. By imposing the boson axis as the vector to which all particles in the final state will be projected onto, the 1-jettiness achieves sensitivity to effects that deflect the final state away from the axis, which would otherwise leave the invariant mass the same. Examples of this include the intrinsic $k_T$ of the struck parton inside the proton, or groomed away soft radiation against which the groomed final state recoiled. Presenting the two observables simultaneously should permit constraining which effects are contributing at what parts of the event shape spectra, thus allowing precision tuning of MCEGs or greater understanding of analytic predictions.
\section{Event Reconstruction}
\label{Sec:HFS}
\subsection{Event Trigger}
When an event occurs, a specific set of requirements must be met in order for the event to be recorded. These requirements are known as triggers, and the specific sub-trigger used in this study is known as S67. Trigger S67 relies on a high-energy cluster in the LAr, along with a veto on signals which are out of time with the expected $e+p$ interactions from HERA, and signals which are likely caused by beam halo or beam-gas interactions. The cluster can be produced by either the scattered electron or the hadronic final state. In high $Q^2$ NC DIS events, the ability to trigger on the electron or the HFS provides redundancy to the trigger and results in an efficiency of close to 100\% for scattered electrons with more than 11 GeV of energy. A fiducial volume cut is applied to the LAr to reject events where the scattered electron impacts in a crack or malfunctioning cell of the LAr. To make sure that signals in the various subdetectors are coming from the same bunch crossing, a measurement of the event start time ($T_0$) is required from either the LAr or the CIP. The CIP and the LAr can each be used to verify the efficiency of one another. The resulting efficiency for a successful $T_0$ measurement is $\ge 99.5\%$. The non-$ep$ veto is provided by the CIP, ToF, and the iron streamer tubes. If an extremely large track multiplicity is seen by the CIP, and the tracks seem to be originating from a vertex located beyond -50 cm\footnote{"beyond" here means the vertex is further in the electron-going, i.e. closer to the SpaCal.}, the event likely resulted from proton-nucleus collisions with gas in the HERA beampipe. The ToF could also be used to reject out-of-time events, which typically come from beam-gas or beam-wall collisions. Muons produced upstream of the detector could traverse the LAr and deposit a large enough signal to trigger the LAr. If hits were registered in the iron streamer tubes, but not in the CIP, the event was likely caused by a muon produced by the beam halo and was rejected.\par
\subsection{Electron Reconstruction}
Efficient detection of the DIS scattered electron is crucial for almost all kinematic reconstruction techniques. For scattered electron polar angles $\ge 151\deg$, the scattered electron is within the acceptance of the LAr. In this analysis, only events in which the scattered electron is reconstructed within the LAr are considered.
\begin{figure}[ht!]
    \centering
    \includegraphics[width=13.9cm]{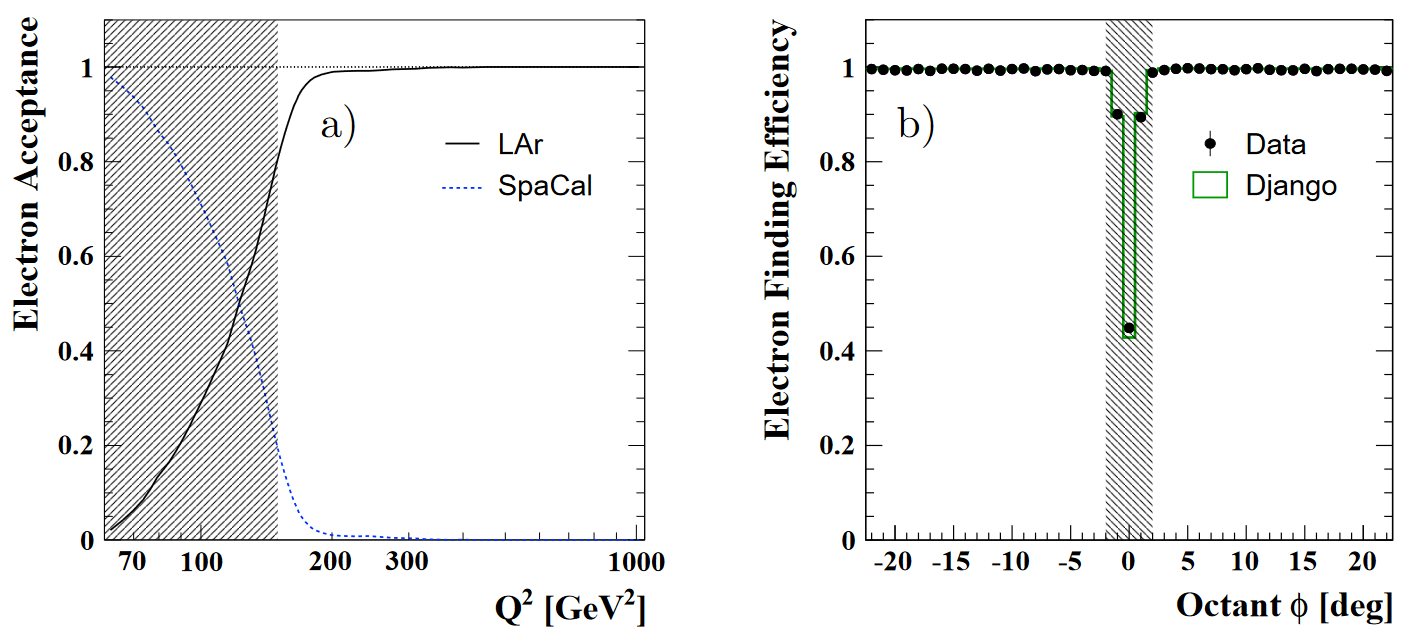}
    \caption{Geometrical acceptance for the scattered election as a function of $Q^2$ (left) and azimuthal angle within an octant of the LAr (right). The transition region from the SpaCal to the LAr is smeared by the vertex distribution of $\sim\pm35$ cm. The shaded regions are not used in this analysis. The geometric acceptance of the LAr for $Q^2 = 150$ GeV$^2$ is around 0.8, rising to 1 at $Q^2 = 200$ GeV$^2$. Figure from~\cite{Kogler:2011zz}}
    \label{fig:ElecEff}
\end{figure}
Electrons are identified out of clusters in the electromagnetic portion of the LAr with a primary vertex fitted track pointing to them. The requirement of a track rejects backgrounds from photons and $\pi^0$s. A large fraction $\ge 94\%$ of the cluster energy should be contained within the electromagnetic portion of the LAr, to ensure that the shower is from an electron and not a hadron that interacted in the electromagnetic portion. The electron cluster is also required to be compact and isolated from other event activity, to reduce the likelihood of jets faking the electron signal. Tracks which point to the electron cluster are stored for later usage (see Sec.~\ref{Sec:HFS}). After these selections, the efficiency for finding an electron is $\ge 99\%$.
\subsection{Energy Flow Reconstruction}
The hadronic final state at HERA consists of charged and neutral particles at energies from zero to hundreds of GeV. Ideally, the four-vectors of every particle would be reconstructed. In general, due to the large range of energies that particles can carry and the presence of neutral particles, doing this with precision requires combining information from the tracking detectors and calorimeters. The technique of combining track and calorimeter information is generically known as ``particle flow". In H1, the algorithm for reconstructing energy flow objects in high $Q^2$ DIS events is called Hadroo2. \par
The low-level objects which enter into Hadroo2 are tracks and clusters. Clusters arising from electronic noise in the LAr are suppressed by requiring multiple neighboring cells of the LAr to show energy above a threshold energy that depends on the location in the LAr~\cite{Kogler:2011zz}. Clusters from beam halo muons are suppressed by removing clusters that appear to be parallel to the beamline. Clusters from cosmic showers and coherent noise are removed by requiring energy deposits in the hadronic portion of the LAr to have a corresponding signal in the electromagnetic portion and/or a track pointing to the cluster. After these noise suppression algorithms are applied, the efficiency to reconstruct low energy signal clusters is determined to be around 99\%. The location of a cluster is determined to be the weighted average of the cells in the cluster, where the weighting factor is $\sqrt{E}$. Clusters are given a likelihood of originating from electromagnetic or hadronic energy, and the energy of the cluster is reweighted based on this assessment to correct for the calorimeter non-compensation.\par
For the purposes of Hadroo2 and therefore this analysis, a track is defined as either forward, central, or combined, depending on which tracking detectors contribute the to measurement. Due to some issues with the forward tracking, the vast majority of tracks reconstructed are central tracks. The cuts applied to measured central tracks to decide if they will enter the energy flow algorithm are:
\begin{enumerate}
  \item Track fitted to a vertex, either primary or secondary
  \item $p_T > 120$ MeV
  \item $20\degree < \theta < 160 \degree$
  \item $|$DCA'$|$ $\ge$ 2 cm
  \item $R_{\text{Start}} \leq$ 50 cm
  \item $R_{\text{Length}} \ge$ 10 cm for $\theta \leq 150\degree$
  \item $R_{\text{Length}} \ge$ 5 cm for $\theta > 150\degree$
  \item $N_{\text{CJC Hits}} \ge 0$
\end{enumerate}
Where DCA' is the distance of closest approach of the measured track to the vertex in the $x-y$ plane at the nominal vertex $z$ location.\par
\begin{figure}[ht!]
    \centering
    \includegraphics[width=9cm]{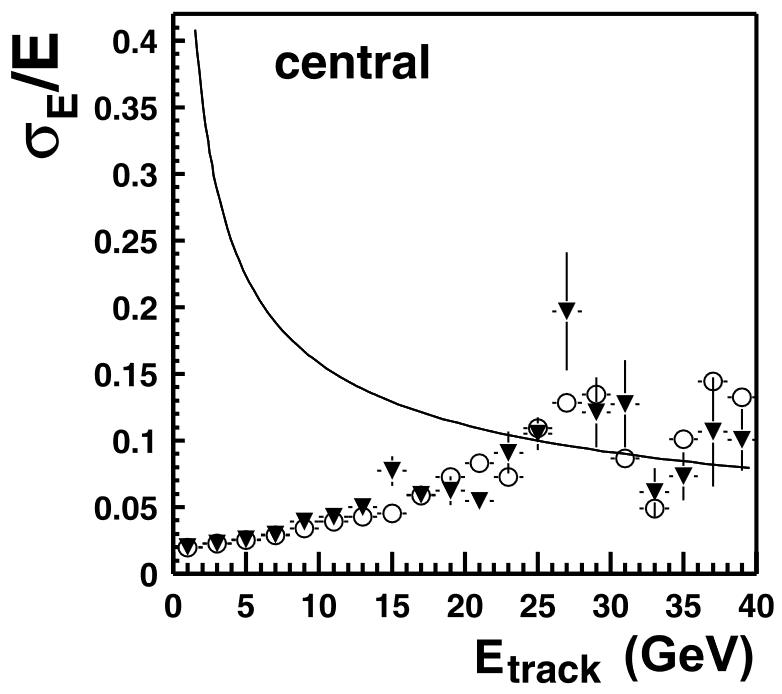}
    \caption{Energy resolution of tracks from MC (open circles) and data (closed triangles) against the energy resolution of the LAr for the same track. The point of equivalent resolution sits around 25 GeV for these objects.}
    \label{fig:CaloVTracker}
\end{figure}

To obtain a unique energy flow object, the double-counting of tracks which produced clusters needs to be eliminated. Despite the CJC providing some information on the particle mass via dE/dx, for the purposes of Hadroo2 all tracks are assumed to be pions. The energy resolution of the tracker is:
\begin{align}
    \frac{\sigma_{E,Track}}{E_{Track}} = \frac{1}{E_{Track}}\sqrt{\frac{p^2_{T,Track}}{\text{sin}^4\theta}\text{cos}^2\theta \cdot \sigma_{\theta} + \frac{\sigma^2_{p_{T}}}{\text{sin}^2\theta}}
    \label{TrackerERes}
\end{align}
where $\sigma_{p_{T}}$ and $\sigma_{\theta}$ are the respective resolutions on $p_T$ and $\theta$ neglecting the correlations between the two measurements. Given this uncertainty, an estimate is then made of the uncertainty on the calorimeter energy measurement for that track. For a first estimate, the uncertainty of the calorimeter on the measurement of a hadron such as a pion is:
\begin{align}
    \frac{\sigma_E}{E} \sim \frac{0.5}{\sqrt{E_{Track}}}
    \label{CaloERes}
\end{align}
The energy flow object is constructed out of the track if:
\begin{align}
    \frac{\sigma_{E,Track}}{E_{Track}} < \frac{\sigma_{E,Calo}}{E_{Calo}} 
    \label{CaloEResComp}
\end{align}
Which is typically the case for particles with $E<25$ GeV, as can be seen from Fig.~\ref{fig:CaloVTracker}
In the forward direction, if a track is found, it is typically preferred over the calorimeter measurements for $E < 12$ GeV. If a track is the preferred measurement, the track is projected to the surface of the LAr and the energy in the LAr in the vicinity of the track is excluded from the HFS to remove double counting. In the case that the measured track energy is significantly larger than the cluster energy, the cluster energy is used instead to suppress erroneously large $p_T$ tracks. If the track points to two nearby clusters of different energy, one of which is of similar energy to the track, the track and the second cluster are kept, with the assumption being that the second cluster likely came from a different neutral hadron. Electrons and muons which are isolated from other tracks or clusters are excluded from the above treatment, as other algorithms handle isolated lepton signals. After this construction of the energy flow objects and the application of additional calibration factors, the Hadroo2 algorithm achieves a jet energy scale resolution of $1\%$. This enables high precision DIS measurements in the high-$Q^2$ regime.\par
\section{Inclusive Kinematics}
\label{Sec:IK}
The kinematics of DIS events can be characterized by the inclusive kinematic variables $Q^2, x,$ and $y$ described in Sec.~\ref{DISKinematics}. The ability to precisely determine these kinematics from the topology of events is one of the key strengths of DIS for studies of QCD.
\begin{figure}[H]
    \centering
    \includegraphics[width=13.9cm]{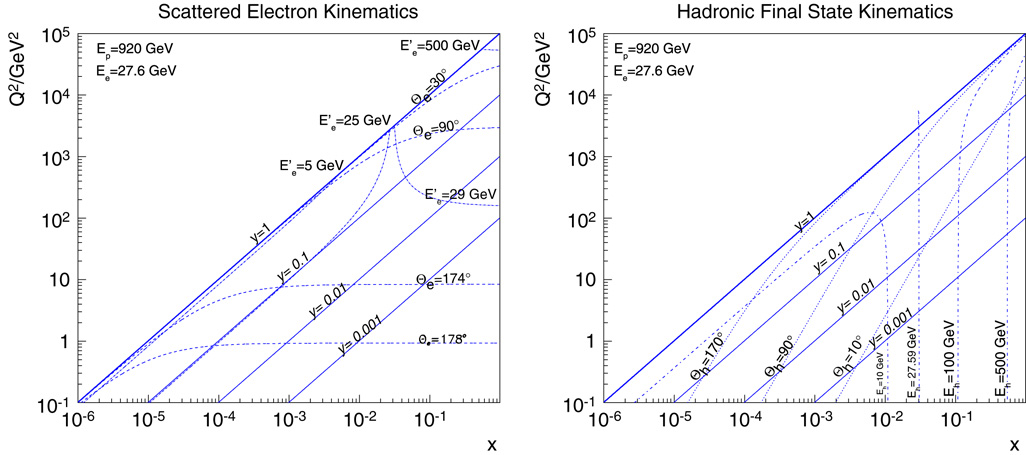}
    \caption{The kinematics of the scattered electron (Left) and hadronic final state (Right) in DIS events as a function of $x$ and $Q^2$. The dashed lines represent contours of constant scattering angle for the electron ($\theta_e$) and hadronic final state ($\theta_h$). The dashed-dotted lines represent contours of constant electron- or HFS energy. Figure from Ref.~\cite{Klein:2008di}.}
    \label{fig:HERAKinematics}
\end{figure}
Since both the scattered electron and the hadronic final state carry enough information to individually uniquely define the event kinematics, the measurement is overconstrained. This fact gives substantial freedom in the choice of the optimal combination of measured quantities to use in reconstructing the kinematics. The method which had been used most frequently since the earliest DIS experiments is the so-called ``electron method", which uses only information from the scattered electron. Given the electron initial-state energy $E^e$, electron final-state energy $E$, and final-state electron scattering angle $\theta$, in the electron method, the kinematics are defined as below:
\begin{align}
    Q^2_e &= 4E^eE\text{cos}^2\frac{\theta}{2} & y_e &= 1-\frac{E}{E^e}\text{sin}^2\frac{\theta}{2} 
    & x_e &= \frac{Q^2}{ys}
    \label{ElecMethod}
\end{align}
The electron method has good precision over much of the phase space by virtue of the fact that the electron can in general be measured precisely. However, at low-y, small changes in the electron energy and scattering angle produce large changes in the reconstructed kinematics, as can be seen in Fig.~\ref{fig:HERAKinematics}. Additionally, the electron method suffers badly from QED radiation off the electron. The naive expectation for a given QED vertex is that radiation of a photon occurs with a frequency proportional to $1/\sqrt{\alpha_{EM}} \sim 0.08$. In the kinematic region considered in the following analysis, the electron undergoes some form of initial- or final-state QED radiation in around 14\% of events. At higher values of $y$, this effect becomes more extreme. Initial-state radiation (ISR) modifies the values of $E^e$ and $s$ in Eq.~\ref{ElecMethod}, thereby causing the kinematics to be reconstructed incorrectly. Photons emitted in the initial-state typically are not reconstructed in the detector, as they are produced at small angles with respect to the incoming electron beam. The result is that the electron method has good precision for the majority of events, but has the potentially to be catastrophically modified in the presence of QED ISR. Final-state radiation in principle has a similar effect on the reconstructed angle and energy of the electron, however final-state radiated photons are typically at small angles with respect to the scattered electron, and thus are generally within the same cluster in the calorimeter. For the purposes of this analysis, which is on high-$Q^2$ data, the electron energy is typically reconstructed more precisely in the calorimeter than the tracker, and thus mostly calorimeter information is used. In this case, final-state radiation has little effect. However, if mainly tracking information is to be used for the reconstruction of the scattered electron, as could be envisioned at the Electron-ion Collider (EIC), where the electron beam energies are lower than at HERA, final-state radiation will have a similar event-by-event effect on the electron method reconstruction quality as initial-state radiation.\par
The above deficiencies mean that the electron method is typically not the method of choice for inclusive analyses. Quantities measured from the hadronic final state are typically also be utilized in reconstruction methods. The angle of the HFS, denoted $\gamma_h$, and the energy of the HFS, denoted $F_h$, must combine to balance the transverse momentum of the scattered electron. At Born-level in the naive quark-parton model, $\gamma_h$ and $F_h$ are the angle and energy of the struck quark. Two quantities can be defined using the measured hadronic final state that can be used to reconstruct the kinematics, namely the HFS longitudinal momentum in the electron direction, denoted $\Sigma$ and transverse momentum, denoted $T$, defined below:
\begin{align}
    \Sigma &= \sum_h(E_h-p_{z,h}) & T &= \sqrt{(\sum_hp_{x,h})^2 + (\sum_hp_{y,h})^2}
    \label{HFSQuants}
\end{align}
Where it should be noted that $p_{z,h}$ is a lab-frame, signed quantity, i.e. negative in the backward direction and positive in the forward direction. A particle emitted in the proton beam direction will have $E\sim p_z$, and thus contribute $E-p_z\sim 0$ to the value of $\Sigma$, while a particle in the electron beam direction will contribute $E-p_z\sim 2E$. These quantities are related to $F_h$ and $\gamma_h$ via:
\begin{align}
    \text{tan}\frac{\gamma_h}{2} &= \frac{\Sigma}{T} & F &= \frac{\Sigma^2+T^2}{2\Sigma}
\end{align}
Using these quantities alone, the kinematics of an event can be reconstructed via the so-called hadron method:
\begin{align}
    Q^2_h &= \frac{T^2}{1-y_h} & y_h &= \frac{\Sigma}{2E^e} 
    & x_h &= \frac{Q^2}{ys}
    \label{HadMethod}
\end{align}
In general, parts of the hadronic final state are not measured. Event-by-event, there will typically be HFS particles that are missed by the detector, such as low-$p_T$ hadrons which curl before reaching the tracker, or particles which are beyond the detector acceptance in the forward or backward directions. In Born-level DIS events, a color string connects the struck parton to the proton remnant, which repeatedly fragments and distributes hadrons at all rapidities between the proton beam and the struck parton. Thus, a large amount of hadronic energy is generally produced at small angles with respect to the proton beam remnant, at $\eta > 4$. This energy cannot be measured by the detector. For this reason, the resolution of a realistic detector with limited acceptance in the forward region on the HFS transverse momentum $T$ is typically suboptimal. Thus, the hadron method is typically only used when necessary, e.g. in charged-current events where the information from the electron is unavailable.\par
The $\Sigma$ method~\cite{Bassler:1994uq} is a complementary alternative to the electron method which achieves superior precision in the low-y region. The quantity $\Sigma$ is mostly insensitive to the dynamics of the forward-going hadrons mentioned above. The $E-P_z$ of the electron beam is a conserved quantity, denoted here as $\Delta = \sum_{Event}(E-P_z) = \Sigma + (E_e'-p_{z,e'}) = 2E^e$, where $E_e'$ and $p_{z,e'}$ are the energy and longitudinal momentum of the scattered electron, respectively. Deviations of $\Delta$ from $2E^2$ arise from the detector resolution, but also particles being missed the detector, such as a QED ISR photon. The kinematics as defined by the $\Sigma$ method are:
\begin{align}
    Q^2_{\Sigma} &= \frac{E^2\text{sin}^2\theta}{1-y_{\Sigma}} & y_{\Sigma} &= \frac{\Sigma}{\Delta}
    & x_{\Sigma} &= \frac{Q^2}{ys}
    \label{SigmaMethod}
\end{align}
It is important to note that for the $\Sigma$ method, both $Q^2$ and $y$ have no dependence on the electron beam energy, $E^e$, and thus are insensitive to QED ISR. The $\Sigma$ method retains good resolution over much of the kinematic phase space, including in events with QED radiation. A variant of the $\Sigma$ method which was used extensively by H1 is the so-called electron-$\Sigma$ or $e\Sigma$ method, which uses $Q^2_e$, $y_{\Sigma}$, and $x_{\Sigma}$. In general, a variety of ``mixed" methods also exist, which utilize different methods for different variables.\par
Another of the standard reconstruction methods used at HERA is the double-angle or DA method. The DA method uses only the angles of the scattered electron and the hadronic final state, thus becoming insensitive to the relative energy scale uncertainties between the electron and the HFS. In the DA method, the kinematics are defined as:
\begin{align}
    Q^2_{DA} &= 4E^{e2}\frac{\text{cot}\theta/2}{\text{tan}\gamma/2+\text{tan}\theta/2} & y_{DA} &= \frac{\text{tan}\theta/2}{\text{tan}\gamma/2+\text{tan}\theta/2}
    & x_{DA} &= \frac{Q^2}{ys}
    \label{DAMethod}
\end{align}
In H1, the double-angle method was used to calibrate the jet energy scale, and a precision of roughly 1\% was achieved. \par
All the reconstruction methods mentioned thus far reconstruct $x$ as $\frac{Q^2}{ys}$. This means that the errors on $Q^2$ and $y$ (as well as $s$ in the case of QED ISR), all compound in the measurement of $x$. Precision in $x$ is particularly important for measurements performed in the Breit frame, which will be discussed at length in the next section. A modification can be made to the $\Sigma$ and DA methods to remove the explicit use of $s$ from the determination of $x$, and thus render them almost completely insensitive to QED ISR. The so-called $I\Sigma$ method uses the same $y_{\Sigma}$ and $Q^2_{\Sigma}$, but reconstructs $x$ instead as:
\begin{align}
    x_{I\Sigma} = \frac{E}{E_p}\frac{\text{cos}^2\theta/2}{y_{\Sigma}}
\end{align}
Where $E^p$ is the proton beam energy, which is effectively constant at HERA. The DA method has a similar modification, the IDA method, where $x_{IDA}$ is defined as:
\begin{align}
    x_{IDA} = \frac{E}{E_p}\frac{\text{cot}\gamma/2 + \text{cot}\theta/2}{\text{cot}\theta/2+\text{tan}\theta/2}
\end{align}
The $I\Sigma$ and IDA methods are both found to have superior resolution on $x$ in the high $Q^2$ phase space considered here. The $I\Sigma$ method is chosen as the preferred kinematic reconstruction method for this analysis due to its good resolution throughout the entire phase space. The IDA method was studied as well and would have been a feasible choice. However, it was not chosen in order to avoid any potential biases arising from the prior use of the DA method in the calibration of the HFS.
\begin{figure}[H]
    \centering
    \includegraphics[width=13.9cm]{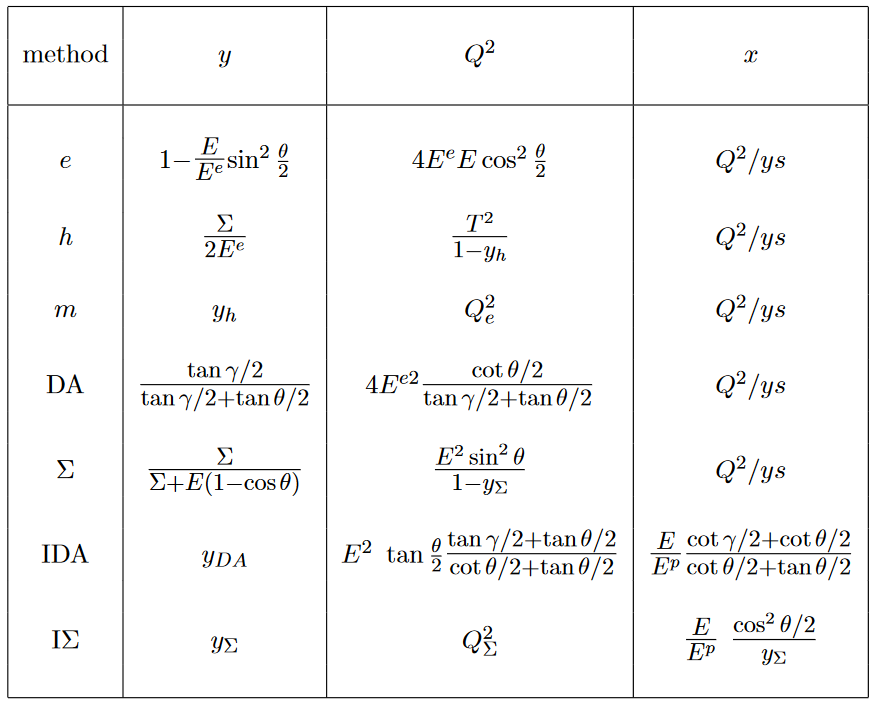}
    \caption{The various methods of reconstructing the DIS kinematic variables from final state particles. Figure from Ref.~\cite{Bassler:1994uq}.}
    \label{fig:Bassler}
\end{figure}
It should be mentioned that the ideal reconstruction method would utilize the full information from both the scattered electron and the hadronic final state. Two techniques for doing this have been developed recently, one based on a Bayesian kinematic fit~\cite{Aggarwal:2022rlk}, and one based on a deep neural network~\cite{Arratia:2021tsq}. Both of these methods achieve improved resolution over the standard reconstruction methods.
\subsection{Event Selection}
Events are selected first by requiring an electromagnetic energy deposit with energy greater than 11 GeV in the LAr. The efficiency of this selection is $> 99\%$ in the phase space considered here. The quality assurance cuts used in this analysis are summarized below:
\begin{compactitem}
\item The measured event vertex is constrained to be within 35 cm of the nominal vertex location. 
\item The total event longitudinal momentum balance $E-p_z$ is required to be $50 < E-p_z < 60$, to suppress badly measured events and events with significant QED initial state radiation. 
\item The $p_T$ of the hadronic final state is required to balance the scattered electron, $0.6 < \frac{p_{T,HFS}}{p_{T,e}} < 1.6$, to ensure the hadronic final state is well measured.
\item The total difference between the $p_T$ of the scattered electron and the HFS is required to be less than 5 GeV.
\item The vector defining the current hemisphere of the Breit frame is required to have a polar angle $0.12<\theta_{c}<3.05$ radians to reduce non-collision backgrounds and ensure the hadronic final state is contained within the detector. 
\item Events where $\theta_{c}>2.6$ and the $\beta$ of the boost to the Breit frame is $> 0.9$ are typically polluted with initial-state QED radiation, and thus are removed. For the same reason, events where the difference between the angle of the current hemisphere vector and the HFS satisfies $\theta_c - \gamma_h > \frac{\pi}{2}$ are removed.
\item Events with $Q^2 < 700$ where the pseudorapidity difference between the hadronic final state and the scattered electron $\Delta\eta > 0.3$ have poor resolution and are rejected.
\end{compactitem}
The above cuts serve to remove backgrounds and poorly measured events, and to ensure that the boost to the Breit frame is well reconstructed. These cuts significantly improve the diagonality of the migration matrix, the importance of which will be outlined in the following section.
\section{Unfolding}
\label{Sec:Unf}
Given a measured distribution, the task of unfolding is to deconvolute the response of the detector from the desired truth distribution~\cite{Blobel:2011fih}. In a typical experiment there will be a vector of measured values $y$, which could be, for example, the number of events in each of $n$ bins. There exists also a vector $x$, which represents the true distribution of the events. The detector is expected to have various effects on the truth distribution that result in the measured one. The detector can misreconstruct events that were in bin $n$ as being in bin $m$, could not reconstruct the event entirely, or could erroneously reconstruct some form of background as having been an event of the desired kind. All these effects are parameterized in the detector response matrix $A$, related to $x$ and $y$ by
\begin{equation}
Ax=y
\end{equation}
For the purposes of high energy physics, the detector response matrix is generally constructed by generating events using a MC event generator that provides a reasonable description of the data and passing them through a simulation, such as GEANT~\cite{Brun:1987ma}, that can determine what the outputs of the detector would be for each event. \par
The simplest form of unfolding, known as bin-by-bin unfolding, approximates $A$ as a diagonal matrix with no off-diagonal components. The inverse of $A$ can then be trivially computed and applied to $y$. The bin-by-bin unfolding procedure effectively multiplies each bin of $y$ by a constant number to reach $x$. This procedure is often used, however strictly speaking it is only correct in the case where there are no migrations between bins, which is very rarely the case in any realistic measurement. When migrations are present, this procedure tends to bias the data towards the model used in the construction of the correction factors.\par
Another method is to determine $A$ including off-diagonal components and perform a matrix inversion to convert $A$ into its inverse matrix $A^{+}$ which can be applied to the measured distribution to reach the truth distribution. This operation is ill-posed in a mathematical sense and often leads to instabilities and fluctuations in the final result. It also requires that the dimensions of vectors $x$ and $y$ are the same. However, a variety of techniques can be used to circumvent these difficulties for the purposes of unfolding. The method which will be utilized here is implemented in the TUnfold package~\cite{Schmitt:2012kp}. The basic method of TUnfold is a minimization of the matrix equation:
\begin{equation}
\chi^2(x) = (Ax-y)^TV_{yy}^{-1}(Ax-y)
\end{equation}
Where $V_{yy}$ is the covariance matrix of $y$. However, this procedure alone suffers from the same instabilities as the standard matrix inversion. To handle these features, a small regularization term is added to the equation:
\begin{equation}
\chi^2(x) = (Ax-y)^TV_{yy}^{-1}(Ax-y)+\tau^2(x-x_b)^T(L^TL)(x-x_b)
\end{equation}
Where $\tau$ is a small but non-zero free parameter governing the regularization strength, $x_b$ is an optional regularization bias term, and $L$ is a matrix defining the conditions of the regularization. The regularization term serves to suppress the fluctuations in $x$ arising from statistical uncertainty on $y$. $x$ can then be determined uniquely from $y$ via
\begin{equation}
x = (A^TV_{yy}^{-1}A+\tau^2L^2)^{-1}A^TV_{yy}^{-1}y
\end{equation}
The matrix $L$ can take various forms, including a unity matrix $L=1$, the so-called ``derivative mode" of $L\sim(-1,1)$, and the ``curvature mode" of $L\sim(-1,2,-1)$. The curvature mode was determined to give the best closure for the measured observables, and thus it is the regularization mode used in throughout the analysis.\par
\begin{figure}[H]
    \centering
    \includegraphics[width=13cm]{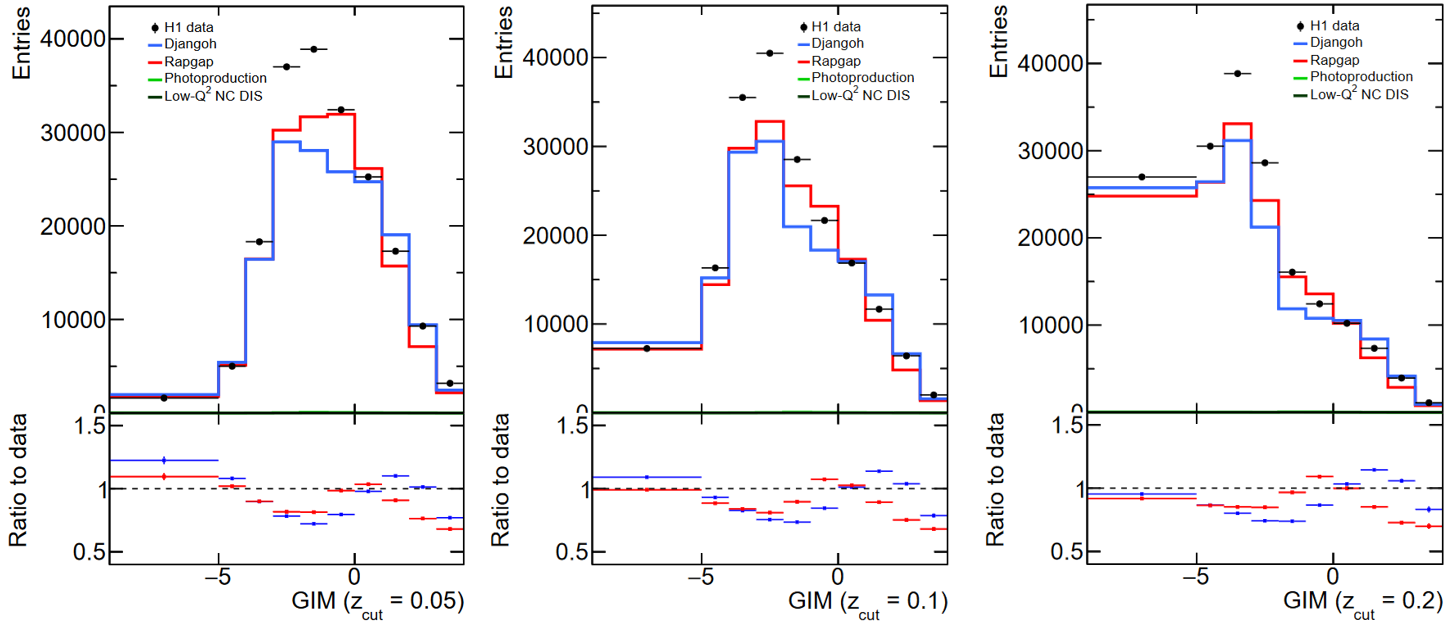}
    \caption{The measured data compared to Rapgap and Djangoh on detector-level for the three values of $z_{cut}$.}
    \label{fig:GIMDet}
\end{figure}
\begin{figure}[H]
    \centering
    \includegraphics[width=13cm]{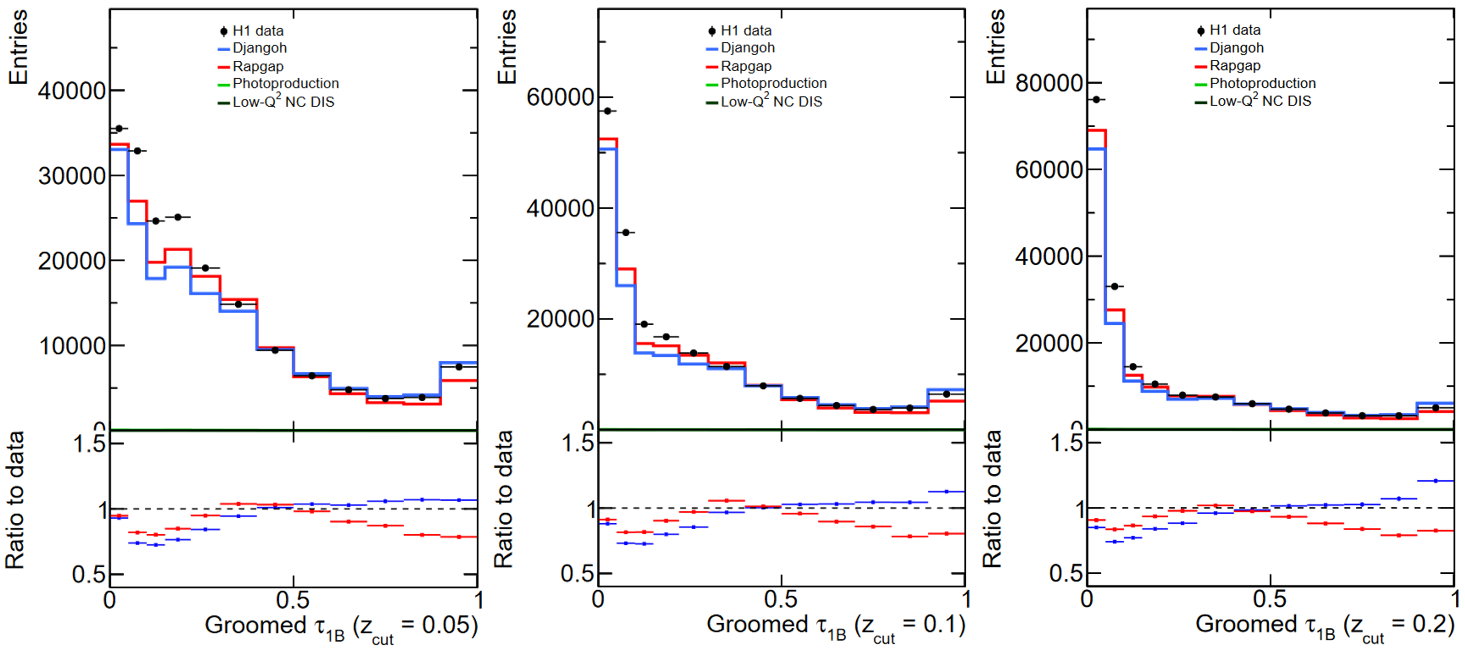}
    \caption{The measured data compared to Rapgap and Djangoh on detector-level for the three values of $z_{cut}$.}
    \label{fig:G1JDet}
\end{figure}
\begin{figure}[H]
    \centering
    \includegraphics[width=13cm]{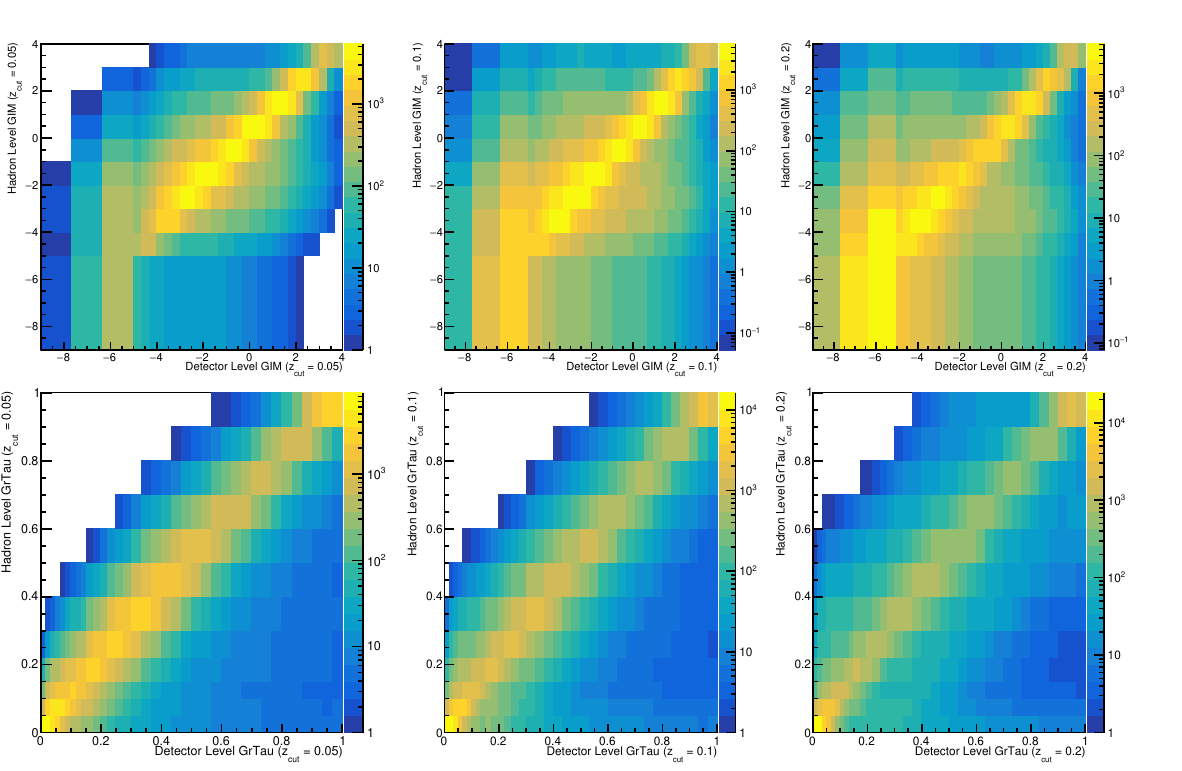}
    \caption{1D migration matrices for the GIM and the groomed \tb. There are three bins at particle-level (y-axis) for each bin in the reconstructed detector-level. At low GIM, the bin size changes dramatically. An artifact of this is that the migration matrices visually appear more non-diagonal than they are.}
    \label{fig:Mig1D}
\end{figure}
\begin{figure}[H]
    \centering
    \includegraphics[width=13cm]{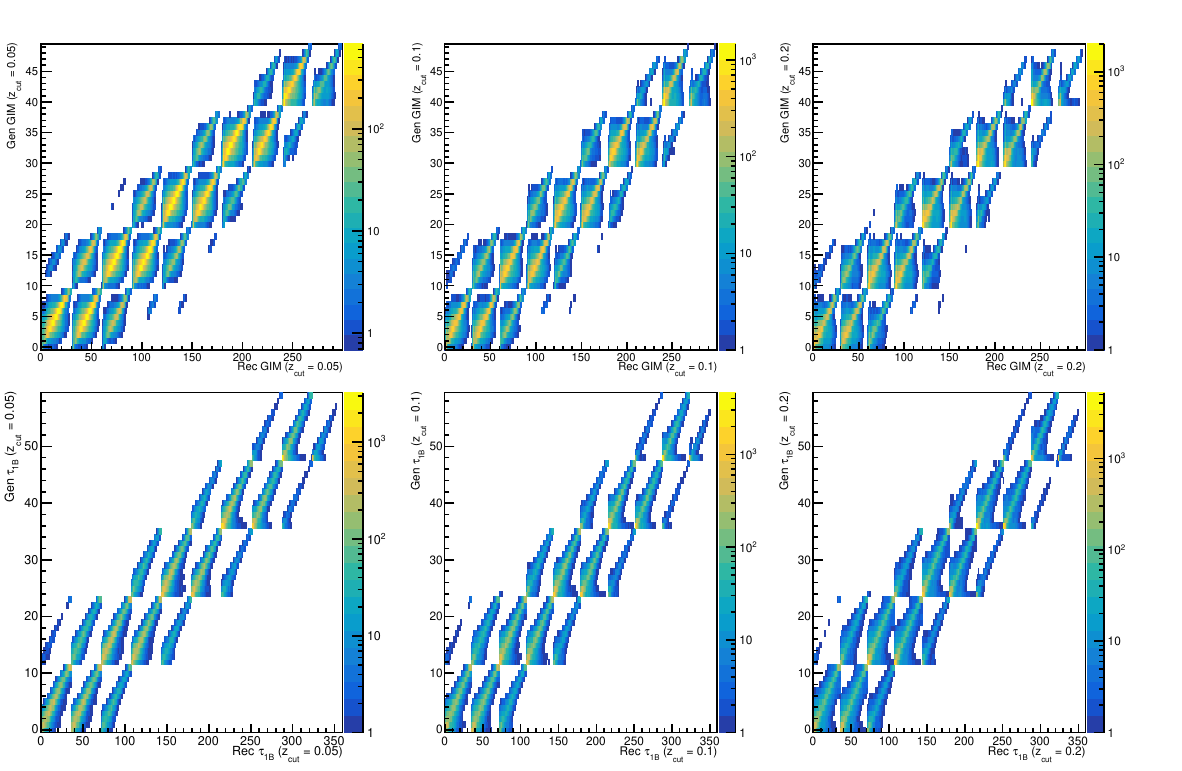}
    \caption{2D migration matrices for the GIM and the groomed \tb. At particle-level, there are three times as many bins in the event shape observables and two times as many bins in $Q^2$.}
    \label{fig:Mig2D}
\end{figure}

The binning in the analysis was chosen to provide the maximum number of bins while still allowing for sufficient statistics and the convergence of the unfolding procedure, i.e. a stable and reproducible determination of $A^+$, the inverse of the detector response matrix. The binning used in this analysis is summarized in table~\ref{tab:binnings}
\begin{table}[b]
  \footnotesize
  \begin{center}
    \begin{tabular}{lc}
      \hline
      Observable & Binning \\
      \hline
     Groomed \tb   &  $[0.0, 0.05, 0.10, 0.15, 0.22, 0.3, 0.4, 0.5, 0.6, 0.7, 0.8, 0.9, 1.001]$ \\ 
     Reduced Groomed \tb   &  $[0.0, 0.05, 0.10, 0.15, 0.22, 0.3, 0.4, 0.5, 0.6, 0.7, 1.001]$ \\
         Ln(\GIM)  &  $[-9,-5,-4,-3,-2,-1,0,1,2,3,4]$  \\ 
         Reduced Low Ln(\GIM)  &  $[-9,-4,-3,-2,-1,0,1,2,3,4]$\\
         Reduced High Ln(\GIM)  &  $[-9,-5,-4,-3,-2,-1,0,1,2,4]$\\
      $Q^2$&  $[150, 200, 282, 447, 1122, 20000]$ GeV$^2$ \\
      \hline
    \end{tabular}
    \caption{Binnings used in the analysis. The reduced binnings are used only in the case of the double-differential measurement, in regions where the event statistics become too poor to maintain a finer binning scheme.}
    \label{tab:binnings}
    \end{center}
\end{table}
The Monte Carlo programs Rapgap~\cite{Jung:1993gf} and Djangoh~\cite{Charchula:1994kf} are used to produce the detector response matrix. Djangoh and Rapgap simulate physics events that are passed through the H1 detector simulation, which utilizes a GEANT3~\cite{Brun:1987ma} model and a fast calorimeter simulation~\cite{Kuhlen:1992ey,Glazov:2010zza}.
Djangoh uses Born level matrix elements for NC DIS and applies the colour dipole model from Ariadne~\cite{Lonnblad:1992tz} for higher order emissions. Rapgap implements Born level matrix for NC DIS and uses the leading logarithmic approximation for parton shower emissions. Both generators use the Lund string model to hadronize the showered partons. The detector-level distributions for the groomed \tb and GIM are shown in Figs.~\ref{fig:GIMDet} and ~\ref{fig:G1JDet}.
With knowledge of the generated and reconstructed values of the observables, a detector response matrix is constructed. The response matrix has three bins in the reconstructed observable for each bin of generated observable. Events which are generated within the phase space and not reconstructed are taken to be an inefficiency, while events generated outside the phase space and reconstructed inside of it are treated as a background. One notable case is the situation where either the detector-level or the particle-level event fails to pass the grooming procedure. The presented cross sections are only the ``visible" cross sections, and thus events which pass the grooming at detector-level and not particle-level are treated as ``backgrounds" and those which pass at particle-level and not at detector-level are treated as inefficiencies. This number of events is only on the order of a few percent for the harshest grooming considered here, and is primarily restricted to affecting the low-\tb low-GIM region. 
The migration matrices produced by Djangoh for the 1D and 2D distributions are provided in Figs.~\ref{fig:Mig1D}~\ref{fig:Mig2D}.

The 2D migration matrices demonstrate that the migration in $Q^2$ is fairly small, and is almost completely restricted to the neighboring bins. This is in part due to the treatment of QED radiation in the kinematic reconstruction. Large off-diagonal migrations in $Q^2$ are typical of QED-sensitive reconstruction methods such as the electron method in events that have QED radiation.
The unfolding is performed at values of the regularization parameter $\tau$, that minimize the influence of the unfolding on the final result while maintaining good closure. Smaller values of $\tau$ induce the large fluctuations described for the standard matrix inversion case, while large values of $\tau$ induce large correlations and bias the data. The choice of $\tau$ is made by looking for the value of $\tau$ that produces the best closure while keeping the global correlation coefficients $< 0.9$ for each bin in the measured observable. The closure of the unfolding procedure is tested by unfolding the detector-level distribution as determined by Rapgap with the response matrix as generated by Djangoh, and vice versa. The output distribution of the unfolding procedure can be compared to the corresponding particle-level distribution to determine whether or not the procedure is returning results close to the expected particle-level. This procedure provides an estimate of the uncertainty associated with the unfolding, which will be described in greater detail in Sec.~\ref{Sec:SysUnc}. Typical values of $\tau$  are $1\cdot10^{-5}$ and $4\cdot10^{-4}$ for the 1D and 2D distributions respectively. The precise values of $\tau$ used in the nominal unfolding are given in Table~\ref{tab:taus}.
\begin{table}[tbhp]
  \footnotesize
  \begin{center}
    \begin{tabular}{llc}
      \hline
      Observable & $z_{cut}$& $\tau$ \\
      \hline
     1D Groomed \tb &$ 0.05$  &  $1.1\cdot10^{-5}$ \\
     1D Groomed \tb &$ 0.1$   &  $1\cdot10^{-5}$ \\
     1D Groomed \tb &$ 0.2$   &  $8.2\cdot10^{-6}$ \\
     1D GIM &$ 0.05$  &  $9\cdot10^{-5}$ \\
     1D GIM &$ 0.1$   &  $7.5\cdot10^{-5}$ \\
     1D GIM &$ 0.2$   &  $8.7\cdot10^{-5}$ \\
     2D Groomed \tb &$ 0.05$  &  $7.9\cdot10^{-4}$ \\
     2D Groomed \tb &$ 0.1$   &  $5.5\cdot10^{-4}$ \\
     2D Groomed \tb &$ 0.2$   &  $4\cdot10^{-4}$ \\
     2D GIM &$ 0.05$  &  $8.7\cdot10^{-4}$ \\
     2D GIM &$ 0.1$   &  $8.3\cdot10^{-4}$ \\
     2D GIM &$ 0.2$   &  $5.5\cdot10^{-4}$ \\
      \hline
    \end{tabular}
    \caption{}
    \label{tab:taus}
    \end{center}
\end{table}
The distribution of the data to be unfolded should ideally be free of backgrounds from other kinds of events. The following backgrounds are subtracted from the measured event yield prior to passing the data through the unfolding:
\begin{compactitem}
\item Low-$Q^2$ NC DIS as simulated by Djangoh for $4 < Q^2 < 60$ GeV$^2$. Events generated and reconstructed between $60< Q^2 < 150$ GeV$^2$ are handled in the migration matrix, and thus are not explicitly subtracted prior to the unfolding.
\item $e+p$ events with $Q^2 < 4$ GeV$^2$, including photoproduction, are simulated by Pythia~6.2~\cite{Sjostrand:1993yb,Sjostrand:2001yu}.
\item QED Compton scattering is simulated by COMPTON~\cite{Courau:1992ht}.
\item Di-lepton production is simulated by GRAPE~\cite{Abe:2000cv}.
\item Deeply virtual compton scattering is simulated by MILOU~\cite{Perez:2004ig}.

\end{compactitem}
However, after the variety of cuts applied to the data, all of the above are on the order of a few events at maximum. The only background which contributes significantly is events with $60< Q^2 < 150$ GeV$^2$, which migrate to higher $Q^2$. This background is on the order of 6\% in some bins, and is subtracted during the unfolding procedure.
\section{Radiative Corrections}
\label{Sec:RadCor}
QED radiative effects influence the measured cross section in a variety of ways. The emission of a real photon off the electron in the initial-state disturbs the measurement of the event kinematics, and occasionally produces an energetic cluster in the SpaCal. Final-state radiation off the electron is typically collinear to the electron and thus produces one energetic cluster in the calorimeter, but occasionally the photon will be produced at larger angle and be resolved. In both cases, these photons need to be removed from the hadronic final state, as they tend to fall at midrapidity in the Breit frame and therefore can significantly disturb the grooming procedure.
\begin{figure}[H]
    \centering
    \includegraphics[width=10cm]{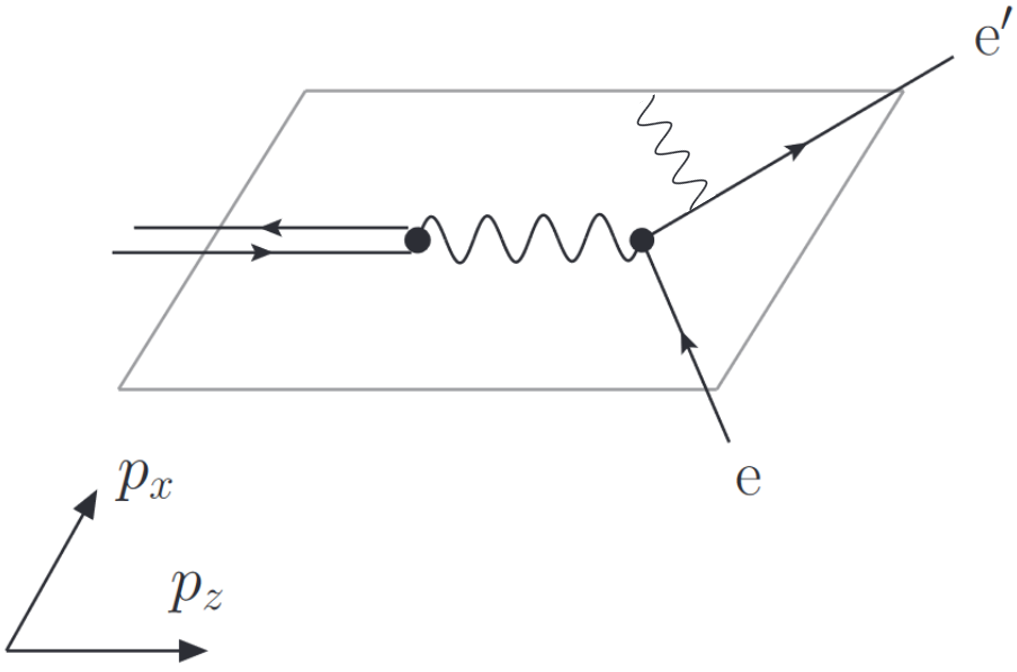}
    \caption{Fig.~\ref{fig:BreitJets} in the case of a photon radiated in the final state. In the Breit frame, this photon is typically emitted at midrapidity, often landing in the current hemisphere.}
    \label{fig:BreitJetsQED}
\end{figure}
Additionally, virtual corrections to the NC DIS process can change the overall normalization and shape of the inclusive cross section. QED effects are included in Djangoh and Rapgap via an interface to the HERACLES~\cite{Kwiatkowski:1990es} program. HERACLES simulates the first-order electroweak corrections to both $e^++p$ and $e^-+p$ DIS, including the photon and $Z$ self-energies, and real photon emission off the lepton. \par
The highest $Q^2$ bin can be seen to have an enhancement in the correction factor, i.e. a decrease in the radiative cross section. This occurs due to electroweak effects, i.e. $Z$ and $\gamma/Z$ interference, decreasing (increasing) the $e^++p$ ($e^-+p$) cross sections~\cite{ZEUS:2009swh}. The sign and magnitude of the $F_3$ term in the DIS total cross section are sensitive to the incoming lepton charge. Since the data presented here include both $e^++p$ and $e^-+p$ DIS, contributing according to the luminosities shown in Fig.~\ref{fig:HERAParameters}, the QED correction factors shown represent a luminosity-weighted average of the two. It is observed that the $e^-+p$ correction factor is roughly flat as a function of $Q^2$, while the $e^++p$ increases slightly at higher $Q^2$. The data are corrected for these effects by applying a bin-by-bin correction factor, $c_{QED}$, which is defined as the ratio between the non-radiative particle-level and the radiative particle-level. \par
\begin{figure}[H]
  \centering
   \begin{overpic}[trim={1.1cm 0 1cm 0},clip,scale=0.4]{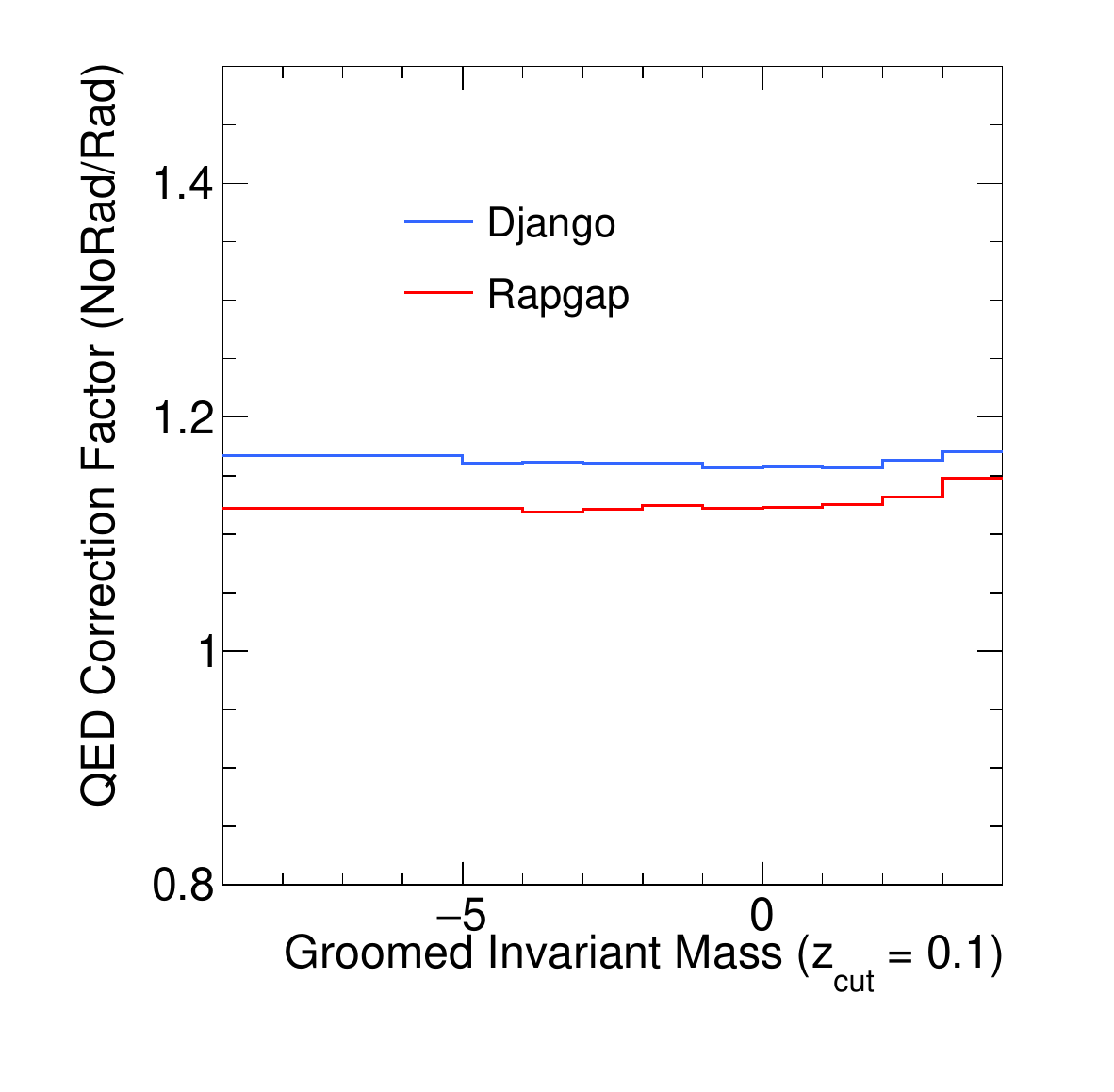}
     \put(89,0){\includegraphics[trim={3.9cm 0 1cm 0},clip,scale=0.4]{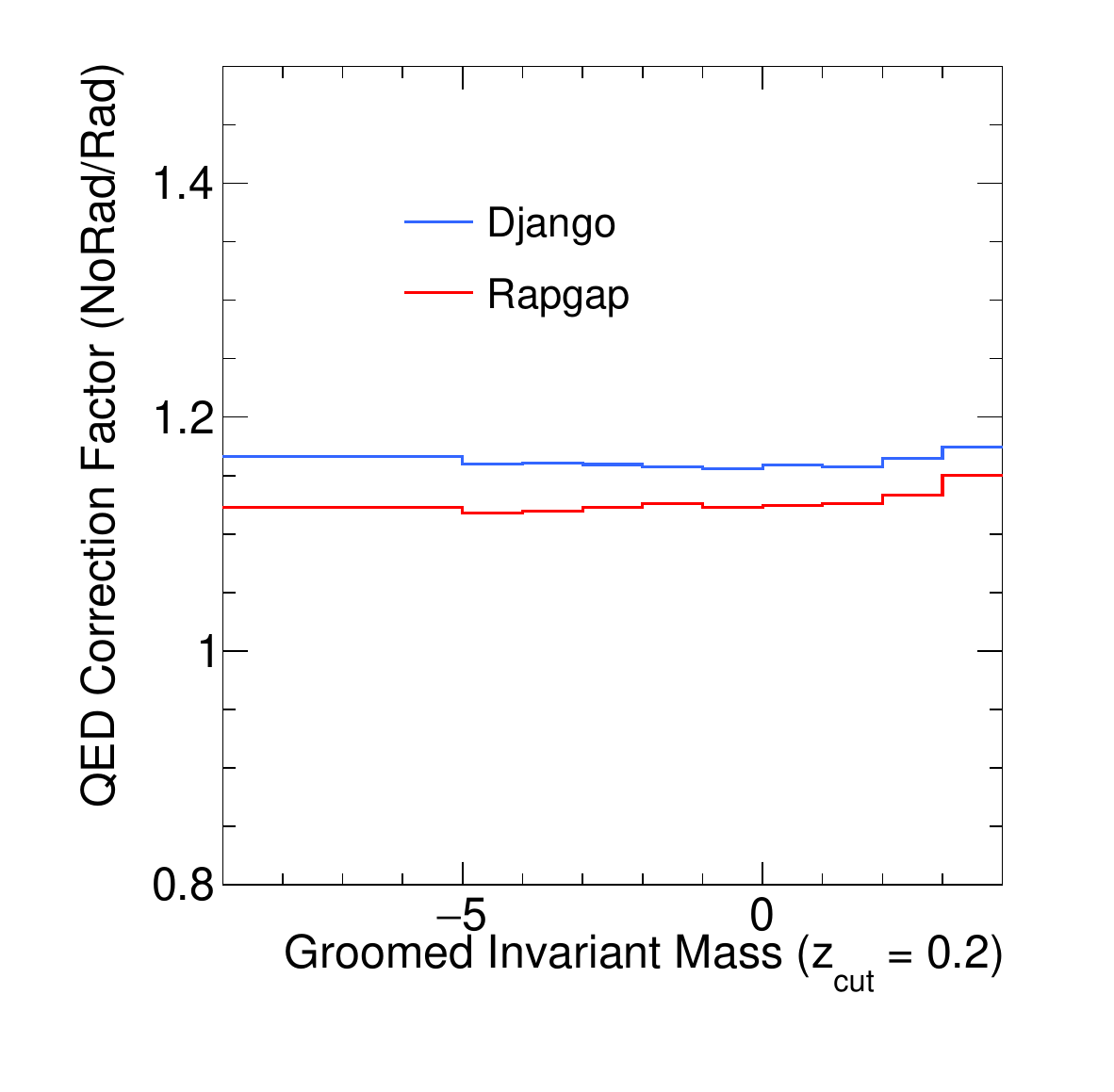}}
     \put(-75,0){\includegraphics[trim={1cm 0 1.5cm 0},clip,scale=0.4]{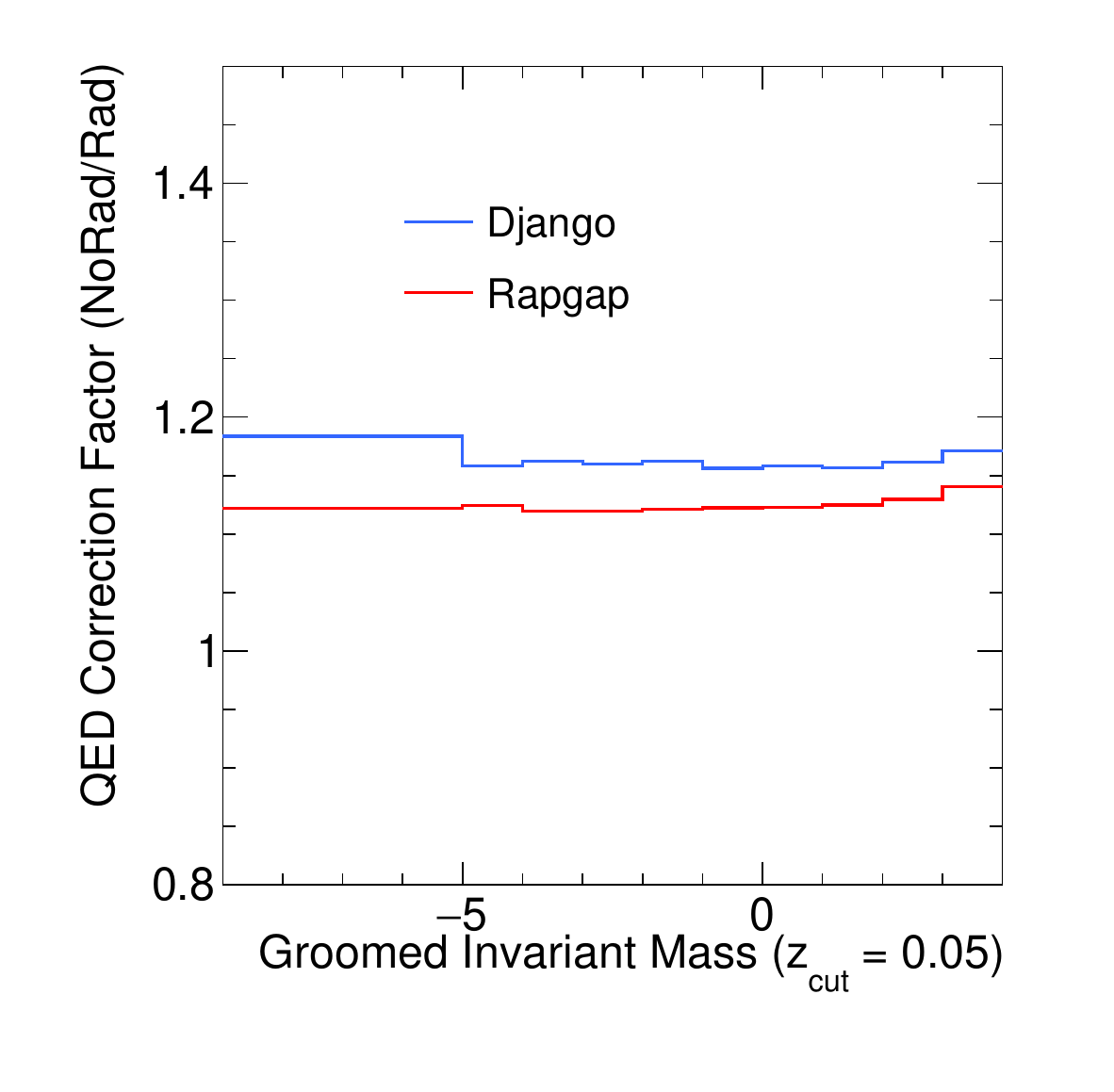}}
  \end{overpic}
\caption{Radiative correction factors for the single-differential GIM.}
\label{fig:RadCor1DGIMs}
\end{figure}
\begin{figure}[H]
  \centering
   \begin{overpic}[trim={1.1cm 0 1cm 0},clip,scale=0.4]{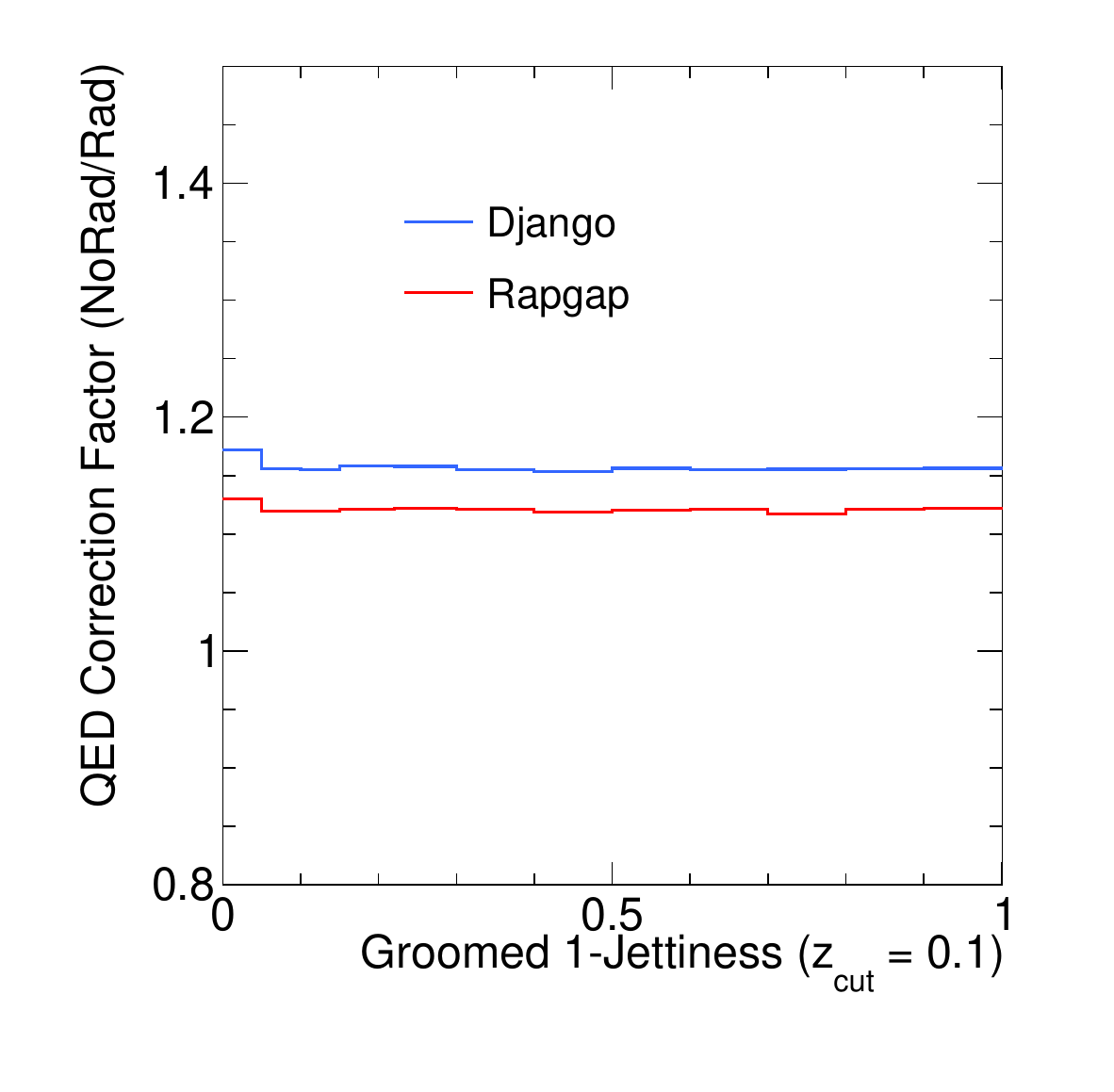}
     \put(89,0){\includegraphics[trim={3.9cm 0 1cm 0},clip,scale=0.4]{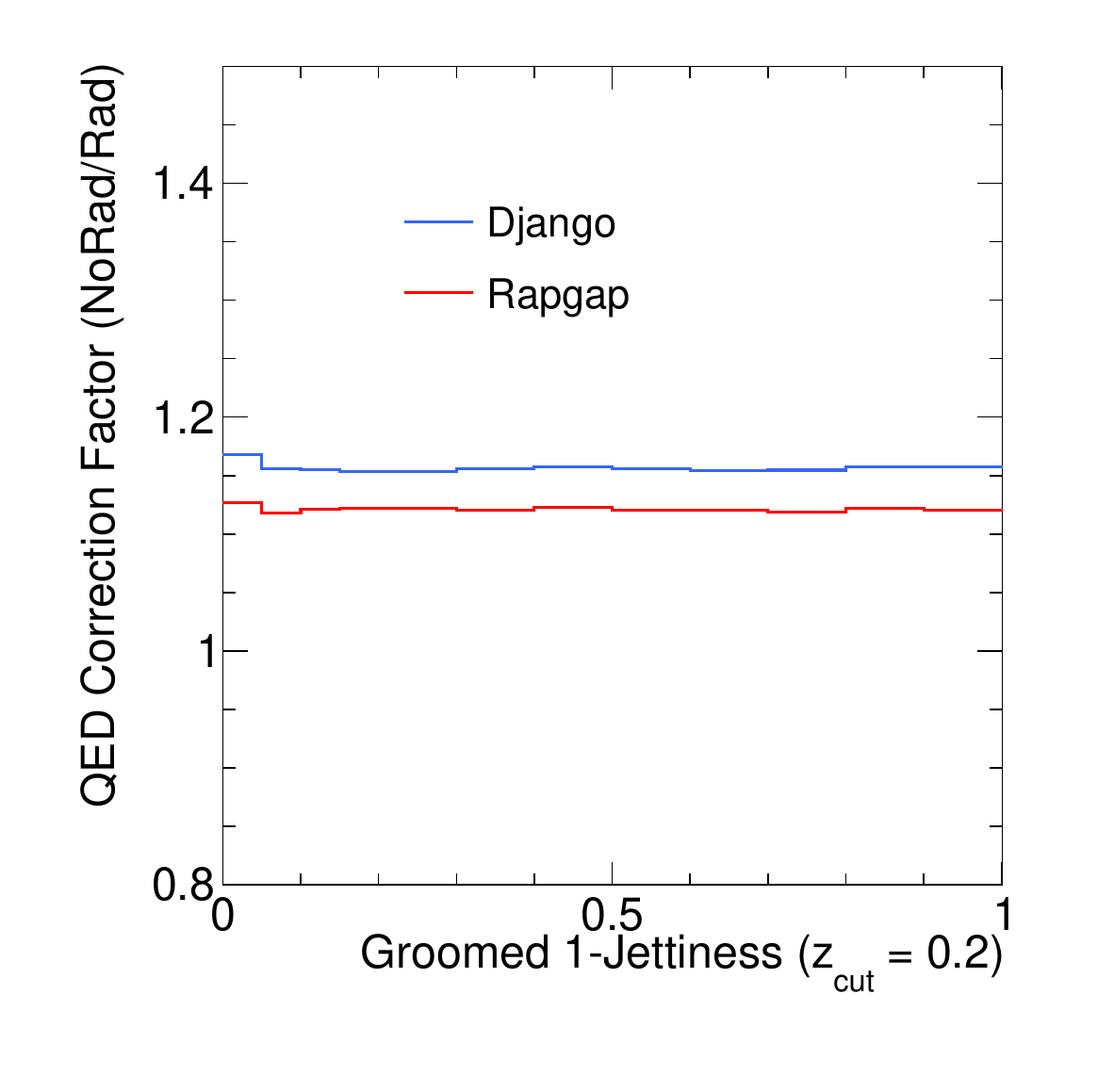}}
     \put(-75,0){\includegraphics[trim={1cm 0 1.5cm 0},clip,scale=0.4]{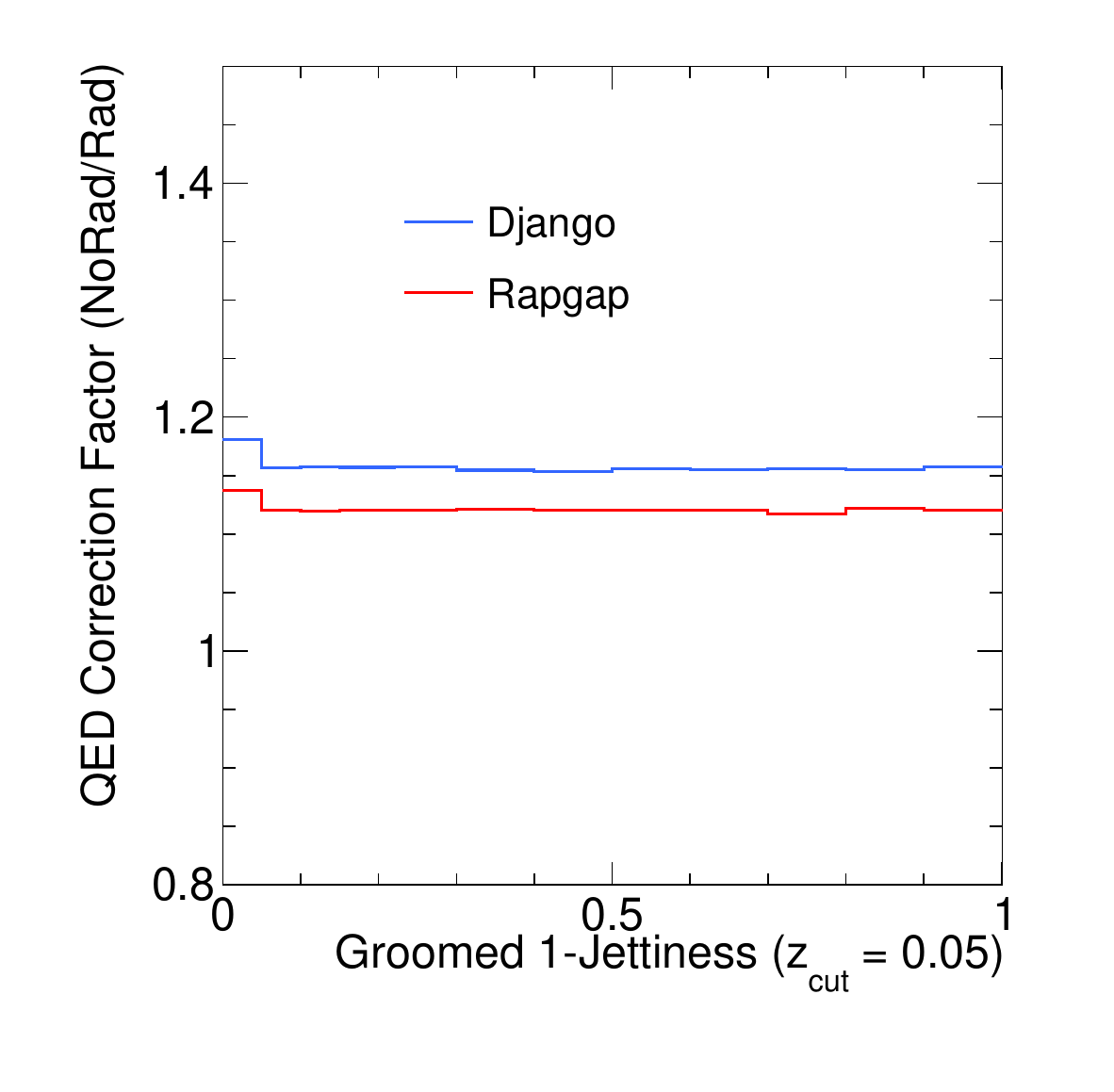}}
  \end{overpic}
\caption{Radiative correction factors for the single-differential groomed \tb.}
\label{fig:RadCor1DGrTaus}
\end{figure}

\begin{figure}[H]
  \centering
   \begin{overpic}[trim={1.1cm 0 1cm 0},clip,scale=0.4]{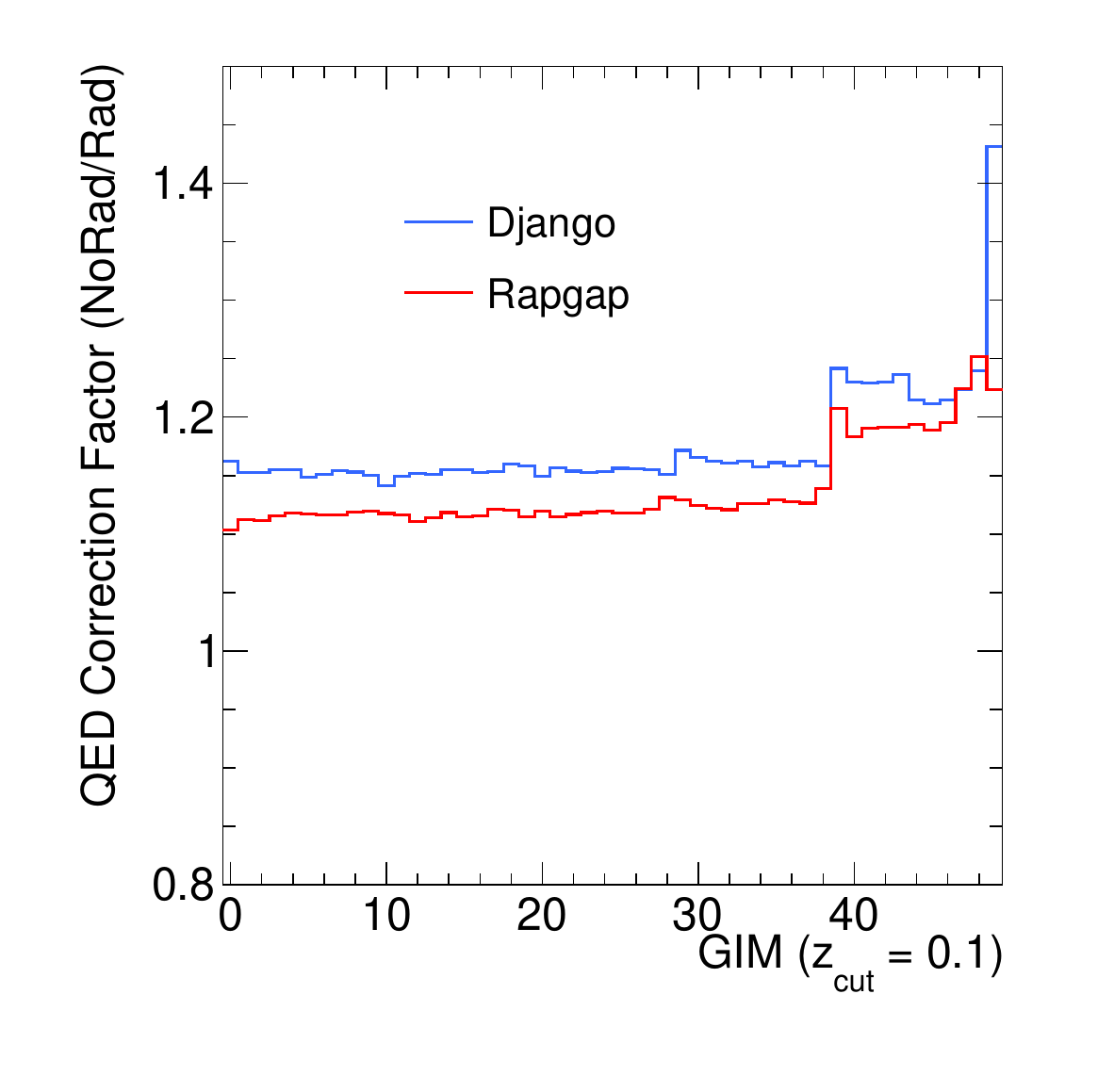}
     \put(89,0){\includegraphics[trim={3.9cm 0 1cm 0},clip,scale=0.4]{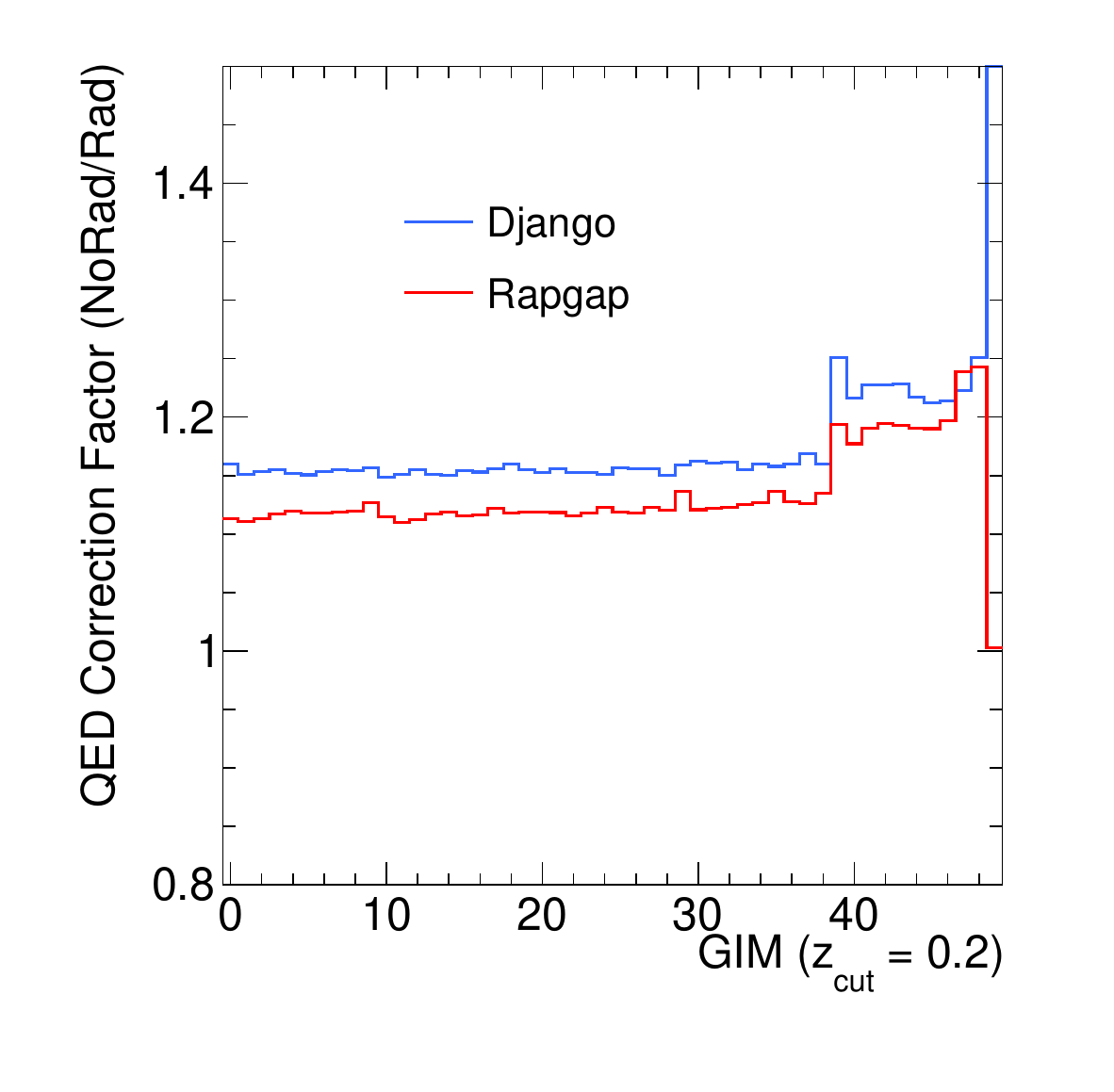}}
     \put(-75,0){\includegraphics[trim={1cm 0 1.5cm 0},clip,scale=0.4]{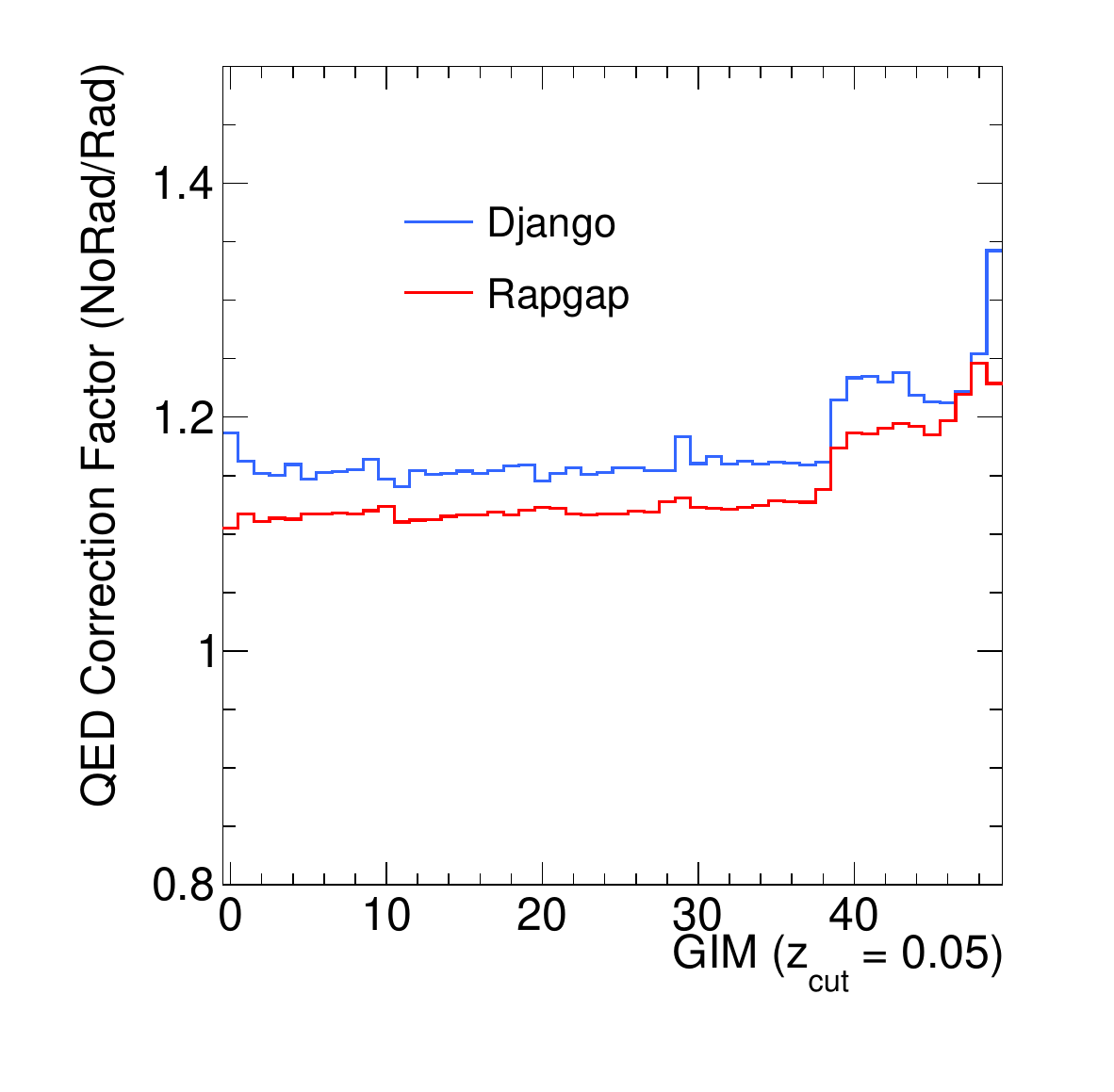}}
  \end{overpic}
\caption{Radiative correction factors for the double-differential GIM. There are 50 bins, corresponding to 10 bins in the observable for each of the 5 $Q^2$ bins.}
\label{fig:RadCor2DGIMs}
\end{figure}

\begin{figure}[H]
  \centering
   \begin{overpic}[trim={1.1cm 0 1cm 0},clip,scale=0.4]{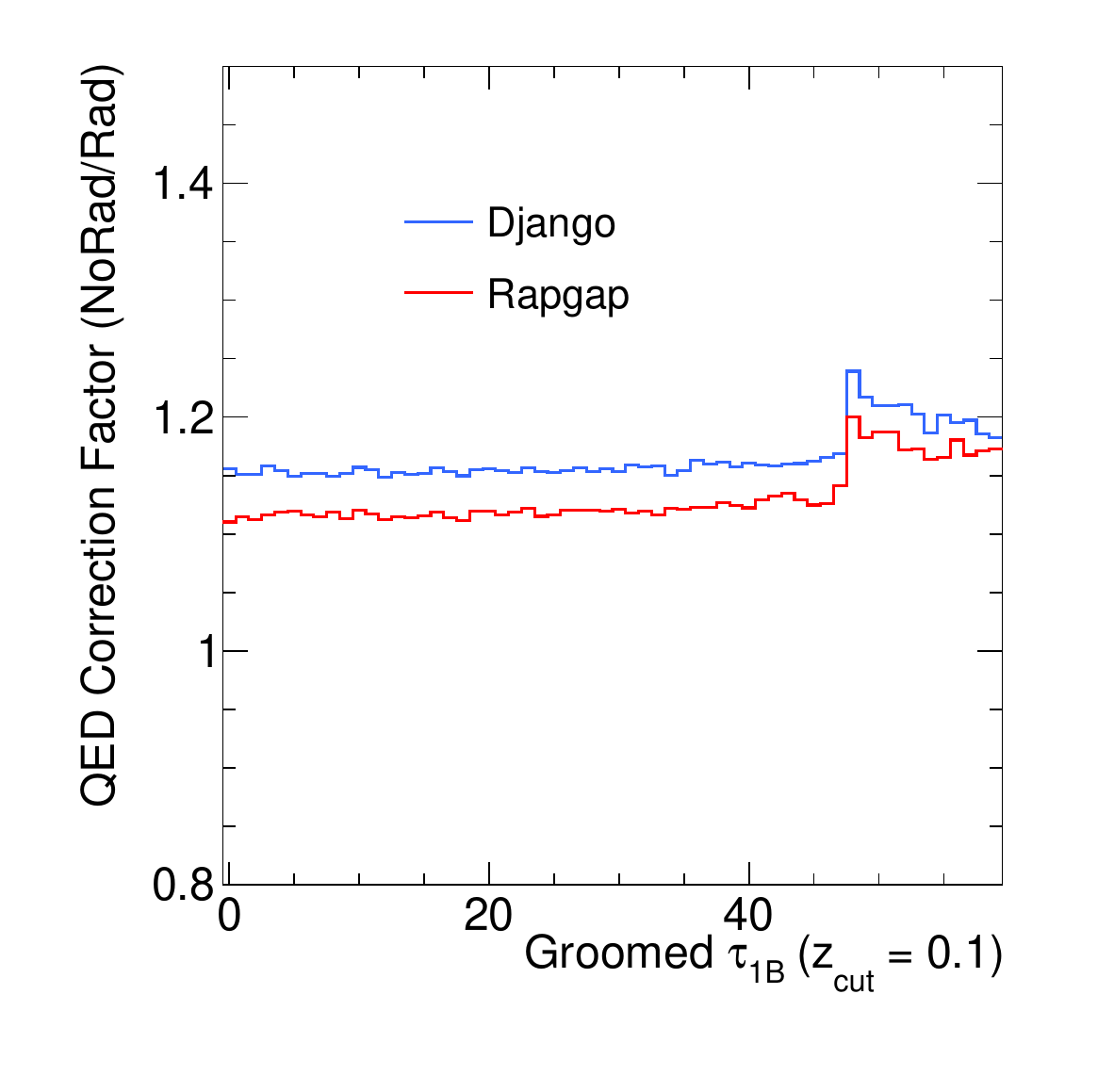}
     \put(89,0){\includegraphics[trim={3.9cm 0 1cm 0},clip,scale=0.4]{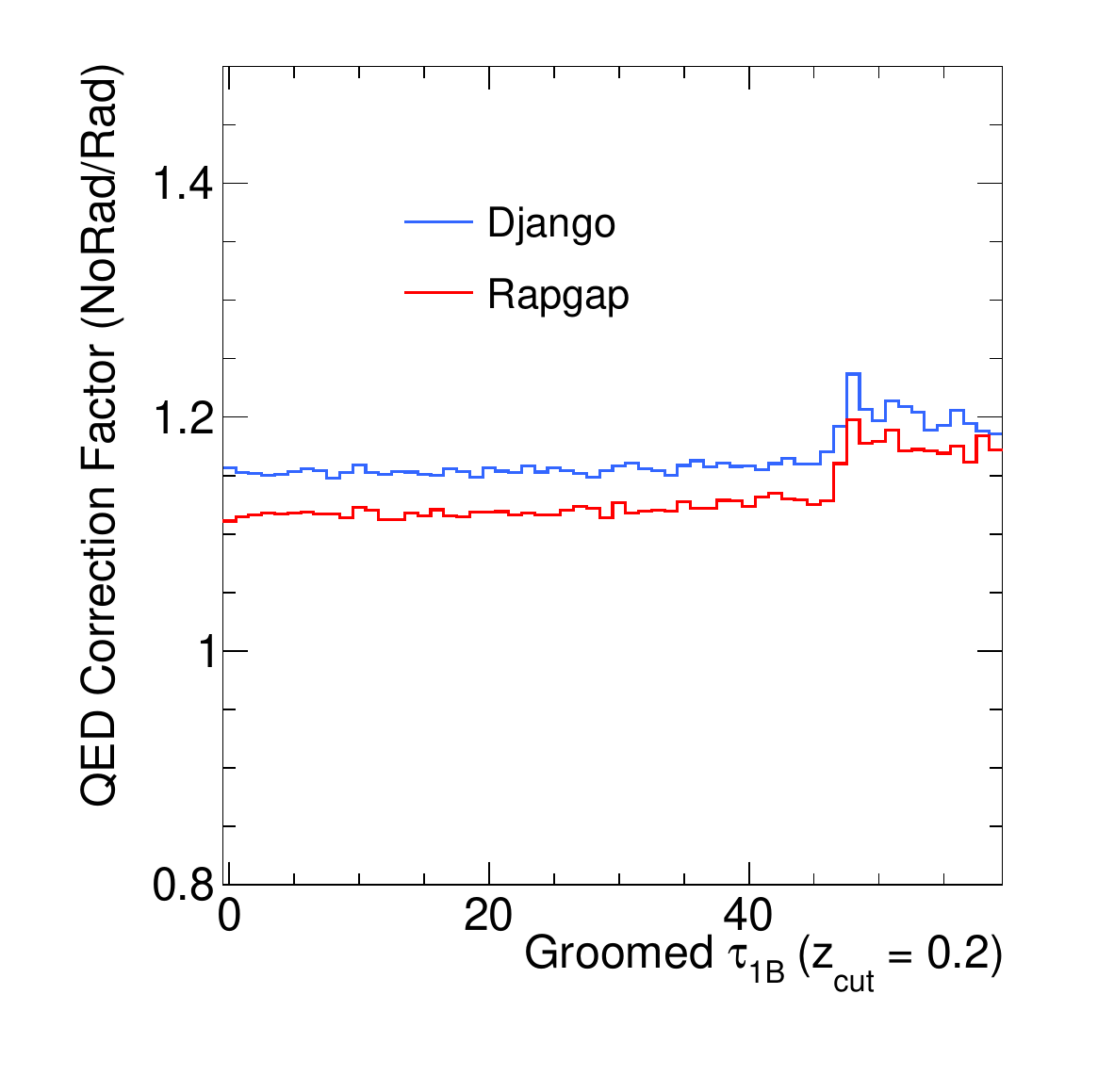}}
     \put(-75,0){\includegraphics[trim={1cm 0 1.5cm 0},clip,scale=0.4]{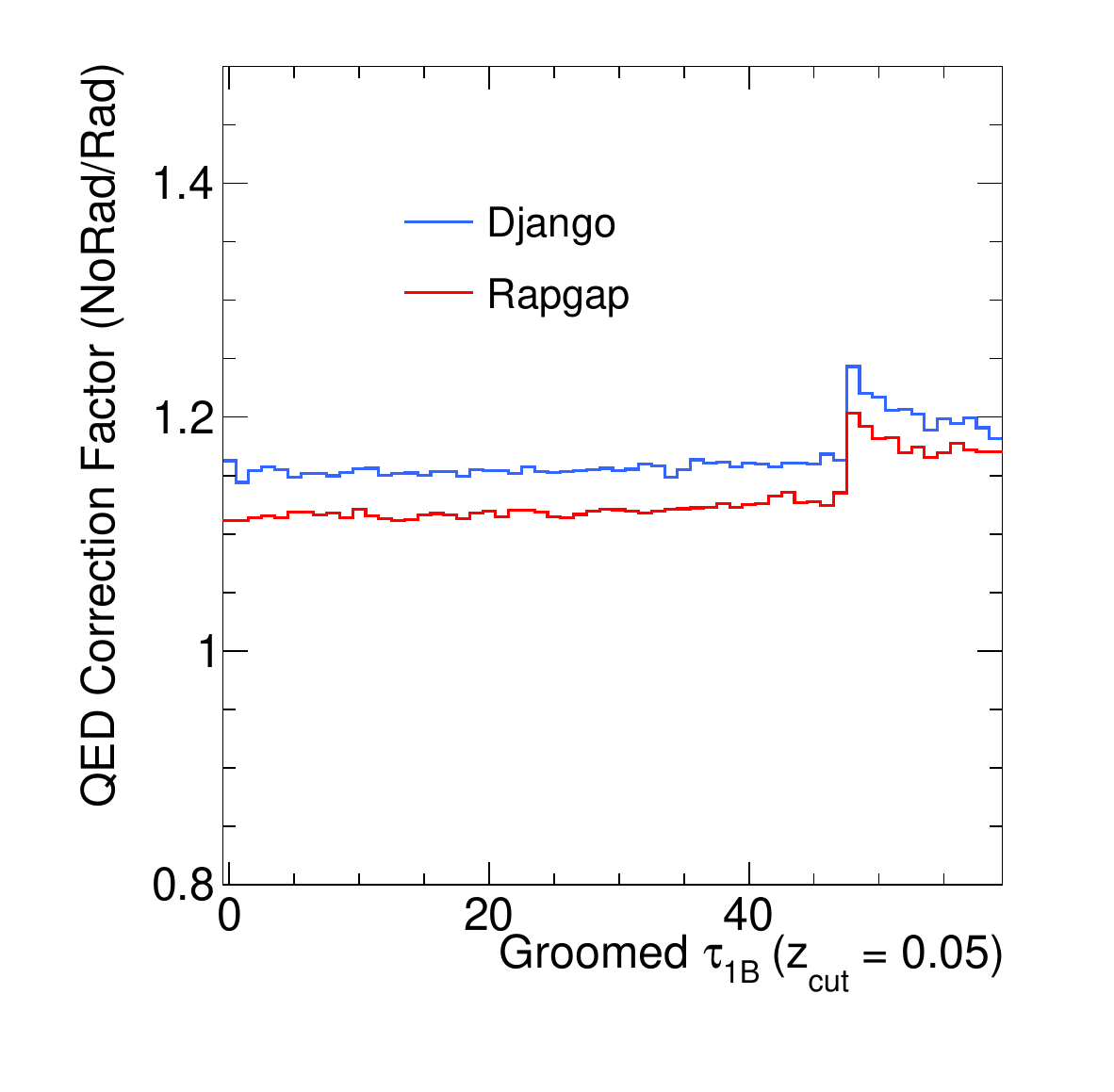}}
  \end{overpic}
\caption{Radiative correction factors for the double-differential groomed \tb.}
\label{fig:RadCor2DGrTaus}
\end{figure}

\section{Systematic Uncertainties}
\label{Sec:SysUnc}
\begin{figure}[!htb]
  \centering
   \begin{overpic}[trim={1.1cm 0 1cm 0},clip,scale=0.4]{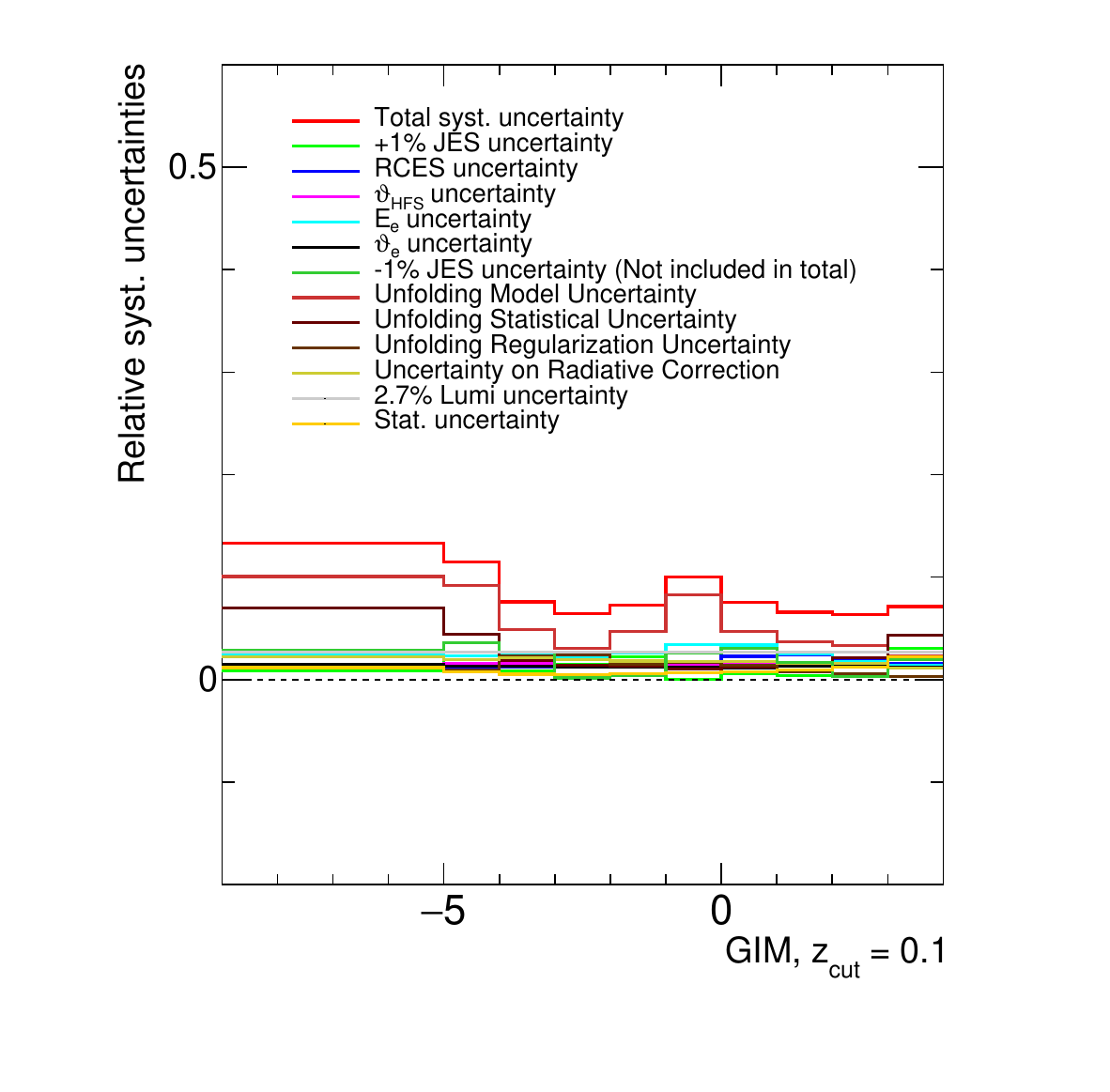}
     \put(89,0){\includegraphics[trim={3.9cm 0 1cm 0},clip,scale=0.4]{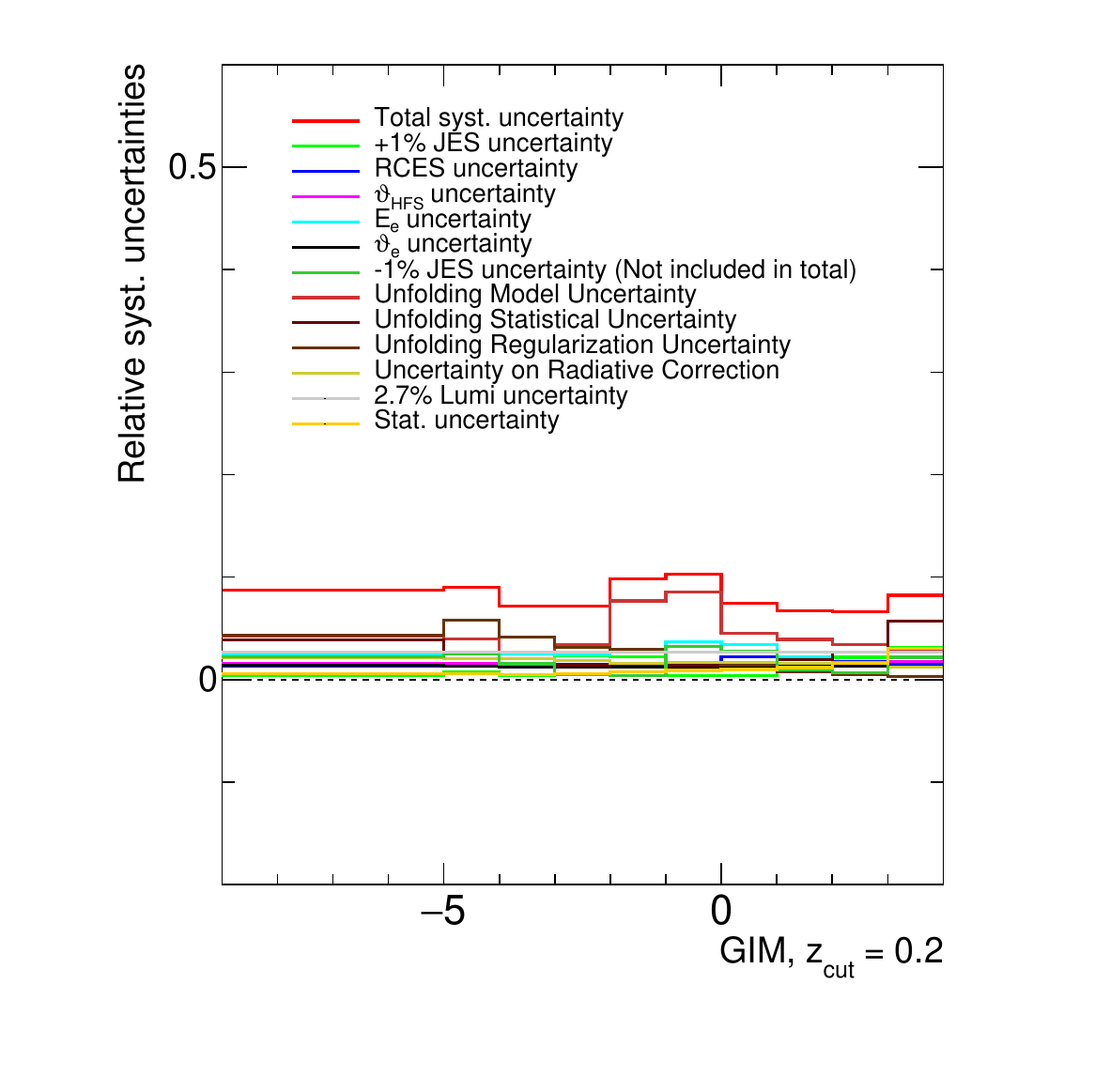}}
     \put(-75,0){\includegraphics[trim={1cm 0 1.5cm 0},clip,scale=0.4]{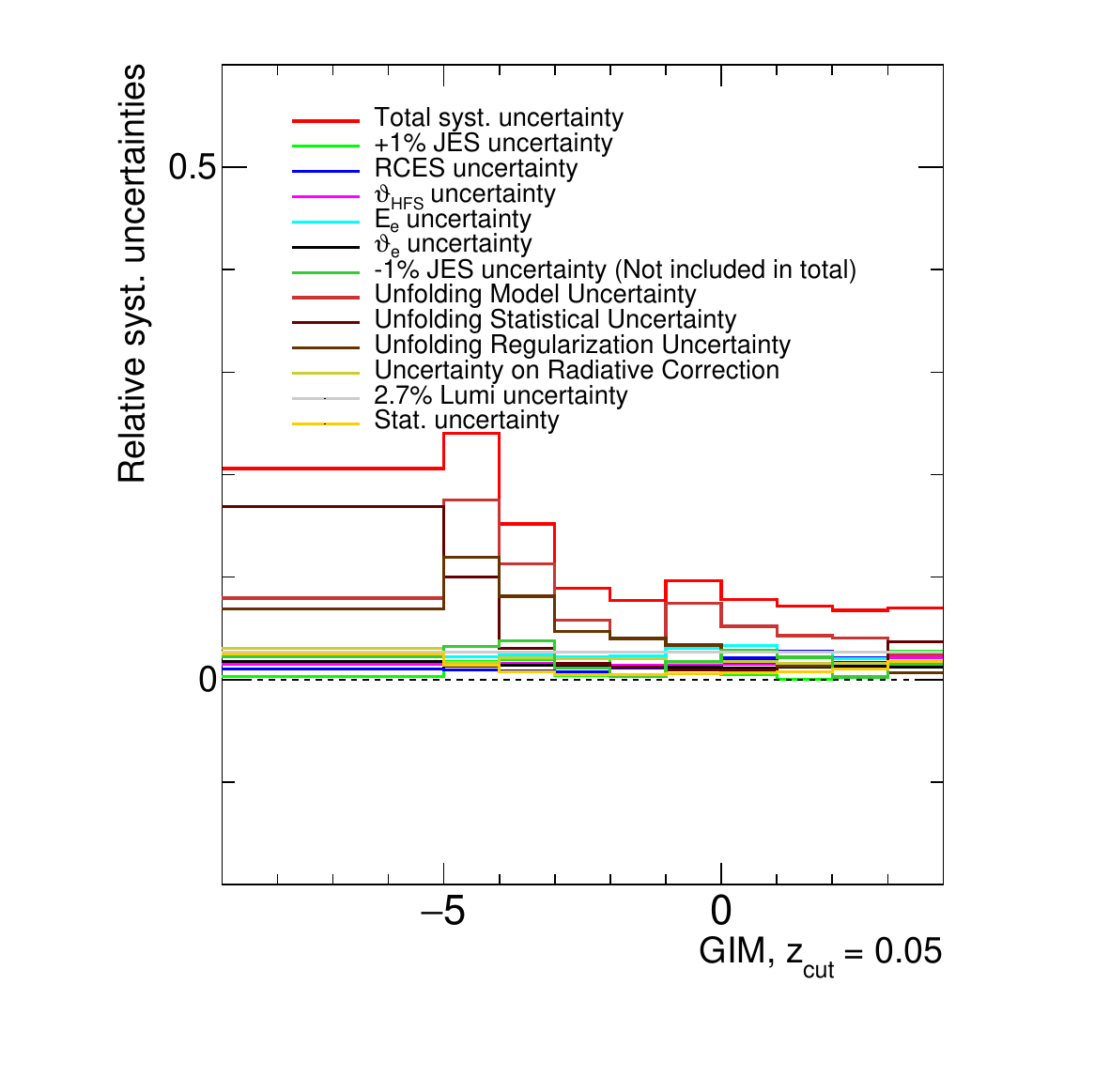}}
  \end{overpic}
\caption{Systematic uncertainties for the GIM at the three values of $z_{cut}$.}
\label{fig:SysUnc1DGIM}
\end{figure}

\begin{figure}[!htb]
  \centering
   \begin{overpic}[trim={1.1cm 0 1cm 0},clip,scale=0.4]{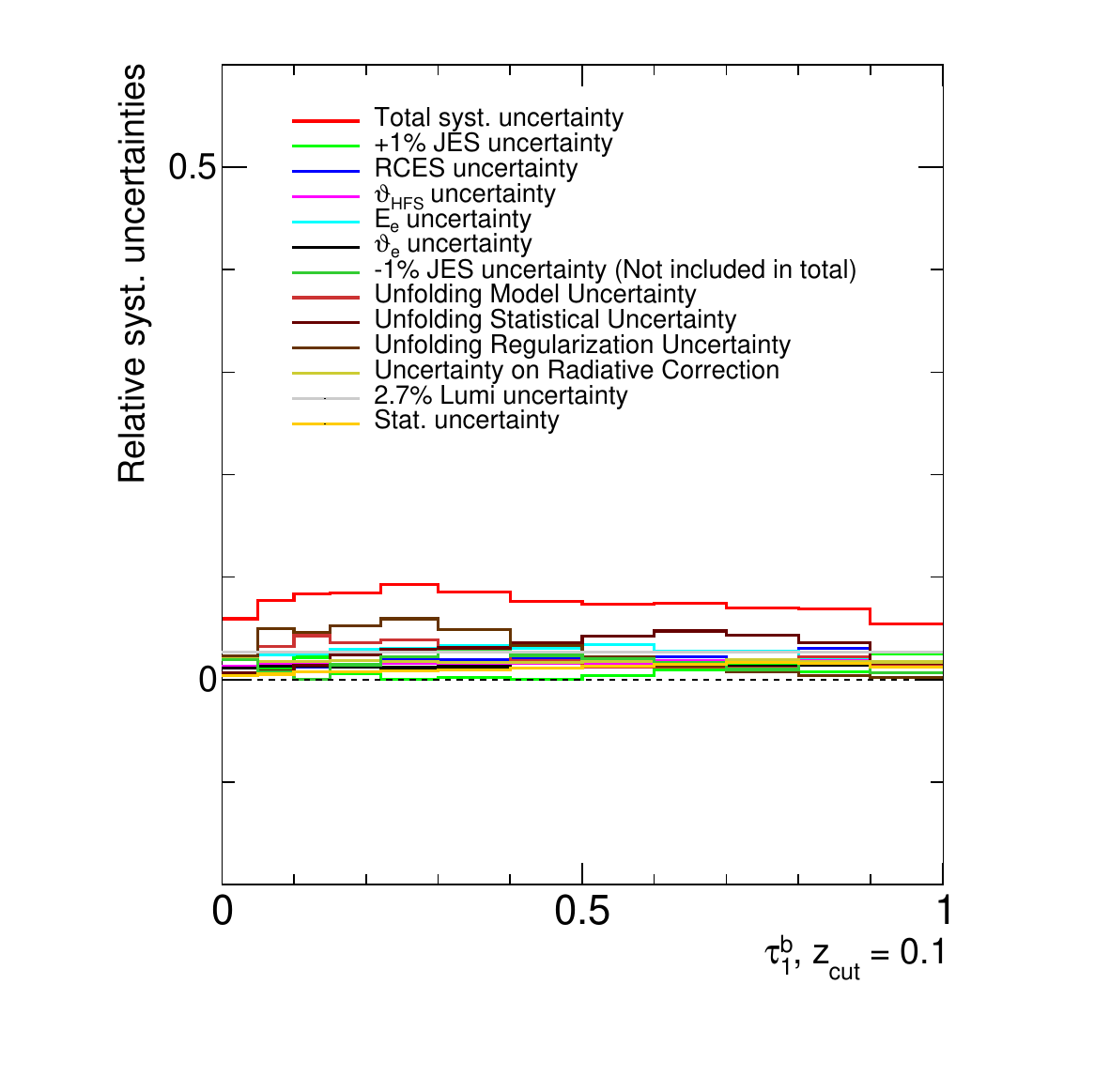}
     \put(89,0){\includegraphics[trim={3.9cm 0 1cm 0},clip,scale=0.4]{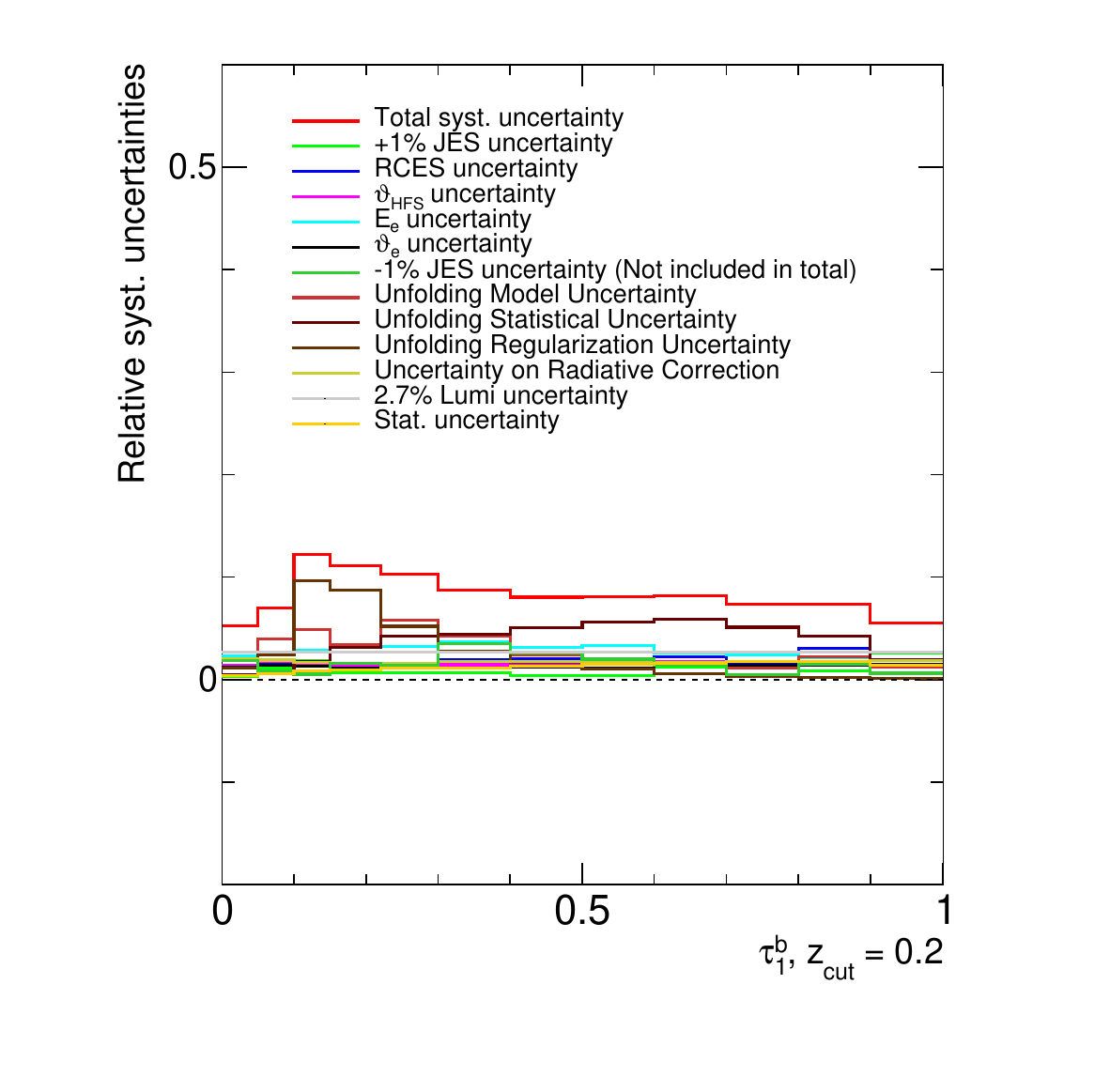}}
     \put(-75,0){\includegraphics[trim={1cm 0 1.5cm 0},clip,scale=0.4]{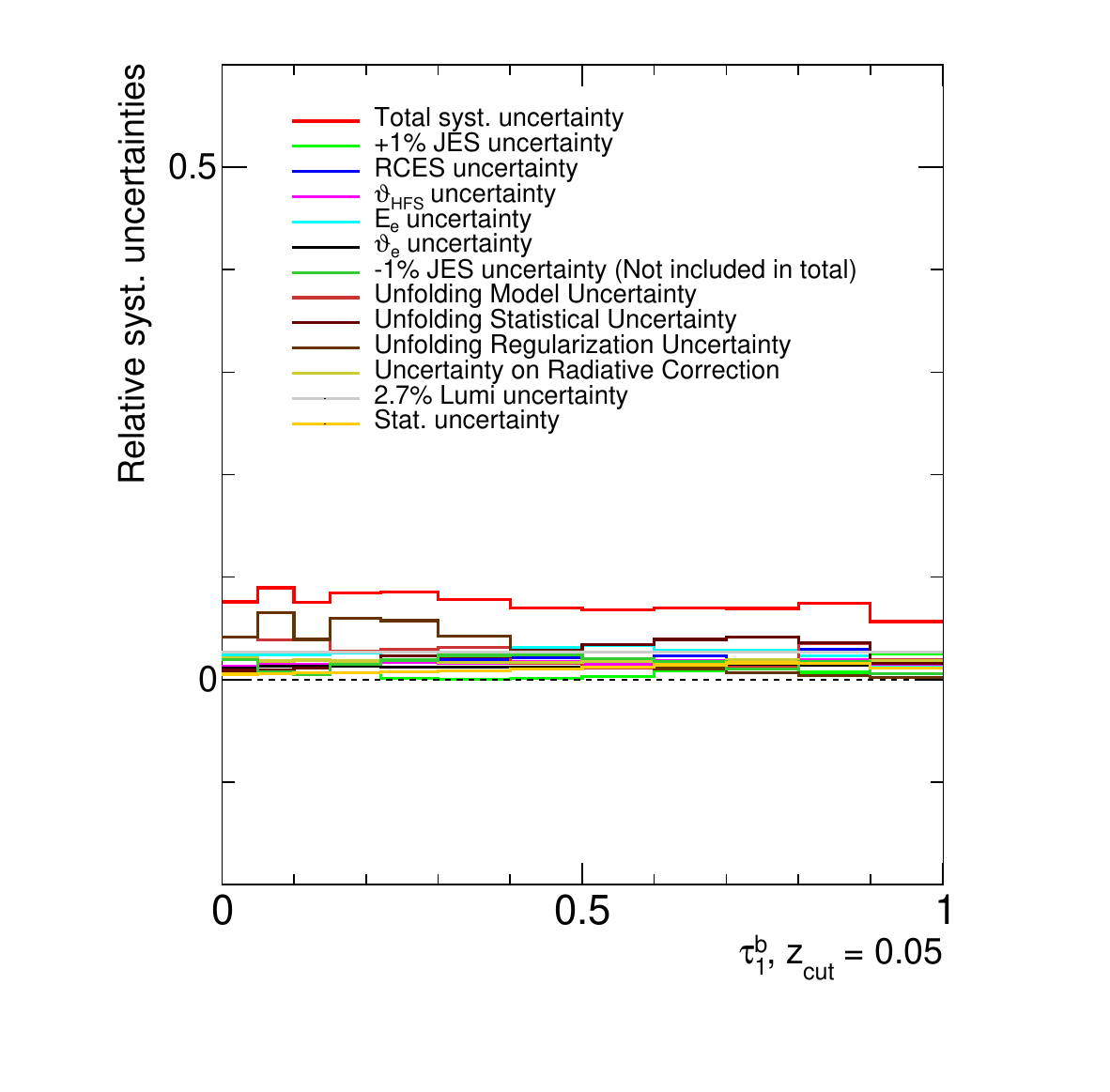}}
  \end{overpic}
\caption{Systematic uncertainties for the groomed \tb at the three values of $z_{cut}$.}
\label{fig:SysUnc1DGrTau}
\end{figure}

\begin{figure}[t]
  \centering
   \begin{overpic}[trim={0 0 0 0},clip,scale=0.7]{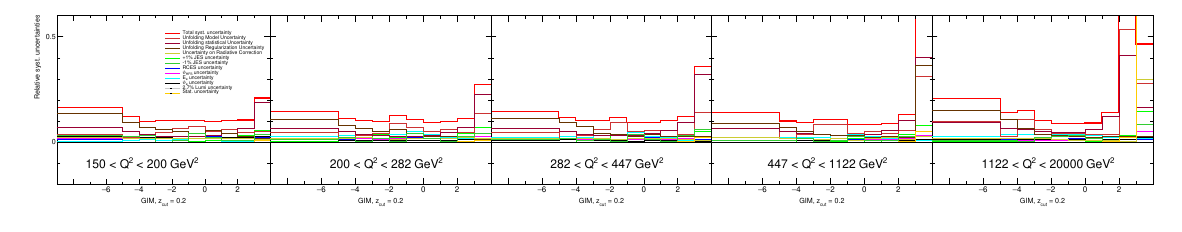}
     \put(0,40){\includegraphics[trim={0 0 0 0},clip,scale=0.7]{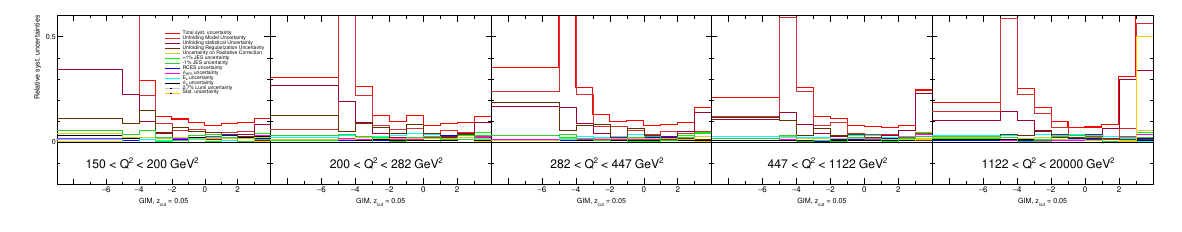}}
     \put(0,20){\includegraphics[trim={0 0 0 0cm},clip,scale=0.7]{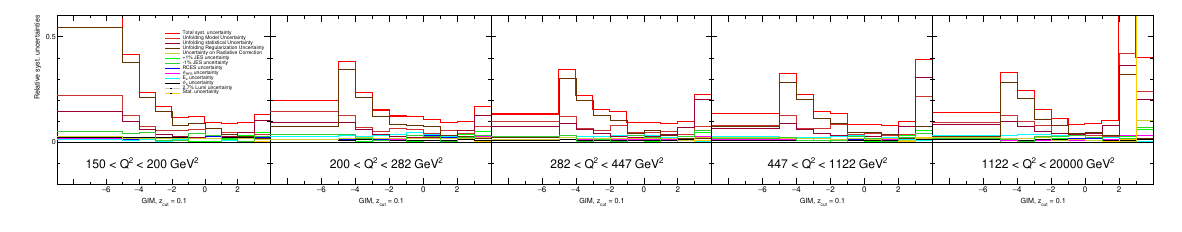}}
  \end{overpic}
\caption{Systematic uncertainties for the three values of $z_{cut}$ in the five bins of $Q^2$.}
\label{fig:SysUnc2DGIM}
\end{figure}

\begin{figure}[b]
  \centering
   \begin{overpic}[trim={0 0 0 0},clip,scale=0.7]{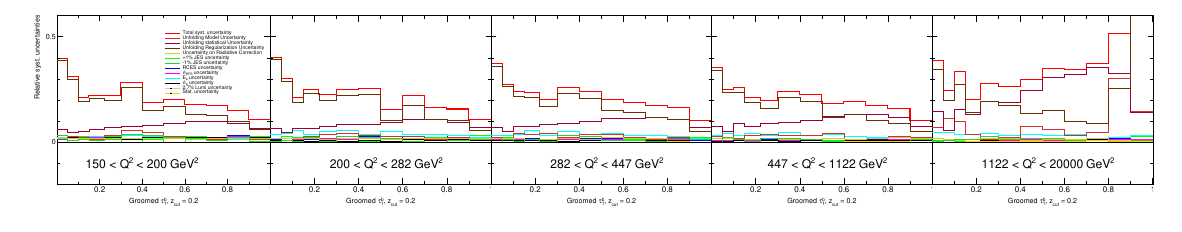}
     \put(0,40){\includegraphics[trim={0 0 0 0},clip,scale=0.7]{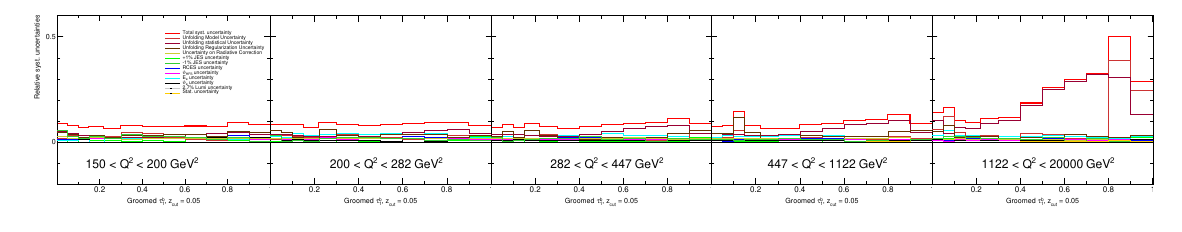}}
     \put(0,20){\includegraphics[trim={0 0 0 0cm},clip,scale=0.7]{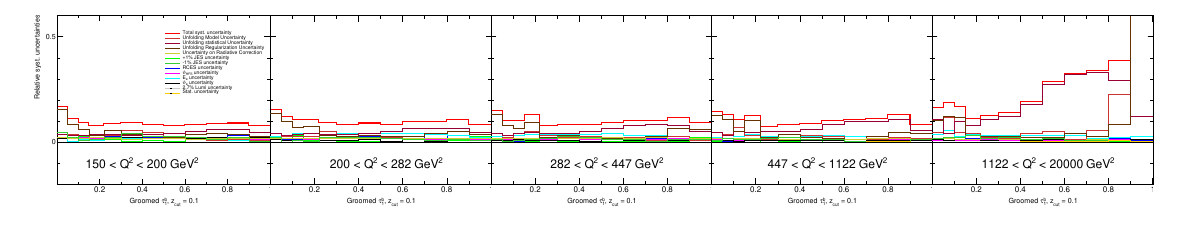}}
  \end{overpic}
\caption{Systematic uncertainties for the three values of $z_{cut}$ in the five bins of $Q^2$.}
\label{fig:SysUnc2DGrTau}
\end{figure}

A variety of uncertainties on the presented cross sections were considered. These uncertainties generally fall into two classes. One class of uncertainties is associated with uncertainties in the physical detector, such as resolutions, alignments, inefficiencies, etc. The other class of uncertainties results from the data correction procedures, in particular the unfolding. The following is a list of all the considered systematic uncertainties; the name of the uncertainty as denoted in the legend of Figs.~\ref{fig:SysUnc1DGIM},~\ref{fig:SysUnc2DGIM},~\ref{fig:SysUnc1DGrTau}, and~\ref{fig:SysUnc2DGrTau} is listed in bold.\par
Three sources of uncertainties originate from the unfolding procedure. They are considered uncorrelated for the purposes of defining the total uncertainty. In almost all bins of the measurement, the unfolding-related uncertainties dominate over the detector-related ones.
\begin{compactitem}
 \item \textbf{Unfolding Model Uncertainty:} Half of the difference between the unfolded results using the migration matrices of Rapgap and Djangoh is taken as an uncertainty on the final result. In many bins of the measurement this is the dominant systematic uncertainty. 
 \item \textbf{Unfolding Regularization Uncertainty:} An uncertainty associated with the arbitrary choice of regularization parameter is defined by varying the regularization parameter by a factor of two in each direction. This uncertainty has the largest impact on the double-differential 1-jettiness at $z_{cut} = 0.2$, where the regularization parameter had to be fairly large to reduce the influence of the unfolding to acceptable levels. For that particular observable, this uncertainty is typically 20\%, while it is sub-leading and typically on the level of a few percent for all other combinations.
 \item \textbf{Unfolding Statistical Uncertainty:} To estimate the unfolding uncertainty arising from statistical fluctuations in the data, the input data were jittered according to the statistical uncertainties associated with each bin and then unfolded. This procedure was repeated one thousand times for each observable, resulting in an uncertainty band that accurately captures the sensitivity of the unfolding procedure to the data statistics. This uncertainty is typically sub-leading except for a few bins in the double-differential cross sections with small statistics.
\end{compactitem}
A smoothing procedure is applied to the unfolding model uncertainty and the unfolding regularization uncertainty to remove unphysically small values of the uncertainties in certain bins. For an uncertainty $\delta$ in bin $n$, the resulting smoothed uncertainty value $\delta_{n,s}$ is 
\begin{equation}
\delta_{n,s} = \alpha\delta_n+(1-\alpha)\delta_{n-1}
\end{equation}
Where $\alpha$ is a free parameter that determines the degree of influence that the previous bin has on the current one. In the case of determining the smoothed value of the first bin of each distribution, $\delta_0$ is taken to be the average uncertainty across all bins. $\alpha$ is chosen to be 0.5 for all distributions, except for the 2D $z_{cut} = 0.05$ GIM distribution, where it is taken to be 0.7 to prevent the large uncertainty in the first bin of the $150 < Q^2 < 200$ GeV$^2$ distribution, which results from the low statistics in that bin, from erroneously affecting the rest of the distribution. In no bin is the smoothed uncertainty allowed to fall below $85\%$ of its nominal pre-smoothed value. The unfolding statistical uncertainty is smooth almost by definition, and therefore does not undergo the smoothing procedure.\par
Uncertainties relating to the physical detector are generally subleading, but for completeness they are are described below:
\begin{compactitem}
    \item \textbf{Luminosity Uncertainty:} The overall integrated luminosity is known to a precision of 2.7\%, and that uncertainty is applied to the final cross sections. This accounts additionally for a variety of smaller normalization uncertainties, including the trigger efficiency, calorimeter noise suppression algorithm, and electron identification.
    \item \textbf{JES, RCES Uncertainty:} A jet energy calibration was performed which independently considered clusters within high-$p_T$ jets, and those outside of them. The uncertainty on the energies of particles inside of jets is known as the jet energy scale uncertainty (JES), and the energies of particles outside of jets is known as the residual cluster energy scale uncertainty (RCES). Both of these energy scales are varied up and down by a factor of $1\%$, and the resulting distributions are passed through the unfolding procedure. The difference between the varied results and the nominal results is considered as the corresponding uncertainty, which is typically around 1-$2\%$. 

    \item \textbf{$E_e$ Uncertainty:} The measured energy of the scattered electron is known to a precision of $0.5\%$ in the backward and central regions of the detector, and to
    $1\%$  precision in the forward region of the detector~\cite{H1:2012qti}. The energy of the scattered electron is varied by the corresponding factors and passed through the unfolding procedure to determine the resulting uncertainty, which is at maximum $\sim3\%$.

    \item \textbf{$\theta_{HFS}$, $\theta_{e}$ Uncertainty:} The polar angle alignment of the tracking detectors to the liquid argon calorimeter is known to a precision of 1 mrad~\cite{H1:2012qti}. This produces an uncertainty on all objects in the hadronic final state as well as the scattered electron. The HFS and electron polar angle uncertainties are considered separately, and each is passed through the unfolding procedure. Both are typically $\sim 1\%$.
    \item \textbf{Uncertainty on Radiative Correction:} The uncertainty on the QED correction factor for a given bin is determined as the bin-by-bin difference between the factor $c_{QED}$ produced by Rapgap and the $c_{QED}$ produced by Djangoh. The magnitude of this difference is typically 1\%.
  \end{compactitem}
The uncertainties on the 1D groomed \tb are presented in Fig.~\ref{fig:SysUnc1DGrTau}. The uncertainties are fairly flat as a function of the value of \tb and on the order of 5-10\%. For a few bins in the near-peak region of the $z_{cut}$=0.2 distribution, the unfolding exhibited a fairly large dependence on the regularization parameter, and the resulting uncertainty is somewhat larger. For the GIM, the maximum uncertainty resides in the low-mass region of the $z_{cut}$ = 0.05 distribution, which has fairly low event statistics. For the 2D distributions, the uncertainties are somewhat larger, as is expected. The uncertainties generally grow as a function of $Q^2$, as a result of the unfolding operating on regions with lower event statistics. The groomed \tb is observed to have higher uncertainties, which are likely also caused by the lower statistics in the tail region, as well as the fact that the normalization factor is $Q^2$ instead of $Q^2_{Min.}$ as in the case of the GIM.\par
It is worth noting that passing the distributions varied in accordance with the detector systematics through the unfolding procedure did not affect the resulting unfolded distributions by an amount substantially larger than the variation in the cross section that they produced. If the unfolding were unstable, a small change in the cross section induced by, for example, the variation on $E_e$, could cause the unfolded cross section to change significantly. Since the above uncertainties are defined as the difference between the nominal and varied \emph{unfolded} cross sections, the smallness of the uncertainties demonstrates that this is not the case. This therefore provides additional confidence in the robustness of the unfolding procedure. \par
The dominance of the unfolding uncertainties is in part attributable to the fairly large differences between the migration matrices produced by Djangoh and those produced by Rapgap. For future measurements seeking to attain high precision, such as those to be performed at the EIC, it should be checked in advance that the generators give a very accurate description of the data at detector-level. The MCs used for this analysis were reweighted in the past to improve the description of jet cross sections. Using the same reweighted MCs for this analysis, the detector-level descriptions of the groomed \tb and GIM were at worst around 20\%. additional reweighting of the MCs could be used to improve the situation somewhat, however, if underlying physics is missing, no amount of reweighting will be able to describe the data universally. For this reason, additional improvement of MCEGs for DIS over the current state-of-the-art is becoming a pressing issue for the upcoming EIC.
\clearpage

\clearpage
\clearpage
  
\section{Results}
\label{Sec:Results}
    \textbf{Authors Note, July 2024: The results provided in this thesis are superseded by the H1 publication ``Measurement of groomed event shape observables in deep-inelastic electron-proton scattering at HERA'', published in Eur. Phys. J. C \textbf{84}, no.7, 718 (2024).}
  \subsection{Comparison to MC Event Generators}
  The data are compared to a variety of MC event generators, described briefly below:
\begin{itemize}
  \item Predictions from Djangoh~1.4 and Rapgap~3.1 similarly to
    those described above, but where purely photonic first order
    corrections are not used in Heracles.
  \item The MC event generator Pythia~8.307~\cite{Sjostrand:2014zea,Pythia83} is
    used, where three different parton-shower models are studied:
    i) the  `default' dipole-like $p_\perp$-ordered shower,
    ii) the $p_\perp$-ordered Vincia parton
    shower based on the antenna formalism
    at leading color~\cite{Giele:2007di,Giele:2011cb,Giele:2013ema,Fischer:2016vfv},
    and iii) the Dire~\cite{Hoche:2015sya,Hoche:2017iem,Hoche:2017hno} parton
    shower which is an improved dipole-shower with additional handling
    of collinear enhancements.
    All models use the Pythia~8.3 default for
    hadronisation~\cite{Pythia83}.
    The Vincia and Dire parton shower use a value of 0.118 for the
    strong coupling at the mass of the $Z$ boson.
  \item The MC event generator Herwig~7.2~\cite{Bellm:2015jjp} is
    studied in three variants.
    For the default prediction,  Herwig~7.2 implements leading-order
    matrix elements that are supplemented with an angular-ordered 
    parton shower~\cite{Gieseke:2003rz} and the cluster hadronization
    model~\cite{Webber:1983if,Marchesini:1991ch}.
    The second variant makes use of the MC@NLO method that implements NLO
    matrix element corrections. In addition, a matching with the default
    angular-ordered parton shower is performed~\cite{Platzer:2011bc}.
    The third variant makes also use of NLO matrix elements, but applies the
    dipole merging technique and a dipole parton shower~\cite{Platzer:2011bc}.
    The events generated with Herwig are further processed with Rivet~\cite{Bierlich:2019rhm}.
  \item Predictions are obtained with
    Sherpa~2~\cite{Sherpa:2019gpd,Gleisberg:2008ta}, where multi-leg
    tree-level matrix elements from Comix~\cite{Duhr:2006iq} are
    combined in the CKKW merging formalism~\cite{Catani:2001cc} with
    dipole showers~\cite{Catani:1996vz,Schumann:2007mg} and
    supplemented with the  cluster hadronization as implemented in
    AHADIC++~\cite{Winter:2003tt}.
    As an alternative prediction, the parton-level is supplemented
    with the Lund string fragmentation model~\cite{Sjostrand:2006za}.
    \end{itemize}

\begin{figure}[!htb]
  \centering
   \begin{overpic}[trim={0.85cm 0 1cm 0},clip,scale=0.45]{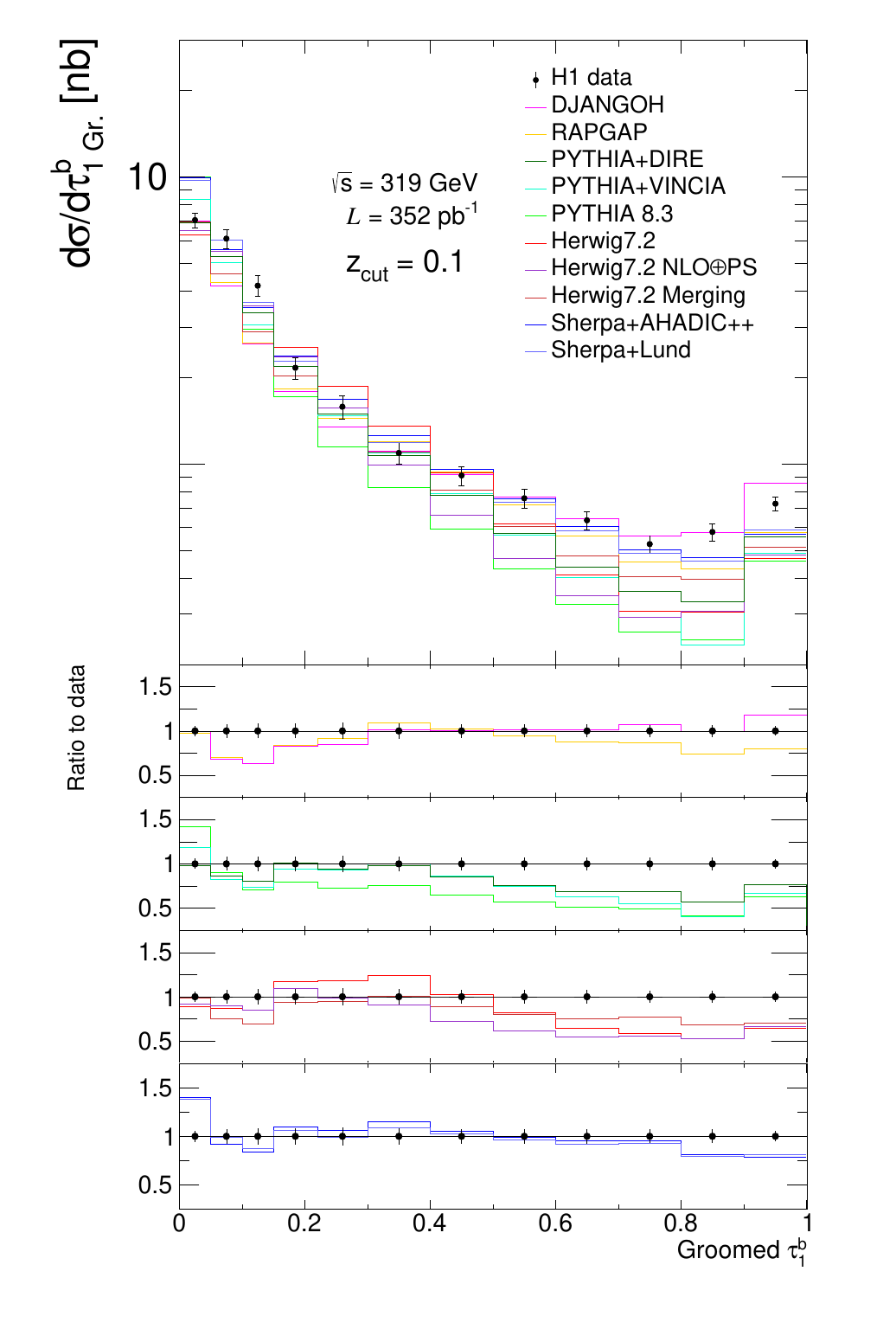}
     \put(60,0){\includegraphics[trim={3.4cm 0 1cm 0},clip,scale=0.45]{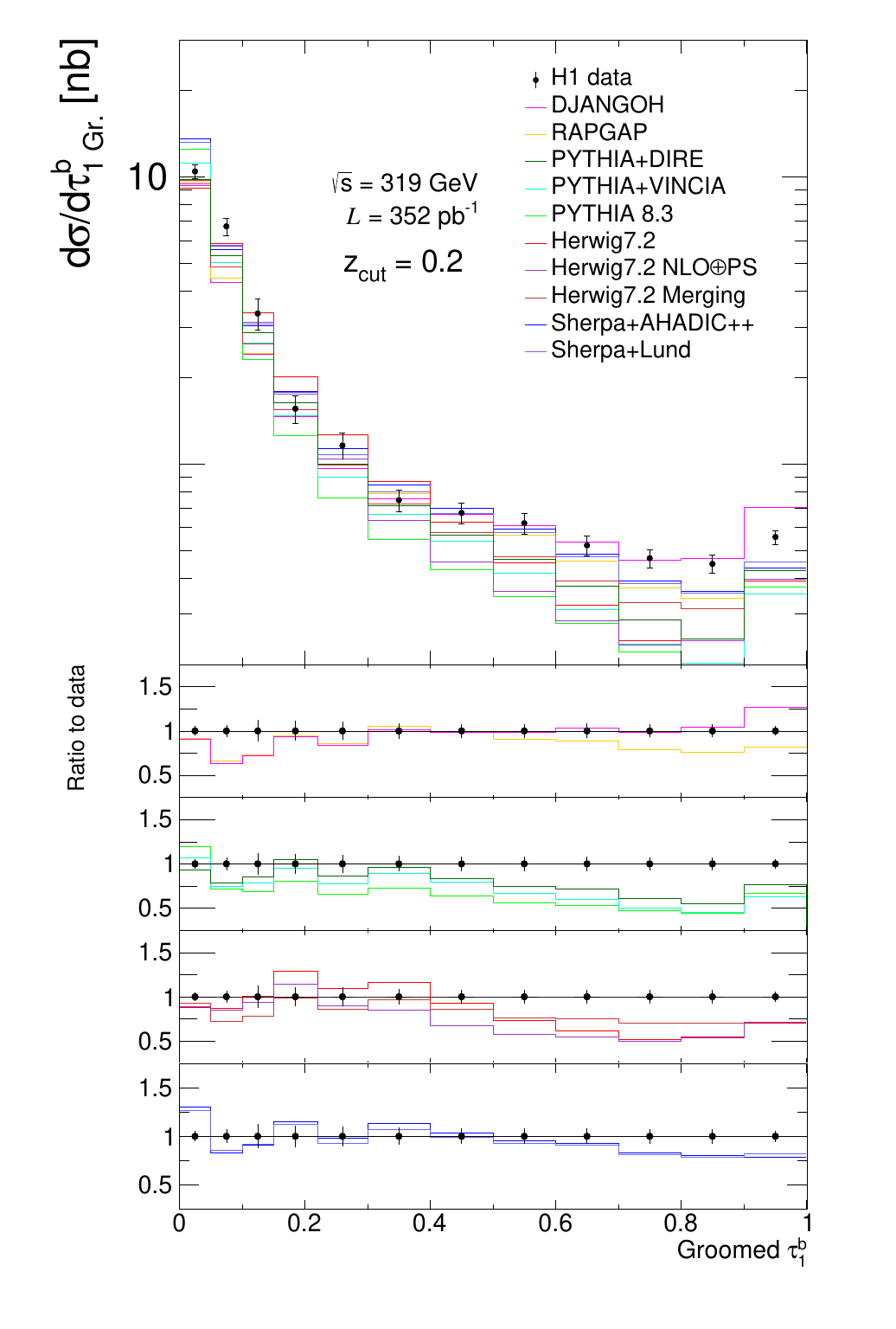}}
     \put(-50,0){\includegraphics[trim={1cm 0 1cm 0},clip,scale=0.45]{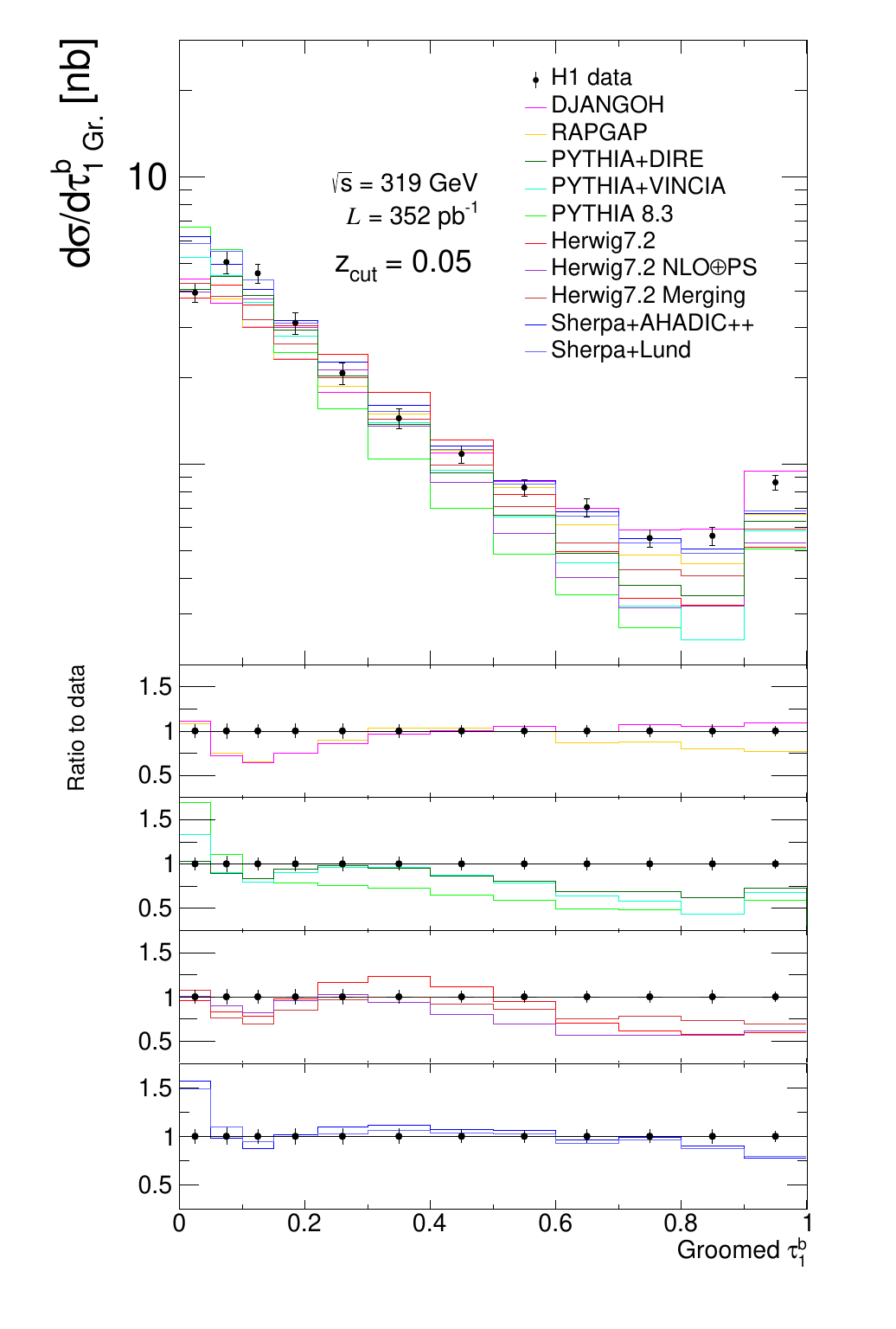}}
  \end{overpic}
\caption{Single-differential groomed \tb at three values of $z_{cut}$, compared to predictions from MC event generators.}
\label{fig:MC1DGrTaus}
\end{figure}
\begin{figure}[!htb]
  \centering
   \begin{overpic}[trim={0.9cm 0 1cm 0},clip,scale=0.45]{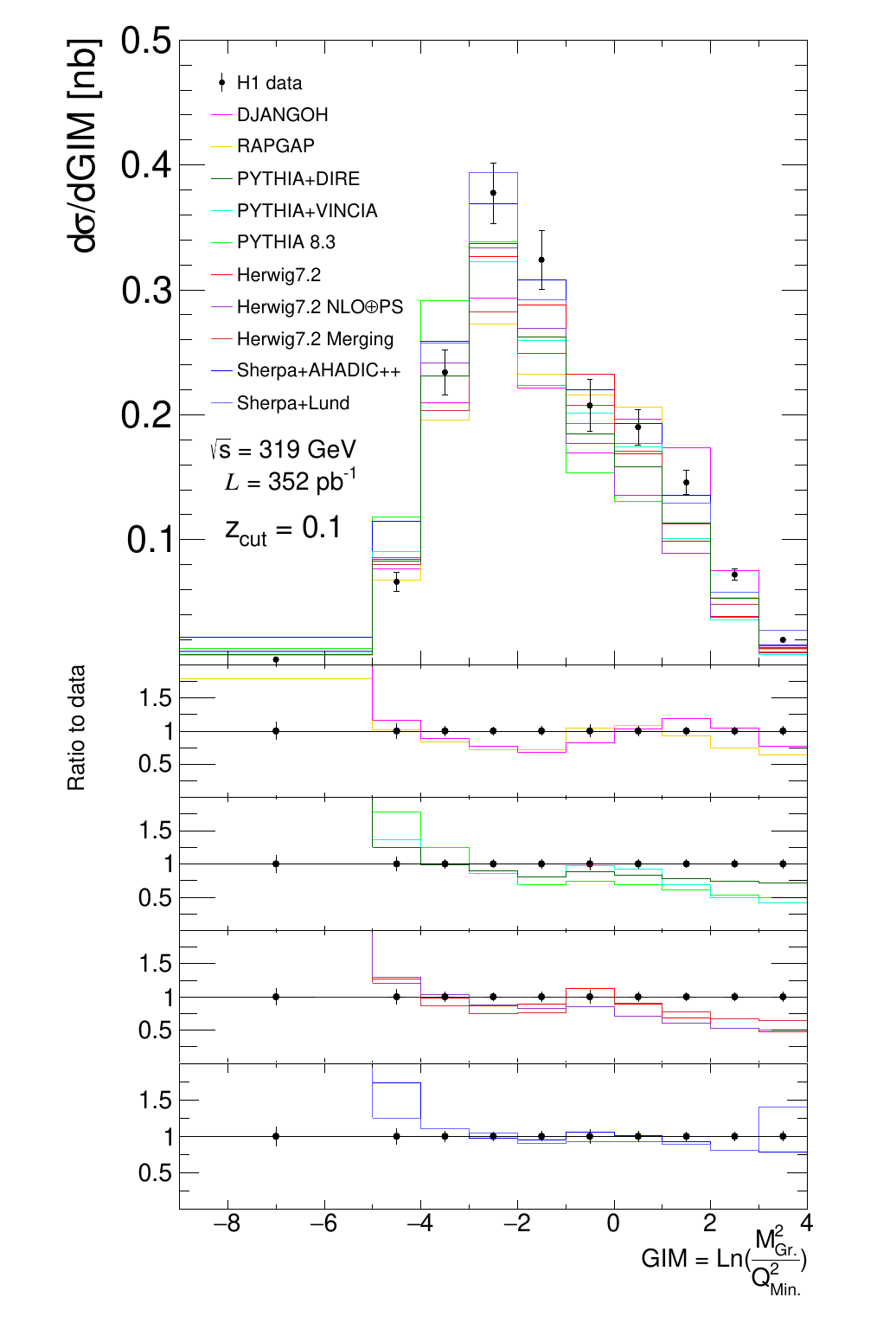}
     \put(60,0){\includegraphics[trim={3.5cm 0 1cm 0},clip,scale=0.45]{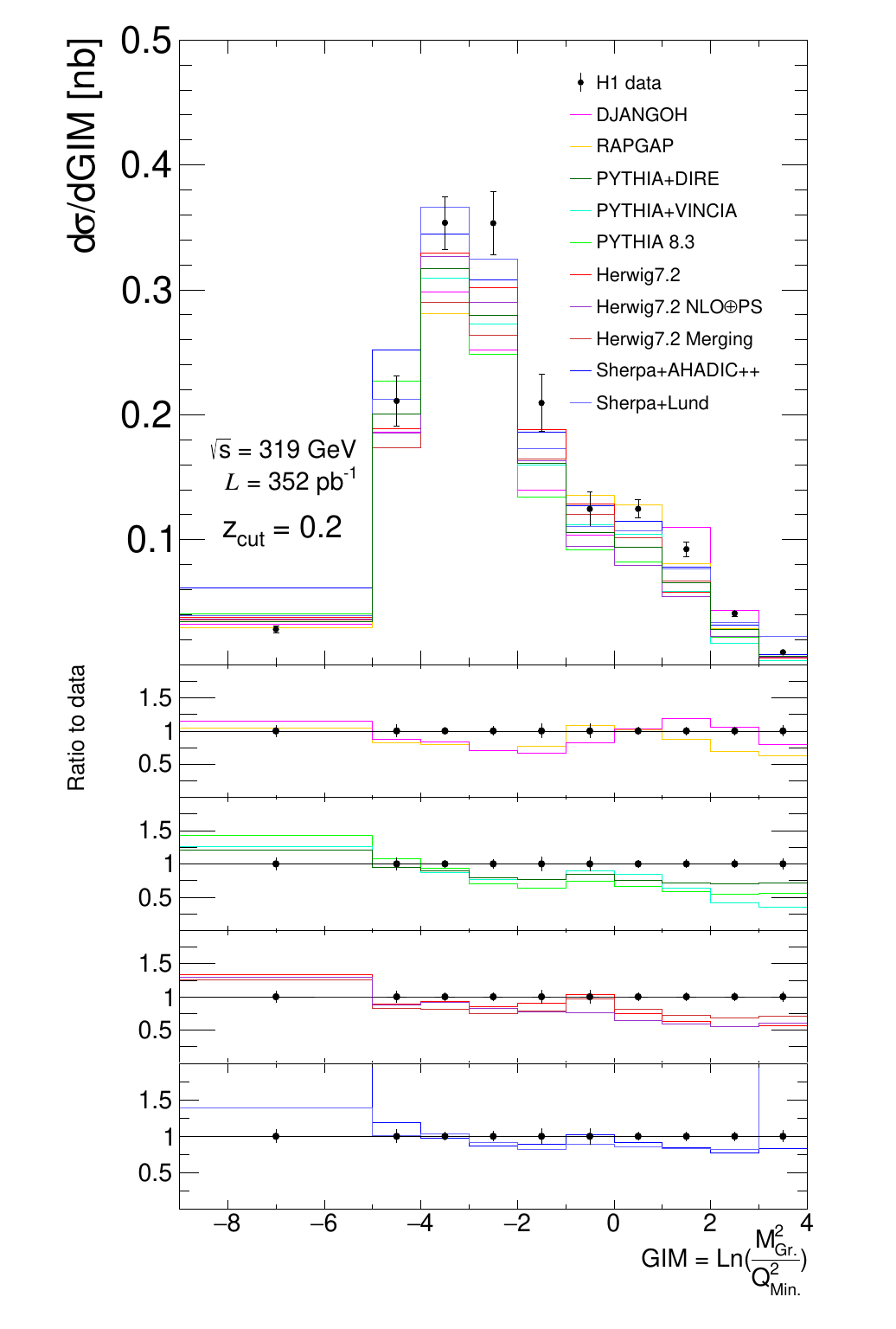}}
     \put(-50,0){\includegraphics[trim={1cm 0 1cm 0},clip,scale=0.45]{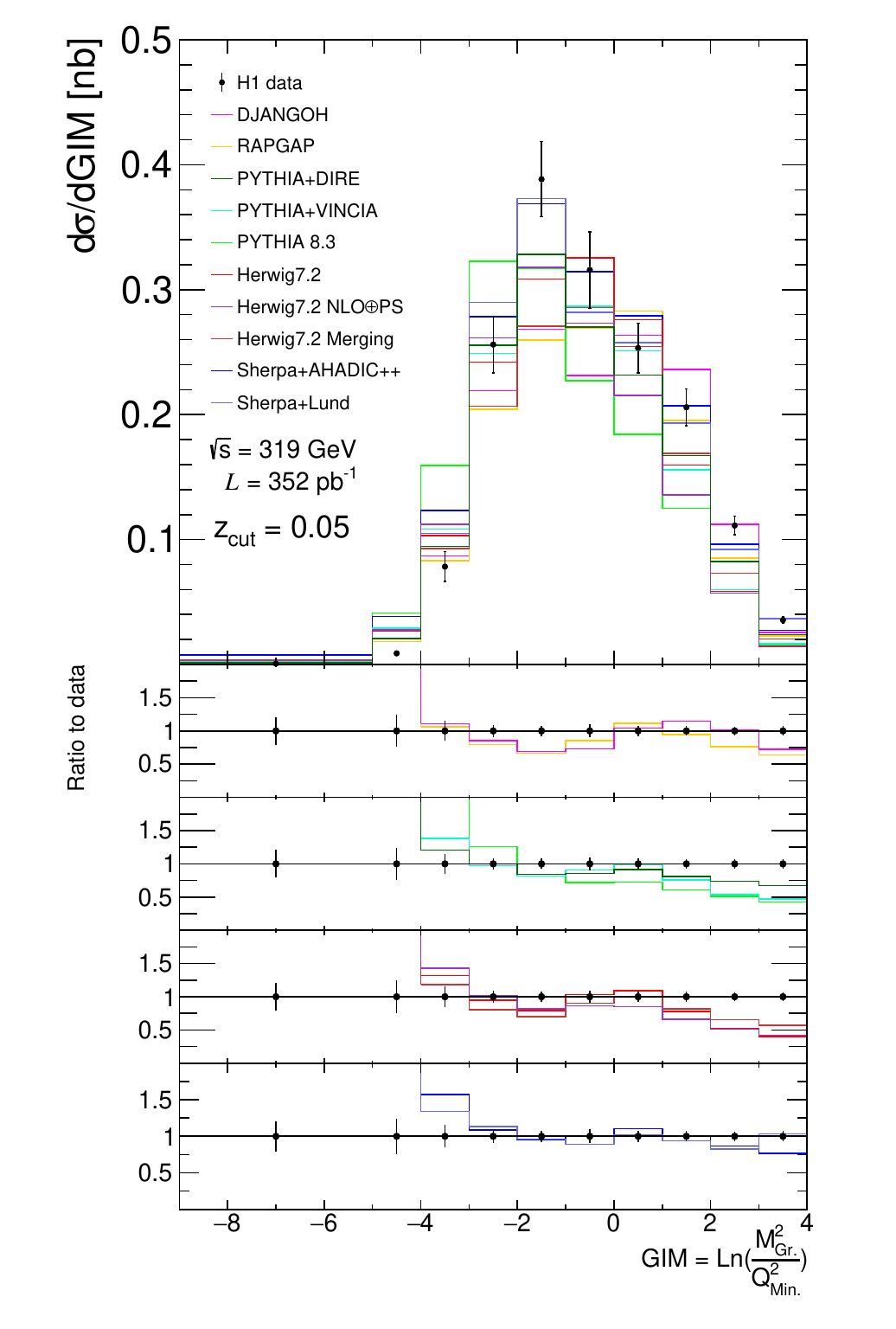}}
  \end{overpic}
\caption{Single-differential groomed invariant mass at three values of $z_{cut}$, compared to predictions from MC event generators.}
\label{fig:MC1DGIMs}
\end{figure}
In the comparison of the data to predictions, it is useful to have a heuristic set of nomenclature for describing the different regions of the observables. The ``tail" or ``fixed-order" region of the GIM and groomed \tb corresponds to large values of the observables, and roughly $\tb > 0.4$ and GIM $> 0$. The ``near peak" region corresponds to $0.2 < \tb < 0.4$, and GIM values between 0 and -2. The ``peak" region will be considered to be $\tb<0.2$ and GIM $<-2$. Roughly speaking, the tail region corresponds to multi-jet events with hard emissions, the near peak corresponds to fairly collimated dijets or broad single jets, and the peak region corresponds to collimated single jets. The GIM is slightly more sensitive to the collimated event topology, in the sense that the first few bins of the GIM distributions correspond roughly to the first bin of the \tb distribution. \par
A few features stand out from the comparison of the data with the MCEGs. The first is the apparent inability of the generators to describe the large values of the GIM and groomed \tb. The default versions of Herwig and Pythia are both run at leading order, and thus this failing is not totally surprising. Sherpa incorporates higher order hard emissions, and thus does a better job at describing the fixed-order region, although the prediction still lies outside the error bars of the data. Djangoh provides perhaps the best description of the tail region, and Rapgap the best of the low-mass region, however the parameters of Djangoh and Rapgap have been tuned to give a reasonable description of the HERA data. The Herwig angular-ordered parton shower appears to describe the low \tb, low GIM region slightly better than the dipole-shower predictions, with the exception of Pythia+DIRE which makes a very similar prediction.\par
Another observation is the overall trend of upward slope in the ratio plots at small \tb and small GIM. This is a result of the data in these regions lying at slightly higher values than the predictions. For the groomed \tb, which compares the HFS explicitly to the Breit frame axis $q_J$, a very interesting possible source of this discrepancy arises from the intrinsic $k_T$ of the struck parton, which is neglected in the definition of $q_J$. Intrinsic $k_T$ will tend to deflect the HFS away from $q_J$, and thus smear the \tb to higher values. The effect of intrinsic $k_T$ on the GIM is less obvious, although since the grooming also utilizes the Breit frame and the nominal direction of the struck parton, it is not inconceivable that there would be a similar smearing to larger masses induced by the grooming preferring to stop earlier in the data. Intrinsic $k_T$ is not present in the predictions from Sherpa 2, and its inclusion in Sherpa 3\footnote{The predictions from Sherpa3 at this point have been received, but the comparison to the data on a plot will be reserved for the H1 publication.} (not shown), along with an improved cluster hadronization model, improves the agreement in the peak region. \par
Another possibility is simply that the Born-level jets at HERA are wider than anticipated, possibly since these MCs have primarily been tuned to LEP and LHC data at higher energies. The fact that the ratio plots for both the GIM and the \tb exhibit similar behavior in the near peak region suggests that this is a more likely scenario than the instrinsic $k_T$ one mentioned above, where the GIM would not be expected to show a substantial deviation, since it does not compare directly to the $q_J$ axis. That the description of this region tends to improve at higher $z_{cut}$ suggests that some aspect of soft radiation is being described poorly, and that the removal of this radiation is what improves agreement. \par
Another observation is that the normalization of the data compared to the MCs is slightly off. Since the presented results are absolute cross sections, the MCs sitting below the data mean that the total cross section for high $Q^2$ DIS is underestimated. Herwig and Pythia underestimate, while Sherpa gives a much better normalization but slightly overestimates. The PDF used in the MC will affect the total DIS cross section. This was studied briefly using default Pythia 8.3, and in general for these kinematics, higher order PDFs produce higher cross sections. All NNLO PDFs tend to give very similar results for the total cross section, while LO PDFs tend to undershoot. In Pythia, the effect of changing the PDF appears mainly in the peak and near-peak regions. Rapgap and Djangoh utilize the CTEQ6L LO PDF, which likely results in the significantly underestimated peak and near-peak regions. Another effect which can change the normalization is the number of events which are groomed all the way down the grooming tree and never pass the grooming procedure. This number of ``failed" events grows as a function of $z_{cut}$. This effect was studied in all the MCs presented here as well as the data, and for the harshest grooming the differences were on the order of a few percent. The events which fail grooming tend to be those which were generated with small values of \tb or GIM This effect is not large enough to explain the normalization issue.
 \subsection{Comparison to SCET Predictions}
\begin{figure}[!htb]
  \centering
   \begin{overpic}[trim={1.1cm 0 1cm 0},clip,scale=0.4]{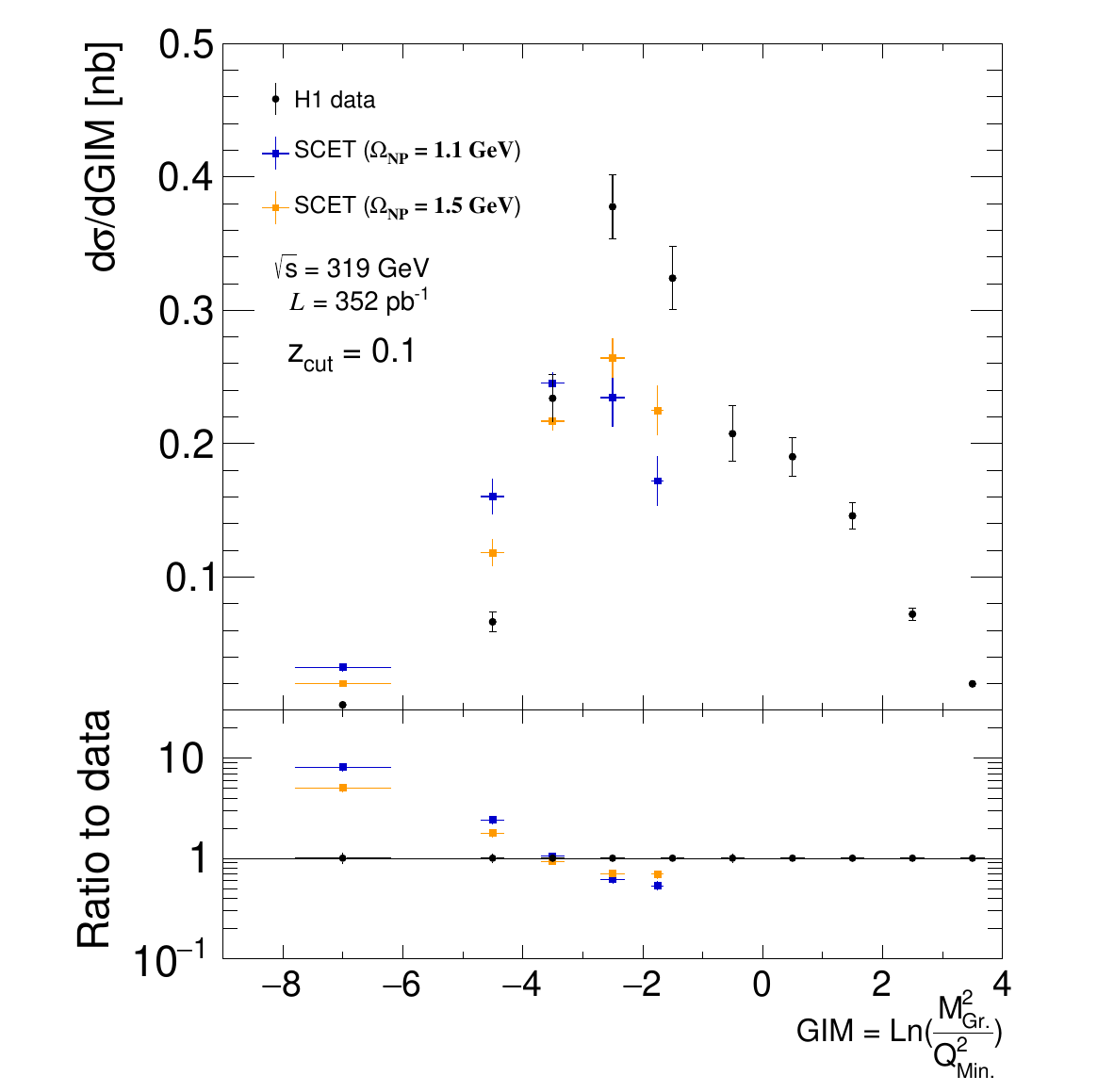}
     \put(89,0){\includegraphics[trim={3.9cm 0 1cm 0},clip,scale=0.4]{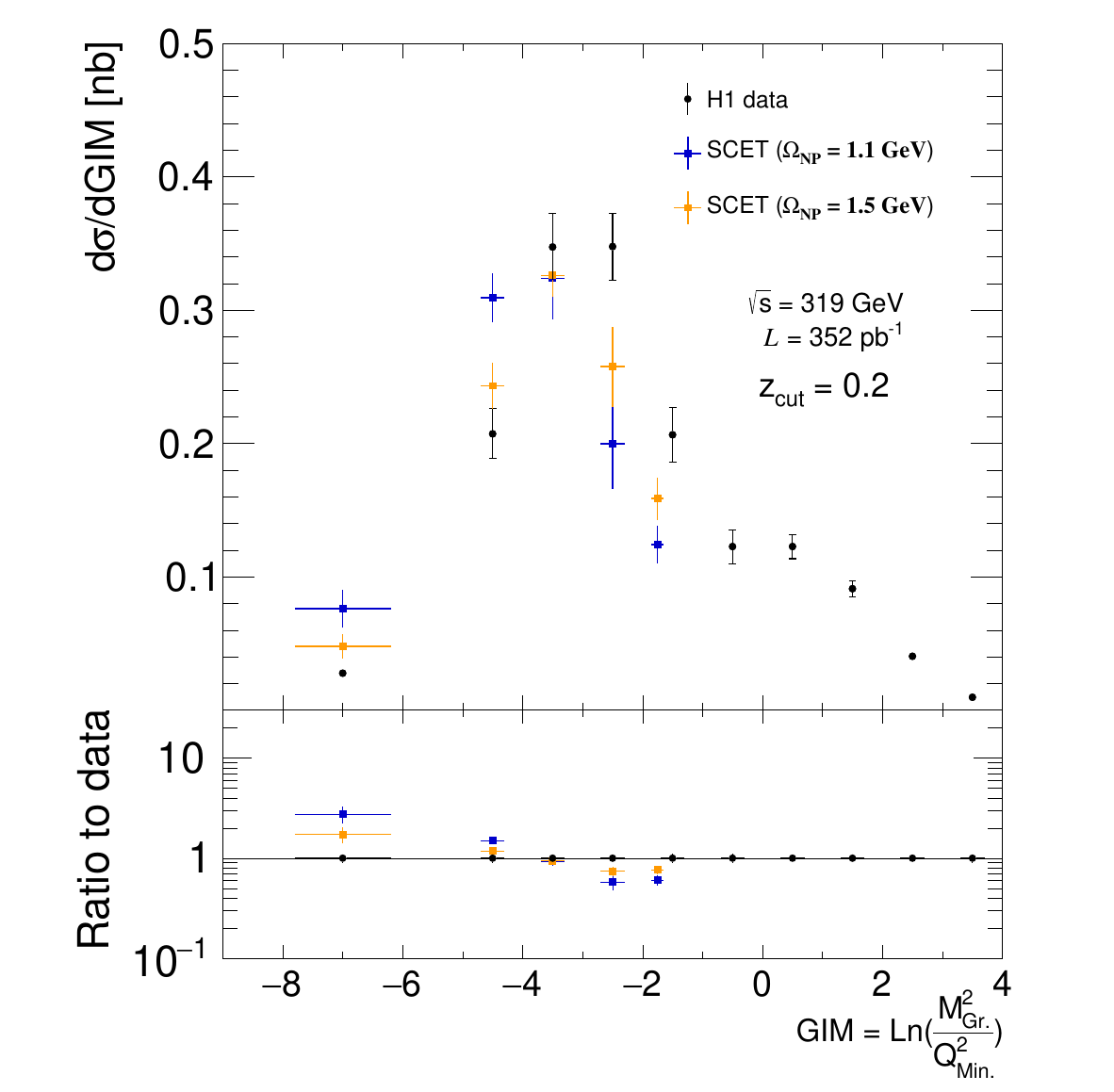}}
     \put(-75,0){\includegraphics[trim={1cm 0 1.5cm 0},clip,scale=0.4]{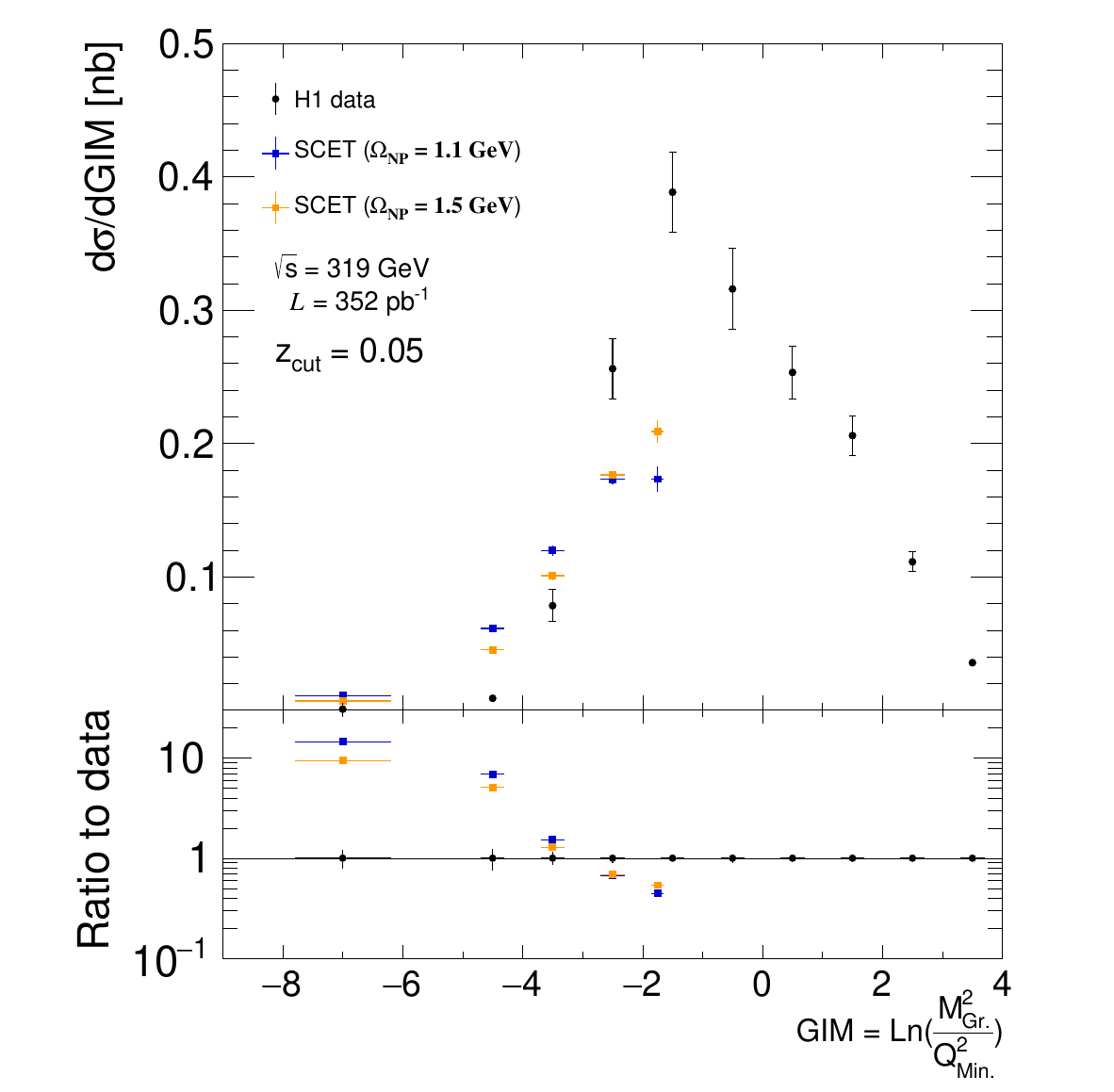}}
  \end{overpic}
\caption{Single-differential groomed invariant mass at three values of $z_{cut}$, compared to predictions from SCET.}
\label{fig:SCET1DGIMs}
\end{figure}

 Explicit analytic predictions from Ref.~\cite{Makris:2021drz} are compared to the data. The predictions are provided at two values of the mean of the shape function, $\Omega_{NP} = 1.1$ GeV and 1.5 GeV. The predictions are provided only for the low mass region, i.e. Ln($m^2_{gr.}/Q^2_{min.}$) $<$ -1.5 = $m_{gr.} > \sim 6$ GeV, and are normalized to the data in that region. The uncertainty on the SCET prediction is produced by varying all scales in the prediction by a factor of 2. The comparison to the data is shown in Fig.~\ref{fig:SCET1DGIMs}.
There exists a clear discrepancy between the prediction and the data. The best agreement is achieved for the $z_{cut} = 0.2$ case with the larger value of $\Omega_{NP}$, although even in this case the prediction disagrees with the data by a few standard deviations. It appears that the data prefer a larger value of $\Omega_{NP}$, and thus a larger impact of the shape function. Interestingly, both the SCET and the MC predictions tend to predict lower masses for the single-jet region. The SCET prediction improves substantially as a function of $z_{cut}$, which is naively expected by virtue of the factorization used (Eq.~\ref{eq:Factorization}), which expects that $z_{cut}$ is large or similar in magnitude with respect to $m^2_{gr.}/Q^2$. For the $z_{cut}=0.05$ case, this expectation is generally not met, and the factorization breaks down. It is therefore likely that at higher $Q^2$, a better agreement between the prediction and data could be achieved.\par
An alternative, although in some sense equivalent\footnote{The prior explanation is a formal one, and the following is a more physical one. The physical manifestation of the breakdown of the factorization scheme likely arises for the reasons listed here.}, explanation comes from the use of the quark jet function in the SCET calculation. The implicit assumption of the calculation is that at low masses, only quark jets contribute. It is possible that even the low mass region of the data contains a substantial admixture of gluon jets, which would tend to produce higher invariant mass final states. In the kinematic phase space considered here, a large fraction of events ($>$30\%) are ``gluon initiated", i.e. a gluon is the parton initially struck out of the proton. This is a natural consequence of the $x$ range, which is roughly $0.001 < x < 0.1$. Additional sources of gluons could be fixed-order emissions, or even fairly hard emissions within parton showers. It is possible that due to the grooming, the fixed-order correction to the cross section contributes substantially even at low masses by mixing gluons into the final state. The inclusion of the fixed-order ingredients provided in the appendix of Ref.~\cite{Makris:2021drz} in the prediction would therefore be very interesting. 

\subsection[Q Independence of Soft Radiation]{$Q^2$ Independence of Soft Radiation}
A prediction is made in Ref.~\cite{Makris:2021drz} that the shape of the groomed invariant mass distribution should be independent of $Q^2$ in the low mass limit, defined by the relation
\begin{equation}
1 \gg z_{cut} \gg \frac{M^2_{Gr.}}{Q^2}
\label{eq:LowMassLimit}
\end{equation}
In this region, the $x$ and $Q^2$ dependence of the cross section is isolated in the portion of the event which has been groomed away, as described in section~\ref{Sec:Obs}. The mass distributions are presented differentially in $Q^2$, normalized to the low-mass region. There are a variety of experimental difficulties that limit the ability to access very small values of $\frac{M^2_{Gr.}}{Q^2}$. These values are sensitive to very small angles, which are difficult to measure in the Breit frame due to the sensitivity to the boost. Additionally, hadron masses begin to play a role in this region. Since H1 lacks hadron PID at high momentum, the particles very close to the boson-going direction are typically not identified. This introduces an unavoidable degradation in the GIM resolution at low values of the GIM. Therefore, in order to reasonably satisfy Eq.\ref{eq:LowMassLimit}, the results are only presented for $z_{cut} = 0.2$. The factor $Q^2_{min.}$ is taken to be the lowest value of $Q^2$ considered in the bin, i.e. 150 GeV$^2$ in the first $Q^2$ bin, 200 GeV$^2$ in the second, etc. 
\begin{figure}[h!]
\begin{center}
\includegraphics[width=13.9cm]{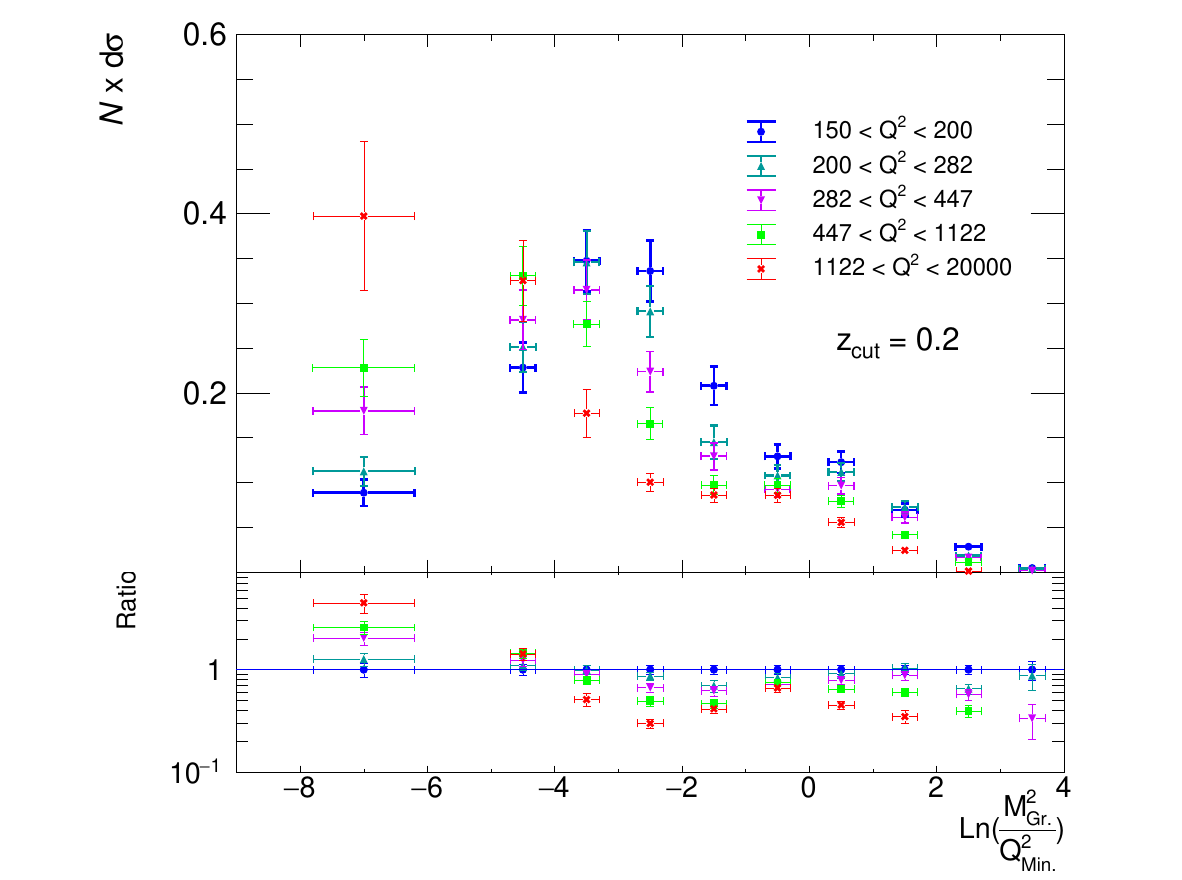}
\caption{Shapes of groomed mass distributions in various $Q^2$ bins, normalized to the region of Ln($\GIM$) $< -2$, i.e. the first four bins of the histogram. The ratio plot in the lower panel shows the ratio of the higher $Q^2$ bins to the $150<Q^2<200$ GeV$^2$ bin. To facilitate comparison of the normalized distributions, the values in each bin have not been divided by the bin width.}
 \label{fig:UniversalityGIM2}
\end{center}
\end{figure}
The shape of the GIM distribution is observed to change dramatically at higher $Q^2$, defying the SCET prediction.
In none of the $Q^2$ bins studied here do the distributions fall on top of each other. A scenario could be imagined where, for example, the $150<Q^2<200$ GeV$^2$ bin exhibits a different shape, but the higher values converge to a universal curve due to Eq.~\ref{eq:LowMassLimit} finally being satisfied. The data seem to suggest that this is not the case, as even the two highest $Q^2$ bins exhibit qualitative disagreement in their shape. A similar interpretation to the one offered above could be considered, that gluon jets populating the low-mass region never disappear, even at high $Q^2$. Going to higher $Q^2$ will change the admixture of gluon jets, but this effect induces a $Q^2$-dependence to the shape of the distribution at all $Q^2$. This could be formally interpreted as fixed-order terms proportional to $Q^2$ contributing everywhere in the GIM distribution. 

\subsection{Double-differential Groomed Event Shape Distributions}
\begin{figure}[h!]
\begin{center}
\includegraphics[trim={0 1cm 0 0},clip,width=13.9cm]{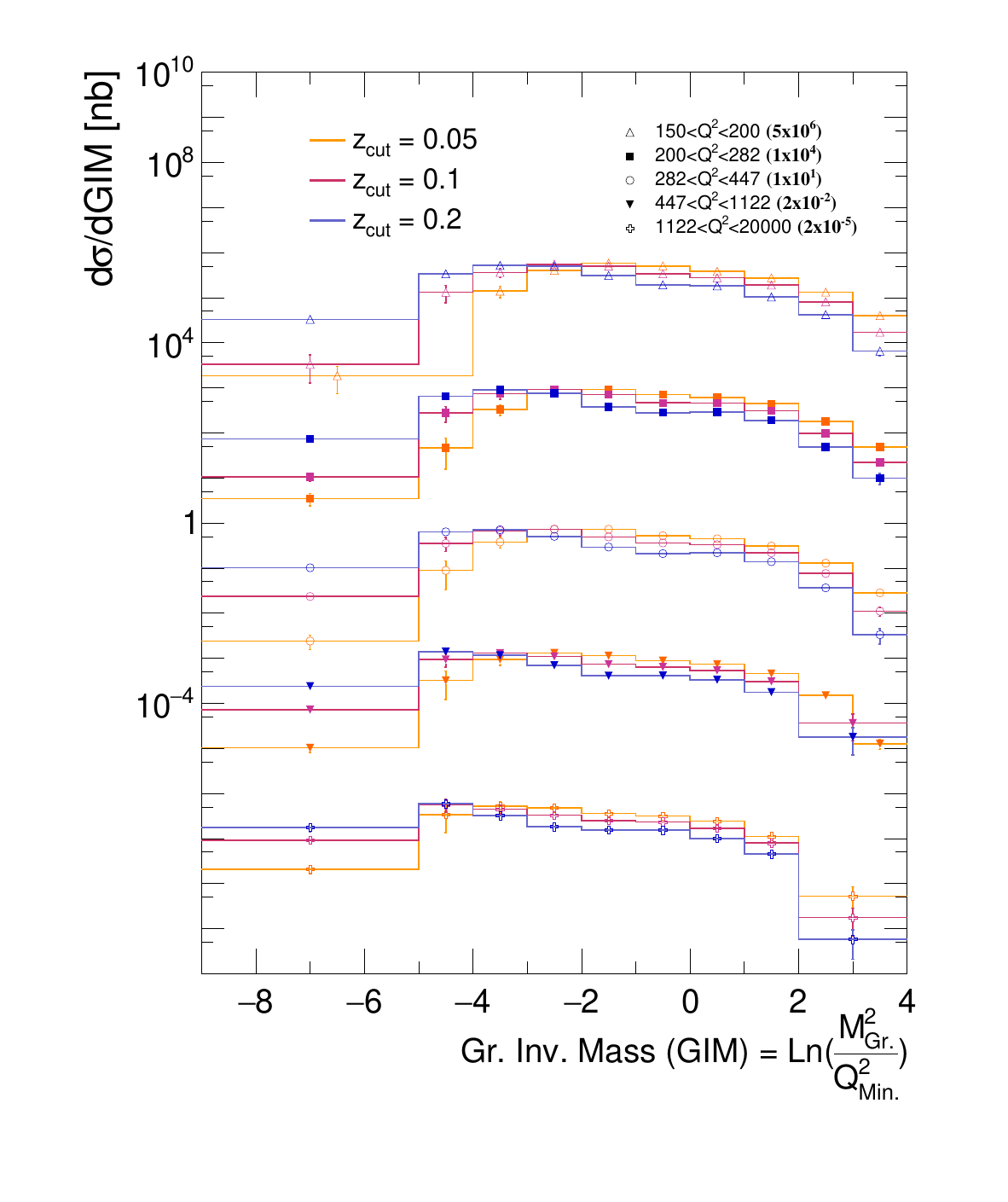}
\caption{Groomed mass at $z_{cut} = 0.05$, $z_{cut} = 0.1$, and $z_{cut} = 0.2$ for five bins in $Q^2$. The trend as a function of $Q^2$ is for the distributions to shift to lower values of the GIM.}
 \label{fig:2DGIM}
\end{center}
\end{figure}

\begin{figure}[h!]
\begin{center}
\includegraphics[trim={0 1cm 0 0},clip,width=13.9cm]{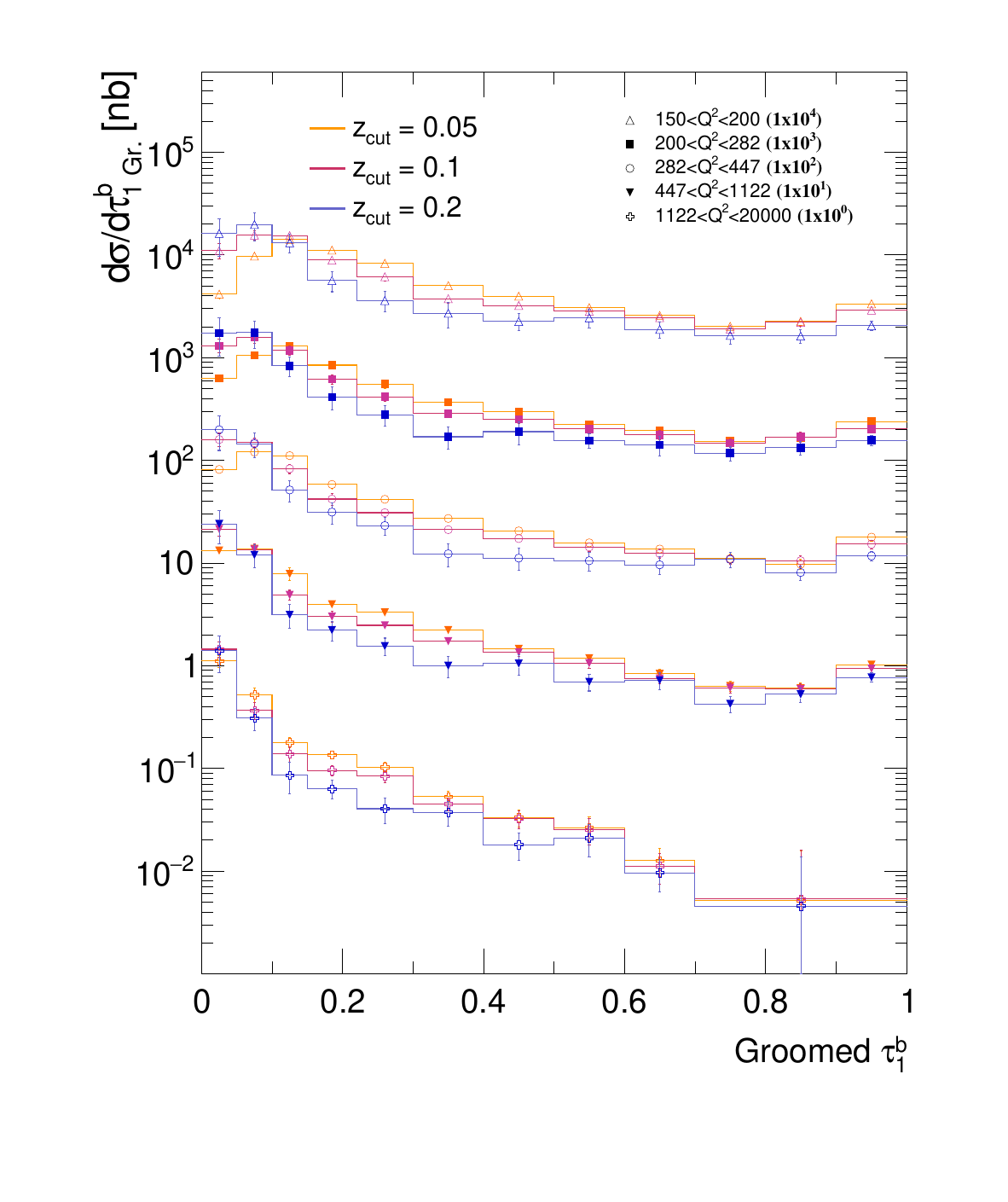}
\caption{Groomed 1-jettiness at $z_{cut} = 0.05$, $z_{cut} = 0.1$, and $z_{cut} = 0.2$ for five bins in $Q^2$. The trend as a function of $Q^2$ is for the distributions to shift to lower values of groomed \tb.}
 \label{fig:2DGrTau}
\end{center}
\end{figure}

For completeness, the measured double-differential cross sections at the three values of $z_{cut}$ and as a function of $Q^2$ are included here in Figs.~\ref{fig:2DGIM} and~\ref{fig:2DGrTau}. The distributions clearly exhibit a $Q^2$ dependence, as expected. The \tb peaks more sharply at higher $Q^2$, and the GIM shifts to lower values. For this reason, the statistics in some bins become too low to perform a reliable measurement, and thus they are combined with neighboring bins. The bins are reduced for the low $Q^2$, low GIM region with $z_{cut}$ = 0.05, where the GIM tends to be large and few events have low masses. Additionally, the binning is reduced in the high $Q^2$, high \tb and high GIM regions, as the combined effects of the distributions shifting to lower values and the lower total cross sections substantially deplete the event statistics. The table of reduced binnings is provided in Table~\ref{tab:binnings}.\par
It is worth noting that the normalization factors also grow as a function of $Q^2$, which is partially responsible for the shifts to lower values. The $Q^2$ dependence of the un-normalized values of $m_{gr.}^2$ and \tb is thus not particularly strong. The grooming likely serves to reduce some of the $Q^2$-dependence of the event shape cross sections. These results can in the future be used for comparison to additional SCET or MC predictions. It would be interesting to use the double-differential data to validate or invalidate the hypothesized sources of discrepancy between the predictions and the data. 
\clearpage

\section{Discussion}
\label{Sec:Discussion}
None of the models studied here accurately reproduce the data in all regions. In particular, three regions of deficiency are highlighted, although they may of course be coupled in various ways. First, the tails of the GIM and the groomed \tb distributions are poorly described by most of the MC models. This highlights the importance of having higher order processes implemented. The tail of the distributions is sensitive not only to the cross section for dijet or trijet production, but also the size of the jets and their angles with respect to each other. In this way, these event shape observables are complementary to standard jet-based measurements. PYTHIA and Herwig in particular underestimate the high values of the event shapes studied here by 25-50\% depending on the tune, suggesting the use of higher order matrix elements is a necessity. There exist also issues with the MC normalizations, which can in part be explained similarly. If processes which occur in the data are missing in the generators, the predicted cross section will tend to be low. The SCET predictions also fail to describe the data. In that case, the key takeaway is that the assumption that after grooming jets in DIS are purely quark jets describable in terms of a quark jet function with dressing of soft-collinear radiation is false, even at high $Q^2$ beyond that which will be accessible at the EIC. Similarly to the MCs, a rigorous treatment of fixed-order physics appears to be vital to accurately describe the data. \par
For the EIC, which is intended to do precision studies of QCD, it will be vital to have MC generators which accurately describe all aspects of DIS events. To first order, a correction for detector effects should not depend on the precise description of the data by the MC which is propagated through the detector model. However, if an event generator distributes particles in a way that varies significantly from the data, the derived detector correction will be deficient. In the analysis presented above, by far the dominant systematic uncertainty arose from the unfolding procedure. This is in large part a result of disagreement in the particle level of the migration matrices of the two MCs available at detector level. In the majority of bins of the measurement, the systematic uncertainties resulting from the detector performance are almost negligible. This highlights additionally the necessity of having more than one MC at detector-level, as it enables the assessment of the systematic uncertainty arising from the detector correction, which can be large. Enormous effort will be placed into the design and construction of a precise detector for the EIC. Based on the results of this analysis, a similar degree of effort must be given to producing precise and accurate DIS MC generators that can be used both to perform detector corrections and aid in the interpretation of the physics results of the EIC.
\chapter{sPHENIX}
\label{Chap:sPHENIX}
\section{History of the Quark-Gluon Plasma}
The possibility of using thermodynamics to predict particle production in high-energy hadron collisions was first realized in the mid 1960s~\cite{Rafelski:2016hnq}. At that time, prior to the pioneering deep inelastic scattering experiments at SLAC that provided the first experimental evidence for the partonic nature of hadrons, the results obtained in high-energy fixed target collisions at the BNL AGS and CERN PS accelerators were fascinating the theory community. The plethora of experimental results were challenging to interpret due to the poorly understood initial state. What was observed in the most violent of these collisions was the production of a large number of particles with a roughly exponential distribution in transverse momentum. Additionally, the large angle elastic scattering cross section dropped precipitously with beam energy. Interleaved with these discoveries was the observation that the mass spacings between hadronic resonances decreased as their masses increased, suggesting the existence of a perfectly continuous spectrum at high masses. This led to the adoption and further study of the so-called ``fireball" model of high-energy hadronic collisions that had been theorized previously to describe multi-particle production in cosmic ray collisions. It was realized that in this model, there existed a universal maximum temperature governing hadron production, now known as the Hagedorn temperature, or $T_{H} \approx 160$ MeV. After the discovery of quarks as the constituents of hadrons, it was quickly realized that at very high temperatures hadrons should ``melt" into their fundamental degrees of freedom. The phase of matter produced by melted hadrons became known as the quark-gluon plasma, and the hunt for its existence began soon after. \par
The first relativistic heavy ion experiments studying the collective dynamics of nuclear matter were performed at the Berkeley Radiation Laboratory (now Lawrence Berkeley National Laboratory). By combining the Bevatron, which had a storied history including the discovery of the antiproton, with the Super Heavy Ion Linear Accelerator (SuperHILAC), Berkeley produced the world's first relativistic heavy ion accelerator, the Bevalac, which began operation in 1974~\cite{PhysToday:1974}. The goal of the initial studies at the Bevalac was to determine the nuclear equation of state by studying the particles produced in collisions of light and semi-heavy ions on heavy fixed targets with energies around 1-2 GeV per nucleon~\cite{Stock:2004iim}. The primary motivation was to produce in terrestrial laboratories conditions similar to those in neutron stars or supernovae. Searches for deconfined quark matter were performed at the Bevalac, with the primary signatures being rapid or discontinuous changes in the equation of state as measured in part by the radial flow of produced particles. However, the energies accessible at the Bevalac proved to be far too low to reach the necessary energy density for deconfinement. \par


The AGS began operation in 1960, shortly after the startup of the similar CERN PS in 1959. The AGS initially serviced a variety of particle physics fixed target experiments, yielding in rapid succession the discoveries of the muon neutrino~\cite{PhysRevLett.9.36}, and CP violation~\cite{Christenson:1964fg}. In 1974, the E598 experiment discovered a resonance in 30 GeV proton on fixed beryllium target collisions in the dielectron channel around 3.1 GeV, consistent with a $c\Bar{c}$ bound system~\cite{E598:1974sol}. These discoveries solidified the legacy of the AGS as one of the primary machines responsible for the foundations of modern particle physics. By the 1980s, the particle physics community had largely moved on from the AGS, in light of the higher energies available for fixed target experiments at CERN and Fermilab, as well as colliders like the $Sp\Bar{p}S$ and the Tevatron. However, the 1980s signified the beginning of the heavy ion program at the AGS, spawning a wide variety of experiments attempting to finally discover the QGP. 
\par
The CERN Intersecting Storage Rings (ISR) collided beams of $\alpha$ particles at 31 GeV/nucleon at in the early 1980s, but the low luminosity of the ISR and the inability to collide heavier species meant that it was generally not the machine of choice for plasma hunters. However, the plethora of studies enabled by the collider environment first provided by the ISR offered tantalizing clues that led the heavy ion community to desire a collider for themselves, capable of reaching higher energies and colliding larger nuclei. \par
The CERN SPS accelerator, which in its tenure in collider mode discovered the W and Z bosons, also proved an excellent machine with which to study high density QCD. The heavy ion program began in 1986, and similarly to the AGS could only inject nuclei with equal number of protons and neutrons, thereby limiting the size of the projectile. The typical beam species were oxygen and silicon, which could be accelerated to 200 GeV/nucleon, reaching a resultant $\sqrt{s_{NN}}$ of $\sim20$ GeV in collisions on various nuclear targets. In 1995, the SPS was upgraded to provide beams of heavier nuclei, including Pb$^{208}$ at 160 GeV/nucleon.
\par
 Although some of the results from the AGS and SPS were indicative of the presence of a QGP in certain conditions, the results were not yet conclusive~\cite{Harris:1996zx}. At that time, a variety of signatures of the QGP had been proposed. A reductive list of them is provided below.
\begin{compactitem}
\item An increase in the number of photons produced.
\item The enhancement of strangeness~\cite{Rafelski:1982pu}.
\item A change in slope of the $p_T$ spectrum of produced particles.
\item The suppression of quarkonia~\cite{Matsui:1986dk}.
\item Changes in the masses and decay widths of hadrons.
\item Hadron abundances in accordance with chemical equilibirum.
\item The quenching of jets.
\item Anisotropic flow of produced particles.
\end{compactitem}
Each of the above signatures individually had a variety of possible non-QGP explanations, and some alternative models, such as the hadron-resonance gas, could still be tuned to describe much of the data. Using the newly upgraded heavy ion beam, the SPS experiments jointly claimed a discovery of a deconfined state of matter in 2000, driven largely by results on hadron abundances, strangeness enhancement, and suppression of charmonium states~\cite{Heinz:2000bk}, which could not jointly be explained by models containing only hadronic interactions. The evidence provided by the SPS was in large part in need of verification at higher energies, where the temperature and energy density of the plasma would be high enough to remove ambiguities that complicated the SPS conclusions and access so-called ``hard probes" of the QGP properties. 
\section{The Relativistic Heavy Ion Collider} In 1963, a summer study was commissioned to determine the feasibility of adding storage rings to the AGS. It was determined that the resulting collider could reach beam energies of 100 GeV for protons~\cite{BrookhavenNationalLabUptonNY:1976soe}. However, in lieu of this proposal it was determined that increasing the intensity of the AGS would be a more cost effective improvement to the BNL accelerator complex. In 1970, with the CERN ISR nearing completion, the idea of storage rings at BNL was revived. In May 1974, the design of the Intersecting Storage Accelerating Facility, or ISABELLE, had advanced to the degree that construction funding was allocated to begin in 1976. The ISABELLE design ambitiously leveraged the experience at BNL with superconducting magnets, aiming at a 200 GeV proton beam energy. Construction began in 1978, but technical issues, competition from the 100 times more powerful Superconducting Supercollider design, and the discovery of the W and Z bosons at the CERN S$p\Bar{p}$S led to ISABELLE's cancellation in 1983.\par
The proposal to build the Relativistic Heavy Ion Collider (RHIC) was submitted in 1984~\cite{RHIC1984}, only 6 months after the cancellation of ISABELLE. Utilizing the recently constructed infrastructure built for ISABELLE allowed for a seamless transition towards heavy ion physics. 11 Experiments were initially proposed for RHIC, all of which were rejected and reconfigured into four experiments, STAR,  PHENIX, BRAHMS, and PHOBOS. \par
The RHIC machine was designed to have a max energy for Au+Au running of 100 GeV/nucleon, corresponding to 250 GeV for protons. The design store-averaged luminosity was 2x$10^{26}$cm$^{-2}$s$^{-1}$ for Au+Au at top energy. Additionally, RHIC required the capability to accelerate ions of many species, from light ions like deuterium up to gold. Asymmetric collision arrangements were also deemed as a vital part of RHIC operations. The wide variety of requirements for the heavy ion program meant that RHIC was to be the most versatile accelerator in the world, a goal which it quickly met. \par
\begin{figure}[ht!]
    \centering
    \includegraphics[width=13.5cm]{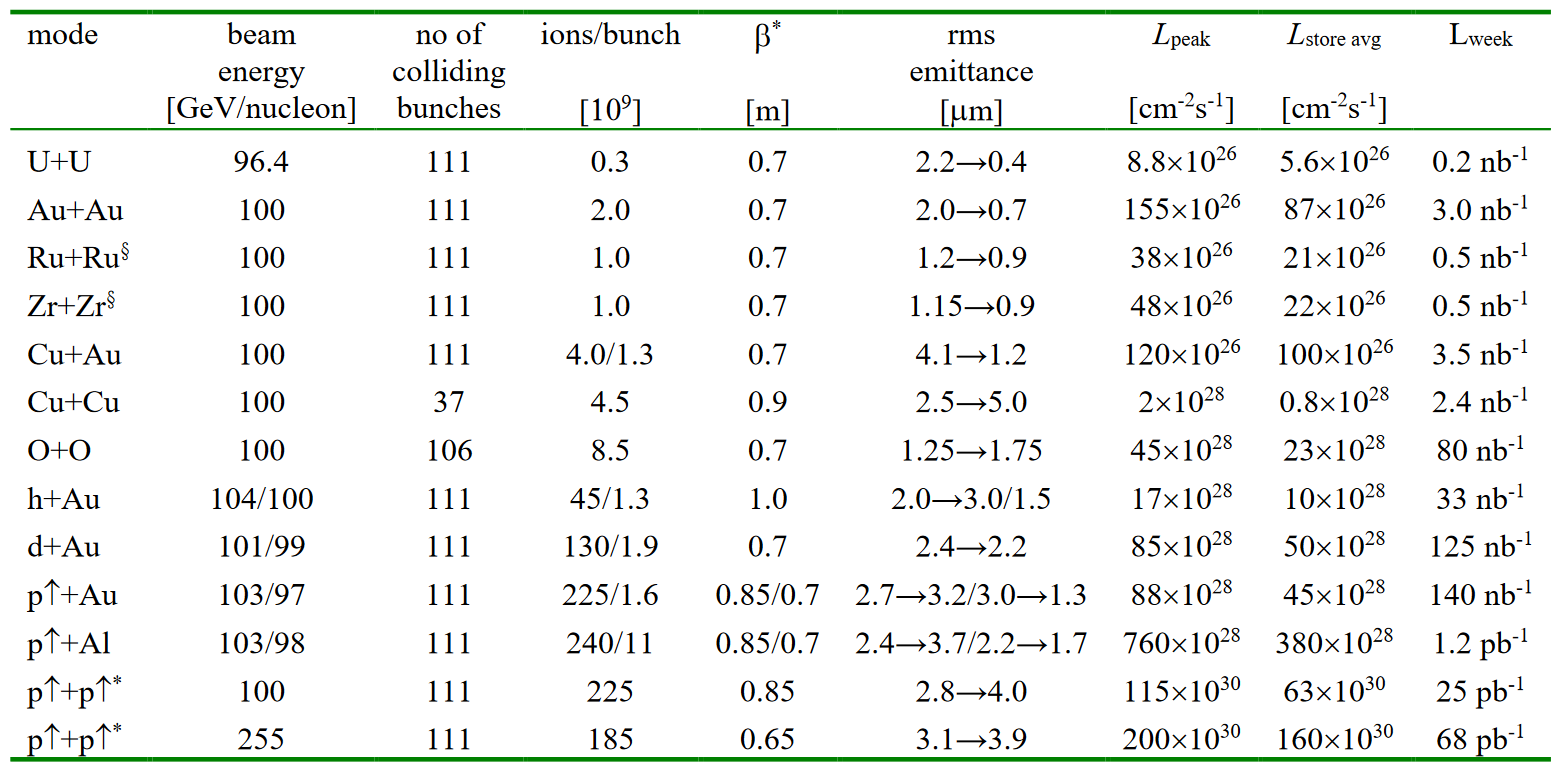}
    \caption{Demonstrated RHIC operating modes as of 2023. In addition, various different energy settings are available for the different collision systems. For lower energy running, the luminosity scales roughly as $E^2$, where $E$ is the ion beam energy. The polarization for the polarized proton running has been measured as 55\%.}
    \label{fig:RHICParam}
\end{figure}

Significant upgrades have since taken place to even further enhance the RHIC physics program, including the ability to collide polarized protons and heavier nuclei (up to uranium), as well as increased luminosity. The demonstrated parameters of the RHIC machine for a variety of collision systems are shown in Fig.~\ref{fig:RHICParam}.
\subsection{PHENIX}
\begin{figure}[ht!]
    \centering
    \includegraphics[width=10cm]{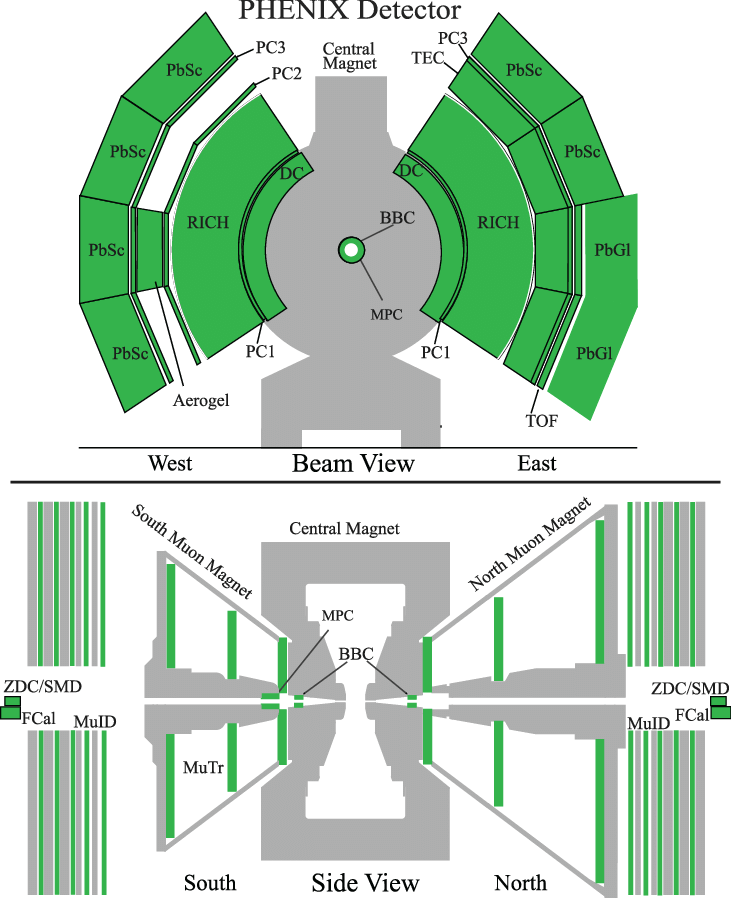}
    \caption{The PHENIX detector as seen from the RHIC tunnel (top) and from the side, looking from inside the RHIC ring (bottom). The two central arms of the detector are slightly offset from 180\textdegree separation.}
    \label{fig:PHENIXDetector}
\end{figure}

The Pioneering High Energy Nuclear Interaction eXperiment (PHENIX)~\cite{osti_10194586} was one of the two major detectors at RHIC, located at interaction point 8 (IP8). PHENIX consisted of two arms at midrapidity ($|\eta|<0.35$) and two muon arms located in the forward and backward directions. The two central arms of PHENIX had a limited acceptance in the azimuth, amounting to a total azimuthal coverage of 180\textdegree. The primary magnetic field was provided by an axial field warm magnet with the field lines oriented along the beamline. The integrated field for particles produced at 90\textdegree was around 1 T$\cdot$m. The tracking detectors, primarily the drift chambers and pad chambers, were located outside the strongest portion of the magnetic field. Particle identification was provided by two RICH detectors and a time-of-flight (TOF) system. At a distance of around 5m from the interaction point sat the electromagnetic calorimeters. 3/4 of the azimuth was covered with a lead-scintillator calorimeter that achieved an energy resolution of around $8\%/\sqrt{E}$ for electromagnetic energy deposits. The remaining sections of the EMCal were made of lead-glass modules from the WA98 experiment, with an energy resolution of $6\%/\sqrt{E}$. The PHENIX data acquisition system was capable of accepting around 15 kHz of Au+Au events, enabling PHENIX to perform high-statistics measurements of high $p_T$ photons and other rare probes. A diagram of the detector is shown in Fig.~\ref{fig:PHENIXDetector}. PHENIX ended data-taking in 2016, in part to allow for a new, upgraded experiment, called sPHENIX to be constructed at IP8.
\subsection{STAR}
\begin{figure}[ht!]
    \centering
    \includegraphics[width=13.5cm]{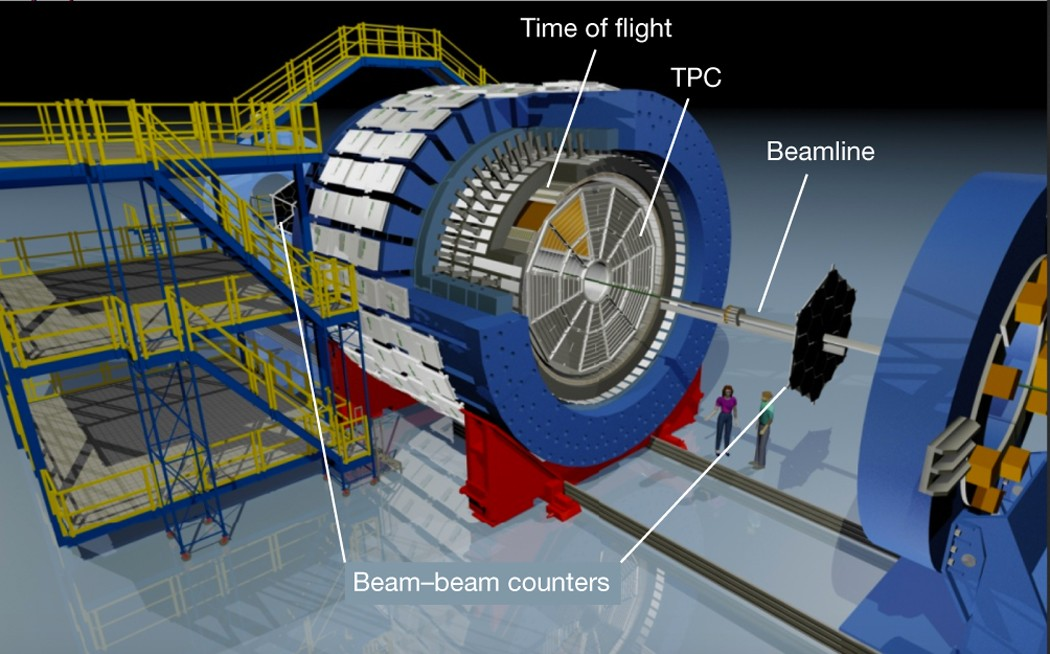}
    \caption{A diagram of the STAR detector.}
    \label{fig:STARDetector}
\end{figure}

The Solenoidal Tracker At RHIC (STAR) is located at interaction point 6 (IP6) of RHIC. STAR utilizes a large TPC as the primary tracking and particle identification system. The TPC covers the whole azimuth, and measures all particles within $|\eta|<1$. The magnetic field of 0.5 T is provided by a warm solenoid with a radius of 2.1 m. The relatively weak magnetic field maximizes the acceptance for low $p_T$ particles, allowing STAR to track and identify particles down to transverse momenta of 150 MeV. This acceptance is crucial for minimally biased measurements of hadron abundances and other soft probes of the QGP properties. The STAR TPC utilizes multi-wire proportional chambers (MWPCs) with pad readout. There generally exists a tradeoff in gas detectors which do both particle identification and tracking; to achieve the optimal dE/dx resolution, position resolution must be sacrificed, and vice versa. The STAR TPC operates in a mode which compromises between the two, allowing for a dE/dx resolution of 7-8\%, and a position resolution of $\sim$ 1 mm per hit. The initial design of the TPC had 45 rows of pads, allowing for 45 individual samples of non-curling track positions and dE/dx. Recently, the inner portion of the TPC was upgraded to bring the total number of pad rows to 72~\cite{Shen:2018pkc}. This improvement in the granularity of the innermost detection plane pushes the momentum threshold for tracking low-$p_T$ particles down to 60 MeV and improves the dE/dx resolution to $\sim$ 6\%. To reduce the amount of ions from the gain stage flowing back into the drift volume, the TPC has a ``gated" readout that can only accept events at a maximum rate of $\sim$ 1 kHz. For further information on TPCs, see Chapter~\ref{Chap:TPC}. To aid in particle identification and triggering, a TOF detector with a time resolution of 87 ps surrounds the TPC. For the measurement of photons and electrons is the barrel EMC, which sits outside the TPC and TOF but inside the magnet coils. A diagram of STAR is shown in Fig.~\ref{fig:STARDetector}.
\section{sPHENIX} \label{sec:sPHENIXPhysics} 
Several key observables that probe the properties of the QGP were unable to be measured precisely by the previous generation of RHIC experiments. For this reason, the sPHENIX experiment was proposed to complete the RHIC physics program by precisely measuring observables which were difficult, or impossible, to measure previously. These observables generally rely on high statistics, precise tracking, large acceptance, or hadronic calorimetry. In general, these fall under the umbrella of ``hard probes" of the QGP properties. The measurements of highest priority for sPHENIX include jets and $\Upsilon$ spectroscopy, which will be outlined in detail below.\par
\subsection{Jets}
As discussed in section \ref{sec:JetPhysics}, jets provide a window into the dynamics of partons. Jets in ``vacuum", such as those produced in $e^+e^-$, $e+p$, and (debatably) $p+p$ collisions, generally exhibit a standard pattern of fragmentation that has been verified by multiple experiments~\cite{ParticleDataGroup:2018ovx}. The explanation for this universality lies in the fact that a hard scattering in these collision systems is not affected substantially by the sparse environment around it, and thus a given parton fragments in all three cases as if it were propagating through the QCD vacuum alone. This standard fragmentation pattern thus provides a standard candle against which to study the effect of the QGP on a parton traversing it. The naive expectation is that a jet passing through the QGP will experience additional soft collisions with the partons of the medium, resulting in the redistribution of the jet momentum to more particles and wider angles. Coupled with the fact that jets in $p+p$ and A+A collisions must be produced in $p_T$ balanced pairs, the disappearance of away-side jets has been one of the most striking pieces of evidence for the existence of the QGP. This concept, known as jet quenching~\cite{Gyulassy:1990ye}, has multiple experimentally observable effects on heavy ion collisions. \par
\begin{figure}[htbp]
    \centering
    \includegraphics[width=10cm]{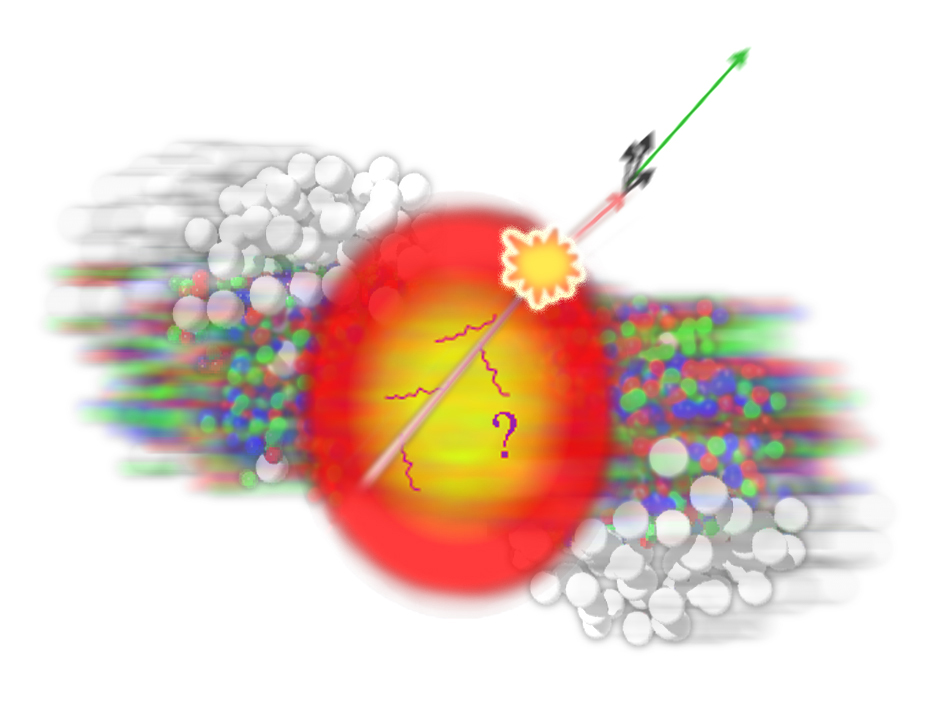}
    \caption{A sketch illustrating the jet quenching process. A hard scattering that occurs on the periphery of the colliding nuclei will produce two jets, one of which may traverse a larger portion of the medium than the other. It is expected that the jet which traverses more of the medium will be quenched compared to its partner.}
    \label{fig:QuenchingSchematic}
\end{figure}
\begin{figure}
     \centering
     \begin{subfigure}[b]{0.45\textwidth}
         \centering
         \includegraphics[width=\textwidth]{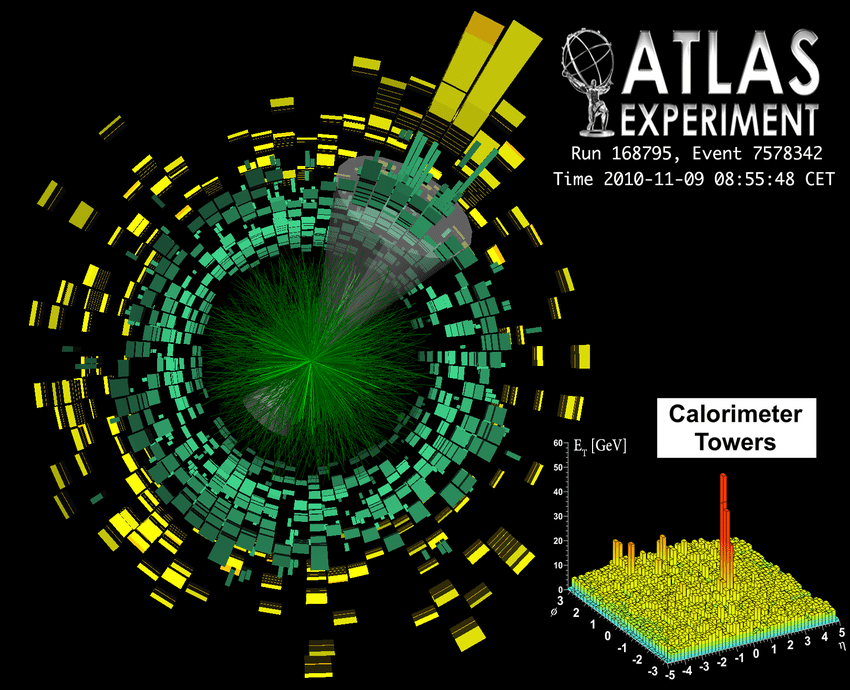}
         \label{fig:ATLASEvent}
     \end{subfigure}
     \hfill
     \begin{subfigure}[b]{0.45\textwidth}
         \centering
         \includegraphics[width=\textwidth]{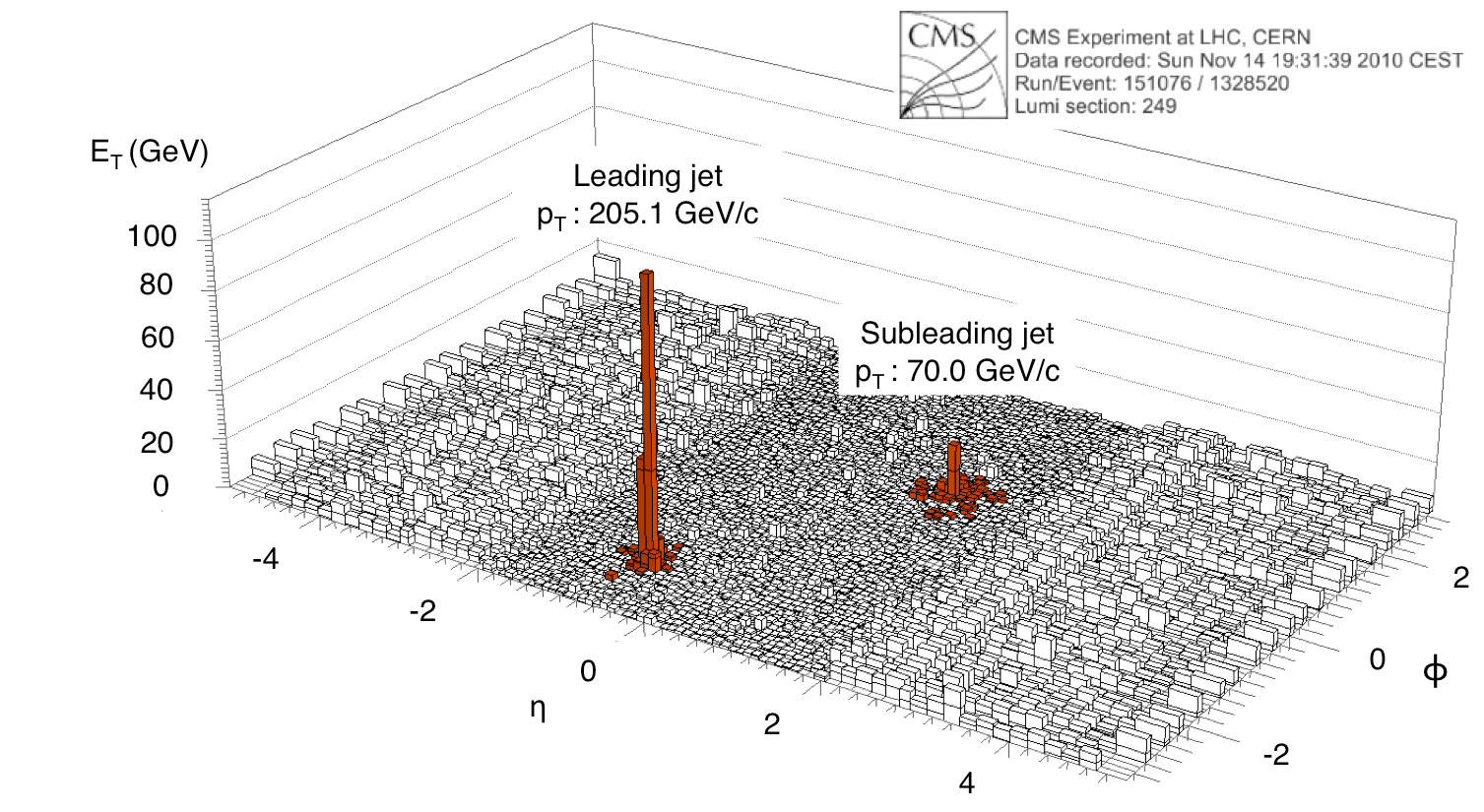}
         \label{fig:CMSEvent}
     \end{subfigure}
        \caption{Event displays from ATLAS (left) and CMS (right) demonstrating the modification of an away-side jet in Pb+Pb collisions at the LHC.}
        \label{fig:JQEventDisplays}
\end{figure}
The first experimentally confirmed effect of jet quenching was the suppression of high transverse momentum single hadrons in Au+Au collisions compared to the expectation from $p+p$ collisions as measured by PHENIX~\cite{PHENIX:2001hpc}. 
\begin{figure}[htbp]
    \centering
    \includegraphics[width=8cm]{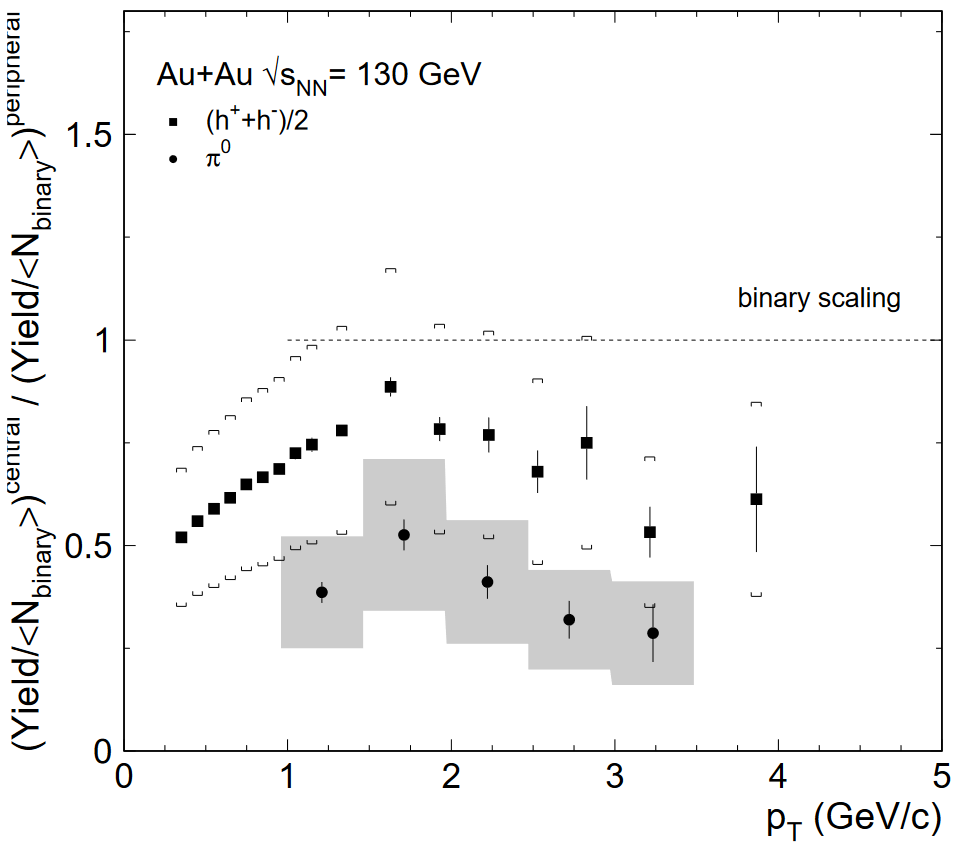}
    \caption{First results on the suppression of high $p_T$ hadrons in central Au+Au collisions compared to peripheral Au+Au collisions at $\sqrt{s} = 130$ GeV, published in 2001 by the PHENIX collaboration. The results from peripheral collisions were found to be consistent with the expectation from an incoherent sum of $p+p$ collisions, while the central case clearly exhibits suppression~\cite{PHENIX:2001hpc}.}
    \label{fig:PHENIXJQ}
\end{figure}
Shortly after the PHENIX result, STAR measured the suppression of back-to-back high $p_T$ hadrons~\cite{STAR:2002svs}, strongly suggesting that the partons produced in hard-scatterings within the QGP were having their momenta reduced by the medium. In both of these measurements, single hadrons were used as proxies for jets, due to the experimental complexity of reconstructing jets in the presence of the enormous background originating from the underlying event in central heavy ion collisions. Single hadrons are useful to gain a qualitative understanding of whether or not partons are quenched, however, for precision studies of the QGP properties, the measurement of jets is preferable. \par
In the decades since the original jet quenching results, substantially more differential measurements have been made, including those at the LHC performed by the ALICE~\cite{ALICE:2008ngc}, CMS~\cite{CMS:2008xjf}, and ATLAS~\cite{ATLAS:2008xda} collaborations. The goal of sPHENIX is to provide similar degrees of precision at RHIC, where the interaction of the initiating parton with the plasma is expected to be stronger. One limit to the precision achieved thus far at RHIC is that previous jet measurements by STAR and PHENIX have been ``biased", either by small detector acceptance (in the case of PHENIX), low rate capability necessitating triggers on e.g. high $p_T$ particles (in the case of STAR), and the lack of hadronic calorimetry. All of these factors result in measurements that contain a bias towards various kinds of events. For example, without hadronic calorimetry, the neutral hadrons produced in jets are typically poorly measured, if measured at all. In this case, jets which happen to have a large component of their energy in neutral hadrons may not be reconstructed as jets at all, thus resulting in a bias towards jets with fewer neutral hadrons. In particular, most prior jet measurements at RHIC have utilized a trigger on a high $p_T$ particle, which biases the selection of jets towards jets with a fragmentation pattern that produces a single high $p_T$ particle. This selection favors, for example, quark jets over gluon jets. These triggers are necessary for experiments which do not have the readout bandwidth to collect all events, and therefore must make selections of which events to keep. sPHENIX intends to collect a large dataset of so-called ``min bias" events, meaning events where the trigger is as unbiased towards any physics observable as possible. With this dataset, effectively all possible jet fragmentation patterns can be studied, and correspondingly the modification of the fragmentation functions between $p+p$ and central Au+Au collisions can be measured in a minimally biased way. \par
\subsection[Upsilon Melting]{$\Upsilon$ Melting}
The temperature at which QCD matter chooses to transition from bound hadrons into a soup of quarks and gluons is anticipated to be around 150 MeV~\cite{HotQCD:2019xnw}. This corresponds to around one trillion Kelvin in SI units ~\cite{Ploskon:2018yiy}. Mapping the QCD phase diagram in the high-temperature regime is one of the primary goals of heavy ion collisions, and various observables exist to measure the QGP temperature. 
\begin{figure}[htbp]
    \centering
    \includegraphics[width=10cm]{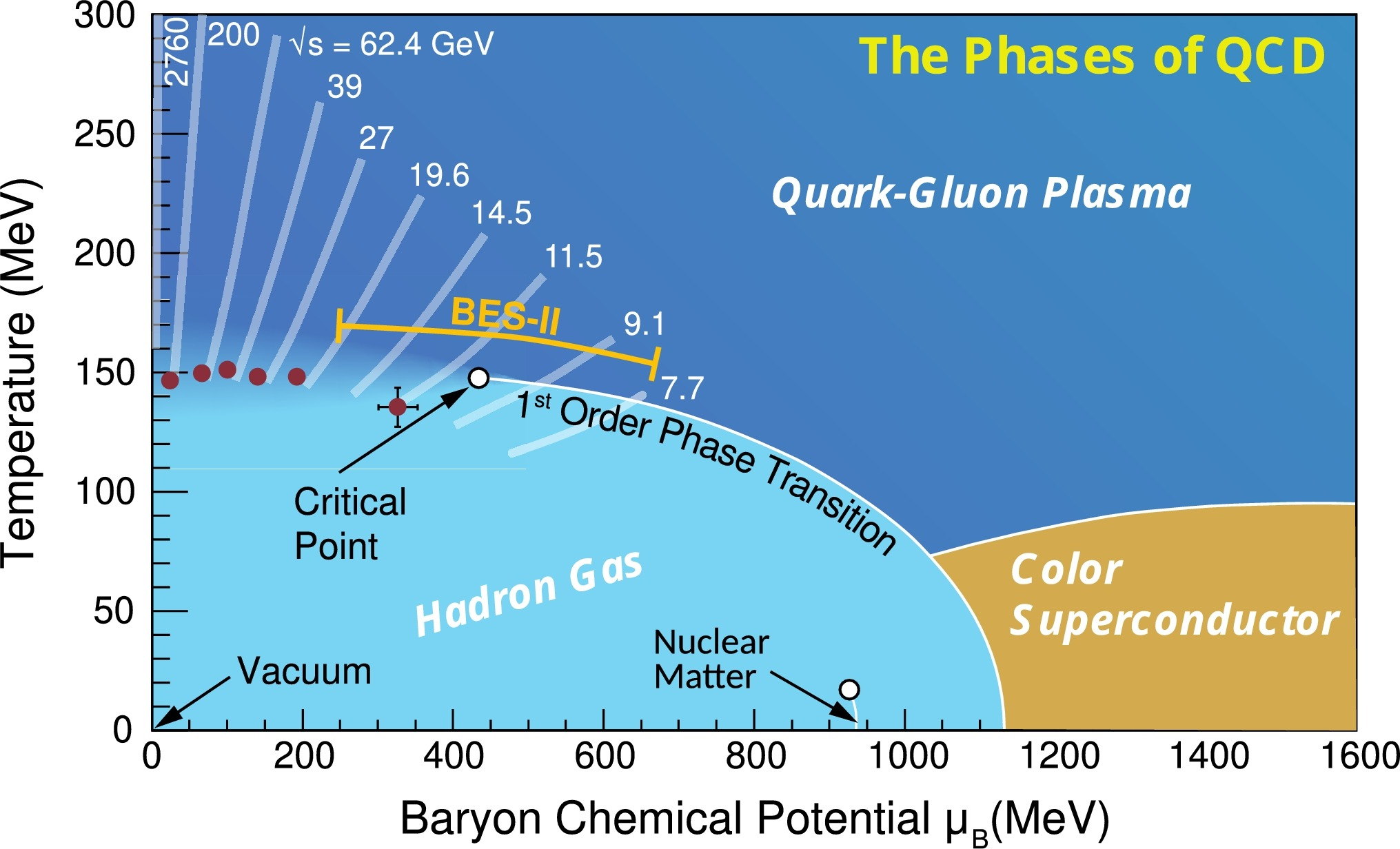}
    \caption{The QCD Phase Diagram. The white lines in the top left corner demonstrate the reach of RHIC and the LHC at different beam energies. Figure from Ref.~\cite{An:2021wof}.}
    \label{fig:Phase}
\end{figure}
Perhaps the most conceptually simple of these is soft direct photons, which are expected to be emitted in a blackbody spectrum that is directly proportional to the temperature. Experimentally, there exist serious challenges to measuring that spectrum, includingbackground photons from meson decays and material-induced bremsstrahlung, as well as photon pair conversion. There currently exists a ``direct photon puzzle", where theoretical calculations are unable to explain simultaneously the spectra and elliptic flow of direct photons. Many measurements of soft direct photons have been performed by PHENIX and STAR, although the results disagree. The experimental situation remains unresolved as of 2023.\par
An experimentally cleaner method is to measure the suppression of heavy quarkonia and their corresponding excited states~\cite{Matsui:1986dk}. Quarkonium refers to the bound state of a quark-antiquark pair. Of these quarkonia, the two most theoretically well-understood, and therefore most relevant for heavy ion collisions at RHIC and the LHC, are the $J/\psi$ (the bound state of a charm and an anti-charm quark), and the $\Upsilon$ (the bound state of a bottom and an anti-bottom). By virtue of the large masses of the charm and bottom quarks, they are expected to be produced almost exclusively in the hard scatterings of individual partons. In vacuum, the $q\Bar{q}$ pair produced in the hard scattering will readily bind into a $J/\psi$ or $\Upsilon$. However, in the SU(3) colored medium of the QGP, the color charges of the $q\Bar{q}$ pair will be screened, thus rendering them unable to ``see" each other and therefore unable to form a bound state. This color screening effect can be understood analogously to the Debye length in classical electrodynamics. As the density and temperature of the charges in the plasma increases, the length over which an individual charge exerts its influence is reduced. Thus, a decrease in the number of heavy quarks which decay as quarkonia is expected in the presence of the QGP.\par
\begin{figure}[htbp]
    \centering
    \includegraphics[width=5cm]{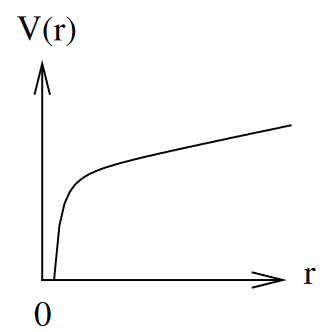}
    \caption{A sketch of the Cornell potential.}
    \label{fig:QP}
\end{figure}

One of the key aspects of heavy quarkonia is their ability to be described theoretically with non-relativistic QCD (NRQCD), which enables quantitative estimates of their size and shape~\cite{Brambilla:2010cs}. This follows from the fact that the masses of the heavy quarkonia are not significantly larger than the masses of the two heavy quarks, therefore the kinetic energy of the quarks within the bound state cannot be enough to make them relativistic. The $J/\psi$ mass is $\sim$3.1 GeV/c$^2$, and the charm quark mass is around 1.3 GeV/c$^2$, while the $\Upsilon$ mass is $\sim$9.5 GeV/c$^2$ compared to the bottom mass of $\sim$4.65 GeV/c$^2$. The actual quark masses are technically scheme dependent, however the smallness of the binding energy compared to the quark masses is clear regardless of which scheme is used. The anticipated velocity of charm quarks within the $J/\psi$ is $\sim0.3c$, while for the bottom quarks in the $\Upsilon$ the velocity is $\sim0.1c$. These small velocities enable convergence of calculations which use the velocity as an expansion parameter. Additionally, the relatively small velocities of the quarks permit the use of the formalism of non-relativistic quantum mechanics, which significantly simplifies calculations. A simple model for the forces between heavy quarks inside quarkonia is the so-called ``Cornell potential", which is coulomb-like ($\propto-\alpha_s/r$) and repulsive at short distances and linear ($\propto r$) and attractive at large distances, as shown in Fig. \ref{fig:QP}~\cite{Eichten:1974af}. The coulomb term accounts for the short distance single gluon exchange between the quarks. The linear term represents the confining potential of QCD, arising from the anti-screening of the QCD vacuum.

NRQCD utilizes the scales of the quark masses $M$, the typical momentum of the quarks $Mv$ and the kinetic energy of the quarks, $Mv^2$. The mass sets the scale of the production and annihilation of the quarks, the inverse of the momentum $1/Mv$, sets the radial extent of the quarkonium, and the kinetic energy sets the scale of the various energy levels accessible for the excited states. Using these quantities and the Cornell potential, the non-relativistic Schr\"{o}dinger equation can be solved to determine the masses and radii of the radially excited states of the $b\Bar{b}$ system, known as the $\Upsilon(nS)$ states~\cite{Satz:2006kba}. The Schr\"{o}dinger equation,
\begin{equation}
[2m_Q-\frac{1}{m_Q}\nabla^2+V(r)]\Phi_i(r) = M_i\Phi_i(r)
\end{equation}
using the aforementioned Cornell potential of the analytic form
\begin{equation}
V(r) = \sigma r - \frac{4\alpha_s}{3r}
\end{equation}
with $\sigma \sim 0.2$ GeV$^2$ being the string tension, $m_Q$ being the heavy quark mass, and $\alpha_s$ being the strong coupling constant, allows for estimates of the properties of the masses and radii of the states to be evaluated directly. The properties of these states are summarized in table \ref{tab:Upsilons}. 
\begin{table}[tbhp]
  \footnotesize
  \begin{center}
    \begin{tabular}{lccc}
      \hline
      $\Upsilon$ State & Binding Energy &Mass & Radius\\
      \hline
         1S&1.10 GeV&9.460 GeV& 0.14 fm\\ 
         2S&0.54 GeV&10.023 GeV& 0.28 fm\\ 
         3S&0.20 GeV&10.355 GeV& 0.39 fm\\
      \hline
    \end{tabular}
    \caption{
      A selection of $\Upsilon(nS)$ states and their properties as estimated by theory. Note that this represents only a small fraction of the known $b\Bar{b}$ bound states.
    }
    \label{tab:Upsilons}
    \end{center}
\end{table}
\begin{figure}[htbp]
    \centering
    \includegraphics[width=13cm]{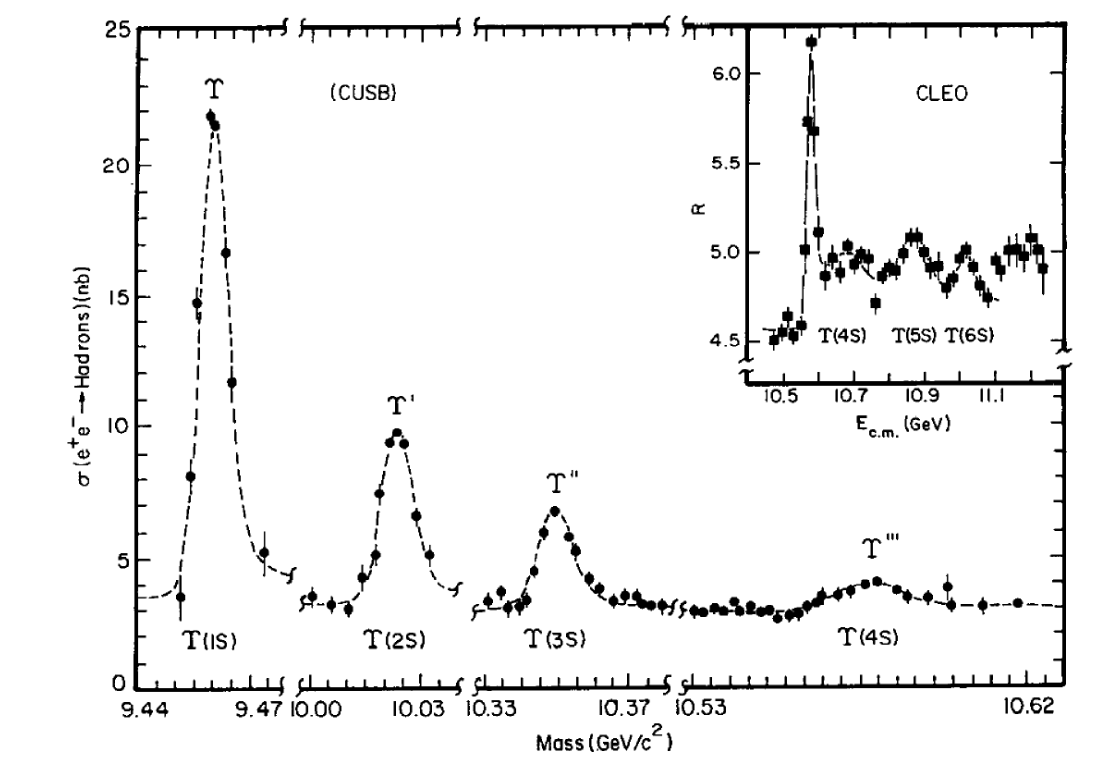}
    \caption{The peaks of the $\Upsilon$ and its corresponding excited states, as measured by sharp increases in the cross section for the s-channel annihilation of $e^+e^- \longrightarrow$ hadrons. The data are from the CUSB and CLEO experiments at the Cornell Electron Storage Ring (CESR).}
    \label{fig:US}
\end{figure}
\par The $\Upsilon(4S)$ state lies above the threshold for production of two $B$ mesons, and thus is often the operating energy of choice for $B$ factory experiment such as BaBar and BELLE\footnote{This feature, along with the small production cross section in $h+h$ collisions, make the $\Upsilon(4S)$ experimentally less commonly mentioned in studies of QGP properties.}. The same procedure can be followed for the $J/\Psi$, although the approximations made in NRQCD are expected to be less valid due to the smaller charm quark mass. 
\begin{table}[tbhp]
  \footnotesize
  \begin{center}
    \begin{tabular}{lccc}
      \hline
      $J/\Psi$ State & Binding Energy &Mass & Radius\\
      \hline
         1S& 0.64 GeV &3.10 GeV&0.25 fm\\ 
         2S& 0.05 GeV &3.68 GeV&0.45 fm\\ 
      \hline
    \end{tabular}
    \caption{
      The two $J/\Psi(nS)$ states and their properties as estimated by theory.
    }
    \label{tab:JPsi}
    \end{center}
\end{table}
Both the $\Upsilon$ and the $J/\Psi$ are experimentally clean signals by virtue of their decays into leptons. The $\Upsilon$ states each have a $\sim 2\%$ branching fraction of $\Upsilon\rightarrow e^+e^-$ and $\Upsilon\rightarrow \mu^+\mu^-$. The $J/\Psi$ has around a 6\% branching fraction for each of the di-lepton channels. Heavy ion collisions produce an enormous amount of hadrons, which generally preclude attempts at reconstructing hadronic decay modes of resonances without further information, such as a displaced vertex decay. The leptonic decay channels of the $\Upsilon$ and $J/\Psi$ produce leptons with fairly high momentum, that can be more cleanly identified.\par
The Cornell potential is expected to describe quarkonia in vacuum, however the introduction of a finite temperature medium is expected to modify the potential. Using finite temperature lattice QCD, a modified potential can be extracted, which can be used to analytically determine the dissolution temperature of the state. Alternatively, using modern methods lattice QCD can directly calculate the dissolution temperature. The expected dissolution temperatures are outlined in \ref{tab:Dissolution}.
\begin{table}[tbhp]
  \footnotesize
  \begin{center}
    \begin{tabular}{lcccccc}
      \hline
      State & $J/\Psi(1S)$ & $J/\Psi(2S)$ &$\Upsilon(1S)$&$\Upsilon(2S)$&$\Upsilon(3S)$&$\Upsilon(4S)$\\
      \hline
         $T_{Melt}$ (GeV)& 0.27& $<$ 0.15 & 0.44& 0.25 & 0.2 &$<$0.15\\ 
      \hline
    \end{tabular}
    \caption{
      The anticipated dissolution temperatures for the states considered here. The $J/\Psi(2S)$ and $\Upsilon(4S)$ states are expected to melt even below the phase transition temperature of $\sim150$ MeV. Table adapted from~\cite{Rothkopf:2019ipj}.
    }
    \label{tab:Dissolution}
    \end{center}
\end{table}
\begin{figure}[htbp]
    \centering
    \includegraphics[width=13.9cm]{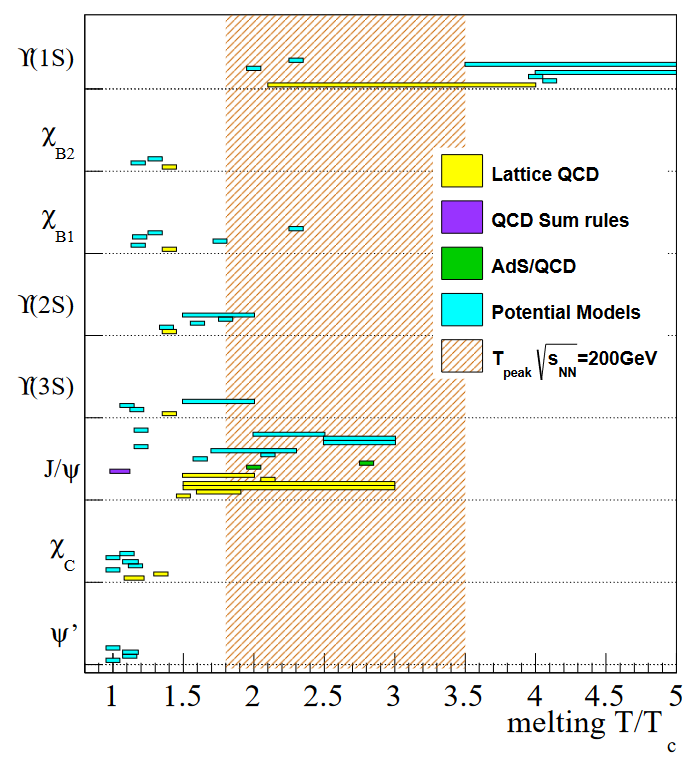}
    \caption{The theoretical predictions for the melting temperatures of various quarkonium states. The observation of suppression of various states correspondingly constrains the red-shaded range of possible $T_{peak}$ values. Figure from Ref.~\cite{PHENIX:2014tbe}.}
    \label{fig:PHENIXMelting}
\end{figure}
It should be mentioned that it is not necessarily the case that the in-medium physical widths of the quarkonia excited states allow them to be uniquely identified~\cite{Larsen:2019zqv}. In vacuum the states are known to be well separated, but in a high temperature medium, thermal effects will cause the peaks to broaden and possibly overlap. This broadening somewhat complicates the extraction of the yields of the individual states in experimental measurements.
\subsubsection{Experimental Status of $\Upsilon$ Melting}
\begin{figure}[htbp]
    \centering
    \includegraphics[width=13.9cm]{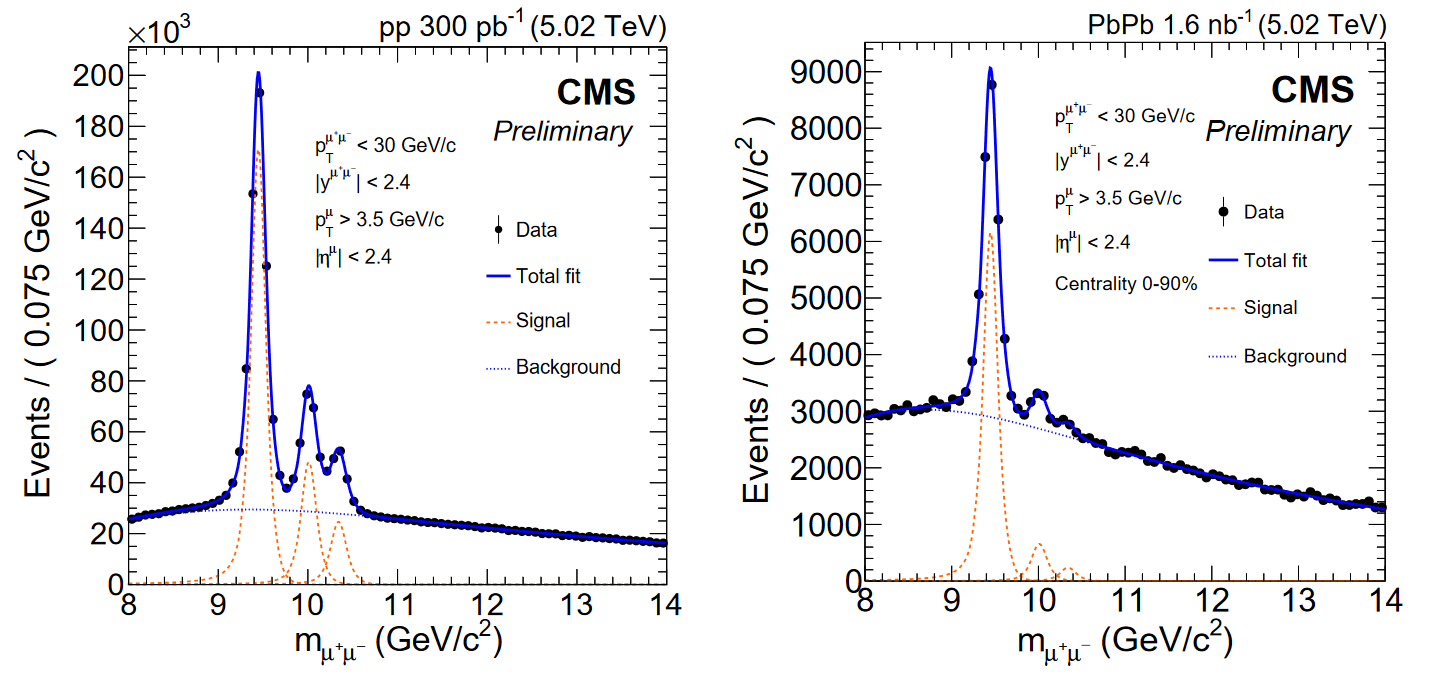}
    \caption{The dimuon invariant mass spectrum in the region of the $\Upsilon$ states, in $p+p$ collisions (left) and in Pb+Pb collisions (right), both at $\sqrt{s_{NN}} = 5.02$ TeV. Figure from Ref.~\cite{Lee:2023ail}.}
    \label{fig:CMSUpsilons}
\end{figure}

\begin{figure}[htbp]
    \centering
    \includegraphics[width=13.9cm]{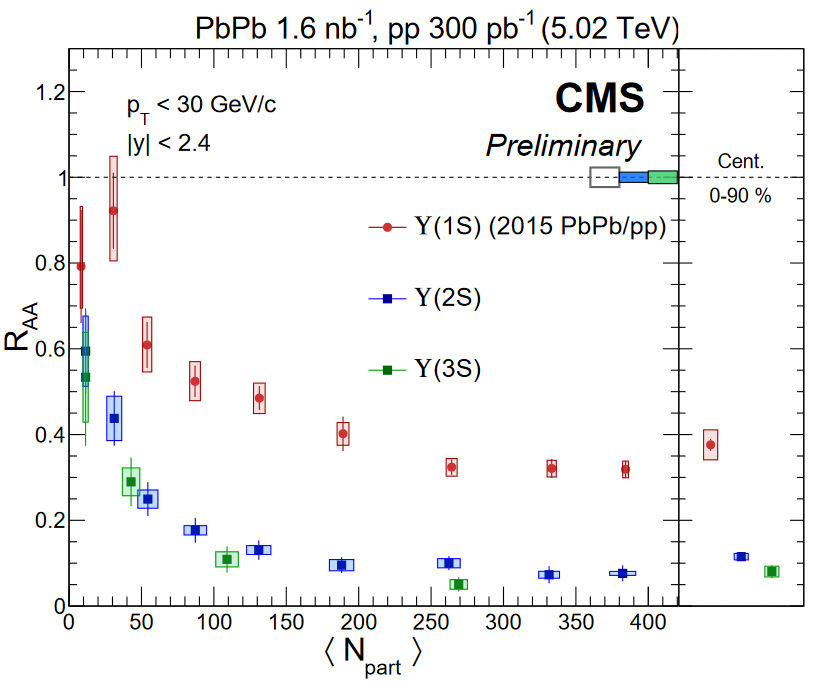}
    \caption{Nuclear modification factor $R_{AA}$ as a function of average number of nucleons participating in the collision. The data show a clearly decreasing trend with $\rangle$N$_{\text{Part.}}\langle$, showcasing the temperature and system size dependence of the suppression. Figure from Ref.~\cite{Lee:2023ail}.}
    \label{fig:CMSRAA}
\end{figure}

As of Spring 2023, the STAR~\cite{STAR:2022rpk}, ALICE~\cite{ALICE:2018wzm}, CMS~\cite{Lee:2023ail}, and PHENIX~\cite{PHENIX:2014tbe} collaborations have published results on $\Upsilon$ suppression in heavy ion collisions. The first observation of $\Upsilon$ suppression was published by CMS in 2014~\cite{CMS:2013jsu}. The capability of CMS to identify and reconstruct muons at high momentum enables very precise measurements in the dimuon decay channel of the $\Upsilon$. The most recent result from CMS is shown in fig.~\ref{fig:CMSUpsilons}.
The results show clear evidence for suppression of the $\Upsilon$ excited states compared to the ground state. Additionally, CMS presented the scaling of $R_{AA}$ with the average number of participants for each of the $\Upsilon$ states, shown in fig.~\ref{fig:CMSRAA}. These results suggest that at the LHC, the temperature of the QGP is high enough to dissolve even the $\Upsilon(1S)$ state in events with large numbers of participants.

\begin{figure}[htbp]
    \centering
    \includegraphics[width=13.9cm]{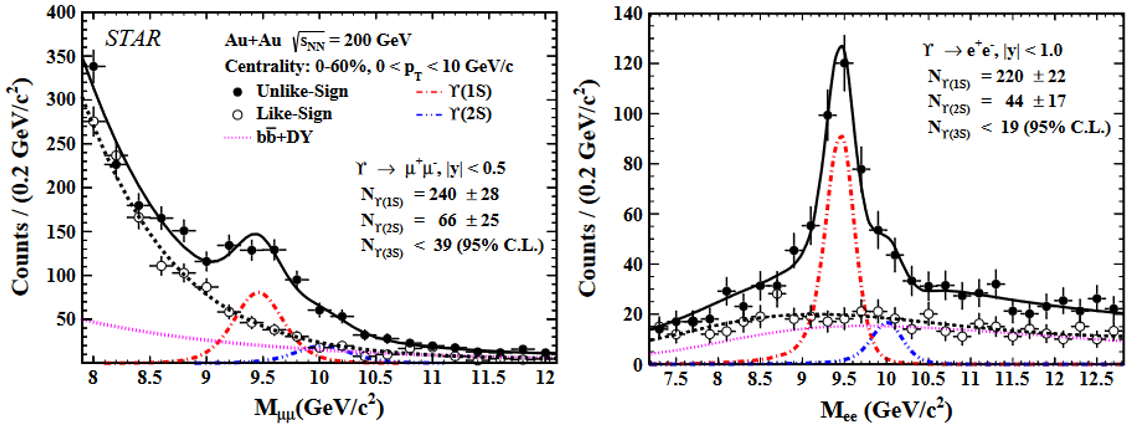}
    \caption{Dimuon and dielectron invariant mass spectra as measured in Au+Au collisions at $\sqrt{s_{NN}}=200$ GeV at STAR. The resolution and background rejection of the dielectron channel is improved by the use of the barrel electromagnetic calorimeter. The $\Upsilon(1S)$ state is clearly resolved in the dielectron case, and the $\Upsilon(2S)$ state can be seen as a shoulder to the $\Upsilon(1S)$. In both channels, an upper limit is set on the possible number of $\Upsilon(3S)$ states produced in the event sample. Figure from Ref.~\cite{STAR:2022rpk}.}
    \label{fig:STARChannels}
\end{figure}

\begin{figure}
     \centering
     \begin{subfigure}[b]{0.45\textwidth}
         \centering
         \includegraphics[width=\textwidth]{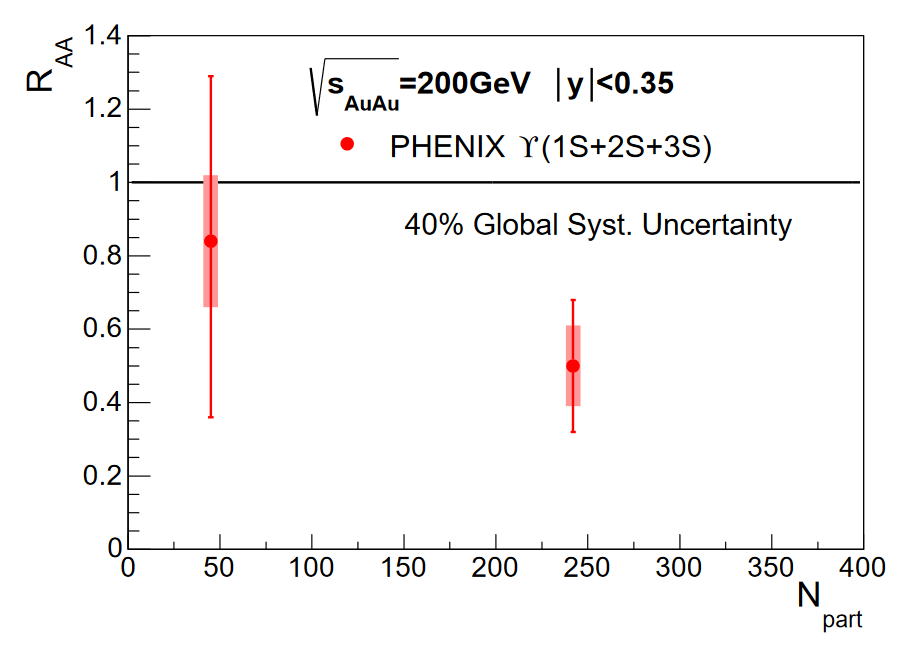}
    \caption{Combined $R_{AA}$ for the $\Upsilon(1S+2S+3S)$ states as measured by PHENIX. Figure from Ref.~\cite{PHENIX:2014tbe}.}
    \label{fig:PHENIXRAA}
     \end{subfigure}
     \hfill
     \begin{subfigure}[b]{0.42\textwidth}
         \centering
          \includegraphics[width=\textwidth]{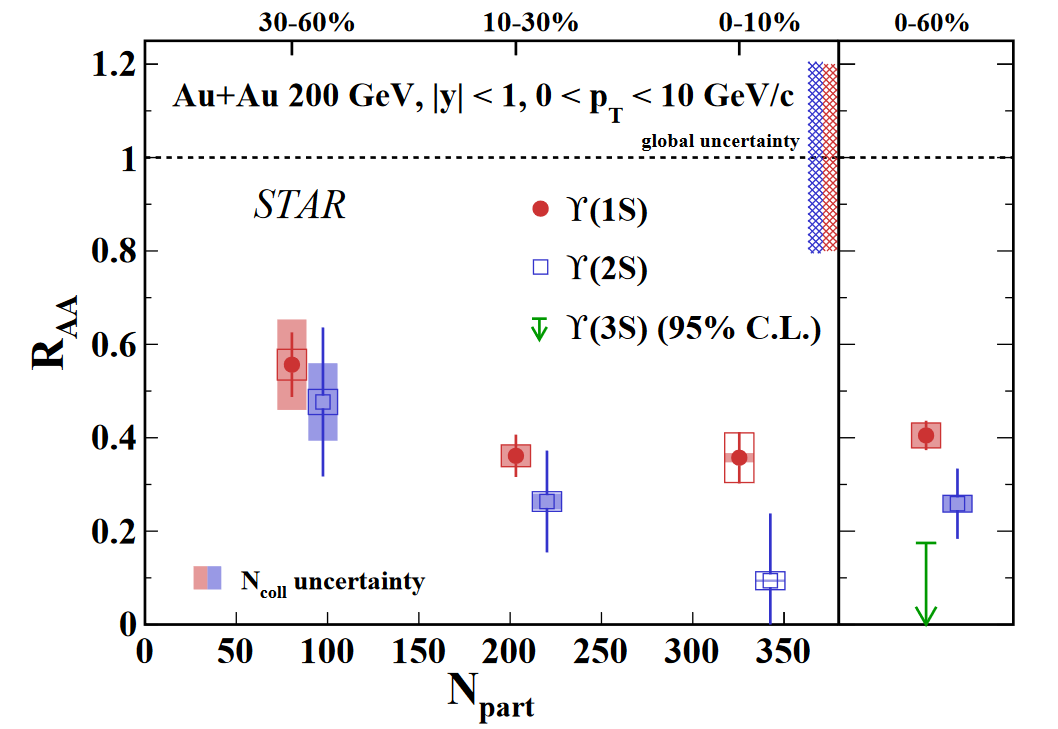}
    \caption{$R_{AA}$ as measured by STAR for the $\Upsilon(1S)$ and $\Upsilon(2S)$, along with an upper bound on the $\Upsilon(3S)$. Figure from Ref.~\cite{STAR:2022rpk}.}
    \label{fig:STARRAA}
     \end{subfigure}
\end{figure}

At RHIC, PHENIX and STAR have both published results on the production of $\Upsilon$ in $p+p$ and Au+Au collisions. The PHENIX result is shown in Fig. \ref{fig:PHENIXRAA}. The fairly small acceptance of the PHENIX central detector makes measurements of $\Upsilon$ challenging, as the large mass of the $\Upsilon$ states means that the produced leptons are typically back-to-back in the lab frame. The forward and backward muon arms were additionally used to determine the rapidity dependence of $\Upsilon$ production. The PHENIX results, shown in Fig. \ref{fig:PHENIXRAA} are consistent with the previous measurements made by STAR and CMS, and suggest suppression of $\Upsilon$ states in events with high activity, however only the results for the integrated $\Upsilon(nS)$ states are presented. STAR measured the $\Upsilon(1S)$ and $\Upsilon(2S)$ in the dimuon and dielectron channels in $p+p$ and Au+Au collisions. The corresponding reconstructed invariant mass spectra are shown in Fig. \ref{fig:STARChannels}

The general consensus from the experimental measurements performed thus far is that the QGP at RHIC and the LHC does indeed suppress all the $\Upsilon$ states, however the degree of suppression of each of the states remains a fairly open question. At the LHC, the results of quarkonium suppression are complicated somewhat by the possibility that the temperature is so high that heavy quarks can be produced directly by the medium. In particular, $J/\Psi$ results have recently been called into question. Since the temperature of the medium produced at RHIC is closer to the critical point, it is expected that $J/\Psi$ and $\Upsilon$ are both still useful probes of the QGP temperature. The interpretation of $\Upsilon$ suppression is less ambiguous than $J/\Psi$ suppression from a theoretical standpoint, and the interpretation of RHIC results are less ambiguous than those from the LHC. However, to use the $\Upsilon$ as a thermometer for the QGP at RHIC, more precise and differential measurements will need to be made. To improve on these and other measurements, the sPHENIX detector is currently being commissioned at RHIC. The design and expected performance of sPHENIX will be described in the following section.

\section{The sPHENIX Detector}
The sPHENIX detector is designed to extend the physics reach of RHIC for rare and hard probes. Hard probes generally result in particles with high energy emerging from the collisions. Examples are high $p_T$ jets, which at RHIC can contain individual hadrons with transverse momenta up to $\sim30$ GeV. For these particles, a strong magnetic field is required to bend their trajectories appreciably to measure their momenta. For this reason sPHENIX has chosen to re-use the superconducting solenoid from BaBar, which has a field strength of $\sim$1.4T at the center of the 2.8m bore. This is almost a factor of 3 higher than the magnetic field used in STAR, and will enable sPHENIX to more easily measure the momenta of high $p_T$ particles. Additionally, sPHENIX will utilize a hadronic calorimeter at midrapidity for the first time at RHIC. The ability to measure high $p_T$ hadrons both in the tracker and the calorimeter will provide valuable cross check and add additional robustness to the sPHENIX jet program.\par
\begin{figure}[htbp]
    \centering
    \includegraphics[width=13.9cm]{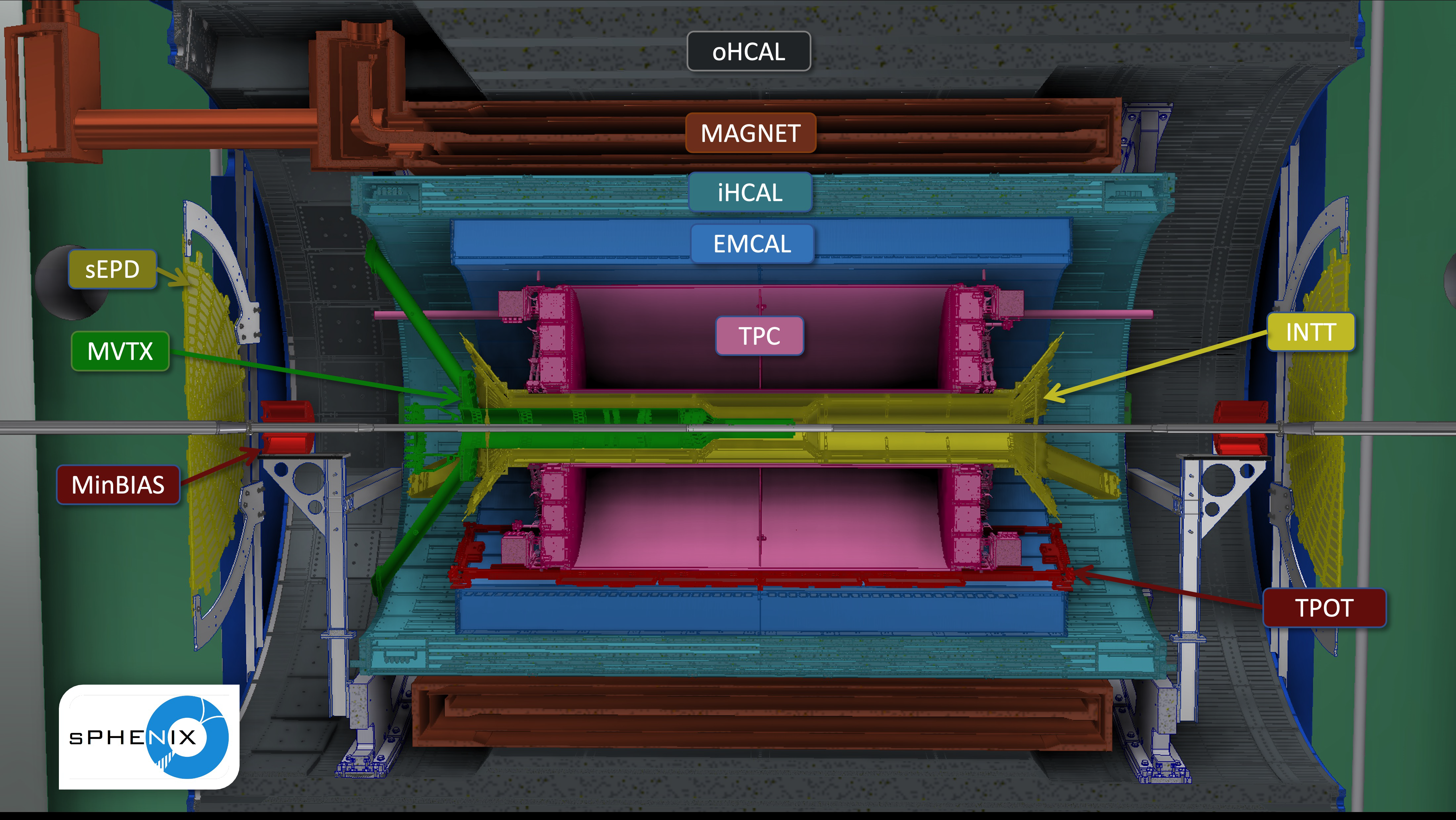}
    \caption{A side view of the sPHENIX detector. The magnet chimney seen on the left side of this sketch is on the south side of the experiment.}
    \label{fig:sPHENIXDetSide}
\end{figure}

Rare probes are statistics hungry, and require large numbers of events to measure properly. The $\Upsilon$ states are an excellent example of a rare probe, as the cross section for their production in heavy ion collisions at RHIC is small. To maximize the event statistics for rare probes, sPHENIX is designed to be intrinsically capable of accepting a large portion of the available luminosity at RHIC. The typical bottleneck in terms of the event rate which can be collected is the tracking system. To this end, sPHENIX will utilize an ungated TPC (described in detail in Sec. \ref{Chap:TPC} as the central tracker. A cross section of the sPHENIX detector is provided in fig. \ref{fig:sPHENIXDetSide}

The detector is largely $\phi$- and z-symmetric. The two main sources of asymmetry are in the magnet, which requires a ``chimney" to transport the liquid helium to the coils to keep them superconducting, and the support of the MVTX. The active detector region of the MVTX is symmetric in z, but it is supported from only the south side of the experiment.
\subsection{Hadronic Calorimeters}
\begin{figure}[htbp]
    \centering
    \includegraphics[width=13.9cm]{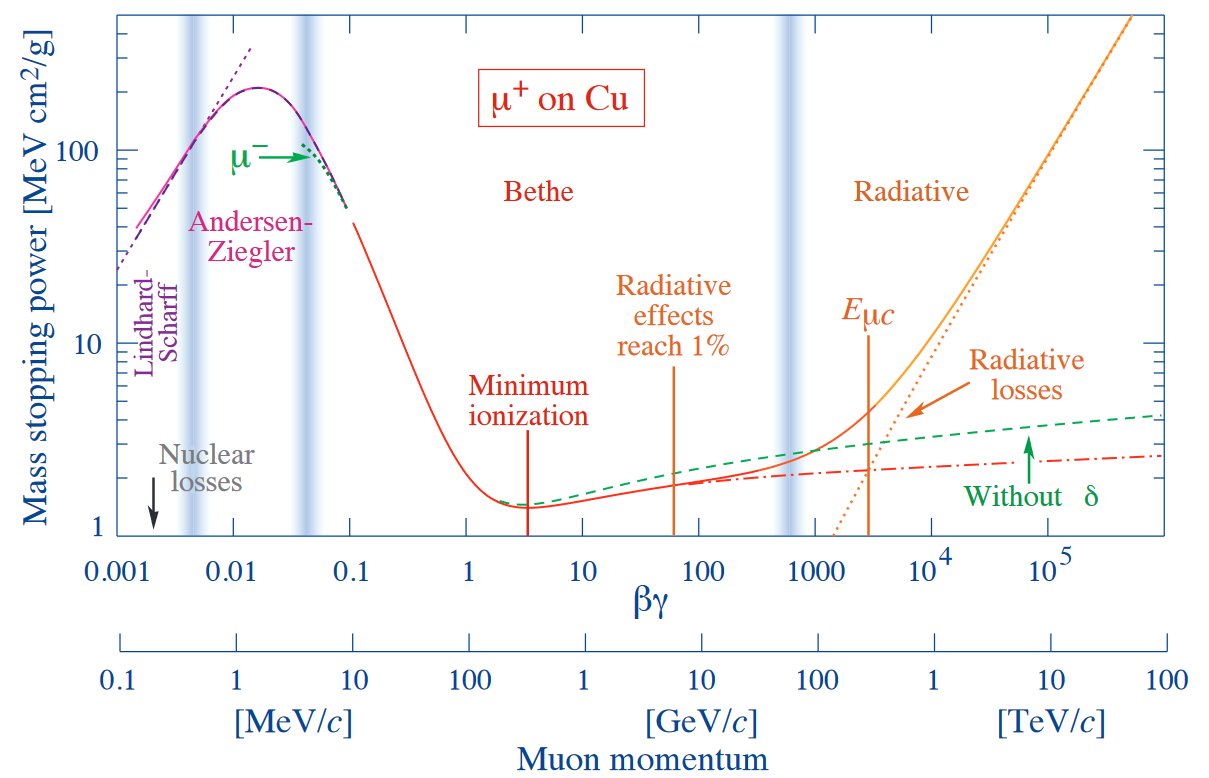}
    \caption{Stopping power ($\langle -dE/dx \rangle$) curve for muons passing through copper. Hadrons at collider energies generally fall into the region of $\beta\gamma\sim 1-1000$, where the average stopping power of materials is low. Electrons, due to their low mass, have $\beta\gamma\sim 2000$ at 1 GeV, meaning they predominantly lose energy via radiative effects, and can be stopped in smaller amounts of material than hadrons.}
    \label{fig:PDGdEdx}
\end{figure}

A hadronic calorimeter is, as the name suggests, a detector which measures the energy of hadrons, including neutral hadrons such as neutrons and long-lived $K^0_L$ mesons. The requirement for sPHENIX is that the HCal as a combined system measure single hadrons with a resolution on the order of $100\%/\sqrt{E}$. To measure the energy of a hadron, the energy lost by the particle as it traverses the detector material must be measured. Clearly then, it is necessary for the hadron to lose all of its energy inside the detector volume. The canonical plot describing the energy loss of charged particles is shown in Fig.~\ref{fig:PDGdEdx}. If a charged hadron lost energy only to ionization\footnote{The ionization process is described in detail in Sec.~\ref{Sec:ChPT}.}, in a dense absorber such as lead, the hadron would lose $\sim 20$ MeV/cm of path length. Thus, for a hadron with 100 GeV of energy, the detector would need to be around 40 meters long to (on average) contain the energy. Thankfully, hadrons can also lose energy to inelastic collisions with the nuclei inside the detector. These inelastic collisions produce additional lower energy particles, which can themselves undergo inelastic collisions and deposit their energy via ionization. The result is a cascade of particles with decreasing energy, known as a ``shower". The cross section for the initial inelastic collisions is small, so hadronic calorimeters typically seek to maximize the density of the detector as much as possible.\par
A typical scheme for constructing a hadronic calorimeter is to subdivide the detector into ``absorber" material and ``active" material. Absorbers are dense materials which typically cannot be read out in any way, such as lead or steel. Active materials are capable of measuring a signal, typically either scintillation light or ionization. The goal of the absorber is to cause hadrons to shower, and the goal of the active material is to measure the energy of the shower. The entire energy of the shower clearly cannot be measured directly, as a significant fraction of the energy is lost to the absorber, however the amount of energy deposited in the active material is generally proportional to the energy of the incident hadron. One of the main goals of hadronic calorimetry is to make the response of the detector as linearly proportional to the incoming hadron energy as possible over the necessary energy range.\par
\begin{figure}[htbp]
    \centering
    \includegraphics[width=5cm]{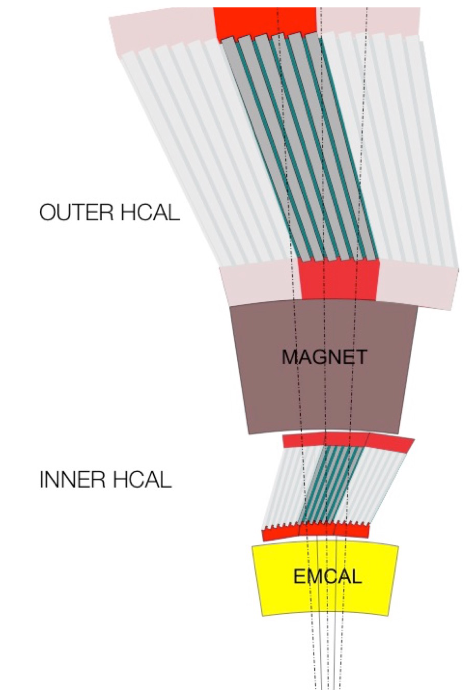}
    \caption{Radial cross section of the sPHENIX calorimeters demonstrating the orientation of the absorbers and scintillators in the HCals. It can be seen that a particle emanating from the interaction point will traverse at least four scintillating tiles in both the inner and outer HCals.}
    \label{fig:sPHENIXCaloStack}
\end{figure}

One major aspect which affects the ability of a calorimeter to have a linear response is the detector thickness, which is typically measured in hadronic interactions lengths (denoted as $\lambda_0$). $\lambda_0$ is the mean free path for a hadron to undergo an inelastic collision with a nucleus in a material. Thus, the probability $P(\ell)$ of having undergone an inelastic hadronic collision as a function of distance travelled in the material is $P(\ell) = e^{-\ell/\lambda_0}$.
The sPHENIX hadronic calorimeter is divided into two sections, inner (IHCal) and outer (OHCal). The inner HCal is inside the bore of the magnet, and is made of aluminum and scintillating tiles, while the outer HCal sits outside the bore and uses steel as the absorber. The inner HCal could not be made out of steel, as the mechanical forces on the steel from the magnetic field would have made integration and mechanical support of the detector challenging. The possibility of using non-magnetic stainless steel for the inner HCal was explored, but it was determined to be excessively expensive. The OHCal is made of 1020 low carbon steel, occupies the radial space 1.82 m to 2.7 m from the beamline, and is around 6.3 m long. The OHCal represents $\sim3.8\lambda_0$, and the combined calorimetry system provides $\sim5\lambda_0$ of active material at midrapidity, where this value is smallest. The steel absorber plates are tapered, and are $\sim 10 - 15$ mm wide at the inner and outer radius respectively. Between the absorber plates are 7 mm thick scintillating tiles. The absorbers and scintillators are tilted in $\phi$ at a 12$\degree$ angle. This ensures that a particle travelling in a straight line from the interaction point will traverse at least four of the absorber plates, as can be seen in Fig.~\ref{fig:sPHENIXCaloStack}.
The largest segment of the OHCal is called a ``module", which is a slice covering $1/32$ of the azimuth. The modules are subdivided into ``towers"; each module contains 48 towers, the segmentation is 24-fold in $\eta$ and 2-fold in $\phi$. Each tower consists of five scintillating tiles, and each scintillating tile has its own silicon photomultiplier. The towers subtend an area of $\Delta \eta \times \Delta \phi \sim 0.1\times0.1$.
\subsubsection{Scintillator}
The active detector elements in the inner and outer HCals are scintillating tiles made of extruded polystyrene plastic, doped with 1.5\% PTP and 0.01\% POPOP scintillator, similar to the ATLAS TileCal~\cite{Adragna:2009zz}. When charged particles traverse the polystyrene, they induce excitations in the polystyrene molecules which emit UV photons around 300 nm via fluorescence. However, these UV photons have a short absorption length in the tile and are quickly absorbed~\cite{KAPLAN2006283}. Some of these photons will be absorbed by the primary fluor, in this case the PTP, and remitted at 350-400 nm. These photons can be once again absorbed, this time by the secondary fluor POPOP, and remitted in the blue-green range of 400-500 nm, where the absorption length is significantly larger. The tiles have a reflective coating designed to contain as much of the scintillation light as possible, to allow it to reach the wavelength shifting fiber that transports the light to the Hamamatsu S12572-33-015P silicon photomultiplier (SiPM).\par
\begin{figure}[htbp]
    \centering
    \includegraphics[width=13.9cm]{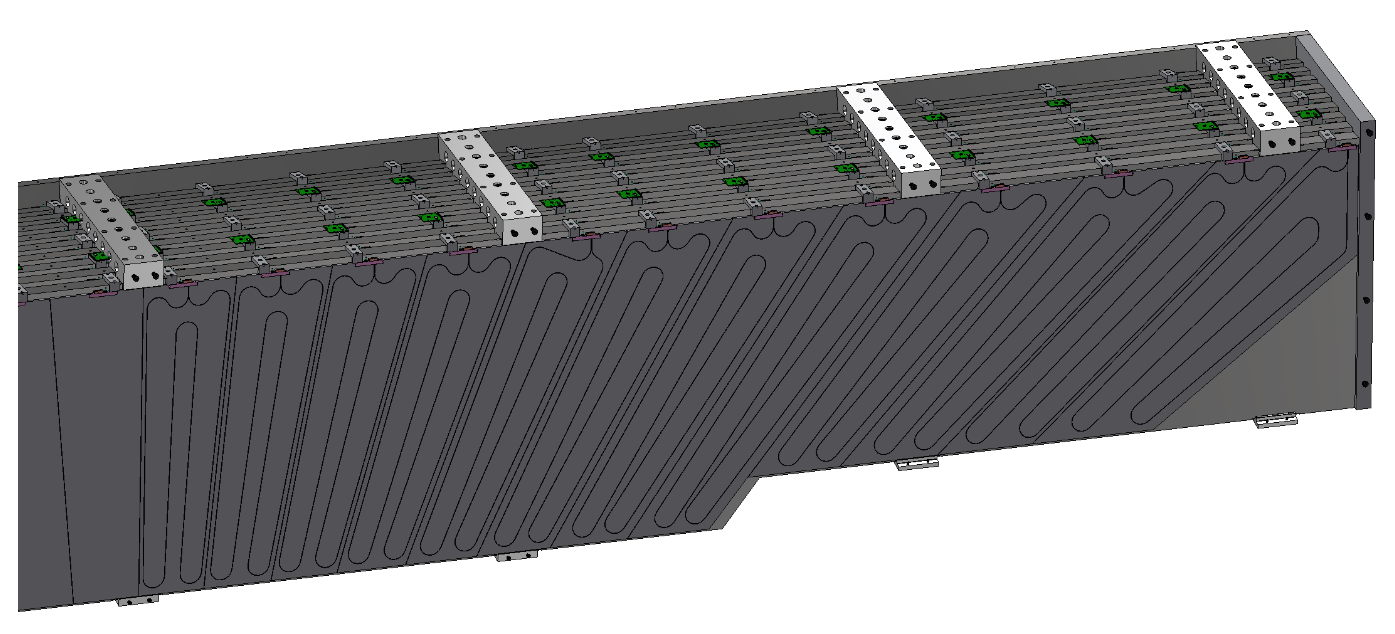}
    \caption{A cutaway of a sector of the sPHENIX OHCal. The tiles are tilted at a 12\degree angle with respect to projectivity, to prevent particles channelling down the scintillator.}
    \label{fig:sPHENIXHcalSector}
\end{figure}

Through the scintillating tiles run wavelength shifting (WLS) fibers of 1 mm diameter from Kuraray. The light produced by the scintillation is totally internally reflected inside the tile until it is either absorbed in the tile or in the WLS fiber. The fibers absorb photons at a peak wavelength around 430 nm, and re-emit around 470 nm. The photons which are re-emitted inside the fiber can be totally internally reflected all the way down the fiber to the SiPM. To ensure a uniform light yield at the SiPM for charged particles incident at different positions on the tile, the fiber is laid inside the tile in such a way that no portion of the tile is more than 2.5 cm away from a portion of the fiber, as can be seen in Fig.~\ref{fig:sPHENIXHcalSector}. The orientation of the tiles, as well as the fibers within the tiles, are shown in Fig.~\ref{fig:sPHENIXHcalSector}. 
\subsubsection{Performance}
\begin{figure}[htbp]
    \centering
    \includegraphics[width=10cm]{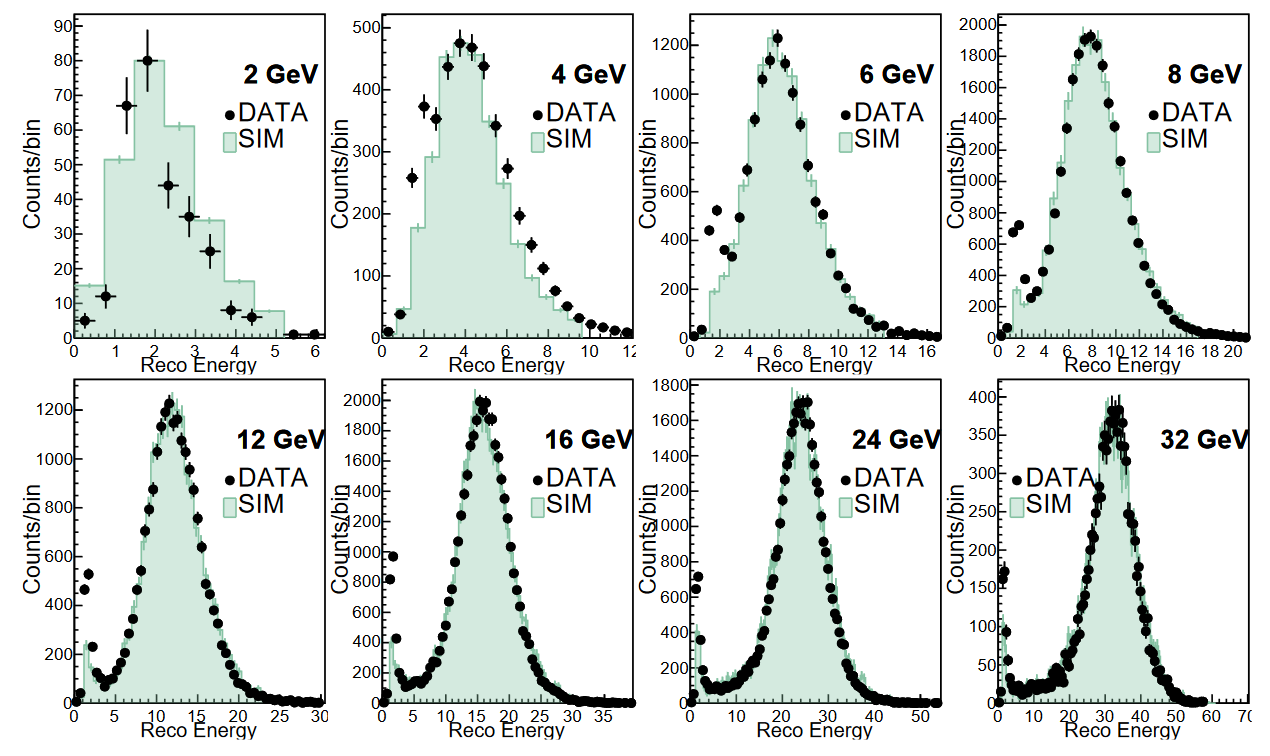}
    \caption{Hadron energies reconstructed for incident pions. A peak can be seen at roughly the value of the incident hadron energy. An additional peak sits near to zero, corresponding to particles which either showered late in the detector or did not shower at all. An admixture of muons in the beam occupy this region, which were not included in the simulation. The agreement between the data and the GEANT simulation is quite good.}
    \label{fig:HCalTestBeam}
\end{figure}

The ideal case is to have the light yield measured at the SiPMs be directly proportional and uniquely mappable to the energy of the incoming hadron. It is generally challenging to predict a priori, either analytically or numerically, the precise signal (i.e. the number of photons) that will be observed in hadronic calorimeters when exposed to hadrons of a given energy. The microphysics of both the hadronic shower itself and the light collection scheme are difficult and computationally expensive to simulate accurately in numerical detector simulation programs such as GEANT~\cite{Brun:1987ma}. For this reason, HCals are preferably studied at test beams where the energy of the incoming particle is known. With knowledge of the incoming particle energy, the characteristics of the final signal measured at the readout can be studied and optimized. This was done for the sPHENIX calorimeter system in 2017~\cite{sPHENIX:2017lqb}. The test was performed at Fermilab with a mixed beam of pions and electrons at various energies. The setup was a prototype of the calorimeter stack shown in Fig.~\ref{fig:sPHENIXCaloStack}. Some pieces of aluminum were placed between the inner and outer HCals to approxmate the material from the magnet. The response of the HCal to pions of various energies without the EMCal in front is shown in Fig.~\ref{fig:HCalTestBeam}.
\begin{figure}[htbp]
    \centering
    \includegraphics[width=12cm]{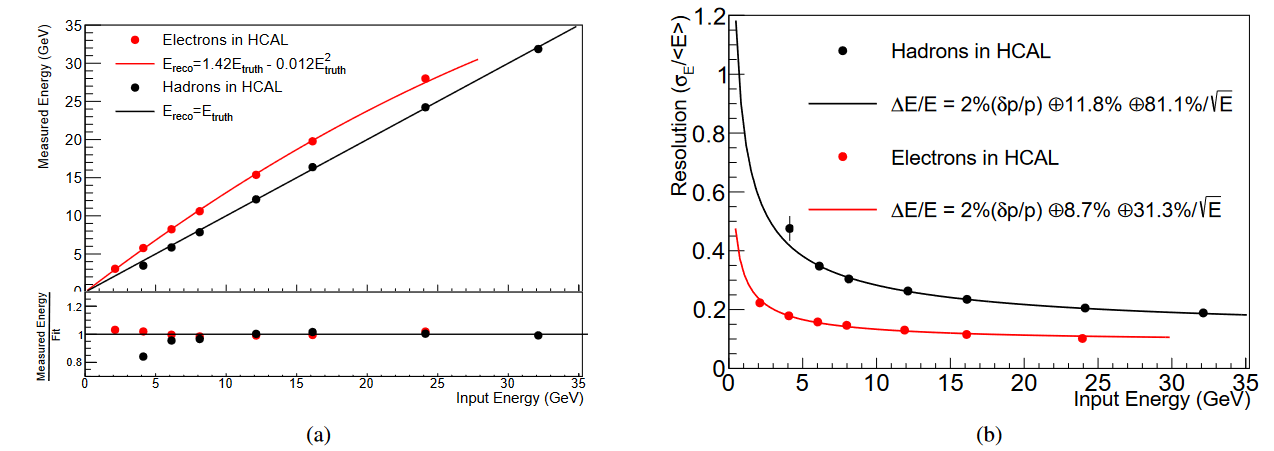}
    \caption{Linearity (a) and energy resolution (b) for the prototype HCal prototype without EMCal in front. The linearity of the detector response to incident hadrons is excellent, and the resolution is better than the $100\%/\sqrt{E}$ needed for sPHENIX. The addition of the EMCal improves the overall hadronic energy resolution to $13.5\%\oplus64.9\%$.}
    \label{fig:HCalERes}
\end{figure}

\begin{figure}[htbp]
    \centering
    \includegraphics[width=10cm]{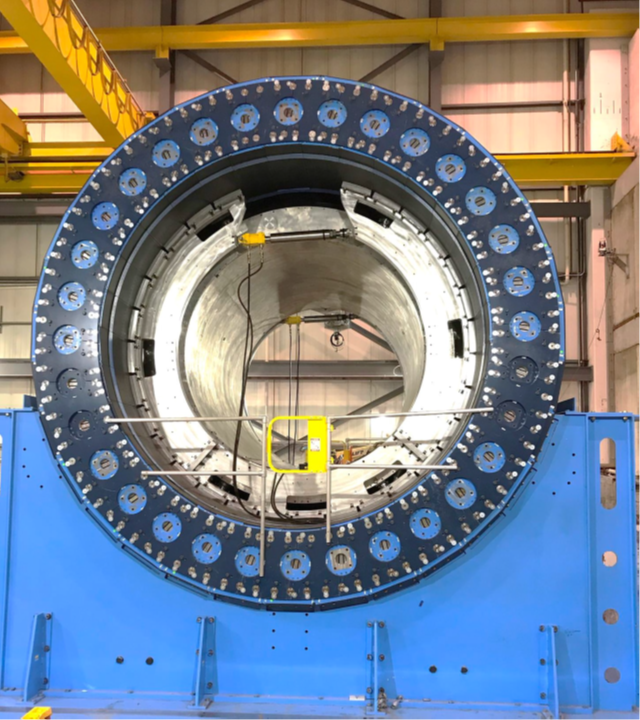}
    \caption{Image of the completed OHCal (blue) with the solenoid (grey) installed.}
    \label{fig:HCalInstall}
\end{figure}
\begin{figure}[htbp]
    \centering
    \includegraphics[width=10cm]{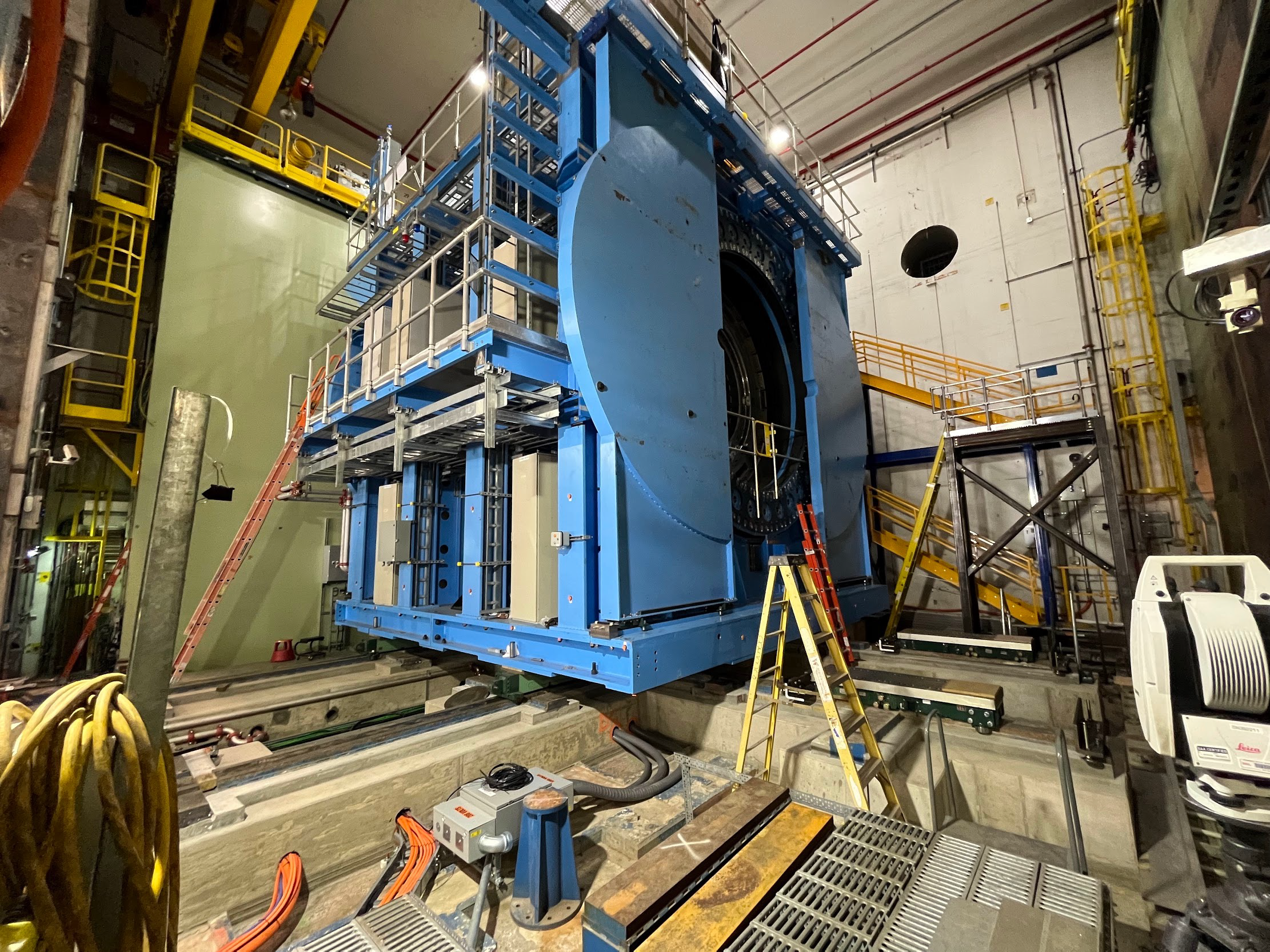}
    \caption{Image of the sPHENIX carriage, including the OHCal, magnet, IHCal, and some sectors of the EMCal in position at IP8. The left flux return door is in its closed position, while the right is in its open position.}
    \label{fig:Carriage}
\end{figure}
\subsection{Electromagnetic Calorimeter}
The goal of the electromagnetic calorimeter is to provide precision measurements of the energies of electrons and photons. For sPHENIX, the desired electromagnetic energy resolution is around $\sigma_E/E = 16\%/\sqrt{E}$, as the resolution is limited to roughly that level by the underlying event background in A+A collisions. The EMCal is vital for the measurement of electrons from $\Upsilon$ decays, as well as measurements of $\gamma+$jet events. In the case of the $\Upsilon$ measurement, the EMCal must be able to provide identification of electrons from the large background of pions at $p_T>4$ GeV. The towers of the EMCal should also be fairly small to geometrically reduce underlying event background to these electrons. Additionally, radial space inside the BaBar solenoid is limited, so the EMCal should be radially compact. All of the above requirements suggest that the density of the EMCal should be as high as possible. To achieve high density, a matrix of tungsten powder and epoxy is used as the absorber. This matrix reaches an average density of around 10 g/cm$^3$, which is higher than that of both lead tungstate and stainless steel. \par
Electrons at collider energies\footnote{This regime can be considered to be where the mass of an $e^+e^-$ pair is essentially negligible. At energies where this approximation breaks down, the primary mode of energy loss becomes ionization and excitation of the surrounding atoms and nuclei. For low energy incident electrons, dedicated absorber materials as described here are unnecessary, and calorimetry can be performed with purely active materials.}, due to their low masses and thus high velocities, generally lose energy very quickly via brehmsstrahlung, as can be seen in Fig.~\ref{fig:PDGdEdx}. The brehmsstrahlung photons are typically of large enough energy that they can be pair converted to $e^+e^-$ pairs, which can also undergo brehmsstrahlung, and so on until all the energy of the initial electron is depleted. Once the energies of the produced electrons fall below the so-called critical energy $E_c$, they predominantly deposit their energy by ionization. The critical energy can be given as:
\begin{equation}
E_c = \frac{610}{Z+1.24}\text{ MeV}
\end{equation}
For solid media, where $Z$ is the atomic number of the medium. This provides a natural cutoff scale At high energies, primary photons incident on materials undergo similar shower dynamics, except they must first pair convert to $e^+e^-$. \par
Analogous to the hadronic interaction length defined above is the radiation length, denoted $X_0$, which is defined as the distance at which a typical electron will have lost all but $1/e$ of its initial energy. The radiation length in a given material can be calculated as:
\begin{equation}
X_0(\text{g/cm}^2)\approx\frac{716\text{~g$\cdot$cm}^{-2}A}{Z(Z+1)\text{ln}(287/\sqrt{Z})}
\end{equation}
Where $Z$ is the atomic number of the material and $A$ is its atomic weight. For photons, the distance at which all but $1/e$ of the photons will have pair converted is $9/7 \cdot X_0$. To provide good containment for electromagnetic showers, calorimeter depths more than 20 $X_0$ are typically preferred. One $X_0$ of the absorber used for the sPHENIX EMCal is around 7 mm. The length of a block is around 140 mm, corresponding to about 20 $X_0$. An additional quantity of interest is given by the Moli{\'e}re radius, which gives the transverse spread of the electrons at the critical energy after one radiation length. 
\begin{equation}
R_M(\text{g/cm}^2) \approx \frac{21X_0}{E_c} 
\end{equation}
Where $E_c$ is expressed in units of MeV. $R_M$ represents approximately the radius of an electromagnetic shower, and it is useful in determining the size that an EMCal tower made of a given material must be to contain a certain fraction of a shower in the transverse direction. For sPHENIX, $R_M$ is $\sim$ 2.3 cm.\par
Each EMCal block is subdivided into four towers, which each have their own light guide to transport the light to the photodetectors. The blocks are roughly 4 mm $\times$ 4 mm, so each tower corresponds to roughly one $R_M$. The EMCal utilizes 0.47 mm diameter Saint Gobain BCF12 SC scintillating fibers with a peak of emission at 435 nm. The core of the fiber is polystyrene and the cladding is acrylic. The fibers run longitudinally through the absorber blocks, pointing approximately at the nominal interaction region (IR)\footnote{For the purposes of this thesis, the ``interaction region" is defined as the center of the detector, i.e. $z=0$. In reality, collisions will be concentrated within a region of around $|z|<$10 cm. RHIC achieves this reduced spread in the collision vertex locations by introducing a 2 mrad crossing angle to the beams. The overlap of the beams will produce a diamond-shaped region where collisions occur.}. The light guides were produced by injection molding of a UV transmitting acrylic. The light from the scintillator is readout by a Hamamatsu S12572-015P SiPM. Each light guide is viewed by four SiPMs, making 16 SiPMs per block. Since SiPMs are susceptible to radiation damage, The SiPM is located on the inner radius of the EMCal, where the radiation dose will be lower. The reason for the lower dose, despite being closer to the IR, is that this location sees predominantly the radiation of primary particles from the interaction, while the back of the EMCal sees the much larger number of ionizing particles produced from the showers of primary particles. Over time, as the SiPMs accumulate radiation damage, the noise levels will increase. This provides a challenge to the calibration of the detector. One way to stabilize the level of noise is to actively reduce the temperature of the SiPMs in accordance with the level of radiation damage, as cooling SiPMs reduces the noise. Thus, the EMCal is equipped with a water cooling system that can cool the readout electronics and the SiPM.
\begin{figure}[htbp]
    \centering
    \includegraphics[width=12cm]{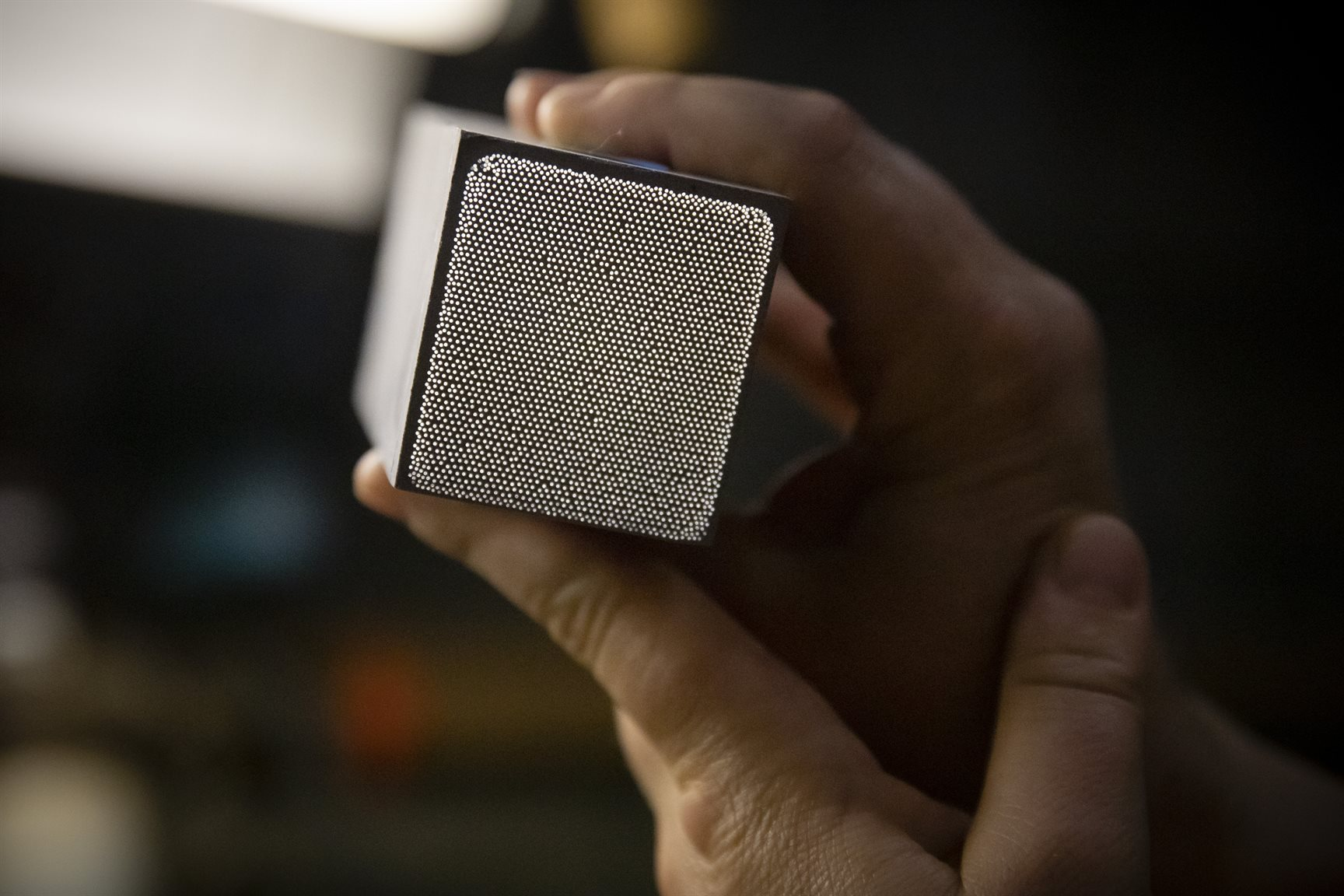}
    \caption{A tungsten-epoxy block of the sPHENIX EMCal. A light from behind illuminates the scintillating fibers.}
    \label{fig:ECalBlock}
\end{figure}
The EMCal sits at 90 cm in radius, and contains 24596 towers each covering a $\Delta\eta$ of 0.025 and a $\Delta\phi$ of 0.025. The design is 2D projective, meaning the towers are tilted approximately at the interaction region. 2D projectivity is necessary to reduce the contribution from the underlying event. For towers which point only radially inward, i.e. toward $r=0$ but not $z=0$, at higher $|\eta|$ showers would naturally be spread over more towers. This would reduce the ability of the EMCal to separate the electrons produced in the $\Upsilon$ decay from hadrons. The nominal pion rejection capability\footnote{Pion rejection is achieved by comparing the measured energy in the EMCal to the energy of the track producing the cluster. Electrons will generally have $E_{\text{Cluster}}\sim p_{\text{Track}}$, while pions will generally have $E_{\text{Cluster}}\ll p_{\text{Track}}$. A cut on E/p will necessarily throw away the tail of the measured electron energy distribution, so typically the rejection factor is cited at a given electron efficiency, in this case 70\%.} of the EMCal is a factor of 100, and it was determined that if the towers were only 1D projective, it would reduce this factor to only 60.
\begin{figure}[htbp]
    \centering
    \includegraphics[width=12cm]{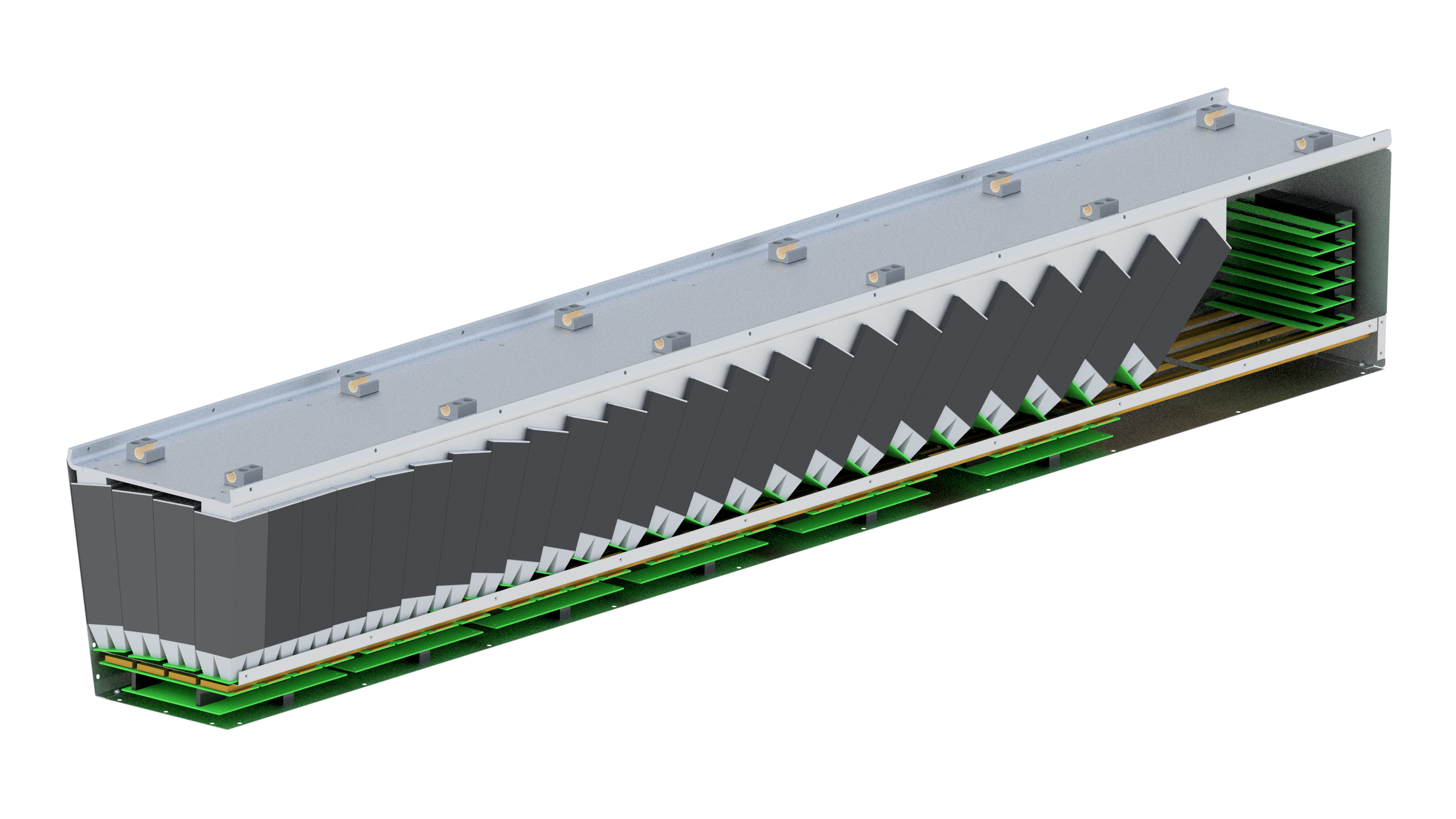}
    \caption{Geometry of an EMCal sector. The SiPMs and readout electronics are located on the inward face of the EMCal, pointing towards the IR.}
    \label{fig:ECalDesign}
\end{figure}
The ``approximate" projectivity of the EMCal is an important point. If the towers were pointed directly at the IR, there would be the possibility that particles could ``channel" directly through the scintillating fibers without hitting the dense absorber, in which case the shower would likely not be fully contained.
\begin{figure}[htbp]
    \centering
    \includegraphics[width=12cm]{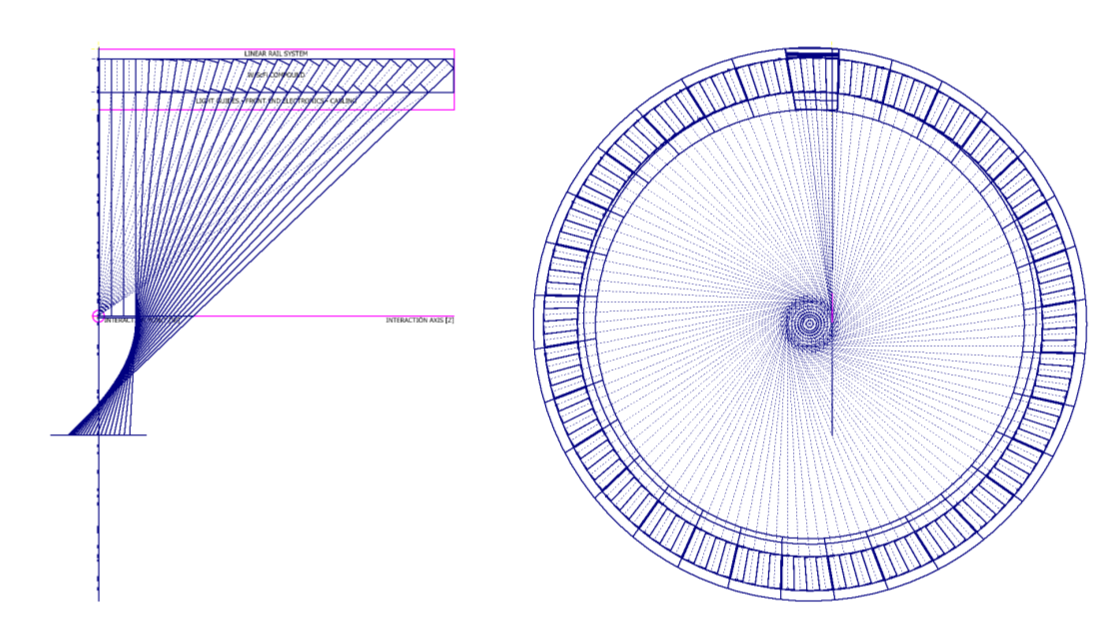}
    \caption{Demonstration of the offsets from full 2D projectivity.}
    \label{fig:ECalProj}
\end{figure}
\begin{figure}[htbp]
    \centering
    \includegraphics[width=12cm]{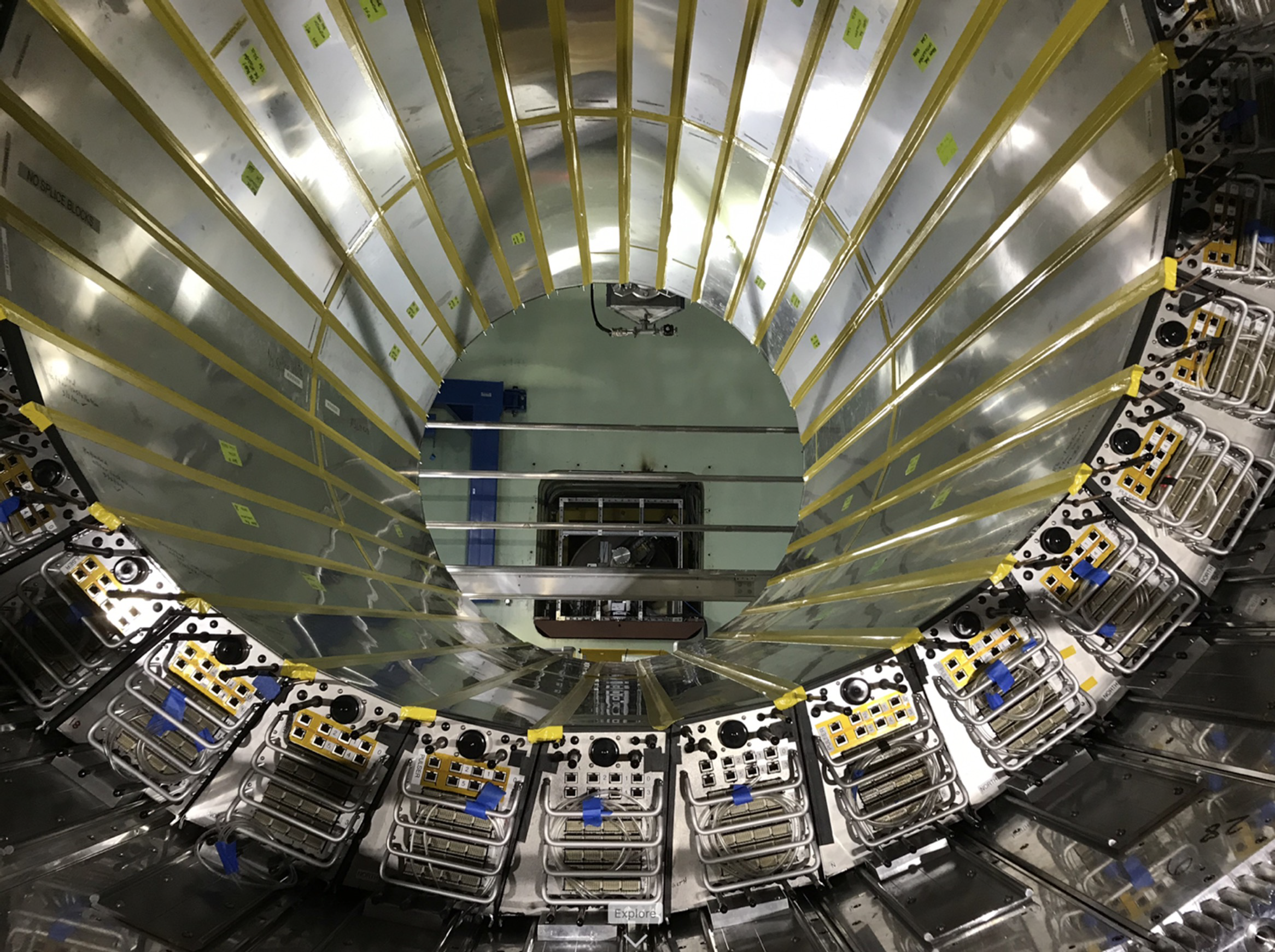}
    \caption{Image of the EMCal before being cabled. The IHCal, from which the EMCal is supported, is visible in the bottom of the image.}
    \label{fig:ECalInstall}
\end{figure}
\subsection{Charged Particle Tracking}
\label{Sec:ChPT}
\begin{figure}[htbp]
    \centering
    \includegraphics[width=10cm]{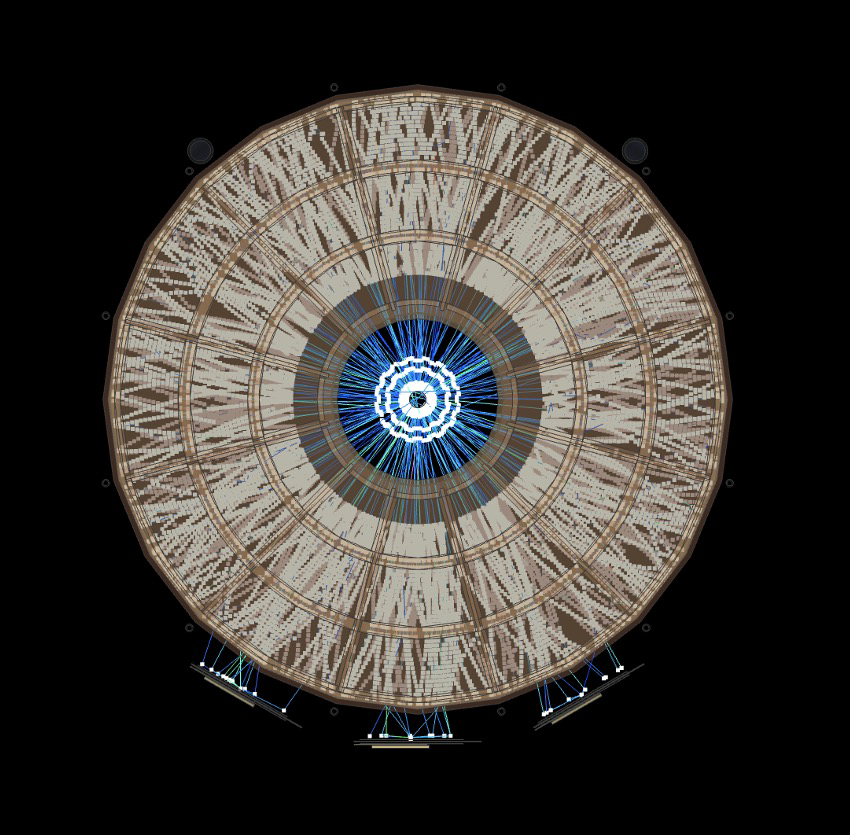}
    \caption{Simulated event display of the sPHENIX tracking system. The innermost white layers are the MVTX, the two jagged white layers are the INTT, the brown is the TPC, and the three planes at the bottom of the image at the TPOT.}
    \label{fig:sPHENIXTracks}
\end{figure}

The sPHENIX tracking system consists of a MAPS-based silicon pixel detector (MVTX), a silicon strip based intermediate tracker (INTT), a Time Projection Chamber (TPC), and the TPC outer tracker (TPOT). An event display demonstrating hits in each of these detectors is provided in Fig.~\ref{fig:sPHENIXTracks}. 
Since the following sections and the upcoming chapter are dedicated to tracking detectors, a brief summary of charged particle tracking is given here. Since charged particles bend in magnetic fields in accordance with their momentum, sequential measurements of the position of a particle as it bends provides a measurement of the momentum of the particle. Many aspects of collider detectors vary depending on the physics goals of the experiment, but tracking the produced particles is almost always required. The classic rule-of-thumb equation for the radius of curvature $R$ of a charged particle with unit charge moving in a magnetic field with momentum $p_T$ in the plane perpendicular to the field is
\begin{equation}
p_{T} = 0.3BR
\end{equation}
Where the units of $p_{T}$ are GeV/c, the units of $B$ are Tesla, and the units of $R$ are meters\footnote{The seemingly miraculous factor of 0.3 comes from the unit conversion, where $c$ is multiplied by $1\cdot10^{-9}$. Thus the oft quoted 0.3 is technically 0.299792..., which is to a good approximation 0.3. There is a certain satisfaction that arises from three apparently unrelated units in experimental particle physics, two SI and one natural, working out in such a convenient way.}. Thus, a measurement of the radius of curvature is a measurement of the momentum in the direction perpendicular to the magnetic field. With the addition of a momentum component in the direction parallel to the magnetic field, the motion of the particle becomes a helix. The uncertainty on the measurement of momentum was studied most famously by Gl{\"u}ckstern in 1963. The resolution on a measurement of $p_T$ can be restated in terms of the measurement of the track sagitta, i.e. the farthest distance between the curve and a straight line connecting its endpoints. In the context of tracking, the sagitta $s$ is generally significantly smaller than the radius of curvature, and thus can be well approximated as:
\begin{equation}
s = \frac{L^2}{8R}
\end{equation}
where $L$ is the path length of the track. The transverse momentum resolution $\sigma_{p_T}/p_T$ is given by
\begin{equation}
\frac{\sigma_p}{p} = \frac{L^2}{8Rs}\frac{\sigma_s}{s} \sim \frac{p\sigma_s}{BL^2}
\end{equation}
Therefore the resolution of a tracker can be improved by increasing $B$ or $L$, or by decreasing $\sigma_s$. In sPHENIX, where the $B$ and $L$ are fixed via the reuse of the solenoid, all the effort has been put into reducing $\sigma_s$ as much as possible. The resolution for the total momentum $p$ in terms of the measurement of $p_T$ and $\theta$ can then be straightforwardly expressed as
\begin{equation}
(\frac{\sigma_p}{p})^2 = (\frac{\sigma_{p_T}}{p_T})^2 + (\frac{\sigma_{\theta}}{\text{sin~}\theta})^2
\end{equation} \par
It is illustrative to consider a particular example of relevance to the sPHENIX physics goals, such as the $\sim$5 GeV electron produced in the decay of an $\Upsilon$. The goal is to separate the $\Upsilon(1S,2S,3S)$ states from one another. The invariant mass of two massless particles is $M^2 = 2p_1p_2(1-\text{cos~}\theta)$, and to a good approximation electrons can be treated as massless. For the simplest case of the $\Upsilon$ decaying roughly at rest, the two electrons will be back-to-back in azimuth and equal in energy, so $1-\text{cos~}\theta\sim2$ and $p_1=p_2$. Therefore, $M \sim 2p$ and $\sigma_M \sim 2\sigma_p $. To keep the resolution $\sigma_M$ on the invariant mass of the di-electron signal around 100 MeV, the momentum of these electrons should be measured with a momentum uncertainty of $\sigma_P\sim50$MeV, or $\sigma_p/p\sim1\%$. Given the constraints of sPHENIX, that $|B|\sim$ 1.4T, and $L\sim80$cm, the radius of curvature of the electron will be $R=p/(0.3\cdot B)\sim10$ m, and the sagitta of that curve will be $s=L^2/(8\cdot R)\sim10$ mm, which needs to be measured to a precision of $1\%$, leading to a sagitta resolution requirement of $\sim$100 $\mu$m. The Gl{\"u}ckstern formula relates the resolution on a measurement of the $r$ and $\phi$ components of position to the resolution on the sagitta in the limit of a large number $N$ of equally spaced measurements.
\begin{equation}
\sigma_s = \frac{\sigma_{r\phi}}{8}\sqrt{\frac{720}{N+5}}
\end{equation}
To achieve the required sagitta resolution with $N=50$, the position resolution $\sigma_{r\phi}$ at each of the $N$ measurement points should be around 200 $\mu$m.
\par
\begin{figure}[htbp]
    \centering
    \includegraphics[width=9cm]{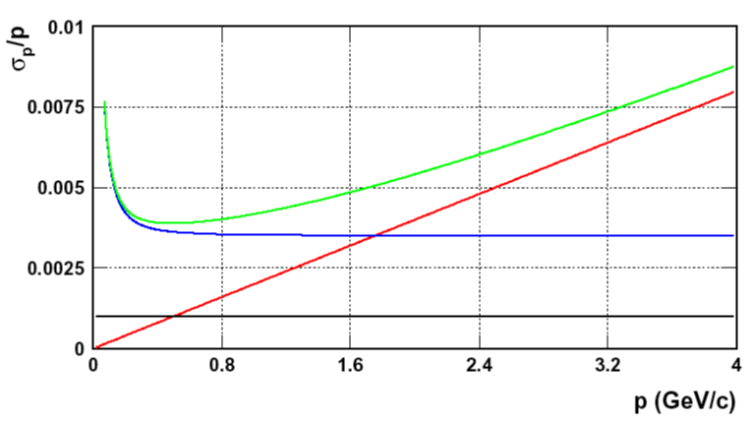}
    \caption{Illustration demonstrating the breakdown of contributions to the momentum resolution of a realistic tracking detector. The green curve represents the total momentum resolution, the blue represents the term coming from multiple scattering, and the red represents the momentum-dependent term from the detector position resolution. The black line represents the term coming from the resolution on $\theta$.}
    \label{fig:PRes}
\end{figure}

The above suggests that the measurement of momentum becomes extremely precise at low momentum. However, another crucial aspect that affects the momentum resolution of a tracking detector is multiple scattering. An excellent discussion of multiple scattering is given in Ref.~\cite{ParticleDataGroup:2018ovx}, and the necessary features are reproduced here. A charged particle incident on an atom will undergo Rutherford scattering on the electromagnetic field of the atom, which will change the trajectory of the incoming particle slightly. The trajectory of a charged particle leaving a material will thus be altered from the incoming trajectory by an angle $\theta$ corresponding to the sum of all the individual scatterings. 
\begin{figure}[htbp]
    \centering
    \includegraphics[width=12cm]{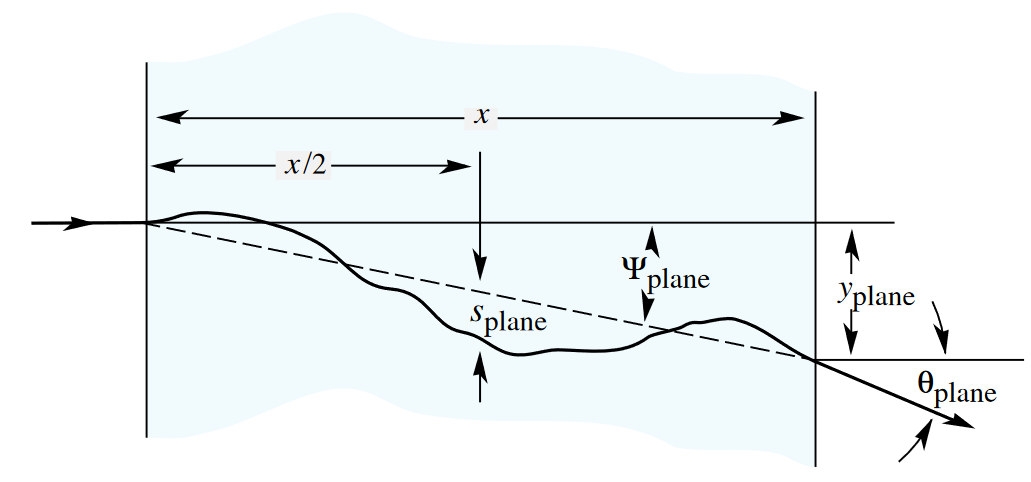}
    \caption{Kinematics of multiple scattering, figure from Ref.~\cite{ParticleDataGroup:2018ovx}. Particles undergoing multiple scattering experience longitudinal displacement ($y_{plane}$) as well as angular displacement ($\theta_{plane}$). For the purposes of tracking, the angular displacement is generally the most relevant effect.}
    \label{fig:PDGMS}
\end{figure}
This sum can, to a good approximation, be described as a Gaussian distribution with mean width of
\begin{equation}
\sigma_{\theta} = \frac{13.6\text{~MeV}}{\beta c p}\sqrt{\frac{L}{X_0}}~[1+0.038\text{ln}(\frac{L}{X_0\beta^2})]
\end{equation}
 where $\beta c$ is the velocity, $L$ is the distance travelled through the material, and $X_0$ is the radiation length. The charge of the particle is assumed to be unity. The contribution to the momentum resolution from this angular smearing can be given as
 \begin{equation}
\frac{\sigma_{p}}{p} \propto \frac{\sigma_{\phi}}{\phi} = \frac{13.6\text{~MeV}}{\beta c p}\sqrt{\frac{L}{X_0}}~[1+0.038\text{ln}(\frac{L}{X_0\beta^2})]\cdot\frac{R}{L}
\end{equation}
Where the substitution $R/L$ has been made in place of $1/\phi$. A further substitution of $R=p/(0.3B)$ yields
\begin{equation}
\frac{\sigma_{p}}{p} \propto \frac{1}{\sqrt{LX_0}B\beta}
\end{equation}
Which contributes a term roughly independent of $p$ to the resolution. A variety of conclusions can be drawn from this result. Clearly this term dominates the uncertainty on the measurement of low momentum particles, and tames the unphysical improvement in the resolution at vanishing momentum. At low momenta, the $\beta$ in the denominator tends to lead to significant worsening of the resolution, as can be seen schematically in Fig.~\ref{fig:PRes}.
The typical strategy employed to reduce the importance of this term is to reduce the distance particles must travel through materials with high $X_0$. Detectors are often arrayed in roughly cylindrical configurations for simplicity of mechanics and integration with solenoidal magnetic fields. In such detectors, there will be an inevitable increase in the distance $L$ traversed through detector material at larger values of $|\eta|$. For planar detectors, the position resolution also typically degrades at large incident angles due to the longer ionization deposit. These effects, coupled with the reduced component of the momentum transverse to the magnetic field, and thus the bending power, typically leads to a substantial degradation in resolution at higher angles. For these reasons, experiments seeking good resolution in the forward regions of $|\eta|\gtrsim1$ typically utilize detectors arrayed perpendicular to the beam.
\par
The above discussion was limited to an idealized set of circumstances that permit quasi-analytic treatment, uniform magnetic fields, materials, measurement uncertainties, etc. Certain approximations are made, the validity of which can reasonably be questioned. For the inevitably more complex situation in a real experiment, MC simulations are typically employed to determine the momentum resolution of a given tracking configuration. \par
With the ``theoretical" groundwork of tracking in place, we turn now to a discussion of the experimental and technological aspects. In general, charged particles can be measured more precisely than neutral particles due to the additional interactions with matter induced by their electromagnetic fields. There exists a broad phenomenology of useful effects that emerge from the interactions of relativistic charged particles with matter that can be leveraged for learning the velocities, energies, and momenta of charged particles. Cherenkov radiation, ionization, transition radiation, and brehmsstrahlung are some of the additional effects that can be used to gain information about a charged particle. Of these effects, tracking detectors almost~\cite{Baur:1993mw,Baur:1993mz} always prefer to use ionization. \par
Ionization is the process of freeing one or more electrons from an atom or molecule. In physics applications, ionization generally occurs in the context of strong electric fields. When the potential from an external electromagnetic field exceeds the potential binding an electron to an atom, the electron can become unbound from the atom. This process leaves behind a free electron and a positive ion. Depending on the field and the atomic binding energies, more electrons can be removed off a single atom as well. It can also be the case that the electron removed in the ionization process itself ionizes an atom, or an atom which is left in an excited state transfers its excitation to a different atom which then ejects an electron. These processes are known as secondary ionization, in contrast to the primary ionization caused directly by the external electric field. In gases, the number of electrons from secondary ionization often outnumbers those from primary ionization. In the context of particle physics, the strong external field that produces the ionization is generally the Lorentz contracted field of a relativistic particle~\cite{Blum:2008nqe}.\par
The use of ionization for tracking charged particles dates back to the cloud chambers invented by Wilson at the end of the 19th century. Ionization of gaseous, liquid, and solid media are all actively being used for tracking in particle physics experiments. The reason for the persisting dominance of ionization in tracking is in part that ionization is plentiful. Even in gases at standard temperature and pressure, the number of electrons freed by a relativistic charged particle can be on the order of 100 per cm of path length, meaning that measurements of position utilizing ionization suffer much less from low statistics than the other methods. The ionized electrons, which are emitted with eV-scale energies, can easily be manipulated by electric and magnetic fields inside the detector, unlike the photons produced by the other effects. Since the electrons from the original ionization can be manipulated, they are also comparatively easy to amplify and detect. Another benefit of ionization is that it is non-destructive, in the sense that particles lose comparatively little of their total energy in exchange for a large detectable signal. For gases, typical ionization energy losses are on the order of keV/cm, leading to an effectively negligible difference between the energy of a particle before and after leaving the tracker.\par
At modern hadron colliders such as RHIC and the LHC, where single A+A collisions can produce hundreds or thousands of particles, an increasingly important aspect of tracking is the ability to reconstruct tracks in the limit of large numbers of hits. An additional challenge is provided by $p+p$ operation, where the rate of collisions and the likelihood to have multiple collisions in the same bunch crossing increase substantially due to the higher luminosity achievable by the accelerator\footnote{The lower luminosity in A+A operation compared to $p+p$ operation is a result of the large Coulomb repulsion experienced by the heavily charged ions inside a bunch. This Coulomb repulsion tends to rapidly defocus the beam, resulting in lower luminosity.}. For this task, the detectors near to the interaction region necessarily must have low occupancy, meaning the number of hits must be substantially smaller than the number of total channels. Only semiconductor tracking detectors, such as the two to be described in the next section, are realistically capable of having such low occupancy in a small area close to the interaction point. These detectors generally use silicon as the ionization medium; in this way they are able to exploit the significant advances in semiconductor technology and nanofabrication. The philosophy of silicon tracking is to maximize the position resolution of a single hit as much as possible, at the cost of having fewer hits. Modern silicon pixel detectors envision pixel pitches on the order of $10\times10$ $\mu$m, providing position resolutions of $10/\sqrt{12}\sim3$ $\mu$m. \par
\begin{figure}[htbp] 
    \centering
    \includegraphics[width=12cm]{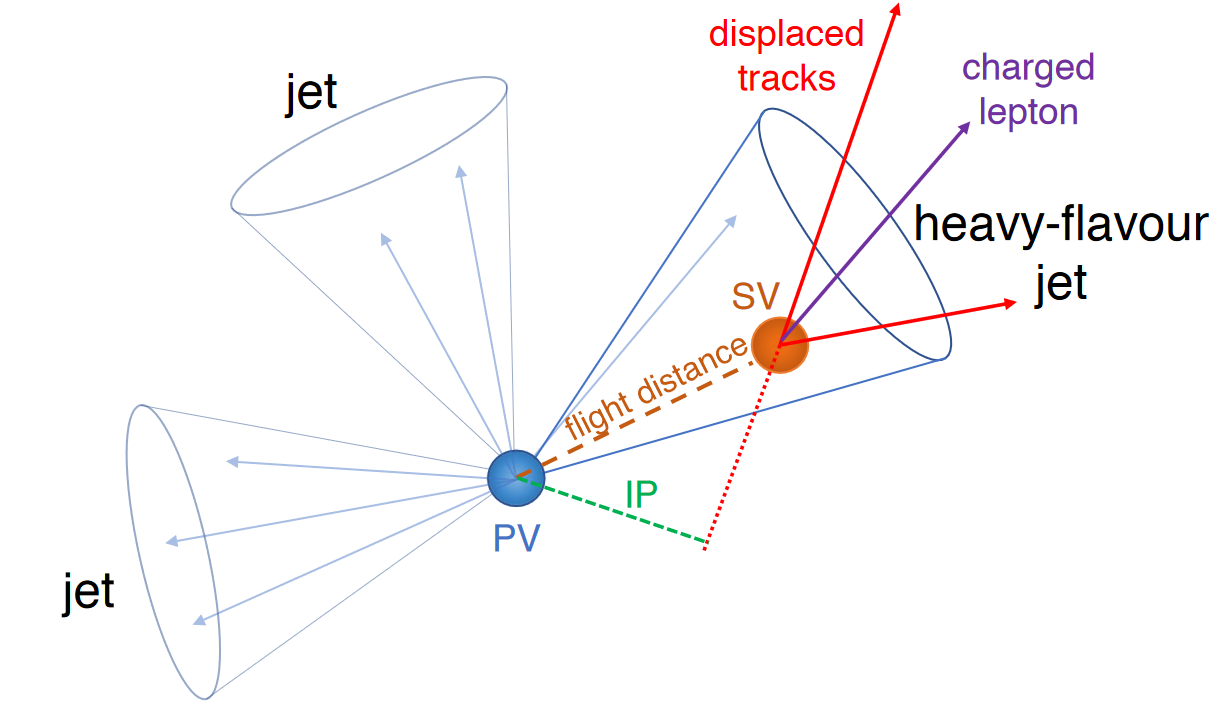}
    \caption{Illustration of the decay of a heavy flavor hadron. The primary collision vertex is denoted PV, and the secondary decay vertex of the hadron is denoted SV.}
    \label{fig:Displaced}
\end{figure}
 An additional benefit of detectors that provide high-precision position measurements placed near to the interaction region is the ability to distinguish not only the interaction vertex location, but also the ever-so-slightly displaced decay vertices of unstable particles with lifetimes on the order of hundreds of femtoseconds. Open heavy flavor hadrons, i.e. hadrons containing one heavy quark and the remainder light quarks, tend to disassemble themselves via the $W$-boson-mediated decay of the heavy quark to a lighter quark flavor. The timescale of that decay can be on the order of picoseconds, long enough that a particle with sufficient velocity can gain a measurable displacement from the primary vertex.\par
\begin{figure}[htbp] 
    \centering
    \includegraphics[width=12cm]{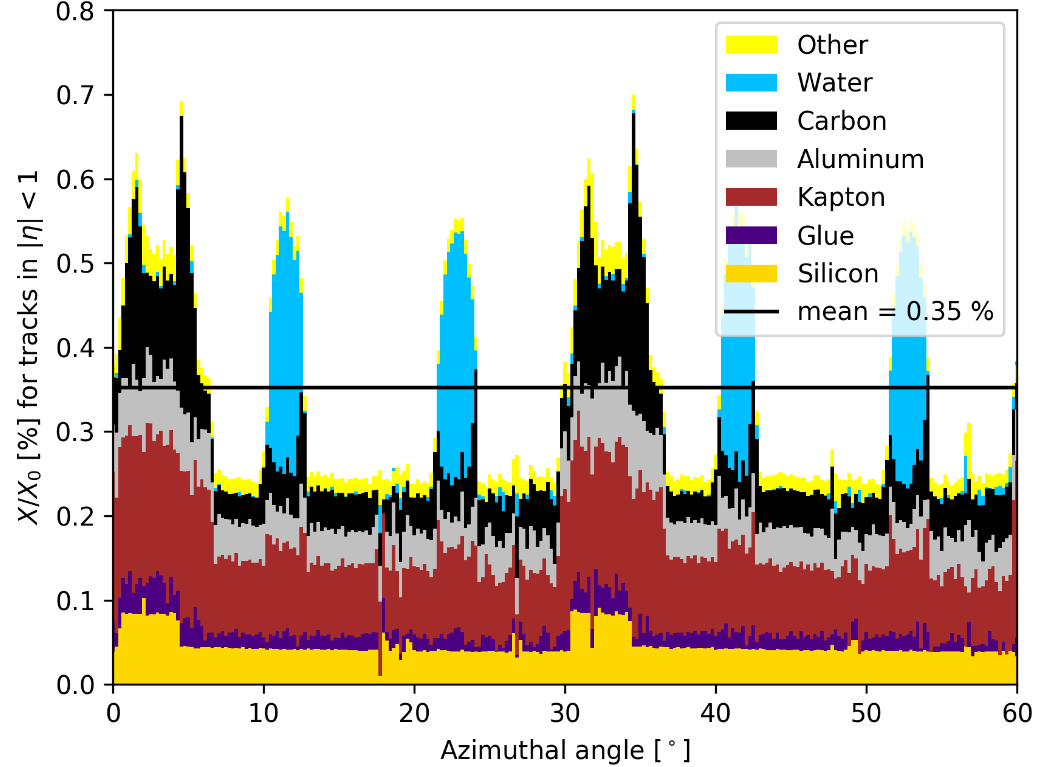}
    \caption{Material budget of the upgraded ALICE inner tracking system (ITS2) design, which represents the current state-of-the-art in terms of material reduction. The ITS2 achieves almost an order of magnitude in material reduction compared to more standard designs. Further R$\&$D is currently ongoing on a new ITS design, which removes essentially all non-silicon material.}
    \label{fig:ITSMat}
\end{figure}

 There are some disadvantages associated with the use of silicon detectors. The first is that silicon detectors are generally very expensive. The cost limits the ability to have a large number of radial measurements of track positions. The CMS tracker~\cite{CERN-LHCC-2000-016}, which represents the most extensive use of silicon detectors in any experiment, achieves around 10 hits for straight tracks at midrapidity. The result of this is that finding tracks out of the hits in the detector becomes a more challenging endeavor. In addition to the resolution considerations discussed above, a clear requirement of trackers is to be able to robustly and regularly find the tracks passing through them. In high multiplicity environments, tracks no longer stand out ``by eye" as they do in detectors with large numbers of position measurements. Thus advanced algorithms must be developed to disentangle the combinatorics of which hits can be associated with a given track. Another challenge is the multiple scattering in the detector material. Although the silicon itself can be very thin, the cooling systems and readout cabling generally increase the material budget substantially, as can be seen in Fig.~\ref{fig:ITSMat}.
The fact that this material is non-uniform and concentrated at discrete points in radius further complicates the tracking situation.\par
The difficulty of pattern recognition can be assuaged via the design employed by ATLAS, which utilizes an extremely precise silicon tracker in conjunction with a gaseous detector that provides a large number of hit points with moderate resolution. A similar technique is employed in sPHENIX, where the TPC nominally provides an additional 48 measurements of particle locations that can be used to find tracks. At RHIC the situation is less challenging than at the LHC for a variety of reasons, the particle multiplicities in events are smaller, the event rates are lower, and the number of pileup events is smaller. Pileup events are the case where multiple collisions occur within a single beam crossing, and in $p+p$ running at the LHC the mean number of pileup events in the general purpose detectors can be as high as 50.
\subsection{Intermediate Tracker}
\begin{figure}[htbp] 
    \centering
    \includegraphics[width=8cm]{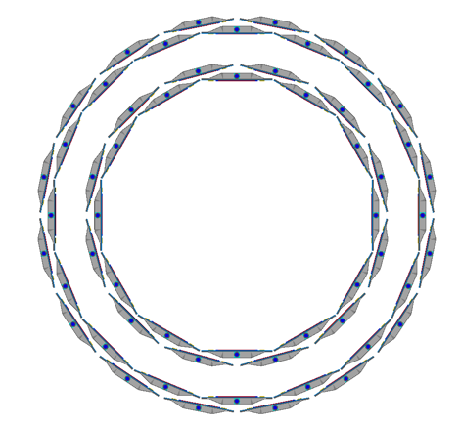}
    \caption{Layout of the INTT staves in sPHENIX.}
    \label{fig:INTTLayout}
\end{figure}

The Intermediate Tracker (INTT) is a multi-layer silicon strip detector. The outer radius of the INTT sits at 10 cm, and the sensors are arrayed into four overlapping layers as shown in Fig.~\ref{fig:INTTLayout}.
For straight tracks, typically the INTT will provide two measurements of position. The 16 outermost layers sit nominally at $r=10.26$cm\footnote{Measured at the distance of closest approach of the stave to the IR; the edges of the staves will sit slightly farther since the staves themselves are flat.}, and the 16 partner layers sit at $r=9.68$cm. The inner two layers of 12 staves each sit at $r=7.73$cm and $r=7.19$cm respectively. The length of the INTT active area is 46.55 cm.
\begin{figure}[htbp] 
    \centering
    \includegraphics[width=12cm]{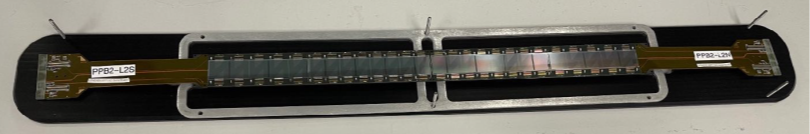}
    \caption{Single stave of the INTT.}
    \label{fig:INTTStave}
\end{figure}
Two pixelization schemes are used, both of which are much finer in $\phi$ than in $z$. The sensors spanning $|z| > 12.8$ cm are $78~\mu$m$\times20$ mm, while those with $|z| < 12.8$ are $78~\mu$m$\times16$~mm. The finer segmentation at smaller $|z|$ allows for lower occupancy at midrapidity. The detector consists of 360k channels, readout via the FPHX electronics chip, reused from the PHENIX FVTX. The radiation length of the INTT is around 1.1\% $X_0$ per layer. \par
An important aspect of the INTT is that the readout is much faster than either the TPC or MVTX. The INTT is capable of operating at the 106 ns bunch crossing timing of RHIC, while the TPC and the MVTX have $\sim14~\mu$s and $\sim5~\mu$s time delays on their respective readouts. This feature allows the INTT to accurately disentangle which tracks originate from which event, which will be crucial to handle the pileup in both the TPC and MVTX. The INTT will additionally help to reject backgrounds, which are generally out of sync with the beam crossings.
\begin{figure}[htbp] 
    \centering
    \includegraphics[width=12cm]{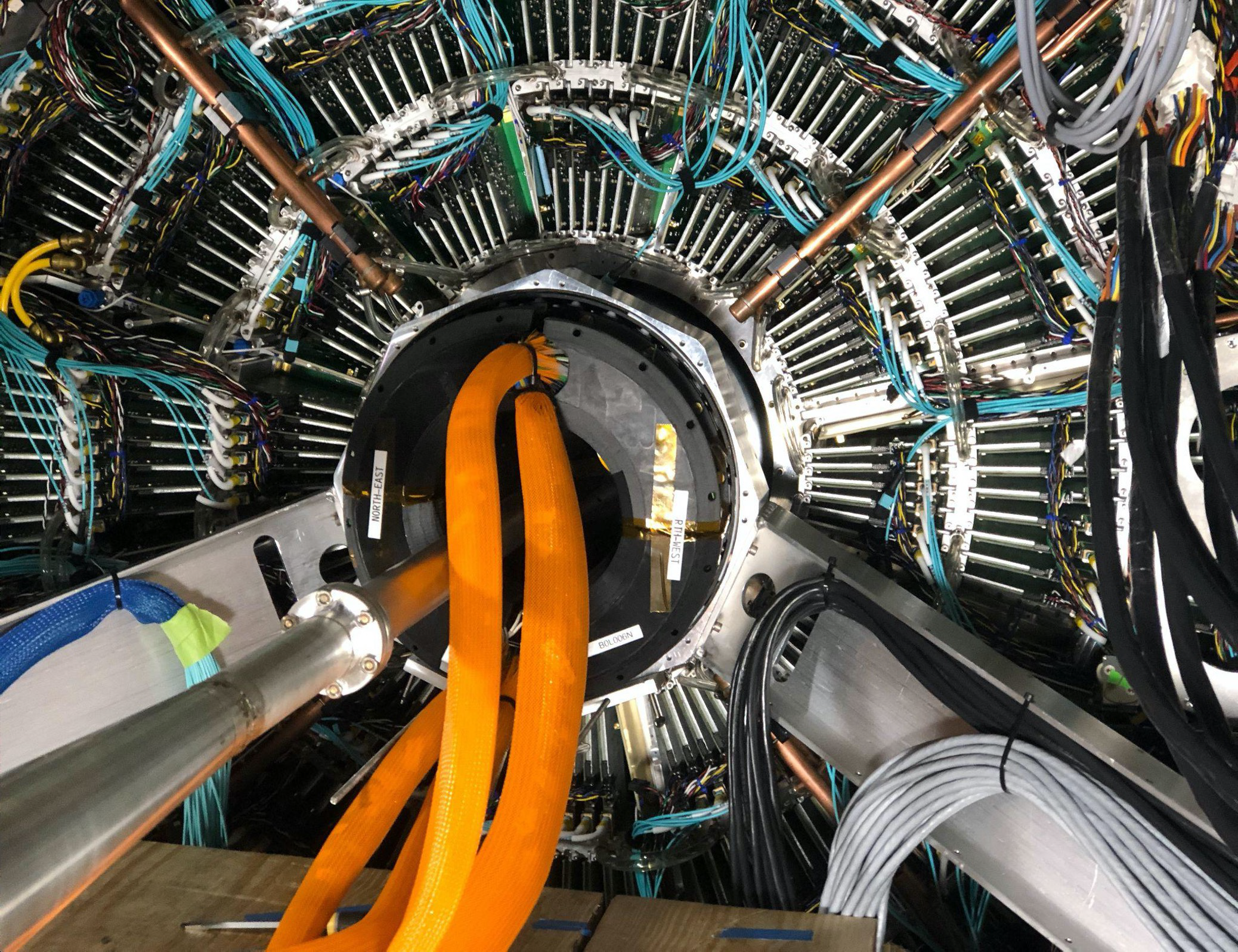}
     \caption{INTT (black) in position in sPHENIX, with the orange umbilical carrying the cables. The aluminum portion of the beampipe can be seen entering the detector from the bottom left of the image.}
    \label{fig:INTTCabled}
\end{figure}
\subsection{MAPS Vertex Detector}
The MVTX is a three layer silicon pixel detector with 27 $\mu$m $\times$ 29 $\mu$m pixels. The detector is a monolithic active pixel sensor, or MAPS, detector that takes advantage of the 180 nm TowerJazz CMOS technology. The design of the MVTX is based on the ALICE ITS2~\cite{ALICE:2013nwm}, and utilizes the same ALIPDE chips. The detector is 27.1 cm long, and the three layers, defined in the same way as the INTT layer locations, sit at $r_{\text{min.}}=$ 2.461 cm, 3.198 cm, and 3.993 cm respectively. Each sensor has an active area of 1.5 cm $\times$ 3 cm, There are in total 48 MVTX staves, each containing 9 ALPIDE chips. 
\begin{figure}[htbp] 
    \centering
    \includegraphics[width=12cm]{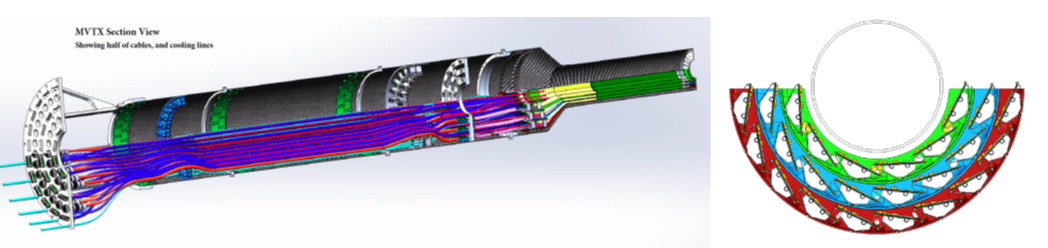}
     \caption{Left: Layout of the MVTX structure and services. Right: Cross section of the MVTX showing the layout and positioning of the staves within the three layers.}
    \label{fig:MVTXLayout}
\end{figure}
The radiation length per stave is only $\sim$0.3\%. This exceptionally low material budget is achievable due to the extremely low power consumption of the sensors and readout. The readout of the MVTX takes around 5 $\mu s$.
\begin{figure}[htbp] 
    \centering
    \includegraphics[width=12cm]{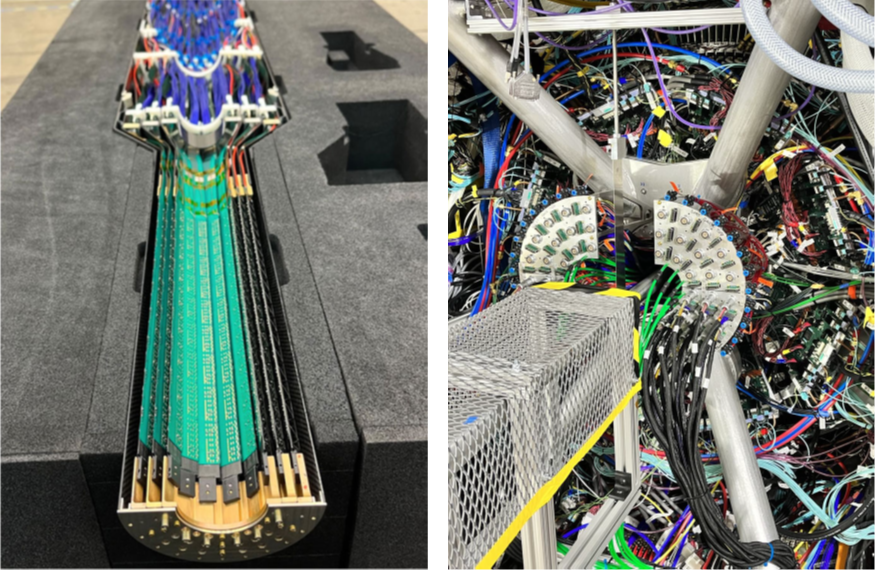}
     \caption{Left: A completed half-barrel of the MVTX. Right: The installed MVTX, with cabling underway. Since the image is taken on the south side of the experiment, the cable attachment locations and aluminum ``X-wing" support structure can be seen. Fig.~\ref{fig:MBDInstalled} shows the corresponding view from the north side.}
    \label{fig:FinalMVTX}
\end{figure}
The MVTX allows for precision measurements of displaced vertices, as discussed in Sec.~\ref{Sec:ChPT}. The precision on the distance-of-closest-approach (DCA) of particle trajectories to the primary vertex is around 40$~\mu$m for particles with $p_T>0.5$ GeV, allowing for precise measurements of, among others, decays of $B$ mesons, which have lifetimes on the order of 1.5 femtoseconds, corresponding to a $c\tau$ of $\sim450~\mu$m. The detector sits as close as possible to the beryllium beam pipe to improve the DCA resolution. The MVTX was tested extensively prior to installation, and was installed at the end of March 2023. After installation, quality assurance tests were performed that ensured that greater than 99.99\% of the channels on each stave were live, to be compared to the sPHENIX requirement of 99\% live channels.
\subsection{Additional Detector Subsystems}
In addition to the subsystems above, sPHENIX has three more detectors which aid in the event reconstruction. \par 
\subsubsection{Min-Bias Detector}
\label{subsec:MBD}
The first is the min-bias detector (MBD), which provides the trigger for the experiment. The MBD is a reuse of the PHENIX beam-beam counter, and consists of quartz cherenkov radiators and mesh-anode PMTs. It is segmented into 64 $\sim$ 1 inch hexagonal elements, each of which are optically isolated. The primary goal of the MBD is to determine if and when an event has occurred. In central A+A events, a large number of charged particles are inevitably produced in the forward direction. These forward particles can be used to determine whether an event occurred, roughly how ``violent" i.e. how central the collision was, and, due to the good time resolution of the PMTs, the location of the event vertex in z. 
\begin{figure}[htbp] 
    \centering
    \includegraphics[width=12cm]{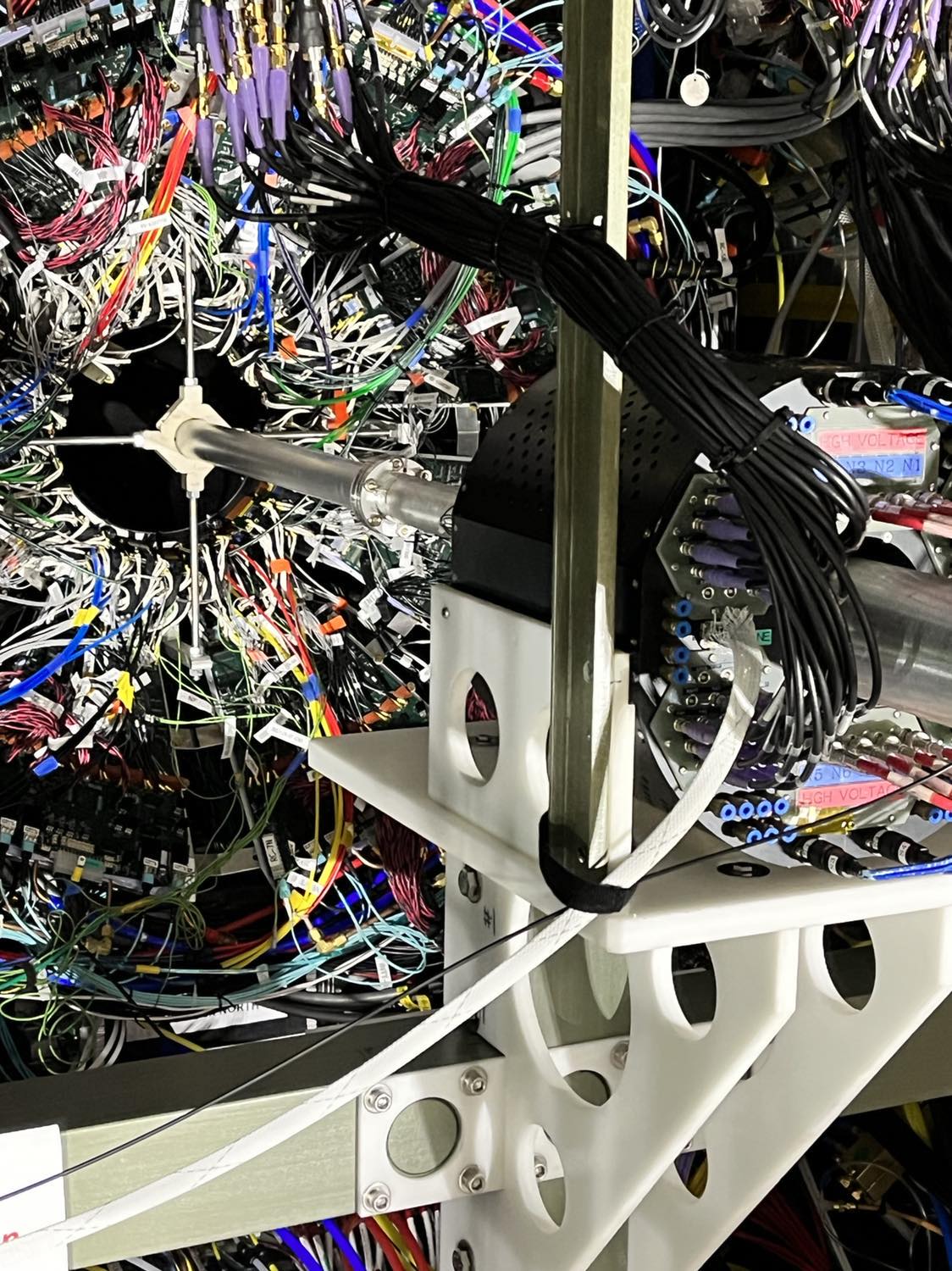}
     \caption{The north-side MBD (black, right) installed around the beampipe. The MBD support, and the cables originating from the PMTs can be observed.}
    \label{fig:MBDInstalled}
\end{figure}
The MBD covers $|\eta|$ between 3.5 and 4.6. Since the number of charged particles produced in central and semi-central A+A collisions in this pseudorapidity region is high, the MBD has an excellent efficiency for accepting these events.
\subsubsection{Event Plane Detector}
\label{subsec:EPD}
The sPHENIX event plane detector (sEPD) is a series of scintillating tiles, similar in concept to those described above used in the HCal, which provides information about the ``event plane" of heavy ion collisions. It is very rare that two nuclei collide completely overlapped. More often, they collide with a non-zero impact parameter, producing an almond-shaped region of overlap where the QGP can be formed. This shape manifests itself in the directions that the final state particles are produced; due to the lower confining pressure the QGP rapidly expands along the short axis of the almond shaped region, resulting in an azmiuthal anisotropy. The event plane is the axis along which the QGP expands, and measuring it enables performing measurements of a slew of observables ``in-plane" and ``out-of-plane". In particular, this is highly beneficial for learning about how jets interact with the medium. By simply measuring the amount of charged particles that are produced in the forward direction, this event plane information can be extracted.\par
The sEPD consists of two halves sitting at $\pm z=319$ cm , covering $|\eta|$ between 2 and 4.9. The inner radius of the detector is 4.6 cm, and the outer radius is 90 cm. Each half is constructed out of 12 sectors of 1.2 cm thick scintillating tiles, machined into 31 optically isolated tiles, for a total of 372 channels in each half. Each of the 31 tiles has three loops of WLS fiber embedded into them for readout of the scintillation light. The sectors are wrapped first in reflective coating and then black coating for light-tightness. The WLS fiber is readout by an SiPM. A similar detector design was recently implemented in STAR~\cite{Adams:2019fpo}.
\begin{figure}[htbp] 
    \centering
    \includegraphics[width=8cm]{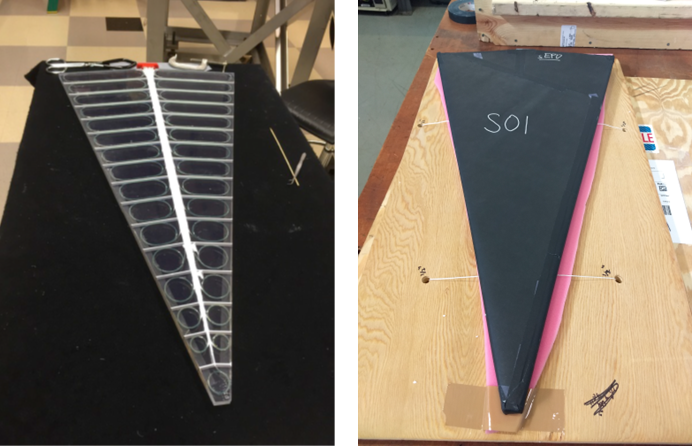}
     \caption{Left: Scintillating tile and readout of the sEPD. The WLS travel down the center of the tile to reach the SiPM readout. Right: Finished sector wrapped in light-tight covering.}
    \label{fig:sEPD}
\end{figure}
\subsubsection{TPC Outer Tracker}
\label{subsec:TPOT}
\begin{figure}[htbp] 
    \centering
    \includegraphics[width=8cm]{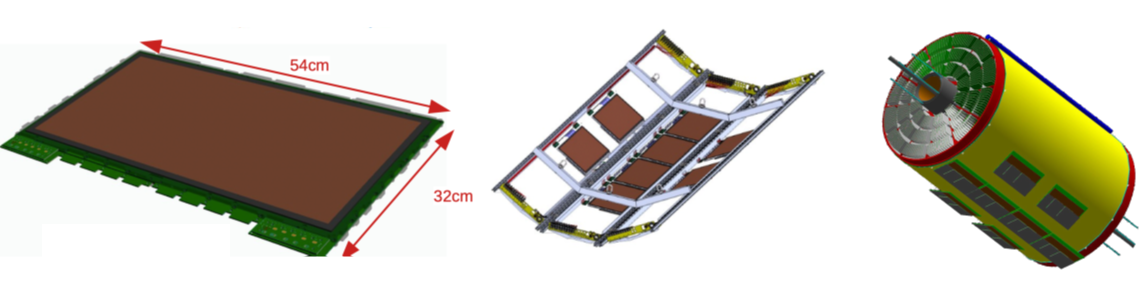}
     \caption{Left: Design of a TPOT module. The modules themselves cover a large area}
    \label{fig:TPOT}
\end{figure}

Outside the TPC sit 8 MicroMegas trackers known collectively as the TPC outer tracker (TPOT). The TPOT provides additional high resolution measurements of tracks at large radius. Since the lever arm of the TPOT is large, the angular error on the positions of tracks is significantly diminished. This allows tracks measured in the TPOT to be used in the calibration procedure of the TPC. The measured positions of ionization deposits in the TPC will not correspond directly with the locations at which they were produced, due to field non-uniformities or space charge\footnote{For more detail on position distortions in the TPC, see Sec.~\ref{Sec:SpaceCharge}.}. The trajectories of tracks measured in the MVTX, INTT, and  TPOT will be well-known, and the distortions in the positions of ionization electrons as measured by the TPC can be extracted. The TPOT modules are arranged as shown in Fig.~\ref{fig:TPOT}.\par
Each TPOT module is a double-sided, planar, bulk MicroMegas\footnote{See Sec.~\ref{subsec:Amplification} for a description of the MicroMegas concept.} detector utilizing 1D strip readout. The two halves of the detector have strips of 1 mm and 2 mm respectively oriented in orthogonal directions, such that the 2D location of incident tracks can be pinpointed. The charge from the mesh of the MicroMegas is avalanched onto a resistive layer, which is detected via image charge in the readout strips. The TPOT gas mixture is 95\% argon and 5\% isobutane. Each module is read out using two of the TPC front end electronics cards, described in detail in Sec.~\ref{subsec:Electronics}.
\chapter{sPHENIX Time Projection Chamber} \label{Chap:TPC}
To achieve the physics goals of sPHENIX, a compact TPC was designed and constructed. The TPC was installed in the sPHENIX interaction region in Spring 2023. This chapter is intended to provide a comprehensive review of the design of the TPC. To give sufficient background, some introduction to gaseous detectors and generic features of collider TPCs are described in Sec.~\ref{Sec:TPC}. Section ~\ref{Sec:DesignConsiderations} discusses many of the driving factors contributing to the design of the sPHENIX TPC, while Sec.~\ref{Sec:construction} presents the design choices made and their implementations in the construction of the detector. 
\section{Time Projection Chambers}
\label{Sec:TPC}
\begin{figure}[htbp]
    \centering
    \includegraphics[width=10cm]{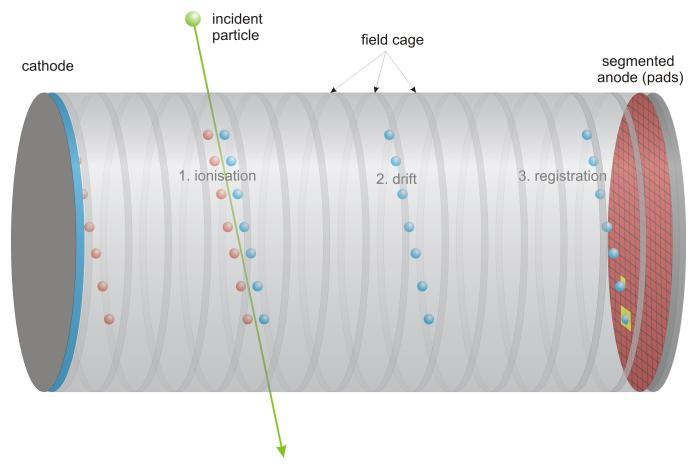}
    \caption{The basic operating principle of a TPC. Ionization electrons from an ionizing charged particle drift a long distance through the medium before being registered at the endcap. The field which causes the electrons to drift is provided by a cathode opposite to the detection plane. The electric field is shaped to point directly towards the detection plane by the field cage. Figure from the LCTPC Collaboration}
    \label{fig:TPCPrinciple}
\end{figure}

TPCs have become a workhorse of modern particle physics due to their versatility and simplicity. TPCs have been used in low-energy nuclear physics, proton decay, neutrino physics, dark matter searches, and as collider detectors. The concept was first proposed in 1974 by Nygren~\cite{Nygren:1974nfi}, for use at the PEP collider at SLAC. The basic principle of a TPC is demonstrated in Fig.~\ref{fig:TPCPrinciple}.
The ``time-projection" moniker is derived from the capability to measure the distance from the readout plane at which the ionization event occurred via a measurement of the ionization arrival time. There exist currently a huge variety of implementations of this concept, and they are in large part beyond the scope of this work. Instead, the following will focus on \emph{gaseous} TPCs as are typically used at colliders. In the collider context, TPCs generally operate in magnetic fields, with the orientation of the magnetic field and the electric field either parallel or anti-parallel. The magnetic field serves to significantly reduce the transverse\footnote{Transverse and longitudinal in this context are defined with respect to the drift direction. In a TPC with electric field in the $z$-direction, transverse diffusion shifts the positions of electrons in the $x$-$y$ plane, while longitudinal diffusion shifts the measured positions in $z$.} diffusion of the electrons, thus enabling precise measurements of the original $r\phi$ location of the ionization event even after the electrons have drifted long distances through the gas. Historically, planar multi-wire proportional chambers (MWPCs)~\cite{Charpak:1970az} located at the endcaps of TPCs have been used to detect the ionization electrons. To understand the advantages of TPCs, it is instructive to compare them with other technologies designed to perform similar tasks.
\par
The most direct point of technological comparison is the drift chamber\footnote{A point of terminology: the generic term for a detector that uses wires to detect ionization is ``wire chamber". A subset of wire chambers, drift chambers utilize the timing information of how long the electrons drifted through the gas to more precisely locate the original ionization location. The distinguishing feature of drift chambers is therefore the capability of the electronics to precisely distinguish time.}; an example of which is given by the H1 CJC described in Sec.~\ref{sec:H1tracking}, which instruments a volume of gas with a large number of long wires. Wires known as ``potential" wires produce an electric field which guides the ionization electrons to the so-called ``sense" wires, which amplify and collect the ionization. A typical drift chamber configuration is divided into ``cells", pockets of space whose electric field lines are shaped to terminate on the same sense wire or wires. The number of potential wires required to shape the electric field in the cell is typically larger, occasionally by factors of up to 20, than the number of sense wires. 
\begin{figure}[htbp]
    \centering
    \includegraphics[width=10cm]{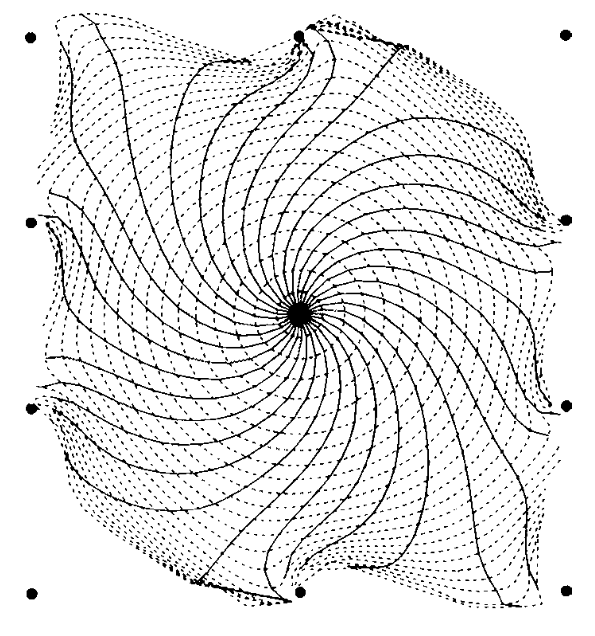}
    \caption{The wire configuration and drift lines of a cell from the ARGUS drift chamber. The wire configuration (Large black dots) is typical, and similar configurations have been used by many experiments. The maximum drift distance in the cell is around 1 cm. The dashed lines represent ``isochrones", lines of constant arrival time. The temporal separation between the dashed lines corresponds to arrival time increments of 15 ns. The solid lines represent the electron trajectories. The achieved position resolution with this configuration was $\sigma_{r\phi}=0.2$ mm. Figure from Ref.~\cite{Blum:2008nqe}}
    \label{fig:ArgusLines}
\end{figure}
The necessity for more potential wires is governed by the requirements on the field uniformity and the required size of the cell. Generally to reach high precision with a drift chamber, the uniformity of the electric fields and the positions of the sense wires must be known with similarly high precision. Both sagging under gravity, and the electrostatic forces between the wires sitting at different potentials must be compensated. The wires therefore must be under significant tension, and this tension must be maintained over time. To support this tension requires highly rigid endcaps. As the number of wires increases, so too do the mechanical stresses on the endcaps of the detectors. Thus, to add more wires and improve the position resolution of the detector, more supporting material must be added, which correspondingly increases the material budget of the detector. Such thin wires also have a tendency to break or detach from their anchor points on the endcap. This can result in large areas of the detector being inoperable until a repair can be made. Long wires are also susceptible to mechanical vibrations which produce a variety of undesirable effects. One of the major advantages of the TPC is the ability to do away with the mechanical challenges imposed by long wires. This comes with the added benefit that the number of active channels can more easily be increased compared to drift chambers, where adding an additional sense wire requires the addition of several more potential wires. As electronics improve in the ability to handle large numbers of channels and decrease in cost, this advantage of the TPC has become a crucial one. The ALICE TPC at the LHC consists of 557568 channels, which, assuming 6 potential wires per sense wire, and 50 grams of tension per wire, would require a mechanical structure capable of supporting at least 165000 kg.\par
Another disadvantage of drift chambers is the so-called ``left-right" ambiguity, which is a result of the fact that a signal arriving at a given time from a sense wire generally cannot be uniquely mapped to one spatial location. Resolving this ambiguity through the design of the detector often involves introducing additional mechanical complexity to the detector design. TPCs, by virtue of the fact that electrons approach the readout from only one side and nominally only one trajectory, inherently lack this ambiguity. This typically reduces the complexity of the pattern recognition techniques that must be employed for tracking in TPCs. TPCs additionally benefit from the typical magnetic field orientation in collider experiments, while it complicates the usage of drift chambers due to the non-zero angle between the electric and magnetic fields. The so-called Lorentz angle is the angle between the electric field and the electron drift direction, which is generally induced by the effects of a magnetic field on the electron drift. An example of this effect can be seen in the ARGUS configuration shown in Fig.~\ref{fig:ArgusLines}, where instead of the naively expected straight electron trajectories, the trajectories exhibit a complicated spiral pattern. Large Lorentz angles tend to spread out the ionization signal in time, decreasing the position resolution for a drift chamber. Since higher magnetic fields tend to result in larger Lorentz angles, there is a competing requirement between wanting to maximize $\Vec{B}$ to increase the curvature of tracks for tracking and wanting to minimize $\Vec{B}$ to reduce the Lorentz angle. In a typical collider TPC, the alignment of the magnetic and electric fields means that the Lorentz angle is essentially zero. This, combined with the fact that higher $\Vec{B}$ fields reduce the amount of transverse diffusion, means that unlike drift chambers, the position resolution of TPCs improves substantially as a function of $\Vec{B}$. \par
An important aspect of almost all TPCs\footnote{The TPC described later in this thesis being a notable exception.} is particle identification. In addition to fulfilling the role of tracking detectors, gaseous detectors such as TPCs and drift chambers can be used for identification of particle species via their energy loss to ionization (dE/dx). The key point is that the amount of primary ionization produced by a charged particle is a function of the velocity of the particle. Thus, a measurement of the amount of primary ionization provides a measurement of the particle velocity, which in conjunction with the measured momentum of the particle can be used to determine the particle mass. The two heavy ion collider TPCs operating currently, STAR and ALICE, measure dE/dx to a precision of $\sim$6-8\% and $\sim$5-7\%~\cite{Yu:2013dca} respectively, allowing for good separation of pions, kaons, and protons up to around 1 GeV, at which point the amounts of ionization produced by the particle species become roughly the same\footnote{The keen reader may recall that the dE/dx curves reach a minimum and then begin to rise again at high momenta. This is known as the ``relativistic rise" region, and many experiments have attempted to exploit this fact to perform PID at high momenta, to mixed success. One major complication is that the ``rise" in total dE/dx comes predominantly from secondary ionization processes that deposit large amounts of energy in a small region, such as knock-on electrons. The standard technique for measuring dE/dx employs a \emph{truncated} mean, i.e. the largest samples are discarded. This is done to prevent large amounts of secondary ionization from biasing the measurement of the primary ionization. When the largest samples of the energy are removed, the effect of the relativistic rise decreases significantly, and the dE/dx at high momentum looks similar to the dE/dx of a minimum ionizing particle (MIP)~\cite{ParticleDataGroup:2018ovx}.}. \par 
One caveat for this technique is the fact that primary ionization cannot be measured directly, since primary and secondary electrons are generally released at effectively the same time and position as the primary electrons. The amount of primary ionization must be inferred from the amount of total ionization, which is a factor of 3.9 and 3.3 larger for the STAR and ALICE gases respectively. This means that a large number of ionization events should be measured to reduce the statistical fluctuations in the number of secondary electrons. The total number of ionization electrons produced per centimeter of path length on average is $\sim$90 in the STAR gas mixture of 90\% argon and 10\% methane (CH$_4$) and $\sim$50 in the original ALICE mixture of 90\% neon and 10\% CO$_2$. This can be compared to the number of primary electrons in each case, which are $\sim$23 for STAR and $\sim$14 for ALICE.\par
For the purposes of tracking, which will be the primary focus of the detector studied in the rest of this thesis, the precise number of electrons produced in the gas is not critical to measure. However, it is crucial that the signals produced by the detector are large enough to be accurately registered in room-temperature electronics. To produce these signals, the amplification of the ionization electrons, described in the following section, is typically necessary.
\subsection{Amplification}
\label{subsec:Amplification}
Essentially all gaseous detectors require amplification of to produce a signal that can easily be distinguishable from noise\footnote{The exception to this rule is the so-called ionization chamber, which directly reads the ionization as a current and can be used in the case of very heavily ionizing particles such as the $\alpha$ produced by Am$^{241}$. This is the operating mechanism of standard Am$^{241}$ smoke detectors.}. Generally in collider experiments the factor by which the signal is amplified, known as the gain, is $O(10^3-10^5)$. Modern collider experiments almost always run gaseous detectors in ``proportional" mode. Proportional mode denotes the regime of amplification where the size of the signal is directly proportional to the number of electrons entering the amplification stage, i.e. one primary electron provides a distribution of signal magnitudes fairly tightly clustered around the mean value. That is to say the gains for each electron are roughly the same, which is vital for particle identification via dE/dx. Even beyond measurements of dE/dx, this proportionality is in general an attractive feature, as it ensures that signals from electrons will not be so small that they are lost to noise or so large that they oversaturate the electronics. Decreasing the necessary dynamic range of the electronics in this way typically allows greater precision and lower costs. \par
\begin{figure}[htbp]
    \centering
    \includegraphics[width=10cm]{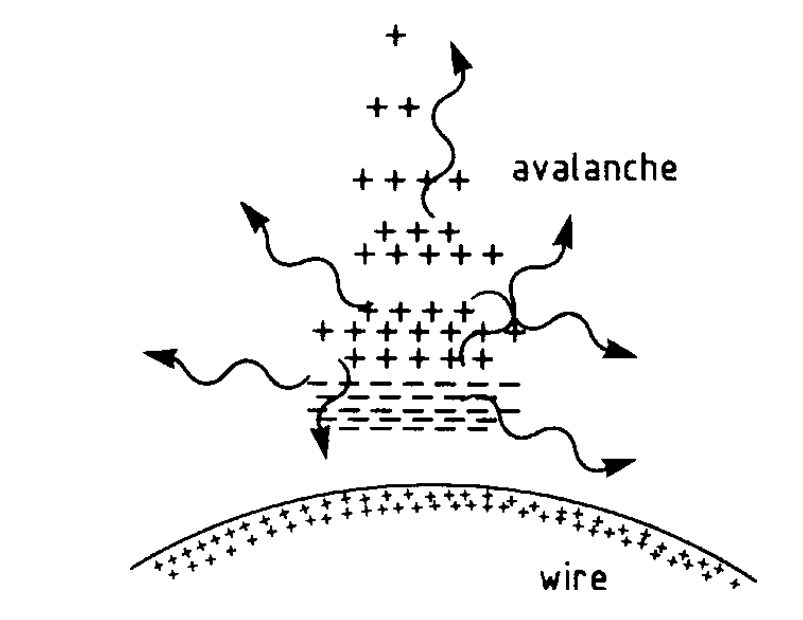}
    \caption{The Townsend avalanche effect. In this case, the avalanche is initiated due to the high electric field near a wire. A consequence of the avalanche is the emission of photons, typically in the UV range. Figure from Ref.~\cite{Blum:2008nqe}.}
    \label{fig:Avalanche}
\end{figure}

The simplest and by far most common technique for amplifying ionization signals in gases is the Townsend avalanche technique, wherein the ionization electrons themselves serve to amplify the signal by further ionizing the gas. In a strong electric field, the ionization electrons can gain enough energy to ionize the gas atoms, producing more electrons which will ionize more gas atoms. The result is known as an ``avalanche" of electrons, and the phenomenon is diagrammatically sketched in Fig.~\ref{fig:Avalanche}. The Townsend coefficient\footnote{Occasionally referred to as the \emph{first} Townsend coefficient, to distinguish it from the \emph{second} Townsend coefficient, which describes electron emission from the impact of an ion on a solid surface.}, denoted often as $\alpha$, describes the number of additional electrons produced by an electron in a given path length. $\alpha$ is a strong function of electric field, simply because the kinetic energy that an electron gains from the field between its collision with the gas molecules can be higher in a higher electric field. The higher the kinetic energy of the electron, the more likely it is that the electron will ionize the next molecule or atom it collides with.
\begin{figure}[htbp]
    \centering
    \includegraphics[width=8cm]{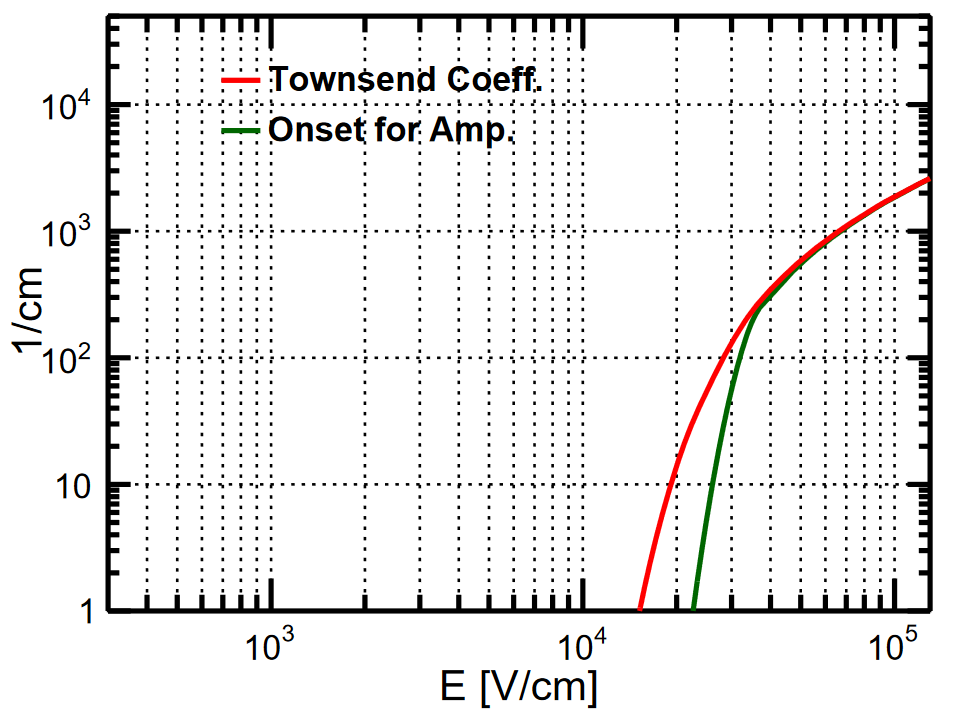}
    \caption{Townsend coefficient for a mixture of 60\% argon and 40\% CF$_4$. The green ``onset for amplification" curve shows the Townsend coefficient minus the number of electrons which re-attach to gas atoms in the same distance. This re-attachment can cause the electric field necessary to actually achieve amplification to be higher. The amplification begins around 20kV/cm.}
    \label{fig:Townsend}
\end{figure}
The electric field close to a vanishingly thin wire scales as $1/r$, producing extremely high electric fields near the wire if even a modest potential is applied to it. 
The qualities of an avalanche depend greatly on the gas mixture, and not all gases can achieve a proportional avalanche in a favorable way. The behavior of the amplification depends heavily on the degree of absorption of the photons produced in the primary amplification. Gases that lack a sufficient amount of ``quencher", a component of the gas with rotational and vibrational degrees of freedom that can absorb photons produced by the avalanche, can have self-sustaining avalanches where the photons produced by the avalanche can travel macroscopic distances and produce additional ionization. Quenching gases are generally molecules such as CO$_2$ or CF$_4$. Fortunately, the addition of a quencher also has favorable effects on the electron transport properties of the gas, as will be seen in the next section. The modes of operation that make use of the photons and their additional avalanches are known as ``streamer mode" or ``Geiger mode", and they tend to produce long-lived signals, which is typically not favorable for collider detectors. Geiger mode is the case where any ionization forms a chain reaction that discharges a large number of avalanches all along the wire. This chain reaction is only stopped by the buildup of positive ions that screen the charge of the wire, effectively reducing the potential that is ``seen" by the electrons attempting to start another avalanche. Since another signal cannot be registered until these ions have dissipated, there is an inherent dead time on the order of 1 ms during which the detector cannot produce another signal. In streamer mode, there is sufficient non-ionizing absorption of the photons to contain the avalanches to a single region of the wire, instead of causing discharges all along the length. The signal size in both of these modes is essentially the total number of avalanches produced, which in the case of streamer mode is controlled by the number of photons from the primary avalanche that are absorbed, and in the case of Geiger mode is proportional to the length of the wire that must be entirely shielded by ions to terminate the amplification. Geiger mode produces signals which are not at all proportional to the number of incoming electrons. The usefulness of Geiger mode is therefore contained to low-rate (by collider standards) applications where a large gain from the amplification stage is a necessity. Streamer mode can have a limited degree of proportionality, but necessarily the utilization of the additional photon-initiated avalanches will come at the price of larger statistical fluctuations, longer signal lifetimes, and longer dead times. Both streamer mode and geiger mode are useful in that they are cost-effective; the signal sizes are generally much larger and do not require sophisticated readout schemes, which are in general expensive. A benefit of proportional mode is that since the signals are fairly small, the recovery time of the detector is fast. \par
\begin{figure}[htbp]
    \centering
    \includegraphics[width=13cm]{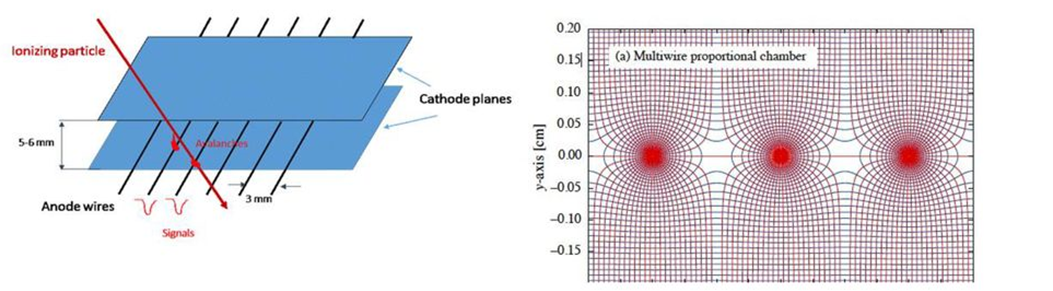}
    \caption{Sketch of the MWPC operating principle (left) and electric field lines (right). The implementation of MWPCs in TPCs is similar to the figure on the left, except one of the two cathodes is displaced by a large distance, on the order of a meter or more for typical collider TPCs.}
    \label{fig:MWPC}
\end{figure}

The proportional aspect of the MWPC made it an ideal choice for use in the first TPCs. The basic operating principle of the MWPC is sketched in Fig.~\ref{fig:MWPC}.
An important feature in gas detectors seeking high precision for position measurements is the mechanism of ``charge sharing". For a single hit registered on a pad of width $w$, the probability distribution for the location of the incoming hit is the continuous uniform distribution, i.e. a constant probability at all points across the width of the pad, from $x=-w/2$ to $x=w/2$. The variance of this distribution is known to be 
\begin{equation}
V(x) = \int_{-w/2}^{w/2} (x-\frac{-w/2+w/2}{2})^2 \,\frac{dx}{w/2+w/2} = \frac{(w/2+w/2)^2}{12} = \frac{w^2}{12}
\end{equation}
Thus for hits on a single pad, the resolution with which the position in the direction of the pad width is known is $\sigma_x=w/\sqrt{12}$. This statement holds for any detector registering a single hit at a given location, including the silicon detectors described in Sec.~\ref{Sec:ChPT}. To improve the resolution of the detector as a whole, one method is to decrease $w$, although this correspondingly increases the number of channels in the detector and thus the cost, cooling required, etc. Another technique is to design the detector in such a way that \emph{multiple} pads are hit by every signal. In an ideal detector with no noise and two pads measuring signals of size $a$ and $b$, if $a\sim b$, then the origin of the signals was centered between the pads. If $a\gg b$, then the origin of the signals was more centered on pad $a$. This allows for the coordinate $x$ to be measured with a precision limited only by the uncertainty in the measurement of the signal magnitudes and the dynamics of how the charge propagates from the origin to the pads.\par
\begin{figure}[htbp]
    \centering
    \includegraphics[width=13cm]{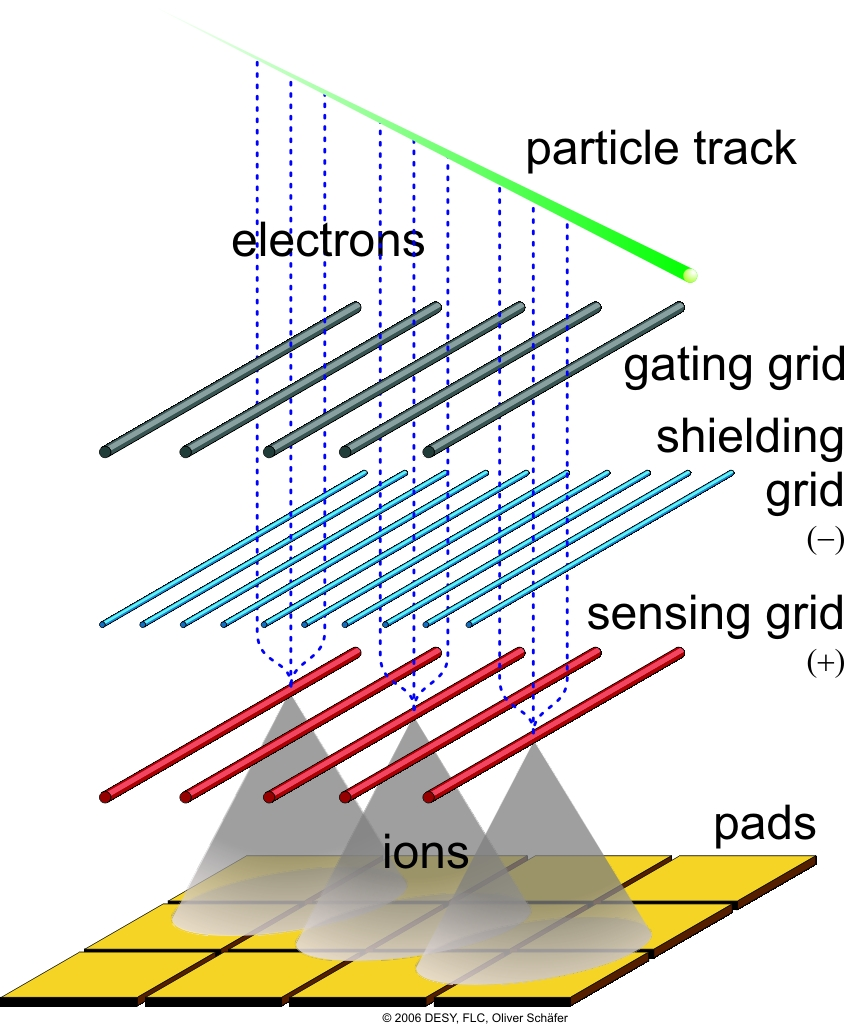}
    \caption{Sketch of the typical implementation of MWPCs in collider TPCs. The gating grid can be ``closed" by introducing alternating voltages on the wires which cause field lines to terminate on the grid. Closing the grid prevents ions from the avalanche from flowing back into the drift volume. The shielding grid protects the sense wires from the field of the gating grid. The cloud of ions surrounding the wire after the avalanche induces an image charge on the pads that allows them to be used as readout in addition to the wires. Image from DESY FLC group.}
    \label{fig:MWPCForTPC}
\end{figure}

One drawback of the specific implementation of the MWPC sketched in Fig.~\ref{fig:MWPC} is that the only position information comes from the wires. This allows the coordinate perpendicular to the wires to be measured precisely, but limits the precision of the measurement in the coordinate parallel to the wires. In collider experiments, since tracking requires measurements of both the $r$ and $\phi$ components of position, generally the cathode is segmented and the image charge produced in the cathode by the avalanche onto the wire is used as an additional measurement of position. A sketch of this design is shown in Fig.~\ref{fig:MWPCForTPC}.
\subsubsection{Beyond Wires}
\label{SubSec:beyondwires}
One challenge in the use of wires for amplification is the aging of the wire after many avalanches. The phenomenon of aging is in general poorly understood due to the complex electrochemistry involved, but the general mechanism involves the buildup of polymers from the dissociated components of the gas. These polymers can create an insulating or conducting layer on top of the wire, both of which decrease the achievable gain. In drift chambers, both the potential wires and the sense wires are susceptible to aging effects. Aging can cause wires to become totally insensitive in certain regions, or cause them to continually draw high currents and/or avalanche. \par
\begin{figure}[htbp]
    \centering
    \includegraphics[width=13cm]{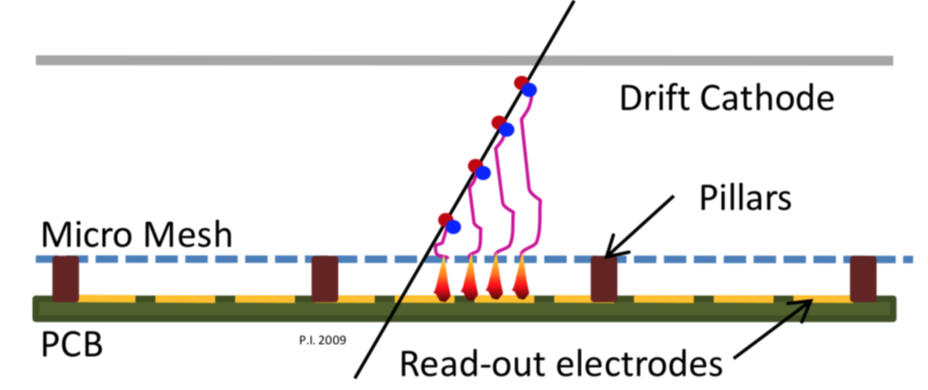}
    \caption{Operating principle of MicroMegas. In the case of operation as a TPC, the drift cathode is far from the mesh and the incident particle typically does not pass through the readout plane. Image from Ref.~\cite{Bhattacharya:2020bqp}}
    \label{fig:Micromegas}
\end{figure}
\begin{figure}[htbp]
    \centering
    \includegraphics[width=10cm]{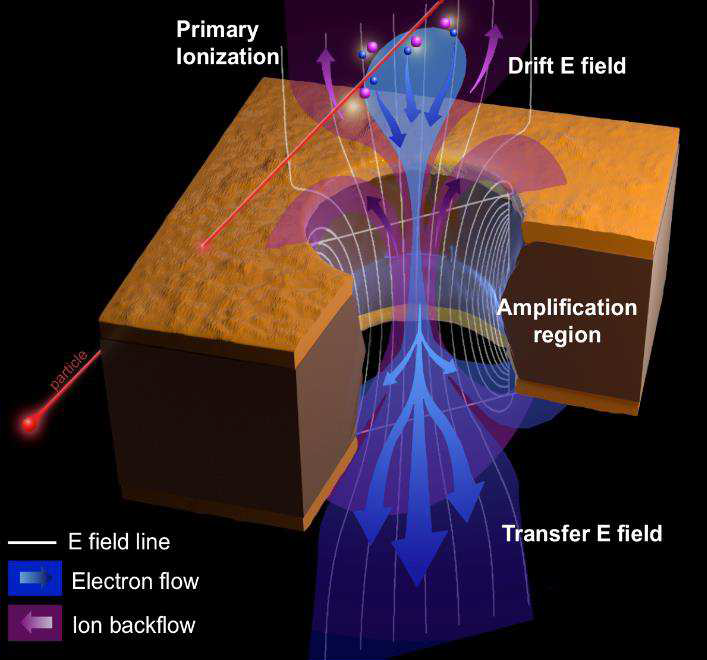}
    \caption{Sketch of a single GEM hole demonstrating the operating principle. Image from Ref.~\cite{Colaleo:2015vsq}.}
    \label{fig:GEMHole}
\end{figure}

Until recently, wires have been the standard tool for amplifying ionization signals in gases. The MWPC was invented by Charpak in 1968, and rapidly became ubiquitous in high energy physics experiments and radiation detection more generally. It was for this breakthrough that Charpak received the Nobel prize in 1992. In the late 1990s two new amplification technologies were invented and quickly overtook wires as the preferred amplification techniques. Both technologies typically avalanche the electrons directly onto pad or strip-based readout structures, allowing for great flexibility in the overall design of the detector and removing the mechanical complexity of wires. The first, invented in 1996 by Giomataris et al., is the micromesh gaseous structure, or MicroMegas. The MicroMegas amplification structure is a mesh placed very close ($O(100)~\mu$m) to the readout pads that produces a very high electric field in the gap between the mesh and the readout. The avalanche occurs in the high field region directly above the readout, which can reach field strengths on the order of 100 kV/cm. The position of the incident ionization electrons can then be inferred via the charge that lands on the readout pads. An alternative scheme avalanches the charge onto a resistive layer above the readout pads, which induces a corresponding image charge signal on the pads. The structure of a MicroMegas detector is sketched in Fig.~\ref{fig:Micromegas}.
The other technology is the gas electron multiplier (GEM)~\cite{Bouclier:1996sw}, invented by Sauli et al. in 1996. The GEM, similar to the MicroMegas, utilizes the fact that two planes separated by a small distance can, with the application of a fairly small voltage, establish an electric field capable of producing amplification. In the case of the GEM, these planes are two planes of copper separated by an insulating material. A series of small holes are etched into the copper and insulator, in which the electric fields can be extremely high. The original GEM design utilized a polymer layer 25 $\mu$m in thickness and voltage differences between the top and bottom conductor of 200 V, producing a field in the hole that reaches values of nearly 40 kV/cm, enough to achieve amplification in most gases used in gas detectors. The principle of operation is sketched in Fig.~\ref{fig:GEMHole}.
The strength of the electric field is controlled by the voltage across the GEM\footnote{``Across the GEM" here and in the following refers to the voltage difference between the two conductors of a single GEM. This is in contrast to the voltages ``between the GEMs" in the case that there are multiple GEM stages.} and the thickness of the insulator, but the voltage cannot be increased indefinitely. Eventually as the voltage across the GEM increases, the two conductors will spark to one another, either through the hole or directly through the insulator. Instead of attempting to achieve large gains in single GEMs, the preferred technique is instead to use multiple GEMs to achieve the desired gain. Thus the required electric field in each GEM can be lower, and each GEM can operate further from the breakdown voltage. In this way a stack of GEMs achieves gain in a multi-stage approach similar to a classical photomultiplier.
\begin{figure}[htbp]
    \centering
    \includegraphics[width=10cm]{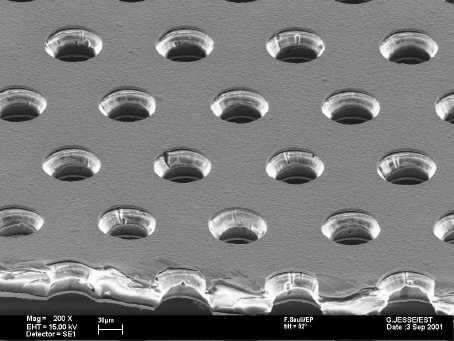}
    \caption{Scanning electron microscope image of a GEM. Image from Ref.~\cite{Colaleo:2015vsq}.}
    \label{fig:GEMSEM}
\end{figure}
The electric field between the multiple GEMs is known as the transfer field, and it can take a wide range of values depending on the application and gas mixture. This versatility is one of the aspects that have made GEMs such a popular choice for gaseous detectors recently, as different applications can be accommodated simply by changing the voltages in the detector. A variety of factors influence the choices for the fields across the GEMs and in the transfer gaps. The voltage across the GEM must be chosen to permit stable operation of the detector while still achieving the required gain. The transfer gap electric fields must be chosen such that electrons transport into the holes of the next GEM in the stack, instead of impacting directly onto the conductor of the GEM below~\cite{Richter:2002hk}. The transfer gap fields should also lie in a regime of low electron attachment. In some mixtures, at fields on the order of thousands of V/cm, 100 or more electrons can be re-attached per centimeter. Since the gains of individual GEMs are typically fairly low, the early GEMs in the stack are susceptible to losing all the signal electrons if attachment is high. Another crucial aspect of the transfer gap is the probability to block ions from returning into the drift volume. This feature will be described in further detail in Sec.~\ref{Sec:GEM}.
\subsection{Electron Transport}
\label{Sec:ElecTransport}
\begin{figure}[htbp]
    \centering
    \includegraphics[width=13cm]{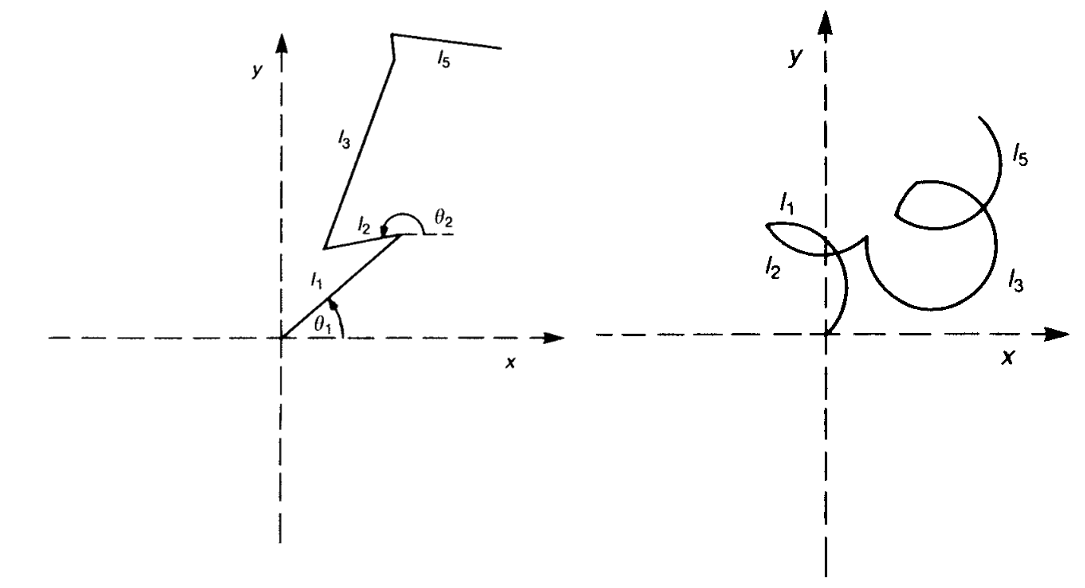}
    \caption{Left: Microscopic diffusion in the $x$-$y$ plane of a single electron drifting in $z$ through a gas with no magnetic field. Right: transverse diffusion of an electron in a magnetic field where $\Vec{E}$ and $\Vec{B}$ are almost parallel. The total transverse displacement of the electron in the case of no magnetic field can be seen to be larger. Figure adapted from Ref.~\cite{Blum:2008nqe}.}
    \label{fig:TransDiff}
\end{figure}

The movement of the electrons in the drift region of a TPC is most generically governed by the Langevin equation, which describes the motion of charged particles with charge $q$, acceleration $\Vec{a}$, and velocity $\Vec{v}$ in electric and magnetic fields $\Vec{E}$ and $\Vec{B}$ in the presence of collisions.
\begin{equation}
m\Vec{a}=q\Vec{E} + q(\Vec{v}\times\Vec{B})-K\Vec{v}
\end{equation}
Where $K$ is effectively a coefficient of friction that parameterizes the random scatterings of the charged particle off the atoms in the gas. It is clear that $K$ will depend explicitly on factors such as the atomic and/or molecular composition of the gas, which controls the scattering cross section, as well as the gas temperature and the gas pressure, which control the likelihood of the scatterings. When electrons scatter off of gas molecules, their direction after the scattering is essentially isotropic due to their small masses. After the scattering, the electric field will once again accelerate them in the drift direction. The timescale at which the acceleration of the electron effectively becomes zero and the particle reaches its terminal velocity is roughly $\tau=m/K$, where $m$ is the particle mass. Electrons, due to their low mass, reach their terminal velocity in realistic gases almost instantaneously. This is a critical point for the operation of TPCs, as the assumption that electrons drift at an effectively constant velocity to the readout is what allows for the measurement of the $z$ position of where the ionization occurred. When the particle has reached its terminal velocity\footnote{Technically this denotes the transition from Langevin dynamics, where acceleration occurs, to Brownian dynamics, where it does not. Brownian dynamics is thus the overdamped limit of Langevin dynamics, which is well motivated in this case due to the large number of collisions experienced by the electron as it drifts.}, the Langevin equation can be solved for the velocity:
\begin{equation}
\Vec{v} = \frac{q\tau|\Vec{E}|}{m(1+\omega^2\tau^2)}\cdot[\hat{E}+\omega\tau(\hat{E}\times\hat{B})+\omega^2\tau^2(\hat{E}\cdot\hat{B})\hat{B}]
\label{eq:DriftVelocity}
\end{equation}
Where $\hat{E}$ and $\hat{B}$ are, respectively, the unit vectors describing the directions of the electric and magnetic fields at the point in space. The term $\omega$ is the cyclotron frequency of the particle, defined as $\omega=q|\Vec{B}|/m$. The cyclotron frequency determines how tightly the particle will spiral in the magnetic field, and thus how ``clamped" in the direction of $\hat{E}$ and $\hat{B}$ it will be. Transverse diffusion manifests itself as a series of kicks that introduce transverse momentum to the drifting electron. In the absence of a magnetic field, this transverse component of the velocity can result in the accumulation of a large displacement from the original ionization location in the transverse direction. This clearly deteriorates the position resolution of the detector, as diffusion is a fundamentally random process, and will produce a gaussian smearing on the positions of the electrons after drifting through the gas. To achieve the optimal position resolution, the diffusion should be minimized\footnote{There are applications where this diffusion is beneficial. The most common case is when the goal is to precisely count the number of ionization electrons for measurements of dE/dx. The spatial displacement given to the electrons when they reach the readout can improve the ability to distinguish individual electrons.}. The thermal energy of the gas mixture determines the magnitude of the effects of diffusion. In the presence of a magnetic field, however, any component of velocity gained in the transverse direction will induce a magnetic force perpendicular to the velocity, causing the electron to spiral, as shown in Fig.~\ref{fig:TransDiff} This is the microscopic picture evoked to explain the decrease in transverse diffusion in a magnetic field. 
\par
It is apparent from Eq.~\ref{eq:DriftVelocity} that achieving a precise determination of the drift velocity of an electron through the gas of a TPC, and thus a precise measurement of $z$, requires the electric and magnetic fields to be known to good precision at all points in the volume. This motivates the meticulousness used in the construction of the sPHENIX TPC field cages, described in Sec.~\ref{Sec:construction}, as well as the utilization of calibration techniques to measure the drift velocity and field components in-situ.\par
Another important aspect of electron transport is attachment. Drifting electrons can be absorbed by atoms or molecules in the gas mixture, resulting in a degradation of the ionization signal. For the purposes of tracking, some attachment is acceptable\footnote{This is not necessarily the case for measurements of dE/dx, which rely on the detected number of electrons being the same as the number of electrons produced in the ionization process. Attachment is a random process, which introduces an additional resolution effect on electron counting measurements.} as long as the electron statistics remain sizable and well separated from the noise in the electronics. The basic process is 
\begin{equation}
e^- + X \rightarrow X^-
\end{equation}
Where $X$ is a molecule in the gas and $X^-$ is a negative ion. The likelihood of attachment to a given molecule depends on the energy of the incident electron, which is roughly proportional to the electric field in a given volume. There is a threshold energy above which an incident electron has the ability to overcome the potential barrier of the neutral molecule and create a negative ion. The gases used in gaseous detectors generally have this threshold energy at the level of several eV, which is larger than the energy of the electrons at the drift velocity in the gas. However, impurities in the gas, or the high electric fields near the amplification region, can still render attachment an issue. \par
Contaminants in gas detectors typically either come from outgassing of materials facing the gas, or inward leaks of atmospheric air\footnote{For this reason, most gas detectors are operated slightly overpressure compared to the surrounding atmosphere.}. In O$_2$, the electron threshold energy for attachment is only 0.5 eV, which means that electrons in the drift volume can easily attach to O$_2$ molecules. The O$_2$ attachment reaction additionally depends on the other components of the gas. In methane mixtures, it was determined that 1 ppm of O$_2$ could produce a loss of 3\% of the drifting electrons per meter of drift. The effects of other contaminants, especially the large molecules that could arise from outgassing of detector materials, are less well studied, and the complex molecular dynamics involved make predicting their effects challenging. It is therefore good practice to limit the amount of contaminants in the gas by using materials which do not outgas significantly and by ensuring the detector volume is well sealed against the surrounding atmosphere.
\subsection{Ion Transport}
\label{Sec:IonTransport}
Naively one would imagine that the transport properties of ions have little impact on the performance of a TPC, by virtue of the fact that ions are not directly detected. However, of the large number of ions produced in the amplification process, some of them will drift back into the electron drift region. These ions can substantially distort the electric field, an effect known as ``space charge". For the performance of the sPHENIX TPC described later, understanding and correcting for this space charge is a necessity. It is therefore warranted to describe in some detail the way that ions move in gaseous detectors. \par
The microscopic picture of ion transport is fairly different from that of electron transport. Since the masses of even light ions are similar to those of the molecules they are colliding with, the scattering of an ion off a molecule is anisotropic; after the collision the ion prefers to continue moving in the direction it was drifting previously. The steady-state drift velocity of the ions can be analytically provided in two regimes, one where the effect of $E$ is small and the resulting ion velocity is thermal (as will turn out to be the case for the drift volume in the sPHENIX TPC), and one where the effect of $E$ is large and the ion velocity is dominate instead by its acceleration in the field. In the first case, i.e. small $E$, the ion drift velocity $v$ can be given as 
\begin{equation}
v = \sqrt{\frac{1}{m}+\frac{1}{M}}\frac{eE}{3kTN\sigma}
\end{equation}
Where $m$ is the mass of the ion, $M$ is the mass of the molecule, $T$ is the temperature, $k$ is the Boltzmann constant, $e$ is the electric charge of the ion, $N$ is the number density of the gas, and $\sigma$ is the scattering cross section of the ion on the gas. It can be seen that the velocity depends linearly on the electric field. At high electric fields where thermal motion can be neglected, the drift velocity is: 
\begin{equation}
v = \sqrt{\frac{m}{M}(1+\frac{m}{M})}\sqrt{\frac{eE}{mN\sigma}}
\end{equation}
which gives a dependence only on $\sqrt{E}$. A dependence on $N$ also appears in both cases, which can be naively interpreted as the pressure of the gas mixture. Historically, to enable extrapolation to different operating pressures, results on the drift velocity have been reported as functions of $E/p$, in units of $\frac{\text{V}}{\text{cm}\cdot\text{torr}}$. The ``high $E$" regime typically sets at values of $E/p\gtrsim100$ $\frac{\text{V}}{\text{cm}\cdot\text{torr}}$, which is not reached in sPHENIX, which is expected to be operated at slightly higher than atmospheric pressure. The low-$E$ regime sets in at $E/p\lesssim20$, meaning in effectively all cases in sPHENIX, the ``low $E$" regime will be a fairly good approximation, and the ion velocity will be linearly proportional to the applied electric field.\par
The above referred only to single-component gases. The ion drift velocity of gas combinations can be determined via Blanc's law\footnote{There exist some gas combinations where Blanc's law fails experimentally due to chemical effects, such as electron transfer from another molecule of the gas to the original ion, turning that molecule into the ion which drifts~\cite{Schultz:1975yd}. The assumption of Blanc's law is that the primary ionized molecule will be the one which drifts, which can be violated in such cases. None of the gas mixtures considered for sPHENIX exhibit this behavior.}:
\begin{equation}
\label{BlancsLaw}
\frac{1}{K_{mix}}=\frac{f_1}{K_{1}}+\frac{f_2}{K_{2}},
\end{equation}
Where $K_{mix}, K_1,$ and $K_2$ represent the ion mobilities for the mixture, gas 1 and gas 2, defined as
\begin{equation}
K=\frac{v}{E}
\end{equation}.
The values of these mobilities have been studied extensively for various gases. Argon and CF$_4$, the mixture considered for sPHENIX, was studied in the context of a TPC for the International Linear Collider in Ref.~\cite{Santos:2018zwk}. The measured ion mobility for the 60\% argon, 40\% CF$_4$ mixture was 1.6 cm$^2$/(Vs). At the sPHENIX drift field of $\sim400$ V/cm, the ion velocity is therefore expected to be $\sim600$ cm/s, substantially slower than the electron drift velocity of $9\times10^6$ cm/s.\par
The slow velocity of the ions means that the magnitude of the magnetic force is substantially reduced. Therefore, to a good approximation, in the TPC field configuration of $E||B$, the motion of ions can be described electric field lines at the drift velocities described above. The differences in the transport properties of ions and electrons can be exploited to reduce the deleterious effects the ions have on the electric fields in the drift volume. As mentioned in Sec.~\ref{subsec:Amplification}, the transfer gaps of GEMs can be tuned to have a high fraction of electric field lines that land on the GEM conductor, instead of going through the holes. Since electrons are influenced more by diffusion and the $B$-field, they can still pass through the GEM holes with a substantial likelihood in such a configuration, while the ions will follow the field lines and be neutralized at the conducting surface.
\section{sPHENIX TPC Design Considerations}
\label{Sec:DesignConsiderations}
The primary advantages of sPHENIX over the previous generation of RHIC experiments are the ability to handle a high event rate and the precision of the tracking. The most demanding physics channel for both of these features is the observation of the sequential melting of the $\Upsilon$, as described in section \ref{sec:sPHENIXPhysics}. The measurement uncertainty on the $r\phi$ position of an ionization deposit can be parameterized as:
\begin{equation}
\sigma_{r\phi} = \sqrt{\sigma_{\text{Pad}}^2+\sigma_{\text{Space Charge}}^2 + \sigma_{\text{Diffusion}}^2}
\label{eqn:totalresolution}
\end{equation}
The intrinsic spatial resolution due to the segmentation of the readout $\sigma_{\text{Pad}}$, the distortions due to space charge $\sigma_{\text{Space Charge}}$, and the diffusion of the electrons $\sigma_{\text{Diffusion}}$, are described in Secs.~\ref{Sec:PosRes},~\ref{Sec:SpaceCharge}, and~\ref{Sec:Diffusion} respectively. The considerations pertaining to the rate capability of the detector are given in Sec.~\ref{Sec:RateCapability}. 
\subsection{Rate Capability}
\label{Sec:RateCapability}
Many factors contribute to the ability of a detector to handle a high rate of collisions. The limiting factor in the rate capability of prior TPCs has been the need to reduce the space charge in the TPC via the use of an active gate, as shown in Fig.~\ref{fig:MWPCForTPC}. The active gate blocks all charge from entering and leaving the readout immediately after an event is registered. This is done to prevent ions from the amplification entering the drift volume, but due to the slow drift velocities of the ions, the gate must stay closed for a time on the order of milliseconds. Thus actively gated TPCs are fundamentally limited to taking data at rates of around 1 kHz. The solution to this issue in ALICE and sPHENIX is the use of a quadruple GEM amplification scheme that passively blocks the ions. 
\begin{figure}[htbp]
    \centering
    \includegraphics[width=13cm]{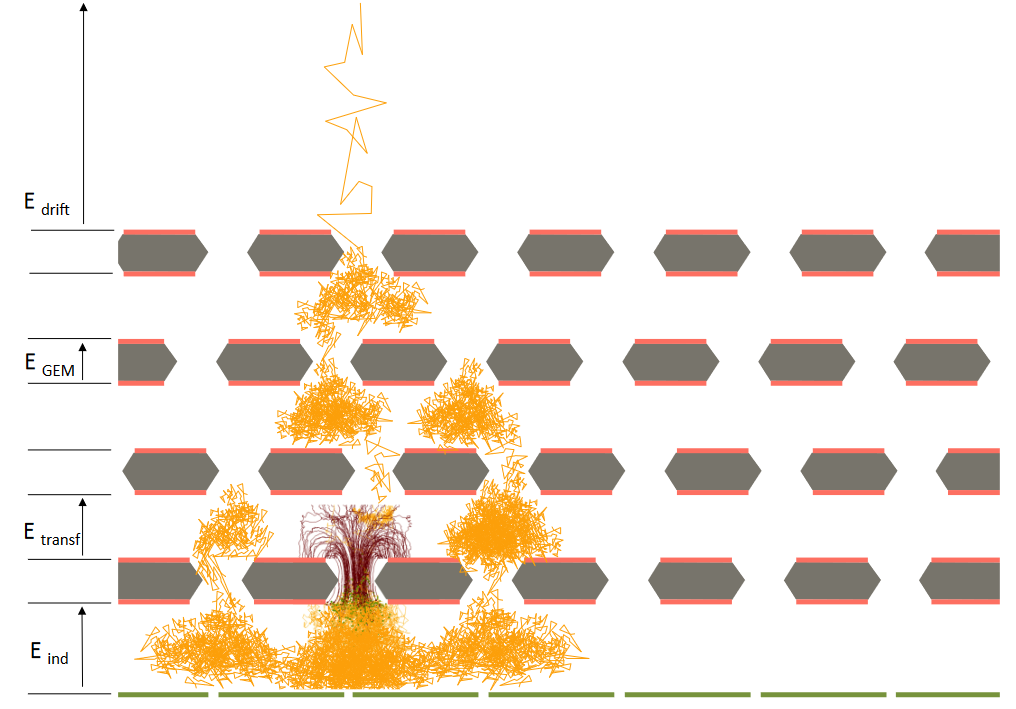}
    \caption{Cross section of an avalanche in a quadruple GEM detector. The paths of the electrons are shown in yellow, and the paths of ions are shown in red (for one GEM hole only). The primary electron enters at the top of the stack and avalanches in the gap between the top and bottom of the GEM.}
    \label{fig:QuadGEM}
\end{figure}
The precise design of the GEMs in sPHENIX is described in Sec.~\ref{Sec:GEM}. The alleviation of the active gate means that the rate capability is limited instead by concerns such as pattern recognition and the speed of the readout, which are more flexible than the ion velocity.\par
The most obvious factor affecting rate capability is the length of time time over which the signals persist in the detector, i.e. how long after an event occurs the detector continues to register hits from that event. In TPCs, this is driven primarily by the drift velocity of the electrons in the drift region. The drift velocity is a function of the gas mixture and electric field throughout the drift volume. Naively, the faster the drift velocity, the higher the rate capability. However, the drift velocity has to be compared with the speed of the electronics peforming the readout of the signal. Electronics generally have a fixed ``shaping time" for incoming pulses that determines the precision on the signal arrival time. In STAR, the electronics shaping time is 180 ns~\cite{Klein:1995fi}, and in ALICE it is 160 ns~\cite{Soltveit:2012jp}. Both ALICE and STAR utilize fairly slow drift velocities compared to those considered for sPHENIX, 2.6 and 5.5 cm/$\mu$s respectively. Slow drift velocities are preferable for optimizing the position resolution in $z$, since they maximize the time over which a given track arrives at the readout, which improves the ability to distinguish tracks that are near to each other in $z$. For sPHENIX, the shaping time of the electronics is 80 ns, and thus a higher drift velocity can be utilized.\par
\begin{table}[tbhp]
  \footnotesize
  \begin{center}
    \begin{tabular}{lcc}
      \hline
      Experiment & Drift Velocity & Shaping Time (ns)\\
      \hline
         STAR~\cite{STAR:1997sav}&5.45 cm/$\mu$s&180 ns\\ 
         ALICE~\cite{Alme:2010ke}&2.83 cm/$\mu$s&160 ns\\ 
         ALICE (GEM Upgrade)~\cite{Lippmann:2014lay}&2.58 cm/$\mu$s&160 ns\\
         sPHENIX& $\sim8$cm/$\mu$s&80 ns\\
      \hline
    \end{tabular}
    \caption{
      A comparison of the operating parameters relevant to the position resolution in $z$ for STAR, ALICE, and sPHENIX.
    }
    \label{tab:RateComparison}
    \end{center}
\end{table}
The nominal resolution on the measurement of $z$ for sPHENIX is therefore $\approx$ 6 mm, which is slightly worse than ALICE and slightly better than STAR. 
\subsection{Pad Position Resolution}
\label{Sec:PosRes}
To reach the necessary position resolution without requiring a substantial increase in the number of channels, charge sharing readout is employed~\cite{Azmoun:2020zwf}. However, for charge sharing to work properly the size of the charge cloud as it lands on the readout plane must be large enough to hit multiple pads. For GEMs in the configuration of sPHENIX, the radius of the charge cloud produced by a single electron is around 0.5 mm. In ALICE, where the transverse diffusion of the gas in the drift volume is larger than in sPHENIX by a factor of $\sim$5, it is typically the case that the several electrons produced by an ionization event spread out enough in the transverse direction to induce a response on multiple pads, irrespective of the size of the cloud produced by the GEMs. The pads of the ALICE readout therefore have widths of 4 mm in $r\phi$ in the inner readout section and 6 mm in $r\phi$ in the outer readout section. In sPHENIX, due to the choice of gas and the high magnetic field, the amount of transverse diffusion is too small to have rectangular pads significantly larger than the GEM charge cloud size, but to produce pads small enough to charge-share reliably would be too expensive. The adopted solution is zig-zag shaped pads~\cite{Azmoun:2018ail}, which interleave significantly such that any charge cloud produced by the GEMs will fall on two or more pads.\par
\begin{figure}[htbp] 
    \centering
    \includegraphics[width=12cm]{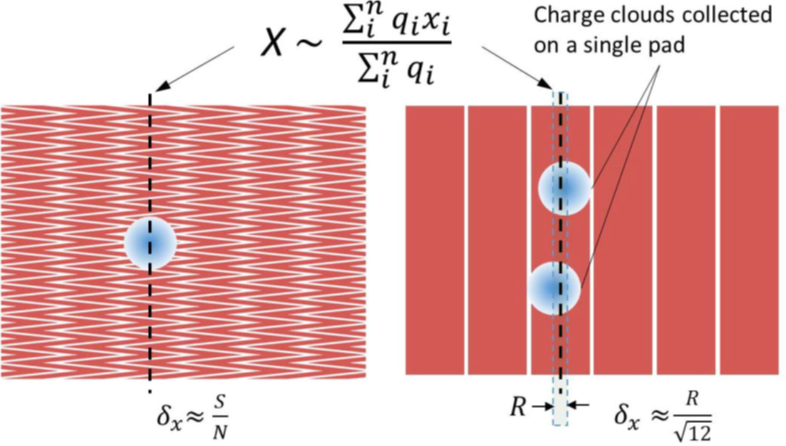}
     \caption{Illustration of the zig-zag charge sharing concept. Right: charge cloud from GEMs falls on standard rectangular pads or strips, the position resolution of the hit is the pad width divided by $\sqrt{12}$. Left: Charge sharing zig-zag readout where single pad hits are effectively eliminated. The position resolution is limited only by the signal-to-noise ratio of the system and the amount of charge lost to dead areas.}
    \label{fig:ChargeSharing}
\end{figure}
An important concept in charge-sharing pad readout is differential non-linearity, or DNL. The ideal case for charge-sharing is that the reconstructed amount of charge on the pads is linearly proportional to the location of the initial electron that creates the avalanche. Pads, by definition, have small dead areas between the conductors. This dead area can introduce a small bias towards the pad locations for incoming primary electrons that happen to be directly above the gap in the pads, leading to a non-linearity that manifests as a degradation in resolution. The DNL can be unfolded to provide a more precise measurement of the incoming electron position, as demonstrated in Refs.~\cite{Azmoun:2018ail,Azmoun:2020zwf,Perez-Lara:2021zji}. Even a naive unfolding algorithm provides a substantial improvement in the resolution. Let us consider the case of a charge cloud that strikes two pads. If pad $A$ at location $x_A$ measures an amount of charge $Q_A$ and pad $B$ at location $x_B$ measures $Q_B$, the standard charge-weight centroid reconstruction technique would place the charge cloud at reconstructed location $x_R$:
\begin{equation}
x_R = \frac{Q_A\cdot x_A + Q_B\cdot x_B}{Q_A+Q_B} 
\label{eqn:Centroid}
\end{equation}
However, if it is known, e.g. from test beam or simulation, that certain values of measured $Q_A$ and $Q_B$ typically correspond to slightly different values of position due to some of the total charge being lost into a dead area, a small $Q_A$- and $Q_B$-dependent correction factor can be applied to recover the true position. Due to their interleaved geometry, zig-zags actually tend to reduce the degree of differential non-linearity when compared to rectangular pads~\cite{Perez-Lara:2021zji}. In sPHENIX, more sophisticated unfolding techniques leveraging the full track information can surely be utilized.\par
\begin{figure}[htbp] 
    \centering
    \includegraphics[width=10cm]{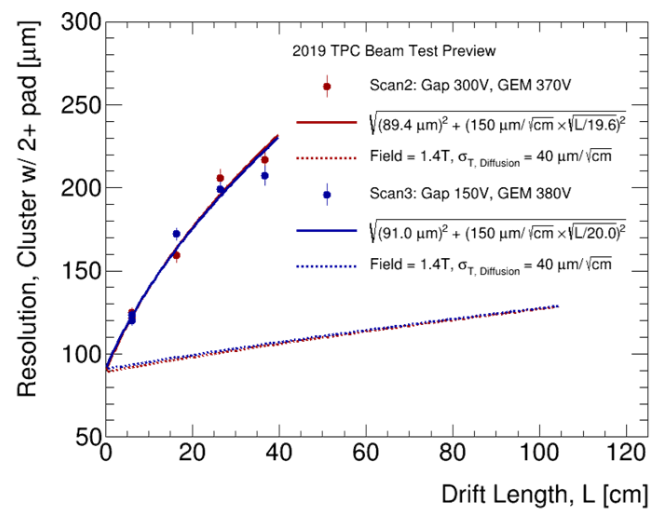}
     \caption{Measured resolution as a function of drift length in the prototype TPC. The intercept of the curves as drift length goes to zero is effectively a measurement of the intrinsic resolution of the readout. The gas mixture in this case was Ne:CF$_4$ 50:50. The solid lines represent a fit to the measured result at two different voltage settings. The dotted lines represent the position resolution extrapolated to the sPHENIX magnetic field, where the transverse diffusion constant $D_t$ (described in Sec.~\ref{Sec:Diffusion}) is estimated to be 40 $\mu$m/$\sqrt{\text{cm}}$.}
    \label{fig:FTBFResolution}
\end{figure}

The intrinsic resolution of the charge-sharing pads in sPHENIX depends on the signal-to-noise ratio of the electronics. Thus, unlike the case of the single-pad hit readout schemes mentioned in Sec.~\ref{Sec:ChPT}, it cannot be estimated from first principles. However, ionization deposits close to the readout were measured at test beam in 2019 with a realistic version of the sPHENIX electronics. Since diffusion and space charge were both effectively negligible in that situation, the intrinsic resolution of the readout could be inferred from the $y$-intercept of the measured resolutions as a function of drift length shown in Fig.~\ref{fig:FTBFResolution}\footnote{The characteristics of the avalanche, which does affect the position resolution, will be different as the gas mixture used in this test beam was different than the final operating gas. However, since the GEM voltages were adjusted accordingly, this is expected to be a relatively small effect.}. 
The intrinsic position resolution of the readout is determined to be around 90 $\mu$m~\footnote{This number was measured using a realistic version of the electronics, but not the final version. The actual intrinsic position resolution of the full detector will depend on the performance of the front end electronics and particularly the TPC cooling system described in Sec.~\ref{subsec:cooling}, which directly impacts the amount of noise seen by the electronics.}. This preliminary resolution is substantially better than the requirements for sPHENIX of 200 $\mu$m, however the transverse diffusion of the ionization electrons and the contribution from space charge also need to be taken into account before the full position resolution of the detector can be properly quantified.
\subsection{Diffusion}
\label{Sec:Diffusion}
Diffusion of electrons in gases is a fairly well understood phenomenon. The diffusion can be calculated both numerically and analytically using a variety of techniques. In general, for arbitrary field configurations, the diffusion is described via the diffusion tensor $D_{ij}$. In the case of a TPC, where $E||B$\footnote{In a realistic TPC, there will always be off-parallel components of these fields, making the use of the diffusion tensor necessary to achieve an accurate and precise description of the diffusion in the TPC volume}, the diffusion takes on only two components, transverse and longitudinal with respect to the drift direction. The measurement of $r\phi$ is affected only by the transverse diffusion, making it the crucial parameter for momentum resolution. The resolution on the transverse position of the drifted electrons scales as $\sqrt{L}$, where $L$ is the length drifted. The proportionality constant $D_t$\footnote{Which should be distinguished from the commonly used diffusion coefficient, often also denoted as $D_t$, that has units of m$^2$/s. The form of $D_t$ used in this thesis is common amongst the TPC community.} describes the amount of transverse diffusion for a single electron as a function of the gas mixture, electric field, and magnetic field, and is straightforwardly related as:
\begin{equation}
\sigma_{\text{T}} = D_t\sqrt{L}
\label{eqn:diff}
\end{equation}
Where the units of $D_t$ are $\mu$m/$\sqrt{\text{cm}}$, and $L$ is given in cm. In the TPC, the resolution on the $r\phi$ location of an ionization deposit is better than that of a single electron by a factor 1/$\sqrt{N_{\text{eff}}}$, where $N_{\text{eff}}$ is a quantity related to the number of ionization electrons measured. The fact that each electron is independently providing a fluctuating amount of charge to the pads and the position is being inferred by the centroid of a charge distribution means that the resolution does not scale directly with the total number of electrons measured. A detailed investigation into $N_{\text{eff}}$ is provided in Ref.~\cite{Kobayashi:2006cp}, but for the gases considered for the sPHENIX TPC it is expected to be around 25. From these considerations it can be seen that the $\sigma_{\text{Diffusion}}$ given in the equation for the total resolution, Eq.~\ref{eqn:totalresolution} can be represented as:
\begin{equation}
\sigma_{\text{Diffusion}} = \frac{D_t\sqrt{L}}{\sqrt{N_{\text{eff}}}}.
\label{eqn:diffres}
\end{equation}
Clearly, $D_t$ should be reduced as much as possible, since it provides an irreducible contribution to the resolution. $N_{\text{eff}}$ should also be maximized, although its effect is only proportional to the square root. These factors heavily influenced the selection of gas and electric field, discussed in more detail in Sec.~\ref{Sec:GasOperating}. \par
Since collisions occur at $z\approx0$ and particles exhibit a roughly flat dN/d$\eta$ distribution within the TPC acceptance, the typical drift length for an ionization deposit can be calculated to be around 75 cm, where the contribution to the position resolution due to diffusion, $\sigma_{\text{Diffusion}}$, is expected to be $\sim70\mu$m. For the full drift length of around 1 meter, $\sigma_{\text{Diffusion}}$ becomes a similar size as the intrinsic resolution of the readout. The combined resolution of the TPC at the full drift length prior to the introduction of space charge is thus around 130 $\mu$m.
\par
Nothing has been mentioned thus far about the deleterious effects of longitudinal diffusion. Longitudinal diffusion contributes to the resolution of the detector on the $z$ location of an ionization deposit, and is to first order independent of the magnetic field. In a TPC, longitudinal diffusion manifests as a time spread in the arriving signals. The longitudinal diffusion exhibits the same length scaling as the transverse diffusion, given in Eq.~\ref{eqn:diff}. Typical values of longitudinal diffusion for the gas mixtures considered for sPHENIX are $D_L$ = 100-200 $\mu$m/$\sqrt{\text{cm}}$. Thus, over the full drift length of the TPC, the spread in the positions will be roughly 1-2 mm. With the $\sim8$ cm/$\mu$s drift velocity in the TPC, this corresponds to arrival time spread of 12-25 ns, which is small in comparison to the SAMPA shaping time of 80 ns. For this reason, the exact value of the longitudinal diffusion is not a critical parameter for the design of the TPC\footnote{TPCs which seek good resolution in the drift direction do, however, have to consider this. In the long drift length liquid argon TPCs envisioned for the next generation of neutrino experiments, the longitudinal diffusion is one of the most important design parameters~\cite{MicroBooNE:2021icu}.}.
\subsection{Position Distortions}
\label{Sec:SpaceCharge}
\begin{figure}[htbp]
    \centering
    \includegraphics[width=8cm]{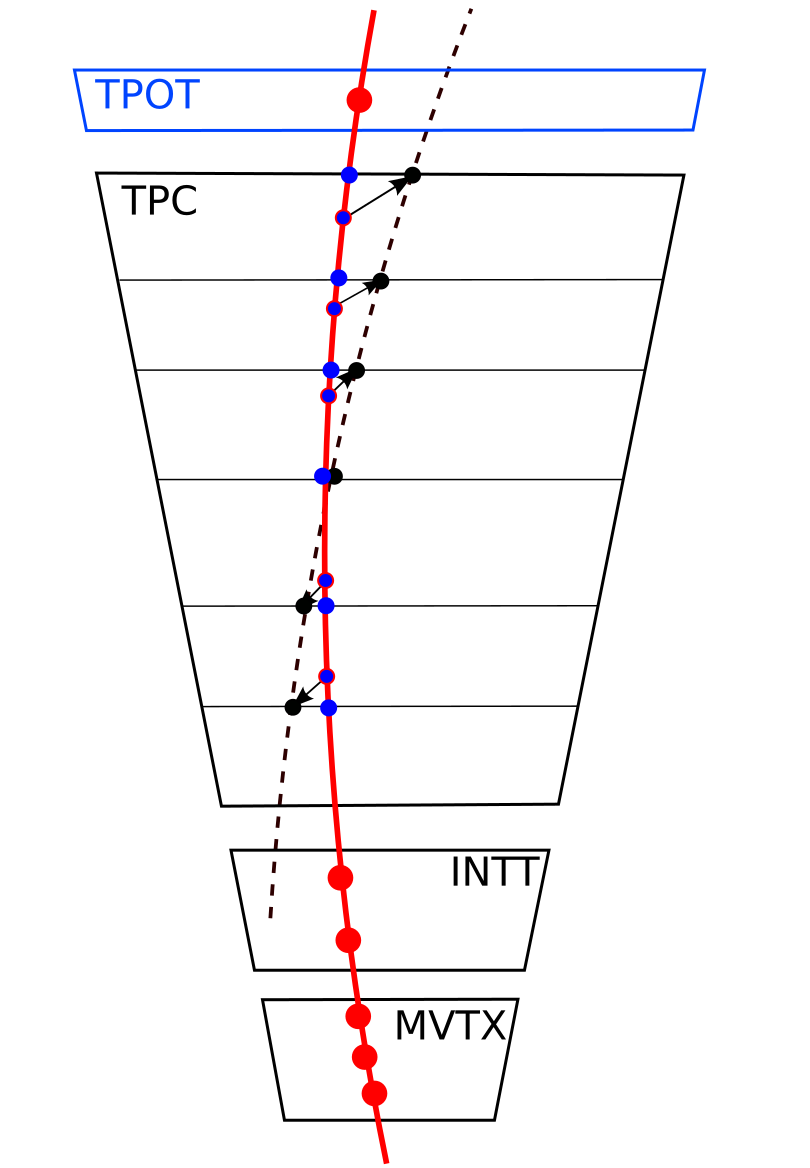}
    \caption{Sketch of the concept of distortion corrections in sPHENIX. The measured electron positions will provide an incorrect trajectory if distortions are not accounted for.}
    \label{fig:Distortion}
\end{figure}

Since the \emph{overall} resolution of the TPC on the $r\phi$ position of an ionization event must be $200~\mu$m or better, the distortions of the measured positions due to space charge and other non-uniformities in the electric and magnetic fields must be correctable to a similar precision. Distortions can be roughly factorized into three components. The first are static distortions, which arise from long timescale field distortions, such as those due to the non-zero $\Vec{E}\times \Vec{B}$ arising from the radial component of the solenoidal field. These distortions are expected to be on the order of 1-2 centimeters in magnitude, but are not strongly time dependent and thus are fairly easy to correct for. Another component of the total distortion is the average space charge distortion, which corresponds to the steady-state occupancy of space charge throughout the TPC volume, and will be time-dependent insofar as it will be proportional to the instantaneous luminosity delivered by RHIC. The final and most challenging to correct for are space charge fluctuations. The average space charge distortions are estimated to be on the order of 1-2 millimeters. The event-by-event fluctuations in space charge due to the topology of the previous events is challenging to correct for and thus needs to be modelled. In a TPC such as sPHENIX, ions drift along the electric field lines at velocities of 1-10 m/s, meaning during 15 kHz operation the ions from up to 15,000 events, each with its own pattern of electric field distortions, could be present in the TPC at a time. The distortions from this rapidly time-varying configuration of ions are estimated to be below 100 $\mu$m. The concept of the position distortions is sketched in Fig.~\ref{fig:Distortion}.
The challenge posed by space charge in particular is that, unlike field distortions caused by mechanical misalignments, space charge produced during operation is highly time dependent. To this end, sPHENIX has implemented various measures to combat and correct for these time-dependent distortions, including readout design choices and various calibration systems.\par
\begin{figure}[htbp]
    \centering
    \includegraphics[width=10cm]{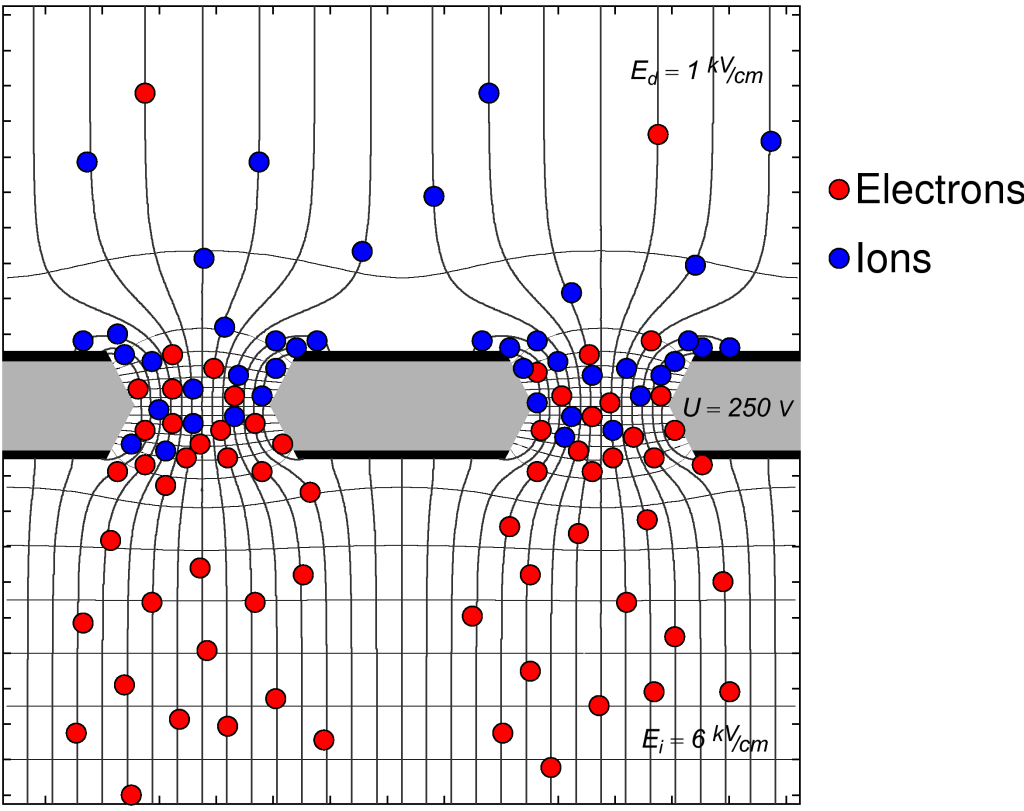}
    \caption{Demonstration of the ion production properties of a GEM. Many ions are produced in the GEM hole, which can then follow electric field lins back into the drift volume. Image from DESY FLC Group.}
    \label{fig:GEMIBF}
\end{figure}

The simplest technique for reducing the time dependence of the electric field distortions induced by space charge is simply reducing the total amount of space charge in the TPC as much as possible. The ions produced directly by the primary ionization process provide an irreducible contribution to the space charge. However, the number of ions produced by the avalanche inside the GEMs is larger than those from the initial ionization process by roughly a factor of the gain. It is these ions that must be prevented from entering the drift region as much as possible. The fraction of these ions drifting back must be kept below 1\%. Even at 1\% IBF, the ions originating in the amplification stage will outnumber the primary ions by a factor of $\sim$20. Fig.~\ref{fig:GEMIBF} demonstrates the process of the ions flowing back into the drift region, known as ion backflow (IBF).
Since the ions tend to follow the electric field lines, some fraction of the ions produced in the GEM hole will necessarily neutralize themselves on the top surface of the same GEM that produced them. This fraction depends on the field strength above the GEM and in the GEM hole, but in general is not large enough to provide the necessary ion blocking. To reduce IBF, it is clear that at the very least, the GEM closest to the drift volume must operate in a regime where it does not produce a substantial amount of ions\footnote{It is important, however, that this GEM has a high probability to collect the primary electrons.}. \par
A rough analytic model for the density of space charge inside a collider TPC has been constructed by the STAR and ALICE TPC groups~\cite{Rossegger:2009tha,Hellbar:2015yqz}:
\begin{equation}
\rho(r,z) = \frac{A-B\cdot z}{r^2}
\end{equation}
Where $A$ and $B$ are extracted from the data. In STAR, with a collision rate of 15 kHz, the values of $A$ and $B$ were determined to be $A=517$ and $B=206$. This simple model captures the 1/r$^2$ dependence resulting from the higher occupancy at small $r$ and the $z$ dependence from the pileup of multiple events. In sPHENIX, despite the TPC volume reaching from $20<r<78$cm, the choice was made only to instrument $r>30$cm to avoid the region of extreme space charge density, where the position distortions due to space charge are estimated to be on the order of a few cm. 
\par
It is worth noting that in ungated detectors, the resolution on dE/dx is inversely correlated with IBF reduction. This is a result of the fact that the first GEM encountered by the drifting electrons must have low gain, since any ions produced by that GEM will have a more direct path into the drift volume. Operating the first GEM in a low-gain mode decreases the primary electron detection efficiency, necessarily resulting in a degradation in the resolution on the number of primary ionization events. This effect was studied extensively by ALICE, where both good dE/dx resolution and ion blocking are desired~\cite{ALICETPC:2018vdw}. In sPHENIX, the reduction of IBF is a more critical parameter than the dE/dx resolution\footnote{This follows from the majority of the sPHENIX physics program relying on \emph{hard probes} which require high precision tracking. PID via dE/dx is typically only reliable at low momentum, and is not required for most of the sPHENIX physics.}, so the operating point of the GEMs is chosen to be the point where IBF is maximally reduced.\par
\begin{figure}[htbp]
    \centering
    \includegraphics[width=10cm]{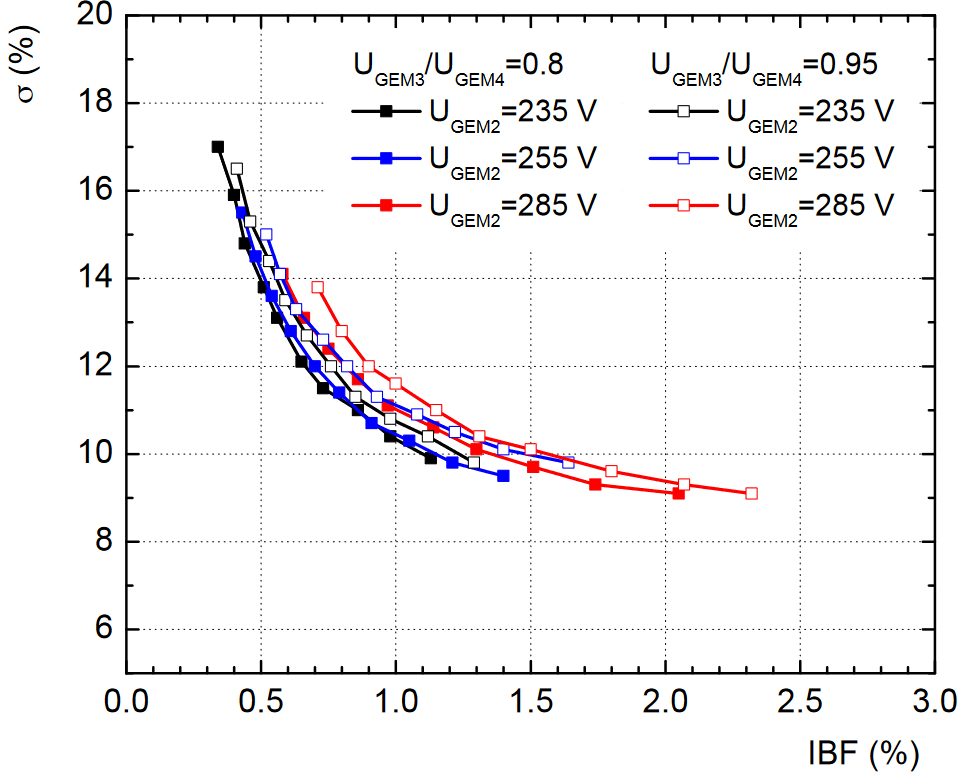}
    \caption{Demonstration of the tradeoff between dE/dx resolution (y-axis) and IBF (x-axis) for different GEM operating voltages. While the Image from Ref.~\cite{Lippmann:2014lay}}
    \label{fig:ALICEGEMIBF}
\end{figure}
Irrespective of the level of IBF reduction achieved by the GEMs, there will be time-dependent space charge in the TPC that must be corrected for. To characterize these distortions, a variety of independent calibration techniques are envisioned. The TPC itself is equipped with two laser calibration systems. They are known as the ``line" laser and the ``diffuse" laser, and their designs are described in Sec.~\ref{lasers}. In addition to the laser systems, using digital currents and tracks measured in the other tracking detectors should enable the construction of a distortion map that can be applied to measured tracks to bring them to their true positions.\par
\begin{figure}[htbp]
    \centering
    \includegraphics[width=12cm]{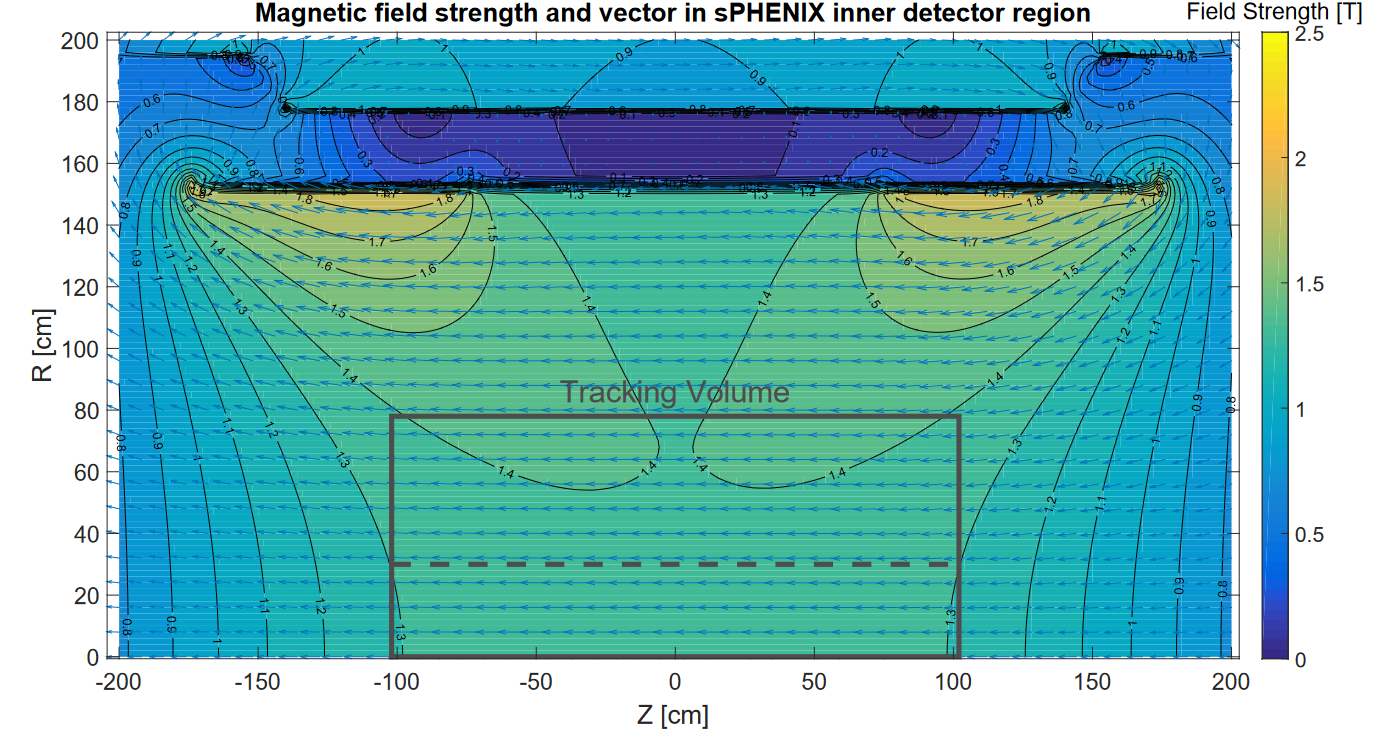}
    \caption{Calculation of the $B$-field strength as a function of position in sPHENIX. The dashed line indicates the $r=30$ cm inner radius of the TPC tracking volume. The field can be seen to be quite uniform in the TPC active volume, with some small deviations at large $r$. The BaBar magnet achieves this extended region of uniform field by varying the coil density as a function of $z$. Note that this is \emph{not} the magnetic field map measured by the magnet mapping effort.}
    \label{fig:BField}
\end{figure}

The line laser is used to characterize static distortions from non-zero $\Vec{E}\times \Vec{B}$. The $B$-field non-uniformity can be seen in Fig.~\ref{fig:BField}.
The line laser produces straight lines of ionization that can be compared to the measured ionization positions at the readout. The measured electron positions will contain the distortions, allowing the magnitude of the distortion to be reconstructed. The line laser has the ability to be ``swept" around the TPC, reaching a large fraction of the TPC active volume. This allows for extraction of the drift velocities of electrons produced as various $z$ locations. Currently operation of the line laser during data-taking is not foreseen, rather it will be used to provide a baseline map of the position distortions prior to the presence of space charge. These static distortions are estimated to be on the order of 2 cm near the outer field cage.\par
The other laser system is the diffuse laser system, which can be used to calibrate the drift velocity and integrated space charge distortions. The diffuse laser system frees electrons from a pattern of aluminum stripes on the central cathode, which drift the full length of the TPC and accumulate distortions along the way. The laser pulses are very short, such that the time the laser photons arrive at the central cathode is known to very good precision. The time of arrival of the electrons at the readout allows for a good determination of the drift velocity over the full length of the TPC.\par
Additionally, the TPOT, described in Sec.~\ref{subsec:TPOT} can be used to calibrate the TPC using tracks projected from the silicon detectors. By having tracking points measured at large radius, the angular resolution on tracks decreases to the point that they can be used to calibrate the TPC. The TPOT extends along the entire length of the TPC for one $\phi$ sector, allowing for an understanding of the distortions to be built out of tracks in that $\phi$ slice. There are additional detectors on either side of that $\phi$ slice to allow a map to be extended and tested at other $\phi$ values.\par
The final technique that can be used to construct a map of the distortions is the so-called ``digital current" technique. This technique assumes that the ion positions leaving the GEMs are the same as the locations where the electrons were registered, and propagates the ions backwards along the electric field lines. This technique is reliable if the electric fields are known well, which is tied in with the operation of the line laser. A challenge posed by this technique is that solving for the electron position distortions given a certain ion density from the digital currents is computationally expensive due to the complexity of building a high granularity electric field map and drifting electrons through it. A fairly high granularity of the map is necessary to prevent aliasing effects due to voxel sizes being too large to properly capture the necessary features of the distributions.
\section{Design, Engineering, and Construction}
\label{Sec:construction}
\begin{figure}[htbp]
    \centering
    \includegraphics[width=10cm]{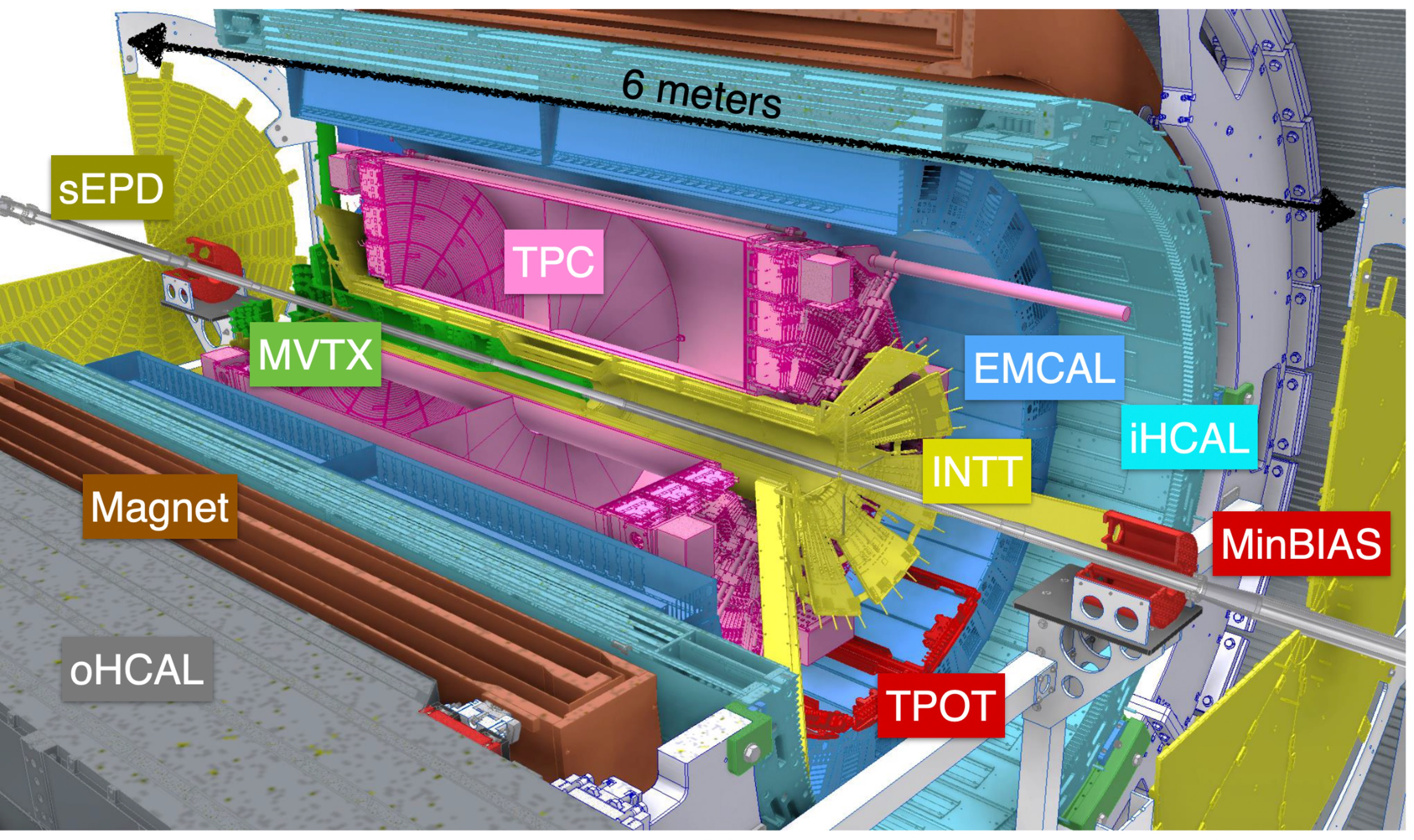}
    \caption{Cross section of sPHENIX showing the TPC in pink.}
    \label{fig:SPHENIXDiag}
\end{figure}

The construction of the TPC took place between 2017 and 2023, primarily at Stony Brook University. Engineering support was provided by BNL. A cross section view of the TPC, capturing essentially all features of the mechanics as constructed, is given in Fig.~\ref{fig:SPHENIXDiag}.
There are many important features in Fig.~\ref{fig:SPHENIXDiag} that will be discussed in detail in the coming sections. On the far side from the perspective\footnote{Which corresponds to the south side of the experiment, as can be determined by the orientation of the MVTX.} given in Fig.~\ref{fig:SPHENIXDiag}, the segmentation of the readout can be seen. Between the two endcaps sit the two field cages and the central cathode. On the near side, one can see the large number of front-end electronics cards, the copper cooling distribution manifolds, the laser benches, and the long beam that connects the TPC to its mounting point on the IHCal. The following sections will discuss many of the choices made in the design and construction of the TPC, although they will by no means be exhaustive.
\subsection{Gas Operating Point}
\label{Sec:GasOperating}
\begin{figure}[htbp]
    \centering
    \includegraphics[width=12cm]{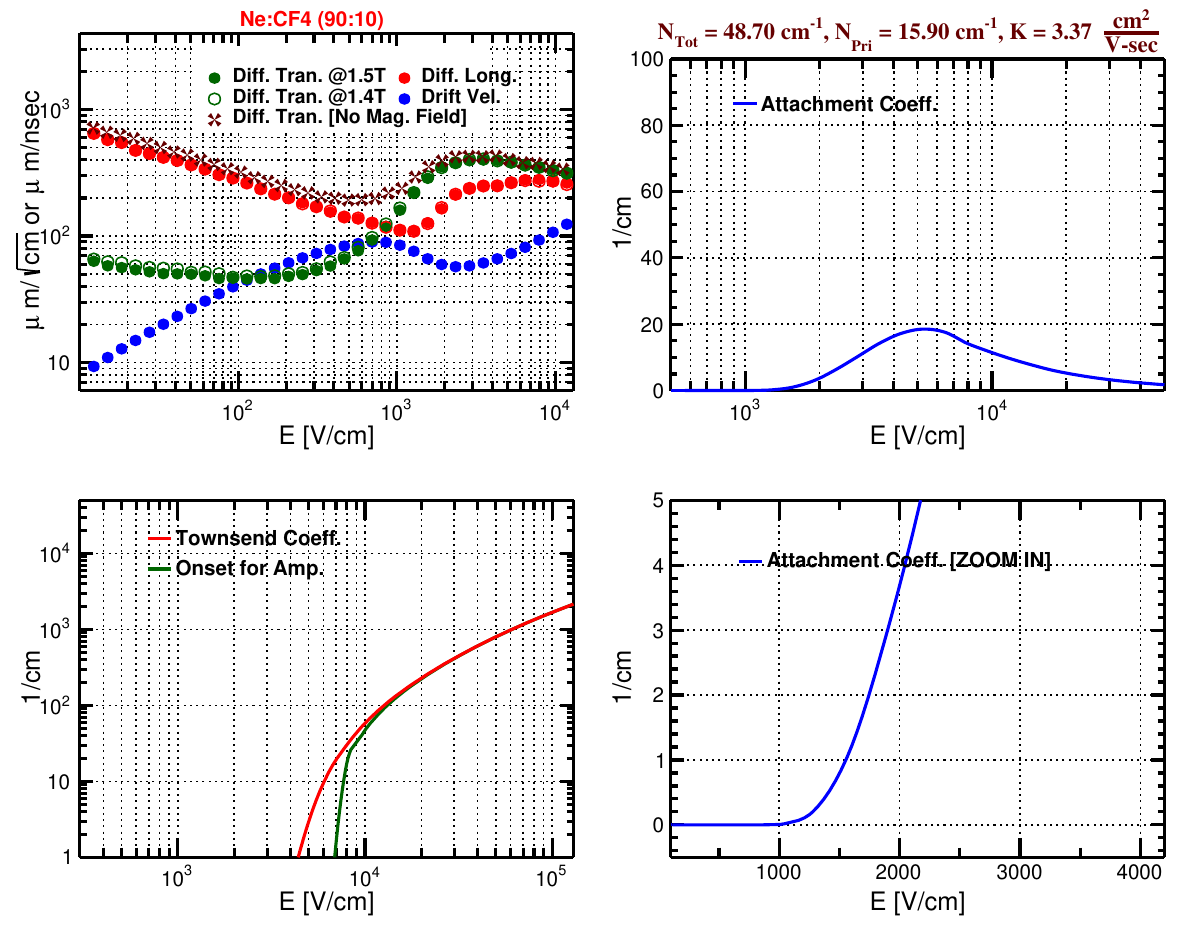}
    \caption{Gas properties of Ne:CF$_4$ 90:10. Top left: transverse diffusion at three different magnetic field settings (green open and filled circles, dark red markers), longitudinal diffusion (red filled circles) in units of $\mu$m/$\sqrt{\text{cm}}$, and drift velocity (blue filled circles) in units of $\mu$m/ns, all as a function of electric field strength in units of V/cm. Top right: Likelihood of attachment in units of incoming electrons lost per cm. Note the text above the plot, which provides the total number of ionization electrons per centimeter, the number of primary ionization electrons per centimeter, and the ion mobility. Bottom left: Townsend coefficient (red) and amplification factor (green) as a function of electric field. Bottom right: zoom in of top right, showing the exact location of the turn on of attachment. Results obtained using Magboltz.}
    \label{fig:NeCF49010}
\end{figure}

In any gas detector, one of the most crucial decisions is the operating point for the gas mixture. All the considerations of Sec.~\ref{Sec:DesignConsiderations} are in some way coupled to the gas mixture. A multi-dimensional phase space of gases was explored. The driving factors in the selection of operating point are the transverse diffusion, the drift velocity, the ion mobility, and the electron attachment probabilities. All the above quantities and more were studied as a function of electric field, to aid in the selection of the drift field. The studies were performed in MAGBOLTZ~\cite{Biagi:1999nwa,Biagi:1989rm}, a Monte Carlo software that provides bulk electron drift properties in gases by numerically solving the Boltzmann equation in the presence of $E$ and $B$ fields. The original choice of gas, which was studied at a test beam in 2018, was 90\% neon and 10\% CF$_4$\footnote{The notation ``A:B X:Y" will be used, where A and B are two gases, X represents the percentage of gas A, and Y represents the percentage of gas B}. The properties of Ne:CF$_4$ 90:10 are shown in Fig.~\ref{fig:NeCF49010}.
Several considerations went into the choice of drift field. Realistically, high drift fields are challenging to maintain over meter scales, due to the high voltage required on the central cathode. Higher voltages require more insulation, and thus more insulating material contributing to the radiation length of the TPC field cages. However, high fields can be seen to produce higher drift velocities, which are preferred to reduce the occupancy in the TPC in high rate operation. It can also be seen that at higher fields, the transverse diffusion increases. These considerations, among others, limited the search for an operating point to fields less than $\sim$500 V/cm. In general, it is also preferable to operate away from electric field values that sit on sharp slopes in the drift velocity or diffusion curves, because small deviations in the electric fields can cause large disturbances in the velocity or diffusion. The the zero-slope points for these parameters are typically at fairly high drift fields. As a compromise between all of these features, $\sim$400 V/cm was chosen as a drift field which maintains the ultra-low diffusion while maximizing the drift velocity. \par
The primary drivers of the choice of Ne:CF$_4$ 90:10 were the low transverse diffusion and very high ion mobility of 3.37 cm$^2$/(Vs) owing to the large concentration of neon, which has an ion mobility of around 4.1 cm$^2$/(Vs). One aspect which was realized during the test beam was the importance of the electron attachment seen in the top right and bottom right panels of Fig.~\ref{fig:NeCF49010}. Inside the GEM stack, the electric fields between the GEMs can be 2000 V/cm or higher in some locations. Compared to the nominal gain of 2000, an attachment of $\sim20$ electrons per centimeter appears almost negligible. However, since the first few GEMs are operated in low-IBF and thus low gain mode, any loss of the small number of electrons traversing the top GEMs becomes problematic. To assuage this issue temporarily at test beam, the GEMs were operated in a higher gain mode that would be unfavorable for sPHENIX running due to the worse IBF performance. An optimal gas mixture would thus have the attachment turn on only at voltages significantly higher than the voltage in the gaps between GEMs. \par
\begin{figure}[htbp]
    \centering
    \includegraphics[width=12cm]{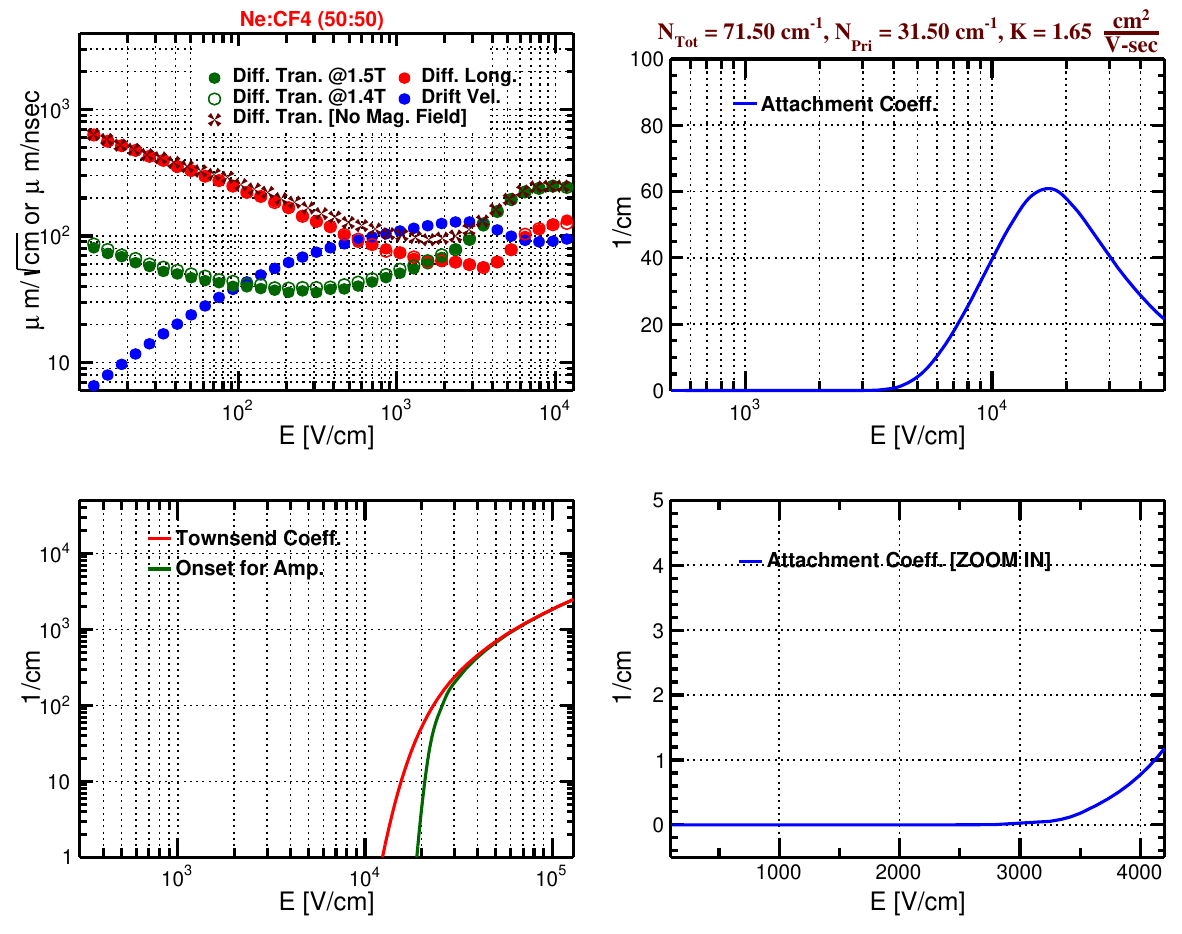}
    \caption{Gas properties of Ne:CF$_4$ 50:50. The primary differences compared to the 90:10 mixture are the ion mobility and attachment properties. The drift velocities are very similar, and the transverse diffusion is slightly smaller in 50:50. The amount of ionization for the 50:50 is also larger, which should increase the value of $N_{\text{Eff.}}$}
    \label{fig:NeCF45050}
\end{figure}
The Ne:CF$_4$ 50:50 gas mixture gave excellent results at a Fermilab test beam campaign in 2019, shown in Fig.~\ref{fig:FTBFResolution}. Since the test had no magnetic field, the position resolution results were extrapolated based on the expected reduction in the diffusion that would be gained by the introduction of the magnetic field. It should be noted that the magnitude of this reduction in diffusion due to the estimated sPHENIX magnetic field is very large, almost a factor of 5. The projected position resolution for ionization produced 10 cm from the readout was $\sim$95 $\mu$m, increasing to $\sim$120 $\mu$m at the full sPHENIX drift length. The structure of the beam at Fermilab, one spill every 60 seconds, meant that no space charge was present when the beam reached the detector, thus the quoted results pertain only to the intrinsic position resolution of the readout and the diffusion in the gas. Since the primary disadvantage of the 90:10 mixture compared to the 50:50 was the ion mobility, the impact of which could not be studied at test beam, these results were not completely unexpected.\par
An additional development in the gas selection originated from the unexpected sharp increase in cost of neon gas throughout 2022 and 2023\footnote{The cost increase was associated with the conflict in Ukraine, as Ukraine was previously the largest producer of neon gas in the world.}. The extreme cost of neon made using it as the operating gas untenable. The decision was therefore made to instead switch neon for argon and once again attempt an optimization. Many different argon-based mixtures were studied, including an attempt to reproduce the ion mobility properties of neon by adding helium to the mixture. Helium alone has a very high ion mobility, around 10 cm$^2$/(Vs), but adding enough helium to have any impact on the mixture\footnote{Since the ion mobilities add in parallel according to Blanc's law, the impact of adding a small fraction of very high mobility gas is small.} as a whole reduced the drift velocity, increased the transverse diffusion, and increased the electric field required for amplification in the GEM hole to a degree that would stress the voltage holding capability of the GEMs. The mixture which was finally chosen is Ar:CF$_4$ 60:40, which reproduces quite closely all the features of Ne:CF$_4$ 50:50 except for the ion mobility, which is worse by about 20\%. It should be noted that Ref~\cite{Santos:2018zwk} measured the ion mobility for Ar:CF$_4$ 60:40 to be around 1.6 cm$^2$/(Vs), higher than the 1.34 cm$^2$/(Vs) predicted by MAGBOLTZ\footnote{Ion mobility is defined as drift velocity divided by electric field. As discussed in Sec.~\ref{Sec:IonTransport}, the assumption of a linear relationship between drift velocity and $E$ only holds for certain electric fields. Thus, depending on the values of $E$ used in the extraction of ion mobility, different values for ion mobility could be measured.}. Despite the normalization difference, it is still assumed that MAGBOLTZ is roughly correct about the relative differences in the ion mobilities between argon and neon.
\begin{figure}[htbp]
    \centering
    \includegraphics[width=12cm]{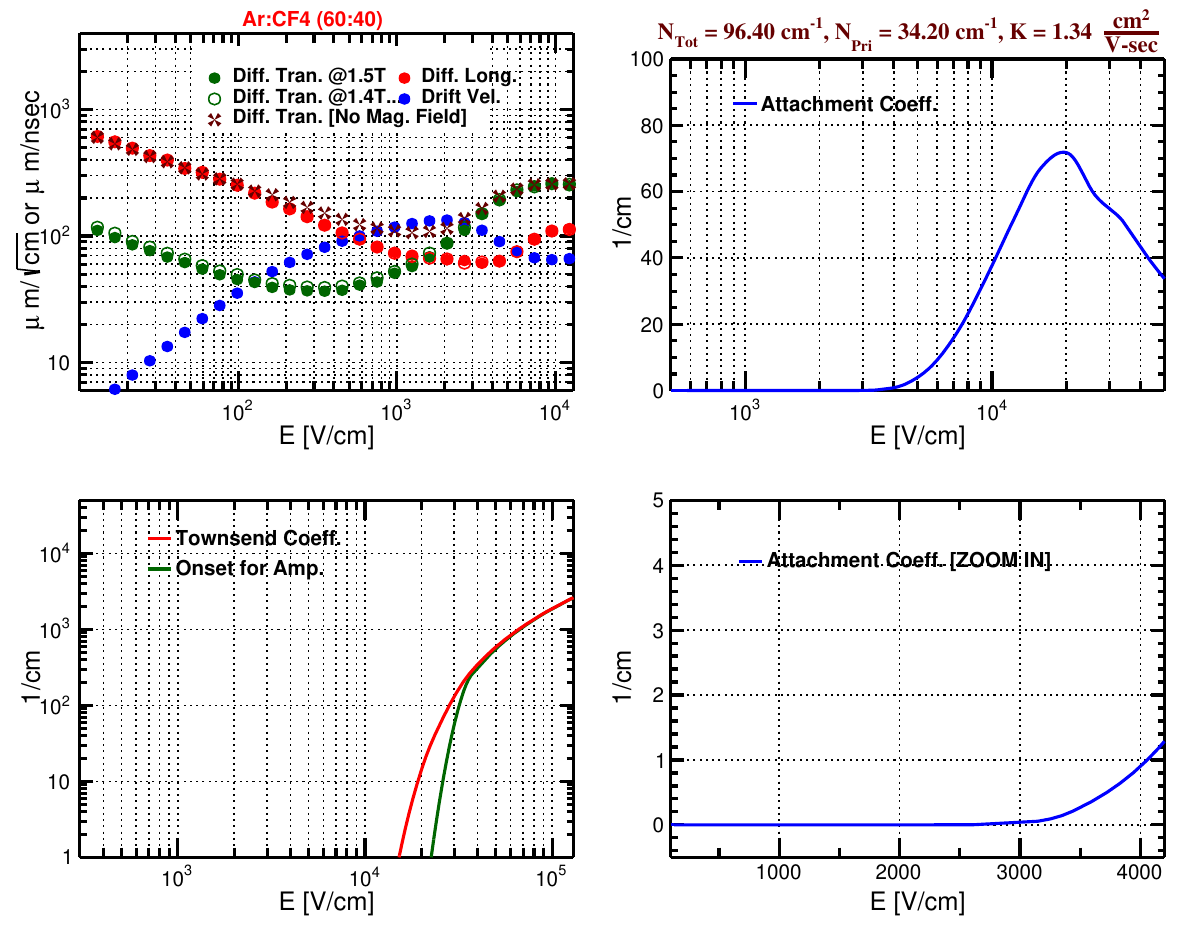}
    \caption{Gas properties of Ar:CF$_4$ 60:40.}
    \label{fig:ArCF46040}
\end{figure}
\subsection{Field Cages}
\label{Sec:FieldCages}
The role of the field cages in sPHENIX is to produce a uniform electric field in the $z$ direction and simultaneously serve as the gas volume. The ideal field cage would be a bulk material that could linearly reduce the voltage from the central cathode to the top of the GEMs, i.e. nominally from $\sim$45 kV to $\sim$5 kV. In lieu of such a material\footnote{These kinds of materials are being studied for use in liquid TPCs, including the DUNE Near Detector.}, small strips of conductor at discrete voltages can approximate a uniform linear decrease in potential. This technique has been used in essentially every collider TPC. In sPHENIX, due to the radial size constraints, the electric field distortions near the field cages should be reduced as much as possible to maximize the fiducial volume of the TPC. For this reason, extremely fine segmentation of the field cage conductor, including backing conductors were designed and employed. The field cages are both 2.1 m long, and the inner field cage (IFC) is 20 cm in radius, while the outer field cage (OFC) is 78 cm in radius.\par 
The main component of the field cages are the printed circuit boards (PCB) which contain field shaping stripes. The purpose of the rest of the assembly is predominantly to maintain the locations and voltage holding capability of these stripes, as well as serve as the gas volume. The field shaping stripes were plated onto 16.555 inch wide flexible PCBs, known as the striped circuit cards. Between each field shaping stripe is a series of resistors that step the voltage The cards for the inner field cage are 53.5 inches long, and the conductor is electroless nickel immersion gold (ENIG). ENIG is an ideal material for this application due to its corrosion resistance. The much longer OFC cards were unable to be produced in ENIG, so copper was used instead for the conductor\footnote{Copper has a tendency to corrode when exposed to air. The OFC cards received from the manufacturer were delivered with some corrosion already present. Various cleaning procedures were implemented to ensure that no corrosion would be present inside the TPC volume. There was, however, some permanent discoloration in the copper that could not be rectified. This discoloration was investigated and determined to have no noticeable effect on the performance.}. The cards for the outer field cage are 189.5 inches long, which stretched the capabilities of industry manufacturing. As a result, various repairs had to be made to the OFC cards, including repairing shorts between the conductors and fixing resistors. The OFC cards additionally had a region of exposed coverlay insulator. This insulator would charge up during the operation of the TPC and cause time-dependent field distortions. Thus they were manually covered by extra stripes. For both field cages, the cards can be thought of as three-layer PCBs, with the top layer being the field shaping stripes, then a layer of polyimide insulator, then a set of ``backing stripes" designed to stop field leakage between the front stripes and create an even more uniform field. The nominal width of the field shaping stripes is 2.3 mm, and the gap between stripes is 0.5 mm. The full field cage consists of 750 front stripes, each different from its neighbor by nominally 120 V to take the 45 kV cathode voltage down to the 5 kV top of the top GEM. 
\begin{figure}[htbp]
    \centering
    \includegraphics[width=12cm]{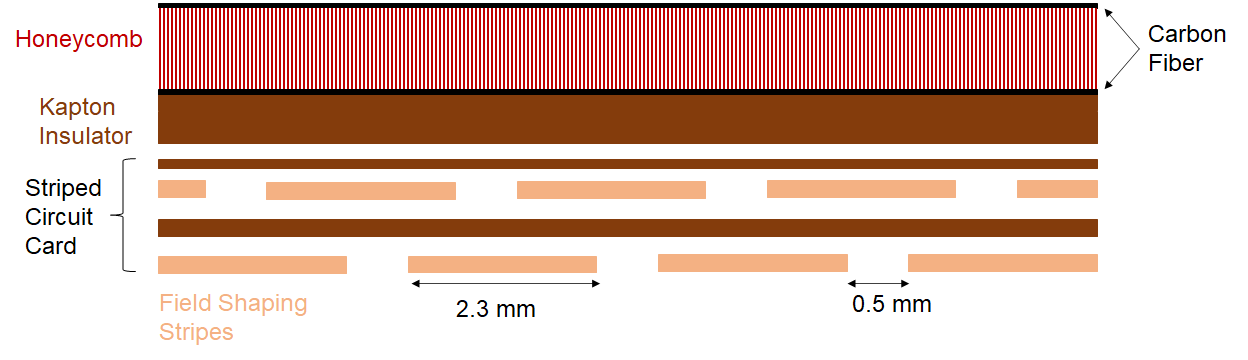}
    \caption{Sketch of the cross section of the field cages, including the striped circuit cards (not to scale). In the IFC, the field shaping stripes are ENIG, while for the OFC they are copper. The use of backing stripes improves the field uniformity near the field cage. The 18 layers of kapton between the conductors and the carbon fiber provide excellent dielectric strength. The carbon-fiber honeycomb sandwich is very lightweight and provides mechanical strength with a small material budget. The IFC was assembled starting from the carbon fiber at the top of the sketch and going down, while the OFC was assembled starting with the striped circuit card at the bottom and going up.}
    \label{fig:FieldCageSketch}
\end{figure}\par
The stripes are connected via four 4 M$\Omega$ HV pulse-withstanding resistors in series and parallel. This design builds in redundancy, such that in the case of a resistor becoming either a short or an open, the field cage can continue to operate with relatively minor field distortions. The resistance values were chosen to reduce the currents flowing through the stripes to a reasonable level. Concerns have been raised in past experiments about the power dissipation of the resistors heating the gas. In sPHENIX, the resistors are gravitationally down on the IFC and gravitationally up in the OFC, in hopes that the heat will tend to conduct into the field cages instead of propagating to the rest of the gas. Behind the resistors is a small piece of FR4 to serve as a stiffener, which prevents the card from flexing underneath the resistors and causing them to detach.\par
The field cages were assembled almost entirely on two mandrels, shaped to precisely the necessary radius and flatness. Since the field stripes sit on the interior of the TPC, the construction procedures for the inner and outer field cages were substantially different. The mandrel for the inner field cage was produced out of a large diameter PVC tube which was machined to the correct radius on a lathe in the Stony Brook machine shop. While the mandrels were to provide a rigid surface on which to build the field cages, equally important was their ability to be disassembled from within without harming the field cages after the field cages were complete. To this end, the IFC mandrel was cut into three independent pieces which could collapse inward to allow the completed field cage to be removed.
\begin{figure}[htbp]
    \centering
    \includegraphics[width=13.9cm]{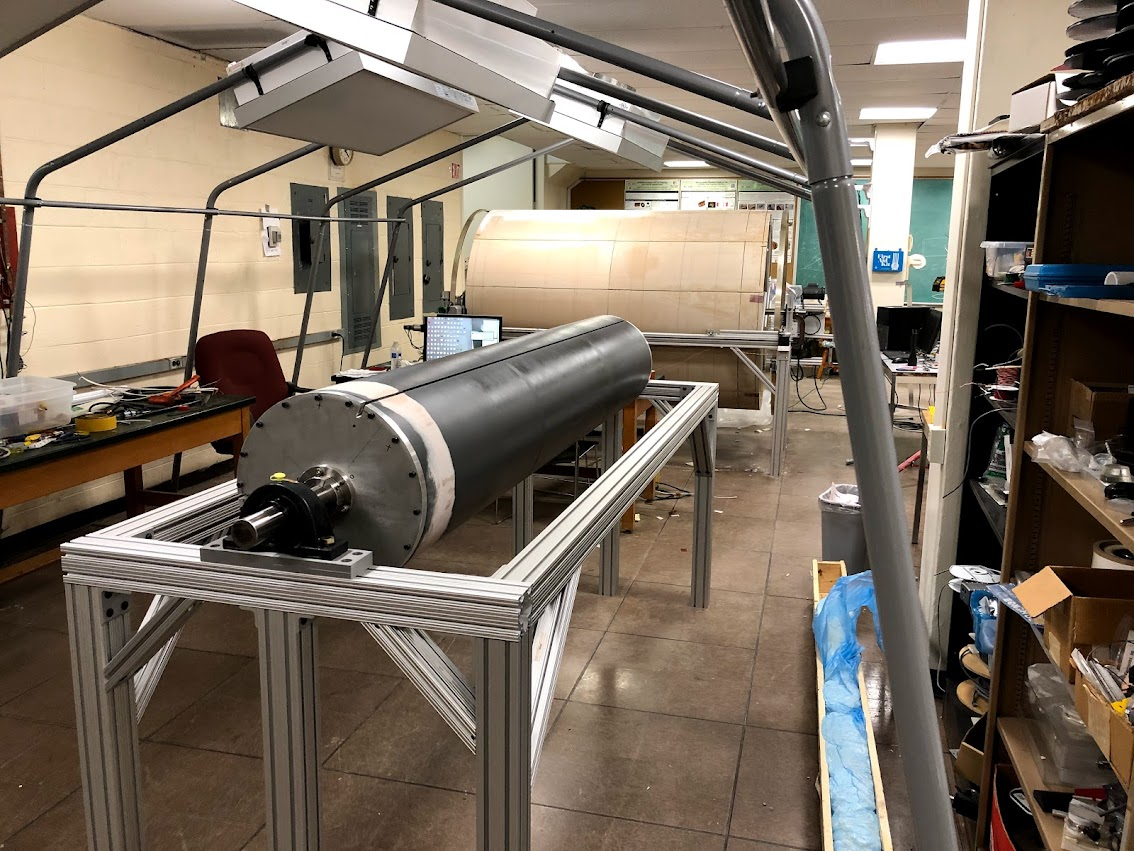}
    \caption{The inner field cage mandrel (near, dark gray) and outer field cage mandrel (far, beige).}
    \label{fig:mandrel}
\end{figure}
The mandrels were assembled on tables constructed of 80-20. Computer controlled motor systems enabled the mandrels to be rotated at fixed or variable speeds. A precisely aligned translating stage with various attachments was able to image and survey the mandrels through the entire process. A microscope mounted to the stage enabled the striped circuit cards to be placed at the proper locations with mil-level precision. \par
\begin{figure}[htbp]
    \centering
    \includegraphics[width=13.9cm]{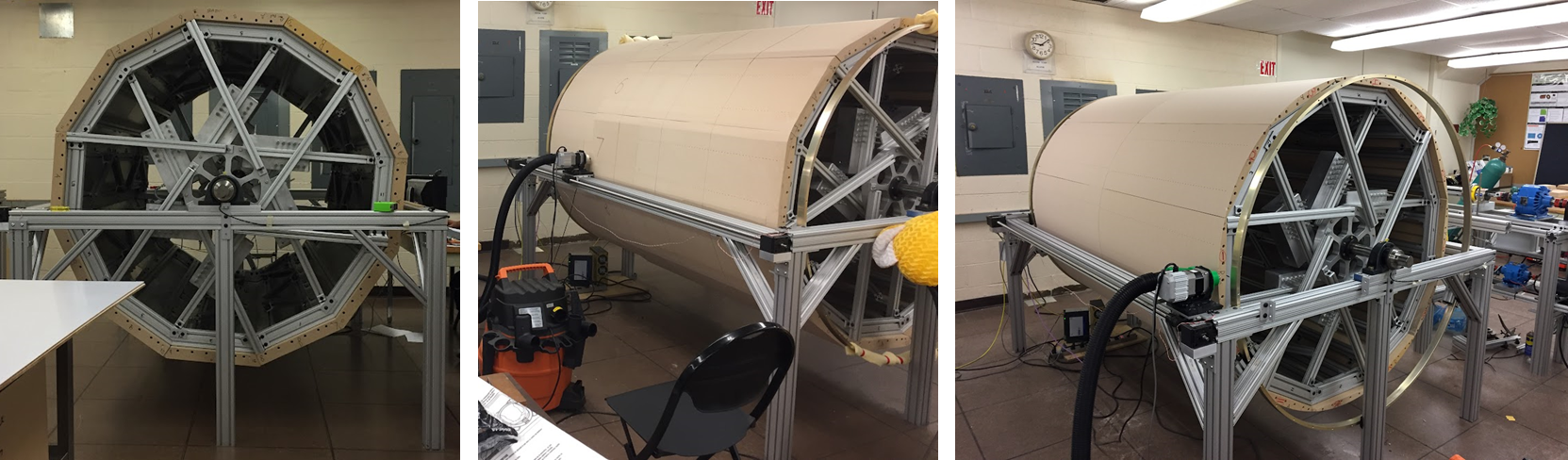}
    \caption{Construction of the outer field cage mandrel. Left: dodecagonal shape prior to foam cutting. Middle: process of foam cutting. Right: cylindrical mandrel after the first round of foam cutting.}
    \label{fig:OFCmandrel}
\end{figure}

For the OFC, where the field shaping stripes had to be positioned face-down onto the mandrel, a vacuum system was devised that could hold the striped circuit cards in place with negative pressure while performing various operations. The challenge was in holding the cards in place, without the use of any material such as glue or tape that would make removal of the OFC from the mandrel impossible, in such a way that the mandrel could freely rotate. The foam boards used in the construction of the OFC mandrel (see Fig.~\ref{fig:OFCmandrel}) had channels machined into them, which could be brought to vacuum by a large capacity vacuum pump. Long, large diameter, flexible vacuum rated tubes were spooled around the motor shaft in a manner similar to a hose reel. These tubes were connected to a set of vacuum distribution boxes that brought the vacuum to a large number of the channels in the foam boards.\par
The decision to merge the field cage and gas volume into one structure was driven by the radial compactness required for sPHENIX. The result is that near the central cathode, the field cage must be extremely well electrically insulated from the grounded materials that are only a fraction of an inch away. The insulator selected was Kapton polyimide from 3M, which has an unmatched dielectric strength of 7700V/mil. In principle, to insulate the 45 kV of the central cathode from ground, it would require only 7 mils of kapton. However, since a dielectric breakdown in the insulator would mean the cathode could never again be brought to full voltage, a substantial safety factor was opted for, amounting to 18 layers of 3 mil thick kapton with 2 mil thick adhesive. Kapton of the $\sim$2 meter width needed to produce the insulating layer of the field cage out of one continuous sheet of kapton was not commercially available, so instead the 18 layers of kapton were applied in lengths of 44 inches, 18 inches, and 21 inches. \par
\begin{figure}[htbp]
    \centering
    \includegraphics[width=13.9cm]{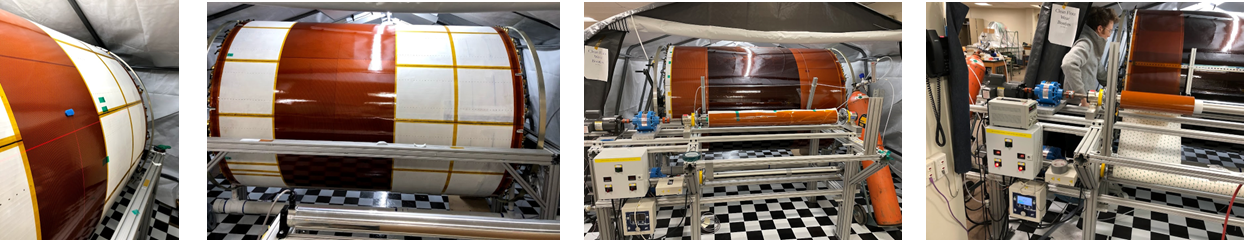}
    \caption{Construction of the outer field cage. Leftmost: Placement of first striped circuit card. Second to left: Two striped circuit cards in place. Second to right: All striped circuit cards in place, first layers of kapton insulator wound around the cards. Rightmost: Procedure of winding kapton onto the mandrel. The tensioner system can be seen in the foreground of the two rightmost panels.}
    \label{fig:OFCConstruction}
\end{figure}
The kapton was applied under tension to remove any air bubbles or wrinkling that could reduce the voltage holding capability. Around 1 lb/in. of tension was applied to the kapton by a tensioner system, shown in Fig.~\ref{fig:OFCConstruction}. Additionally, helium gas was blown underneath the kapton as it was applied. The helium supply tube can be seen under the kapton being applied in the last panel of Fig.~\ref{fig:OFCConstruction}. The reason for this is that helium can diffuse through kapton at a fairly high rate, on the order of hours for the kapton thicknesses used for the TPC, thus producing bubbles of helium which disappear over time. The tensioner system pulled back on the material to provide tension as the mandrels were spun by the harmonic drive motor. On the outer field cage, where the radius was substantially larger, the tension applied by the tensioner produced a torque that was too great for the mandrel motor. This was overcome by offsetting the torque of the tensioner with weights that were manually shifted during the kapton application process. The gaps between the kapton layers were filled with DP460 epoxy, which also has a fairly high dielectric strength of 1100V/mil. The epoxy seams were strategically placed in alternating locations, to prevent any line-of-sight from the circuit cards to ground consisting of only epoxy. \par

\par
To maintain a uniform field, the field shaping stripes must maintain the correct locations and voltages with respect to the central cathode, including any mechanical deflections experienced due to sagging under gravity. To reduce these deflections, the mechanical structures of the two field cages were constructed out of a honeycomb-carbon fiber sandwich which exhibits extreme stiffness and is very low mass. The sheets of carbon fiber are electrically connected to the end rings, which provides a reference ground. The inner field cage additionally has a sheet of aluminized mylar providing the ground beyond the layers of kapton.\par
The 45 kV high voltage is brought to the central cathode through the OFC. The HV cable carrying the cathode voltage passes through the South end ring and between the two layers of carbon fiber, until it approaches the center of the field cage in $z$. At that location, a high voltage ``house" made of PEEK accepts the HV cable with the ground braid removed, as the ground braid needs to be kept far from the exposed conductor. The insulated cable continues through a small hole in the house until it reaches the point in the striped circuit card where it is to be soldered. The HV house has a large opening at this point to allow for manual soldering of the conductor to the striped circuit card. After the soldering was complete, the opening in the house was filled with around 1 inch of DP460 epoxy with dielectric strength of 1100 V/mil. The total dielectric strength is therefore significantly higher than any voltage that would be applied to the conductor. 
\begin{figure}[htbp]
    \centering
    \includegraphics[width=13.9cm]{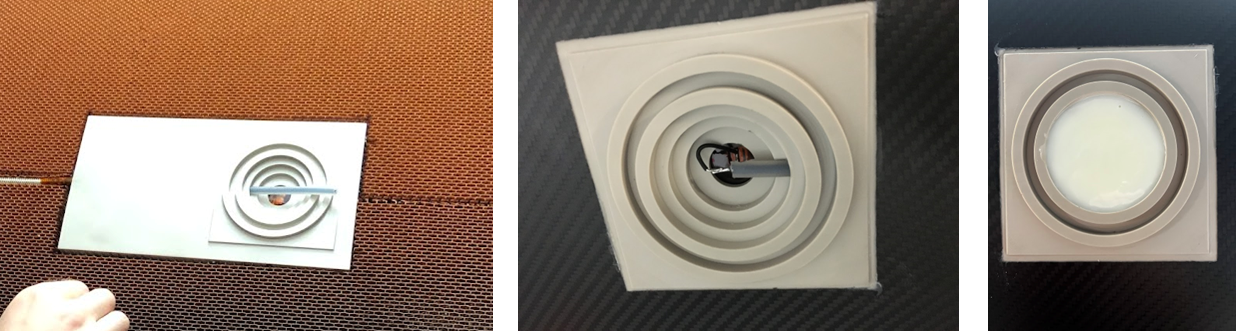}
    \caption{The procedure of bringing the central cathode voltage to the center stripe of the field cage. Left: The placement of the HV house (tan) inside the honeycomb, prior to the second layer of carbon fiber. Middle: HV cable inside HV house, soldered to center stripe of OFC. Right: HV house fully insulated with DP460 epoxy.}
    \label{fig:HVHouse}
\end{figure}

\begin{figure}[htbp]
    \centering
    \includegraphics[width=13.9cm]{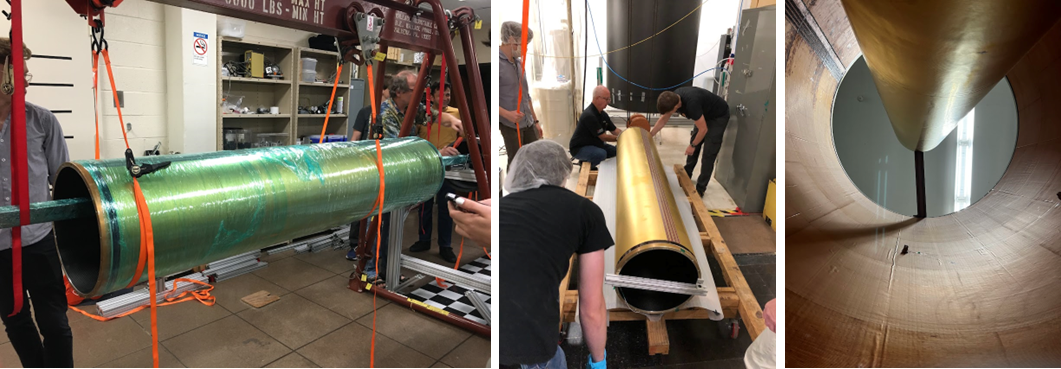}
    \caption{The procedure of removing the IFC and placing it concentric with the OFC. Left: IFC being removed from the PVC mandrel. Middle: IFC prepared to be lifted and placed inside the OFC. Right: IFC centered inside the OFC. }
    \label{fig:IFCInstall}
\end{figure}
The central cathode is soldered to the central stripe of the OFC, which then brings the voltage to the corresponding center stripe on the IFC. The IFC and OFC both have wires soldered to the last stripe that allow moderate voltages to be applied to them. This is necessary to equilibrate the voltage of the final stripe with the top of the top GEM, to prevent field distortions near the field cages. It also allows the TPC to be operated at different field strengths without the danger of a spark between the GEM stack and the field cage. The voltage holding capabilities of the field cages were tested prior to beginning assembly of the full TPC. Both field cages successfully maintained the expected current at the final stripe with 50 kV applied at the center stripe, i.e. where the central cathode is connected, for over an hour. 
\subsection{Pad Planes}
\label{Sec:PadPlanes}
The sPHENIX pad planes are designed to provide good charge sharing capabilities for improving position resolution on the $r\phi$ component of particle locations. Additionally, there should be low noise arising from capacitive coupling of the traces that connect the pads to the connectors. They should have pads which are fine enough to have effectively zero single-pad hits for the charge clouds coming from the GEMs.\par
\begin{figure}[htbp]
    \centering
    \includegraphics[width=8cm]{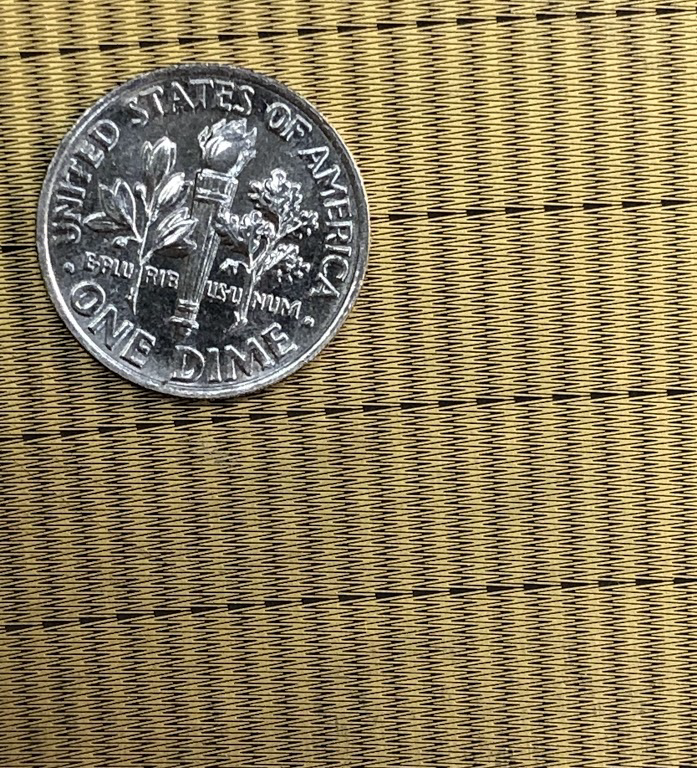}
    \caption{Zig-zag pattern pads on an R2 pad plane, with dime for reference. The width of the conductor is roughly 2mm, and the height of the pad is 1 cm. The radial location where the pads meet produces a small dead area.}
    \label{fig:ZigZags}
\end{figure}

Three different sizes of pad planes were designed and fabricated for the R1, R2, and R3 locations respectively. The pads are zig-zag shaped copper conductor, plated with electroless nickel immersion gold (ENIG). Each pad plane has 16 pad rows in $r$, producing a total of 48 measurements of ionization locations for tracks which traverse the whole TPC.
\begin{figure}[htbp]
    \centering
    \includegraphics[width=13.9cm]{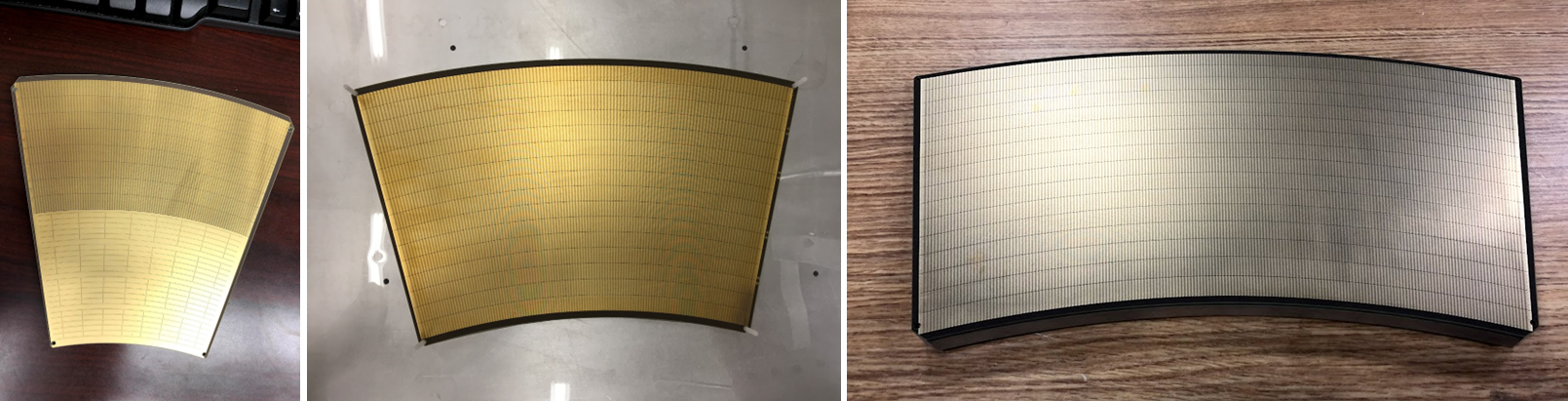}
    \caption{The sPHENIX TPC pad planes. Left: The R1 pad plane, containing 1024 zig-zag pads and several ``antenna" pads that can be used to monitor the amount of incoming charge at the innermost radii. Middle: The R2 pad plane, containing 2048 pads. Right: The R3 pad plane, containing 3072 pads.}
    \label{fig:PadPlanes}
\end{figure}
The pad shapes themselves are unique, and were the subject of extensive R\&D. While so-called ``chevron" shaped pads have been used in the past, true zig-zags with a large number periods and substantial interleaving are being used for the first time in sPHENIX. The width of the conductor producing the zig-zag is around 2mm, and the gap between conductors is 2.7 mil. The gaps are produced by a chemical etching process. Laser etching, which can generally produce much smaller gaps, was deemed too expensive. The industry standard for distances between conductors using chemical etching is 4 mil, so the manufacturer often had to make post-production modifications to remove shorts between pads, broken pads, etc. Since chemical etching cannot easily produce sharp corners, the tips of the zig-zags were blunted, giving them a rounder shape. The blunted tips of the zig-zags may also aid in the electrical stability, as sharp tips have the potential to induce sparks\footnote{This behavior was noticed in some early tests with MicroMegas. A possible explanation is that the sharp tips can lift slightly off the PCB substrate and create high-field regions that spark. No sparks to the pad plane were observed with GEMs, likely because the fields are not as high near the pad plane.}. For R2 and R3, the zig-zags are 1 cm in height, while for R1 the radial pad density is doubled, thus the pads are only 5 mm in height. The density of R1 is larger to help deal with the high occupancy expected at small radii. The radial point where the zig-zags meet on R2 and R3 produces a small dead area that can be seen in Fig.~\ref{fig:ZigZags}. The sPHENIX GEMs carry the same radial segmentation, such that the GEM dead area and the pad plane dead area overlap. The segmentation of the R1 GEMs is the same as R2 and R3, but the pad density is higher. Thus some R1 pads will radially meet one another inside the active area. To reduce the dead area as much as possible, zig-zag pads which almost completely occupy this dead area were designed and implemented.
\begin{figure}[htbp]
    \centering
    \includegraphics[width=10cm]{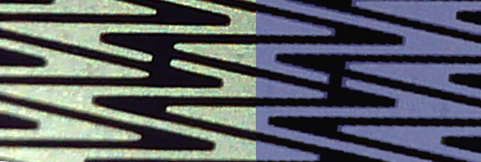}
    \caption{Microscope image of the updated radial termination scheme of the zig-zags, utilized in the R1 pad plane. The right side overlays the CAD design with the finished product, showing quite good agreement.}
    \label{fig:R1Boundary}
\end{figure}
Special pads sit on the $\phi$ boundaries of the pad plane. The zig-zags do not terminate nicely against a radial boundary, so a single large filled pad is used at that location.
\begin{figure}[htbp]
    \centering
    \includegraphics[width=10cm]{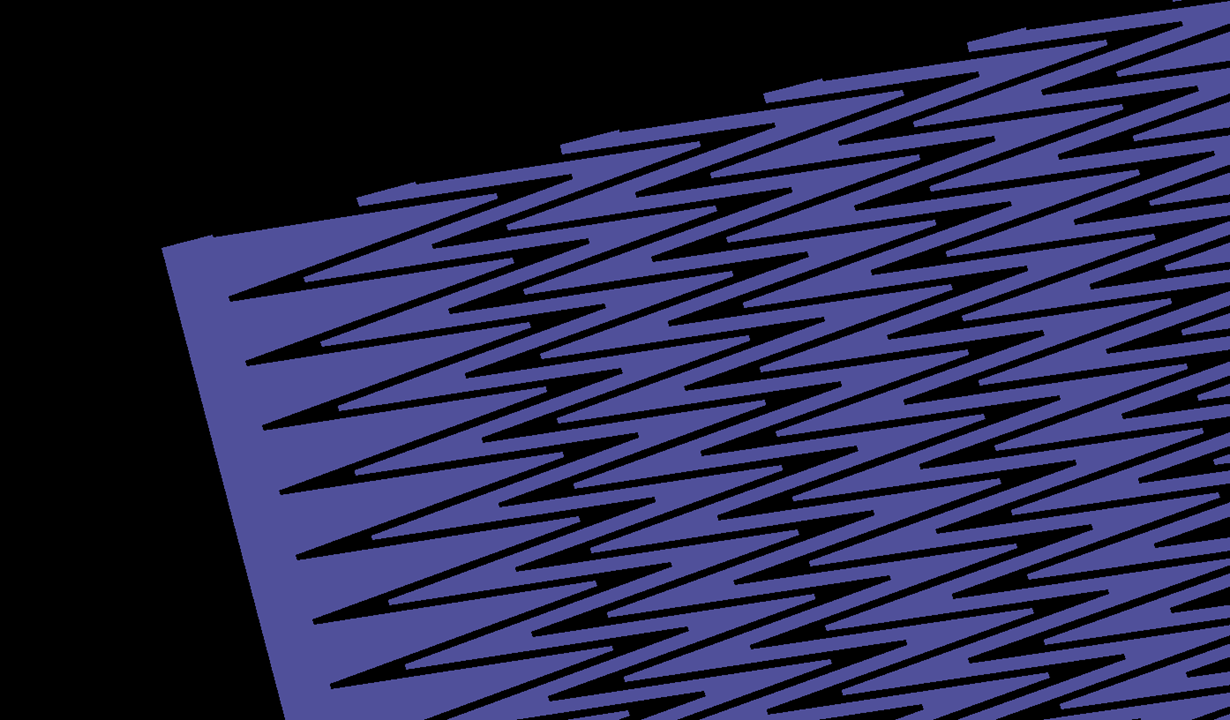}
    \caption{CAD drawing of the radial and azimuthal boundaries of the pads. The large pads on the azimuthal boundary are expected to suffer from lower gain owing to electric field distortions associated with the GEM frame, which is nearby.}
    \label{fig:PadPlaneEdges}
\end{figure}
\par
R1 additionally has 32 ``antenna" pads, visible in Fig.~\ref{fig:PadPlanes}, located at the innermost 10 cm in radius, that can monitor the amount of ionization produced at small radii. The choice was made not to instrument this region with the normal zig-zag pads, as the magnitude of the space charge and other electric field distortions is largest in this region and it is unlikely that signals from this region could be reliably reconstructed. The size of the active area of the antenna pads was chosen such that signals from the much larger antenna regions would not fully saturate the electronics, which were designed for fairly small signal sizes.
Underneath the pads is an 18 layer FR4 PCB. Every other layer is a ground plane, to reduce cross-talk between traces and noise as much as possible. Traces are routed in such a way that they do not approach near to other traces on the same layer. The pads are routed from the top layer down to the bottom layer, where they connect to a pin on a high-density connector produced by Samtec. \par
\begin{figure}[htbp]
    \centering
    \includegraphics[width=10cm]{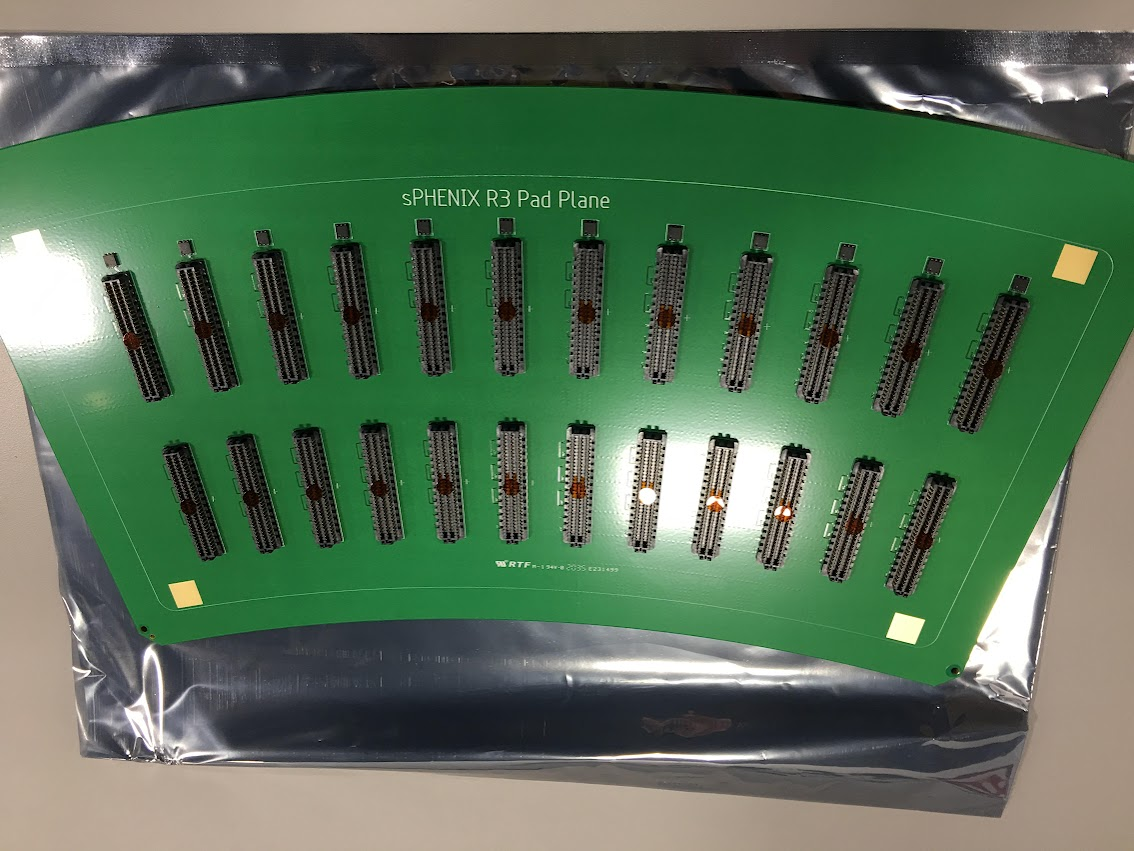}
    \caption{Back side of an R3 pad plane. The Samtec connectors each have 128 pads associated with them. The spacing of the connectors is designed to accommodate the electronics cards.}
    \label{fig:R3Back}
\end{figure}
The pad planes have through holes in the corners which accept nylon threaded rods that align the GEMs on the padplane and maintain their positions. The nylon rods pass through the pad plane and enter the aluminum strongback, to which the padplane is epoxied. The epoxy is applied both to mechanically attach the pad plane to the strongback, as well as to produce a gas seal between the two. A ``trough" in the strongback allows the HV tails of the GEMs to pass from the TPC volume to the outside even after the padplane is epoxied to the strongback. Later on, this trough is filled with DP460 epoxy to finish the gas seal\footnote{It was considered to use some kind of silicone plug that could be removed in the event that a GEM failed and needed to be replaced. However, the GEMs were found to be extremely stable during testing. Even under conditions much more extreme than those in sPHENIX, i.e. bombardment with heavily ionizing $\alpha$ particles, the GEMs did not spark. Only an intention increase in the voltages beyond what would reasonably be used in sPHENIX caused the GEMs to spark. Thus it was deemed an unnecessary risk to use a plug, which could introduce a gas leak. Even with the tails epoxied in place, a repair is still possible, albeit more challenging.}. An alignment pin in the strongback mates with an alignment hole in the pad plane and guarantees the alignment of the padplane with respect to the strongback. The strongback with padplane and GEMs in place then attaches to the wagon wheel, described in Sec.~\ref{Sec:WagonWheels}. The strongback has an O-ring running around the perimeter, which seals when bolted to the wagon wheel.
\begin{figure}[htbp]
    \centering
    \includegraphics[width=10cm]{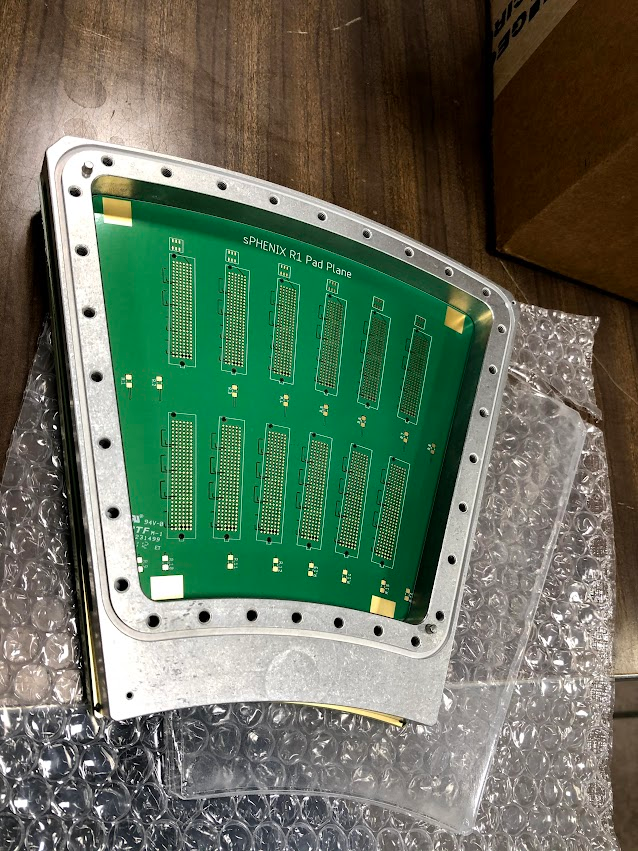}
    \caption{An R1 strongback placed behind an R1 pad plane. The strongback O-ring groove and the pins (top left, bottom right) which allow it to be easily aligned in the wagon wheel can be seen. The pad plane has not yet had the Samtec connectors attached, therefore the individual pads to which the zig-zags are routed are visible.}
    \label{fig:R1Back}
\end{figure}
\subsection{Wagon Wheels}
\label{Sec:WagonWheels}
\begin{figure}[htbp]
    \centering
    \includegraphics[width=10cm]{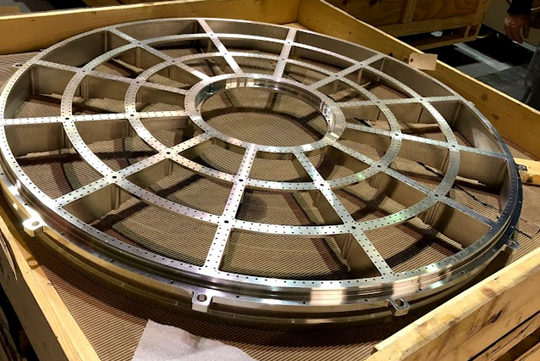}
    \caption{One of the two TPC wagon wheels.}
    \label{fig:WW}
\end{figure}

\begin{figure}[htbp]
    \centering
    \includegraphics[width=10cm]{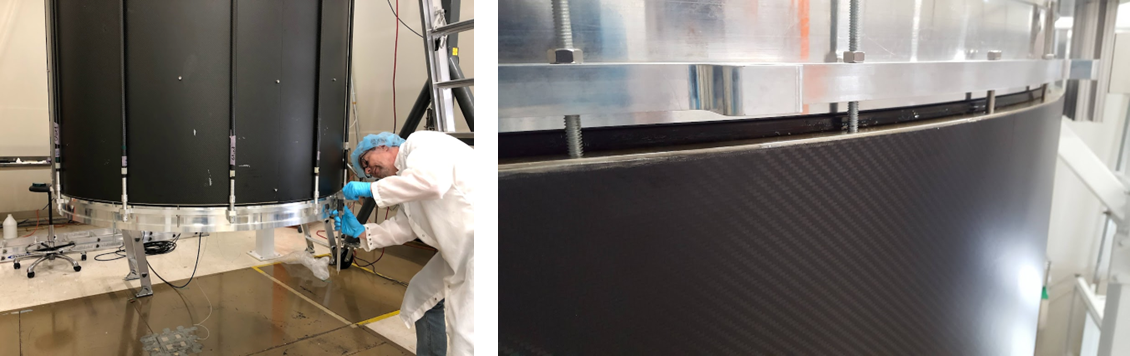}
    \caption{Left: Mounting the final wagon wheel to the field cages. The carbon fiber rods are inserted in the wagon wheel support holes, and the nuts for tightening the rods down are in place. In this image, the wagon wheel is not yet fully mounted to the field cages. Right: The spring energized seal (black) seated on the wagon wheel before making the seal on the end ring. The seal was coated in vacuum grease to lubricate it and allow a better seal.}
    \label{fig:WWRods}
\end{figure}

The large aluminum structures at the endcaps of the TPC which support all the readout modules and maintain the positions of the field cages are known as the ``wagon wheels". They are machined out of a single piece of aluminum to ensure mechanical strength and gas tightness. An image of one of the wagon wheels is provided in Fig.~\ref{fig:WW}.
The wagon wheels have a series of 12 ``spokes" each of which have two NPT ports through which either gas can be flowed, or the laser systems can enter the TPC volume. The spokes are reinforced with thicker aluminum ribs for mechanical strength when supporting both field cages. The spokes contain a large number of holes, to which the strongbacks are bolted. The O-ring of the strongback makes a gas seal to the flat surface of the wagon wheel. The inner and outer field cages were bolted to the inner and outer radii of the wagon wheels. The wagon wheels have grooves at these locations into which spring-energized seals were placed to provide gas seals on the end rings of the field cages. Around the exterior of the wagon wheel are holes which capture long carbon fiber rods that traverse the entire length of the TPC. These rods are used to further increase the rigidity of the TPC while only adding marginally more material. The carbon fiber rods and the way they mount to the wagon wheel can be seen in Fig.~\ref{fig:WWRods}.
Along the outer perimeter of the wagon wheel are the aluminum brackets that strain relieve the high voltage cables that run to the GEMs and field cages.
\subsection{Gas Electron Multipliers}
\label{Sec:GEM}
The design of the GEM readout is very similar to that used in ALICE, with a few key differences. The target gain for the GEMs is 2000, the same as ALICE. Also the same as ALICE, all the GEMs are 5 $\mu$m thick copper plated onto 50 $\mu$m polyimide insulator. Four GEMs with the same pitch and rotation as those used in ALICE are used for sPHENIX. The GEM facing the drift volume of the TPC is denoted as GEM 1, and the GEM closest to the pad plane is GEM 4. The ``pitch" of a GEM refers to the distance between the centers of the holes. Two hole pitches are used, 140 micron and 280 micron. GEMs 1 and 4 are 140 micron pitch, while GEMs 2 and 3 are 280 micron pitch to maximize their ion blocking properties. The GEMs are also ``rotated" 90$\degree$ in their hole patterns, meaning there is no direct line of sight from the drift volume to the pad plane. This pattern of GEM stack was also inherited from ALICE, who performed extensive studies to optimize the ion-blocking capabilities.
\begin{figure}[htbp]
    \centering
    \includegraphics[width=10cm]{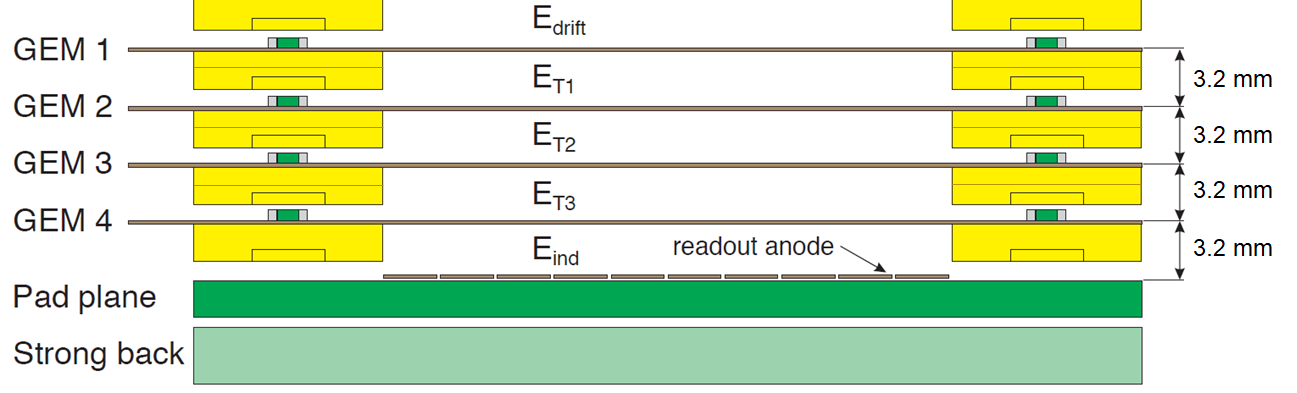}
    \caption{Sketch of the sPHENIX GEM stack configuration. Figure adapted from Ref.~\cite{Lippmann:2014lay}}
    \label{fig:GEMSketch}
\end{figure}
The gaps between the GEMs are nominally 3.2 mm, larger than the ALICE gaps of 2 mm. Also unlike ALICE, the GEMs are glued to frames on both the top and bottom sides of the GEM with a thickness of 1.6 mm.\par
In sPHENIX, the readout at the endcap is segmented 3 times in $r$ and 12 times in $\phi$. The three radial segmentations are denoted as R1, R2, and R3, with R1 being the closest to the interaction region and R3 being the farthest. Since the GEMs all subtend the same angle, the R3 GEM foils are substantially larger than those of R1. The sPHENIX GEMs are small and finely segmented compared to those used in ALICE. Since the orientation, hole patterns, and high voltage ``tail" locations of the four gems are not identical, four unique GEM designs (denoted as "flavors") were required to be manufactured by the CERN MPGD workshop. 
\begin{figure}[htbp]
    \centering
    \includegraphics[width=13.9cm]{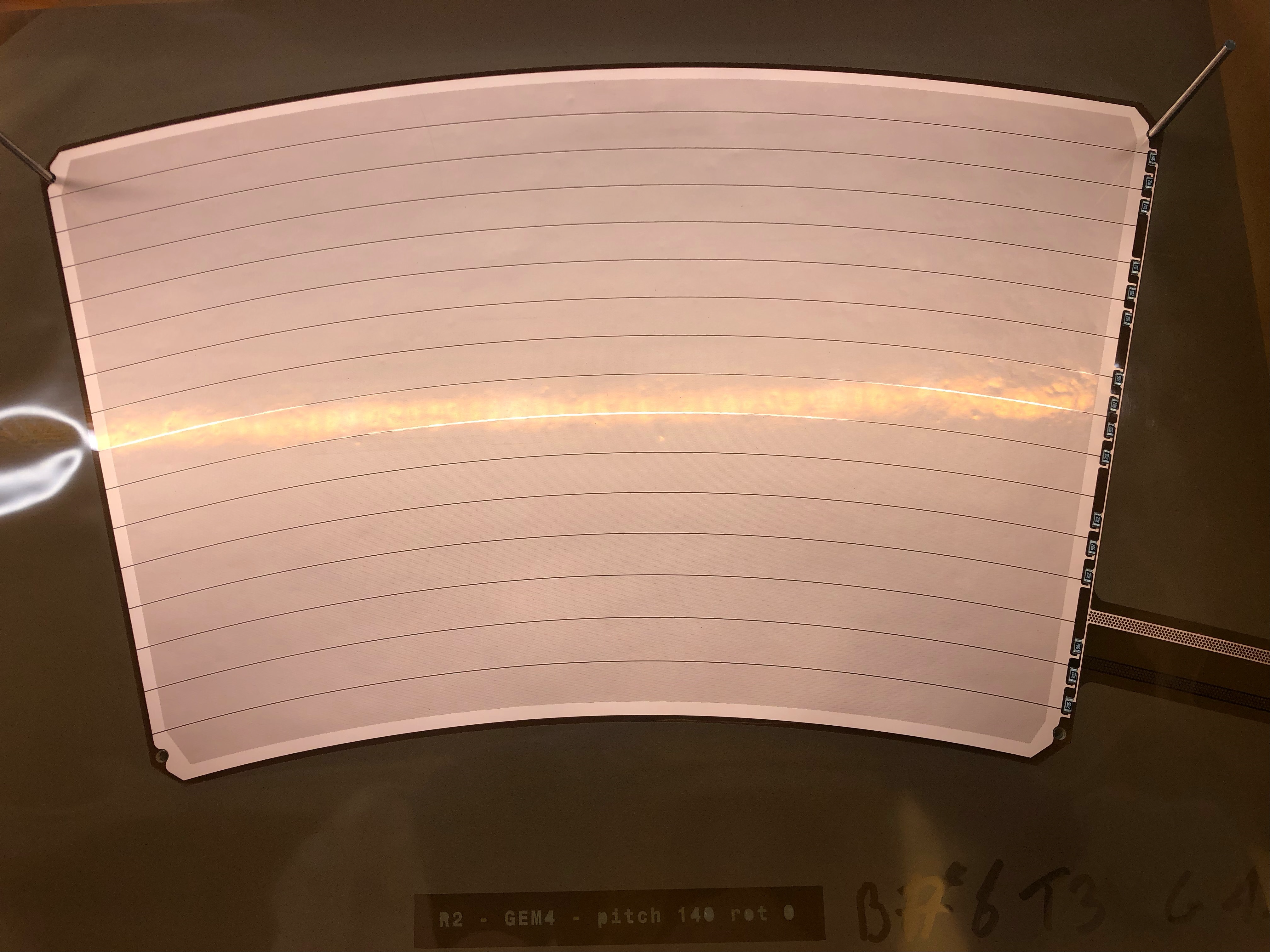}
    \caption{R2 GEM foil before framing. The fine radial segmentation and the built-in resistors (right) are visible.}
    \label{fig:GEMFoil}
\end{figure}
 The segmentation of the GEM copper follows the radial segmentation of the pad plane, i.e. the copper segments are 1 cm wide in the radial direction. The radial boundaries between segments are free of copper and avalanche holes, which disturbs the electric field slightly and produces a dead area slightly larger than the size of the holeless region. This segmentation can be found on all four flavors of GEM foil used in sPHENIX. The fine GEM segmentation reduces the amount of stored energy in any segment of the GEM, which can cause damage to the GEM if released due to a discharge or an unexpected change in voltages. The segments are supplied voltage through a 20 M$\Omega$ HV pulse-withstanding resistor. R3 is additionally segmented down the middle, meaning it has two HV tails, one on either side. This would be dangerous if the R3 GEMs were operated from different power supplies, as a drop in voltage of one half would certainly cause discharges between the two halves. To alleviate this, both halves of the R3 GEM are powered from the same HV cable.
\par
\begin{figure}[htbp]
    \centering
    \includegraphics[width=13.9cm]{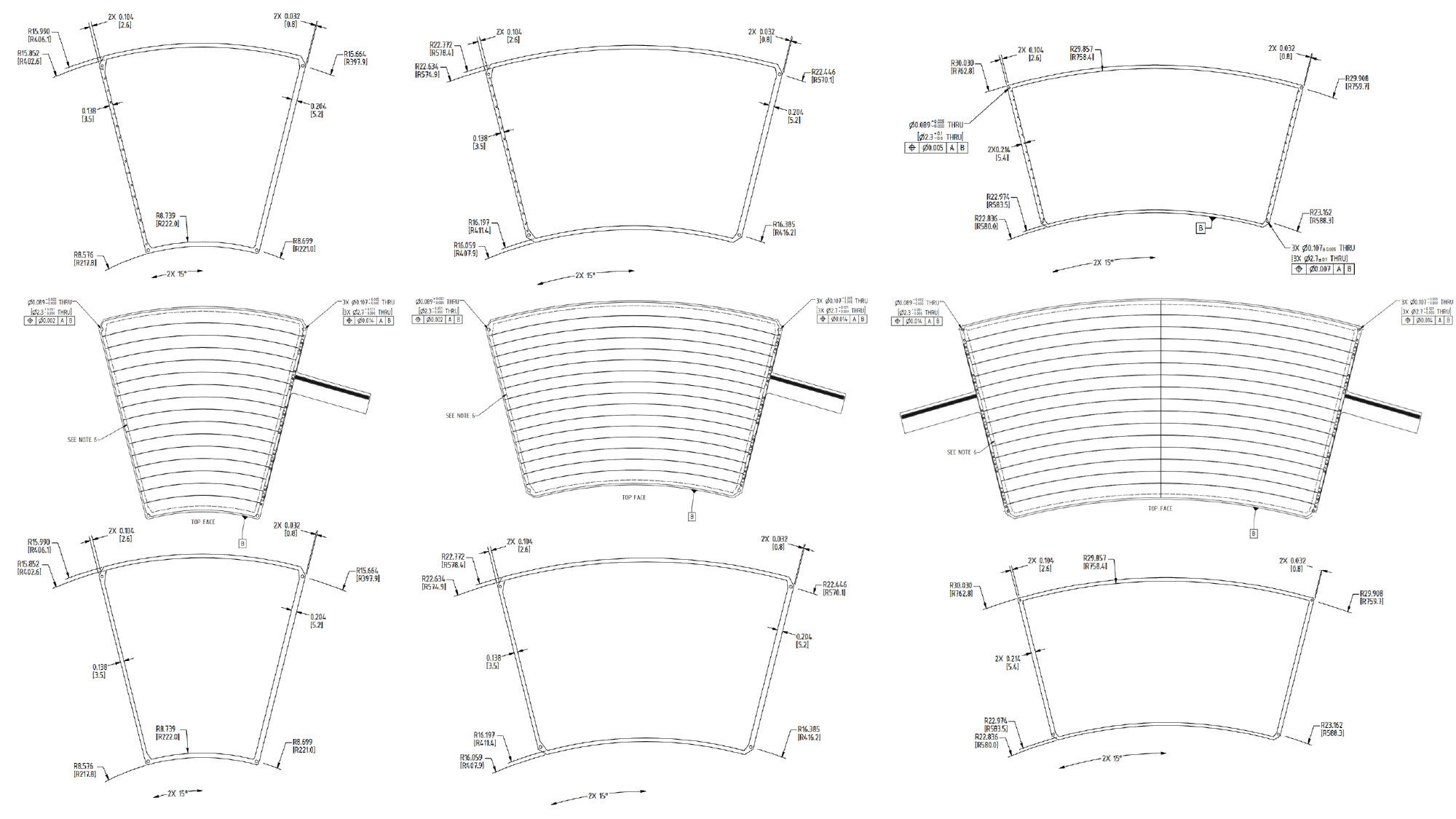}
    \caption{Engineering drawings of the GEM foils and GEM frames for R1 (left), R2 (center), and R3 (right). Dimensions are in inches [mm]. The topmost drawing is the top frame of the GEM, the middle drawing is the GEM foil itself, and the bottom drawing is the bottom frame. Note 6, mentioned in the GEM foil drawings, states that the perimeter of the GEM foil should be free of holes to permit better mechanical strength and adhesion to the frames.}
    \label{fig:GEMFrames}
\end{figure}

The GEMs from CERN were glued to frames, the designs of which are shown in Fig.~\ref{fig:GEMFrames}. Two frames are used, one on top of the GEM and one on bottom. The frames are made of polished NEMA G11/FR5, and are 0.063 inches (1.6 mm) thick. The frames were glued to the GEMs with a thin layer DP460 epoxy. The GEMs were framed at Temple University, Wayne State University, Vanderbilt University, and the Weizmann Institute of Science. For some of the GEMs, the epoxy was mixed with a small amount of alcohol to decrease the viscosity and allow the epoxy to be applied to the thin area around the perimeter of the GEM more easily. 
The framed GEMs were then shipped to Stony Brook for assembly into readout modules\footnote{Throughout the following a ``module" is, at minimum, four GEMs stacked onto a pad plane which is epoxied to a strongback.}. \par
At Stony Brook, the framed GEMs had the excess polyimide material removed and the resistors potted with DP460. Since the modules in sPHENIX are very close to one another, the potential for a spark from a module to the neighboring modules in the event of a HV difference between the modules was a concern. The resistors were potted, i.e. the entire trench containing the resistors and their traces was filled with epoxy, to prevent the exposed copper traces from being a potential spark point for the neighboring module in the event that a module were to be ramped down to ground. A thin sheet of kapton was wound around the finished GEM stack for the same purpose. The HV tails had thin pieces of kapton tape placed over the exposed copper. The GEMs were stored on laminar flow tables inside a clean tent to reduce the amount of particulates on the GEM surfaces. The GEMs were subject to a variety of quality assurance tests, including leakage current tests and visual inspection for deformities, missing holes, wrinkles, tears, significant discoloration, and lifting of the GEM frame, among others. \par
The modules were assembled out of GEMs that passed all QA steps. The GEMs were stacked on top of a pad plane that had already been glued to a strongback. The GEMs are held down with nylon nuts on nylon threaded rods that mate to holes in the pad plane. The stacked module was then HV tested in air to ensure the GEMs would hold voltage while stacked. 
\begin{figure}[htbp]
    \centering
    \includegraphics[width=13.9cm]{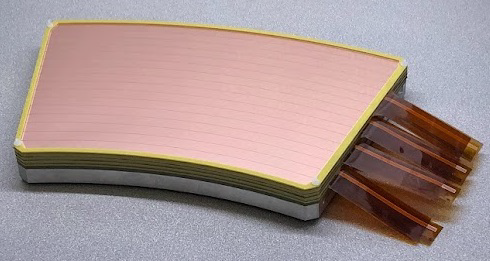}
    \caption{A fully stacked R2 module. The white nylon nuts can be seen in the four corners of the module. The staggering of the HV tails is also apparent.}
    \label{fig:GEMStack}
\end{figure}
After the stack passed this test, the HV tails were tucked through the gap between the strongback and the pad plane and glued in place with DP460. This is the methodology by which the HV tails are made accessible from outside the TPC volume while maintaining a good gas seal.
\begin{figure}[htbp]
    \centering
    \includegraphics[width=13.9cm]{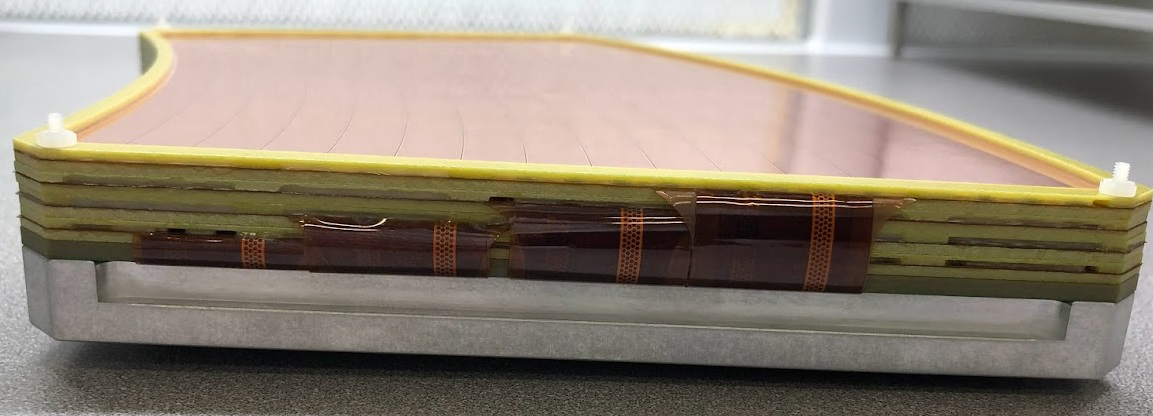}
    \caption{A side view of the module after the tail potting procedure. The tails were kept as close to the module as possible, to prevent them from obstructing the neighboring module during insertion into the TPC.}
    \label{fig:GEMStackTucked}
\end{figure}
The fully assembled module was then tested with an X-ray source. The modules were inserted into the module test box, which is a sealed gas volume with a drift cathode to provide a similar drift field to the setup in sPHENIX. The modules were then transported to the X-ray test stand where they were irradiated with X-rays from an X-ray source utilizing a copper target. The source is mounted to a 2-D translating stage that allowed it to scan all positions on the module surface. The current on the pad plane was measured as a function of the X-ray source position. This produced a map of the relative gain as a function of location on the module. The test box was hooked up to a picoammeter to measure the IBF as a positive current on the drift cathode. In this setup, the voltage on each GEM surface could be varied. This is in contrast to the real TPC, which has fixed value resistors defining the voltages on all the GEM surfaces.\par
Initially, the same set of voltages used in the 2019 test beam were used, however the signal was smaller than expected. The voltages were increased slightly to compensate for the difference in elevation between Fermilab, where the operating voltages had been originally determined, and Stony Brook, at which point the expected signals were observed. The GEM module performances were characterized at various voltages to determine the optimal operating point in terms of IBF reduction and gain. Once this point was determined, fixed resistors were purchased which would most closely produce the desired voltages\footnote{Due to supply chain issues in 2022, fairly few values of resistors were available at the time of construction. To achieve the necessary voltages, some resistors had to be soldered in parallel.}. The final voltage settings for the GEMs are summarized in Table~\ref{tab:GEMVoltages}.
\begin{table}[tbhp]
  \footnotesize
  \begin{center}
    \begin{tabular}{lc}
      \hline
      GEM & Voltage (V)\\
      \hline
        4 Top&4793.0\\ 
        4 Bottom&4623.2\\ 
        3 Top&3375.2\\ 
        3 Bottom&2987.1\\ 
        2 Top&1739.2\\ 
        2 Bottom&1291.6\\ 
        1 Top&1260.4\\ 
        1 Bottom&624.0\\ 
       
      \hline
    \end{tabular}
    \caption{
     GEM voltages implemented in sPHENIX with respect to ground.
    }
    \label{tab:GEMVoltages}
    \end{center}
\end{table}
In the TPC as built, only one HV cable services an entire module. The voltages are distributed to the GEMs via a HV-rated PCB designed to accept the resistors necessary to achieve the GEM voltages listed in Table~\ref{tab:GEMVoltages}. The GEM voltages result in nominal electric fields which are given in Table~\ref{tab:EFields}. 
\begin{table}[tbhp]
  \footnotesize
  \begin{center}
    \begin{tabular}{lc}
      \hline
      Location & Electric Field (kV/cm)\\
      \hline
        Drift Volume &0.4\\
        GEM 4 Hole &33.96\\ 
        GEM4 $\rightarrow$ GEM3 Transfer Gap&3.90\\ 
        GEM 3 Hole &77.62\\ 
        GEM3 $\rightarrow$ GEM2 Transfer Gap&3.90\\ 
        GEM 2 Hole &89.52\\ 
        GEM2 $\rightarrow$ GEM1 Transfer Gap&0.097\\ 
        GEM 1 Hole &124.8\\ 
        GEM1 $\rightarrow$ Pad Plane Transfer Gap&1950\\ 
      \hline
    \end{tabular}
    \caption{
     Electric fields in GEM transfer gaps and GEM holes implemented in sPHENIX.
    }
    \label{tab:EFields}
    \end{center}
\end{table}
\subsection{Electronics}
\label{subsec:Electronics}
\begin{figure}[htbp]
    \centering
    \includegraphics[width=13.9cm]{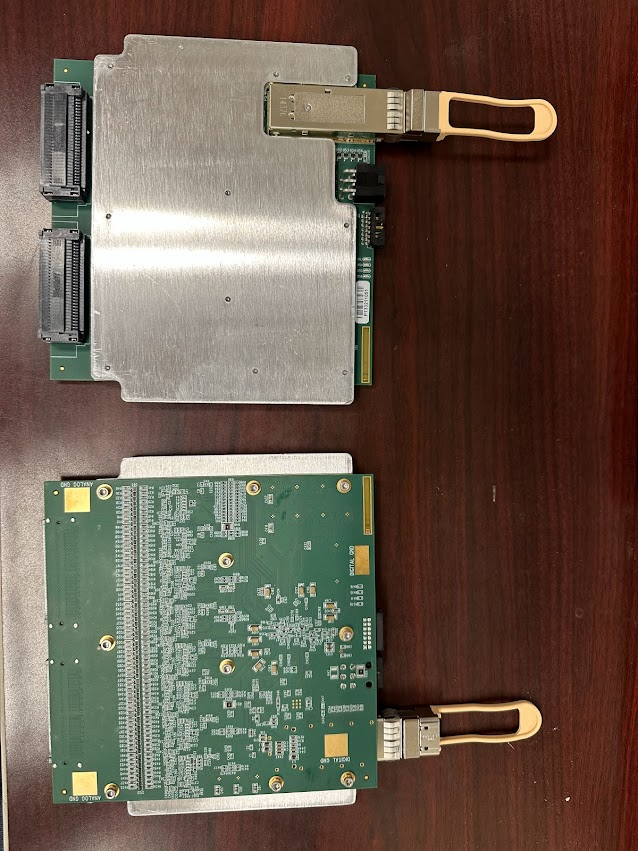}
    \caption{The front and back of a TPC FEE card mounted to a cooling backplate. The entire front face of the FEE is embedded in gap pad to produce good thermal contact to the aluminum plate.}
    \label{fig:FEEBack}
\end{figure}

The data created in the TPC for a single Au+Au event is around 10 megabits. At the nominal 15 kHz interaction rate, this means that the TPC is producing around 20 GB/s The TPC front-end electronics (FEE) are fairly similar to those used by ALICE. The goal of the FEE is to accept the analog signal coming from the pad plane, amplify it, convert it to a digital signal, and pass it along an optical fiber to the data-aggregation module (DAM). The primary component of the TPC front-end electronics is the SAMPA chip~\cite{Barboza:2016ala}, a custom circuit designed for continuous readout data processing. 
\begin{figure}[htbp]
    \centering
    \includegraphics[width=13.9cm]{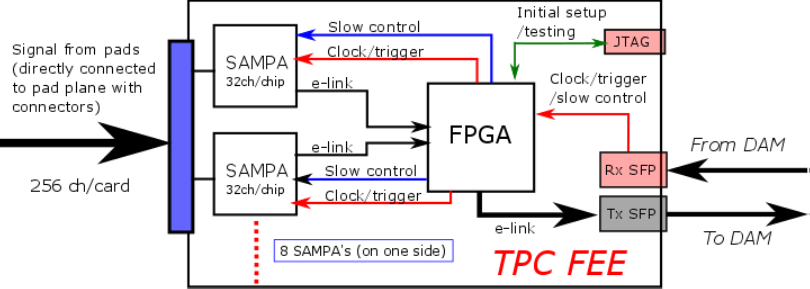}
    \caption{A schematic outlining the features and layout of the sPHENIX TPC FEE.}
    \label{fig:FEEDiagram}
\end{figure}
The SAMPA consists of a charge-sensitive amplifier, an analog-to-digital converter (ADC), a shaper, and a digital signal processor. Each SAMPA chip reads out a set of 32 pads, and each sPHENIX FEE card contains 8 SAMPA chips. The total number of FEE utilized by the TPC is 624. Each R1 readout module takes 6 FEE cards, R2 takes 8, and R3 takes 12. 
\begin{figure}[htbp]
    \centering
    \includegraphics[width=13.9cm]{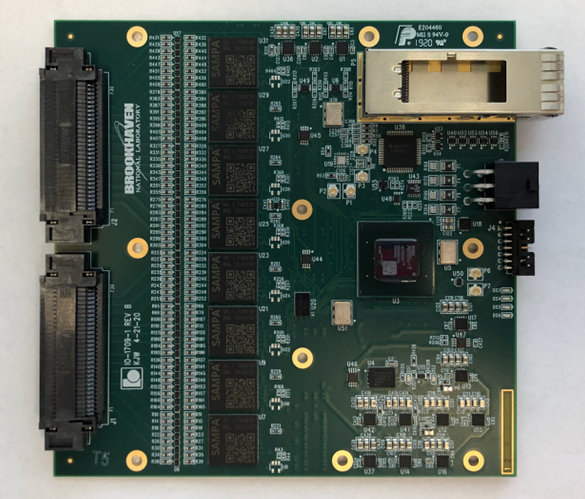}
    \caption{A TPC FEE card before being mounted to a cooling backplate. The 8 SAMPA chips can be seen, along with the fiber optic coupling housing in the top right.}
    \label{fig:FEECard}
\end{figure}
The primary difference between the SAMPA as applied in ALICE and sPHENIX is the shaping time, mentioned briefly in Sec.~\ref{Sec:RateCapability}. The SAMPA chip natively has multiple options for the shaping time of the analog signals, input polarity, and sensitivity. The shaper component of the SAMPA chip can operate at shaping times of 300, 160, and 80 ns. ALICE selected the 160 ns shaping time, while the 80 ns option was heavily preferred for the sPHENIX application. To that end, the University of Sao Paulo re-implemented the 80 ns option in the SAMPA v5 chip\footnote{Now distinct from the version currently being used by ALICE, which is the SAMPA v4.}. The SAMPA amplifier can operate at various gains, including 20 and 30 mV/fC. After the re-tooling by Sao Paulo, the v5 SAMPA chips were tested by Sao Paulo and Lund. The distribution of information to and from the SAMPA chips is handled by an Artix-7 200T FPGA. The power consumption of one FEE is around 10 W. To aid in the dissipation of this heat, the FEE are mounted via thermally conductive adhesive known as ``gap pad" to an aluminum backplate, shown in Fig.~\ref{fig:FEEBack}. This aluminum backplate will be cooled via a cooling system described in Sec.~\ref{subsec:cooling} and will provide fairly uniform temperature across the FEE.
\par
After the signals leave the FEE, they enter the DAM. The DAM aggregates the data and processes it for eventual storage. The DAM also has the capability to send information to the FEE, allowing for slow control, trigger information, and timing distribution. The DAM is essentially a high-bandwidth server, which can interface between the FEE and the data acquisition system. The DAM boards are frontend link exchange (FELIX) cards, developed for the ATLAS experiment~\cite{Ilic:2019yyj}. 
\subsection{Cooling}
\label{subsec:cooling}
\begin{figure}[htbp]
    \centering
    \includegraphics[width=13.9cm]{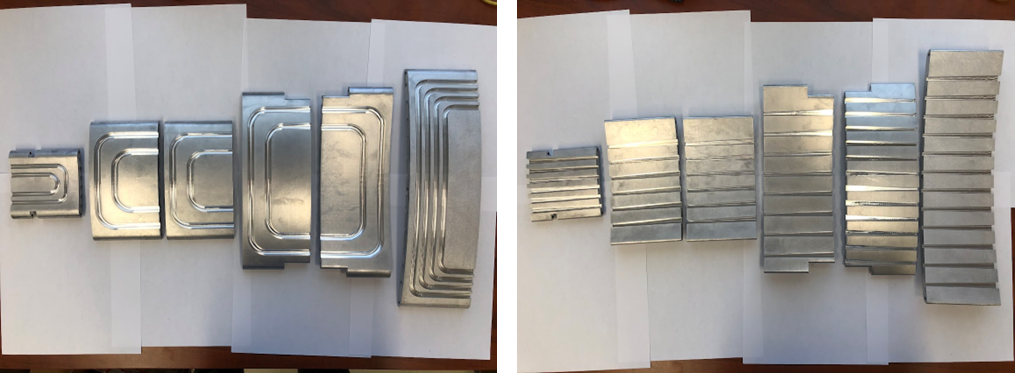}
    \caption{The aluminum cooling blocks. Left: View of the trenches into which the copper tubes are embedded with thermally conductive epoxy. The second and third blocks from the left and the fourth and fifth from the left are the blocks which service modules on both sides, and these two halves will additionally be epoxied together, to produce more thermal uniformity. Right: View of the slots for the FEE cards. The FEE will be inserted along with a wedge-shaped piece of aluminum and a thermally conductive piece of gap pad to ensure that the thermal contact of the FEE backplate to the cooling block is sufficient.}
    \label{fig:CoolingBlocks}
\end{figure}

Since the total power output of the electronics for one side of the TPC is around 3-4 kW, substantial cooling is required to prevent both the FEE cards from overheating, and the temperature of the gas changing via thermal conductivity through the endcap. To this end, the FEE are mounted into water-cooled cooling blocks that sit at the radial boundaries between modules, shown in Fig.~\ref{fig:CoolingBlocks}. The cooling water flows through the cooling blocks via a series of polyflow tubes that connect them to each other, and finally to a larger copper cooling manifold. Each quarter of the TPC face has two manifolds stacked on top of one another, one of which serves as a supply of cool water, and one which serves as the return for water that has been warmed by the FEE. The water flows through the blocks in two directions to ensure uniform cooling. The cooling system is powered by an underpressure chiller from Chilldyne, meaning leaks in the system will introduce air into the water flow instead of leaking water out. This is a significant benefit for a detector like the TPC which cannot be easily accessed. The INTT and MVTX supports and services effectively block access to the TPC, making the ability to fix water leaks that occur on the TPC severely limited.\par
The cooling blocks themselves are made of aluminum to minimize material budget and cost. Inside the cooling blocks, copper tubes are epoxied in place using a thermally conductive epoxy. The copper tubes are responsible for bringing in the cooling water and cooling the aluminum block. Images of these components are shown in Figs.~\ref{fig:CoolingBlocks} and \ref{fig:CoolingBlocks}.

\begin{figure}[htbp]
    \centering
    \includegraphics[width=13.9cm]{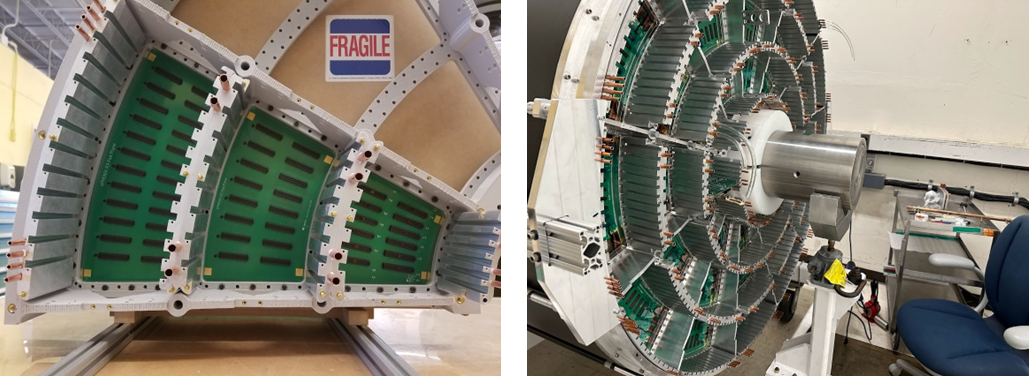}
    \caption{Left: The TPC cooling blocks installed in the 3D printed test sector. There are four cooling blocks, corresponding to the inner radius of R1, the outer radius of R3, the boundary between R2 and R3, and the boundary between R1 and R2. Right: The cooling blocks installed in the TPC.}
    \label{fig:CoolingLayout}
\end{figure}
The cooling blocks have a series of wedge-shaped slots, into which the FEE are mounted. The FEE backplates have extensions that are designed to fit loosely into the slots. This allows them to be installed with the necessary mechanical freedom to ensure a good fit of the FEE connector to the pad plane connector. Strips of gap pad were attached to the FEE backplate extensions. Gap pad only thermally conducts properly once it has been squeezed between the two materials that should be in thermal contact. Once the FEE is situated, a wedge is driven between the FEE and the cooling block, on the side of the FEE backplate that does not have the gap pad. The wedge presses the side of the backplate with the gap pad directly into the cooling block, achieving thermal contact between the FEE and the cooling block. The vertical positioning of the wedge is controlled with a single bolt, which can apply large amounts of force. To avoid mechanical stresses on the FEE, in most cases the bolts were not tightened to the utmost. It was additionally noticed after some power cycles of the FEE that the backplate, wedge, cooling block system became very tightly bound to one another, possibly due to thermal expansion. This unexpectedly good contact between the metal surfaces bodes well for the smooth operation of the cooling system.\par
\begin{figure}[htbp]
    \centering
    \includegraphics[width=13.9cm]{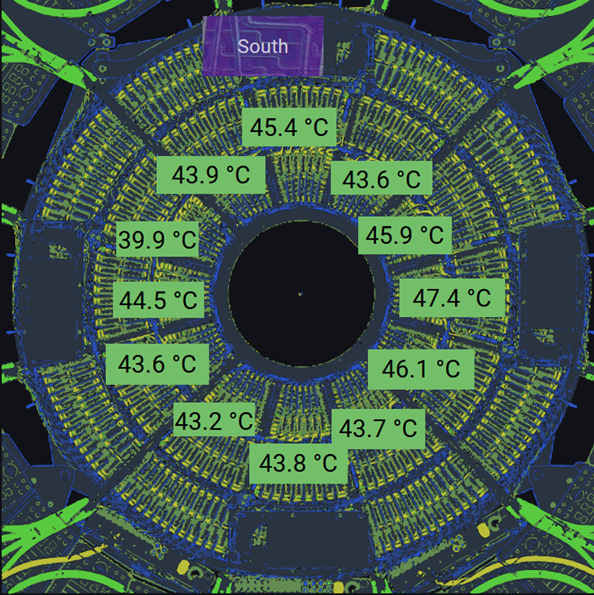}
    \caption{State of the TPC FPGA temperatures in steady-state operation of the FEE without beam. It is likely that during RHIC operation the temperatures will increase slightly.}
    \label{fig:TPCTemps}
\end{figure}

A typical problem in applications such as this is that the coolant heats up as it travels through the system. Thus elements downstream see warmer coolant, and are not cooled to the same degree as elements upstream. In sPHENIX, the coolant cannot be made too cold, or else it will produce condensation on various surfaces such as the cooling blocks, which could damage the nearby electronic equipment. To alleviate this issue, in the TPC cooling system the supply and return of the coolant, which is water with additives to prevent the growth of biological organisms, flow in opposite directions, such that a cooling block near a cooling manifold will see both fresh cold water coming directly from the manifold and some water which has already been heated via contact with the previous cooling blocks in the series. In general, the flow proceeds from block to block by a series of U-shape segments of polyflow tubing clamped to the copper tubes. The exception is at locations where the line laser system interferes with this pattern. At those locations, a ``shark-bite" push-to-connect elbow is used to mate the copper tube to a yellow polyflow tube at a 90 degree angle. 
\begin{figure}[htbp]
    \centering
    \includegraphics[width=13.9cm]{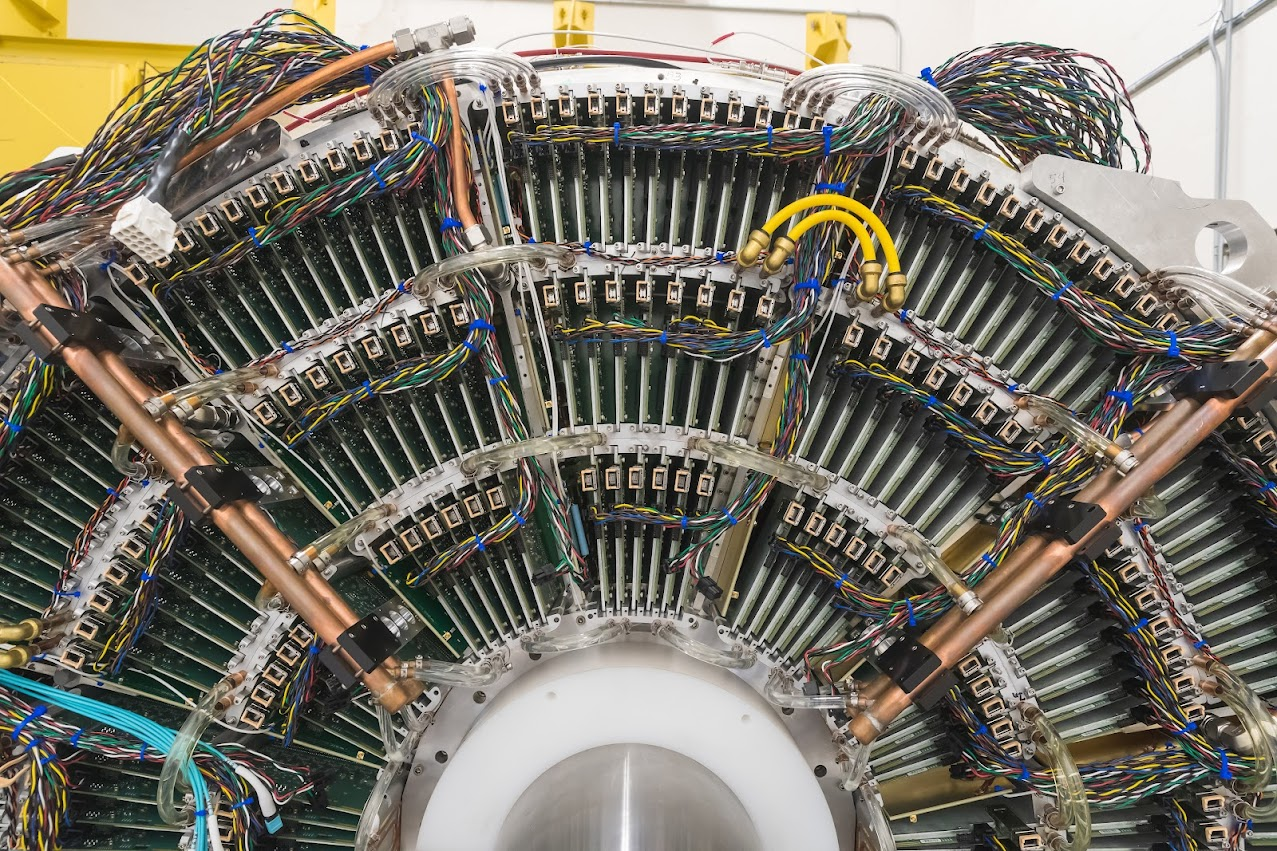}
    \caption{Image of the TPC face demonstrating the line laser avoidance technique. The normal clear cooling tubes can be seen on the left, and the yellow polyflow tubes that travel around the line laser (which is not yet installed in this image) can be seen on the right.}
    \label{fig:SharkBite}
\end{figure}
This polyflow tube bends to circumvent the interference from the line laser and reaches another shark-bite connector that interfaces it back into the normal cooling pattern. Overall, the counter-circulating of the supply and return achieves a very uniform temperature distribution.
\begin{figure}[htbp]
    \centering
    \includegraphics[width=13.9cm]{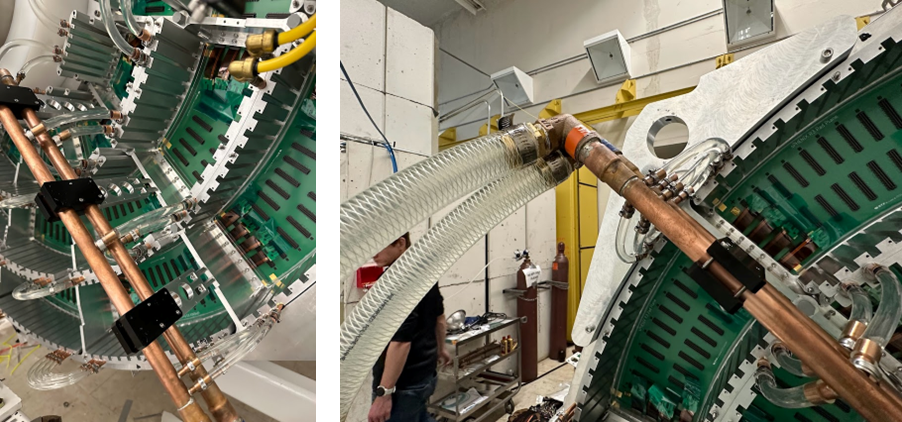}
    \caption{Left: Image of the copper cooling manifold bolted in place to the rib of the wagon wheel. Right: Cooling manifold at the outer radius of the TPC, where the cooling water is supplied and returned.}
    \label{fig:CoolingManifolds}
\end{figure}
The TPC FEE have the ability to report their temperatures to the DAM. If the temperatures begin to climb to unsafe levels, an alarm is thrown and the FEE are powered down. The element on the FEE cards which typically runs at the highest temperature is the FPGA. With the full cooling system in place, the FEE on a full endcap can be turned on and readout in essentially the same situation as when the experiment is operating. The highest temperature reached at the FPGA during long-term operation of the FEE with the cooling system operating is around 50\textdegree C, while the FEE components are rated to 85\textdegree C. Typical temperatures are shown in Fig.~\ref{fig:TPCTemps}.
It is thus clear that the cooling system as built is capable of handling the sizable heat load produced by the TPC FEE. 
\subsection{Central Cathode}
\label{subsec:cathode}
The TPC central cathode is a lightweight honeycomb-PCB sandwich that is resistant to mechanical vibrations, and is designed to sit at 45 kV to provide the drift field of 400 V/cm. Previous TPCs such as STAR, ALEPH, and ALICE, have utilized a stretched mylar ``membrane" coated in conducting material as the central cathode. While the thinness of the stretched membrane maximizes the acceptance around $z\sim0$, it also introduces the ability for vibrations, gas pressure differences, or electrostatic forces to produce macroscopic deflections in the cathode, which consequently affect the drift field at all locations in the TPC. These deflections can be challenging to pin down and correct for during a running experiment, as they can be time-dependent\footnote{In the case of mechanical vibrations, seismic activity and other man-made environmental vibrations are generally on the order of 0.1 to 200 Hz. Depending on the tension with which they are pulled taut, membranes are susceptible to picking up resonances in these frequencies, producing electric field variations that are often too rapid to calibrate away with slow calibration methods, but are slow compared to the electron drift times.}. The losses in tracking due to a central cathode occupying more space in $z$ were studied in simulation, and determined to be negligible. The central cathode is formed of a single piece of large-cell 20-mil thick honeycomb with 18 thin, ENIG-coated PCB ``petals". 
\begin{figure}[htbp]
    \centering
    \includegraphics[width=13.9cm]{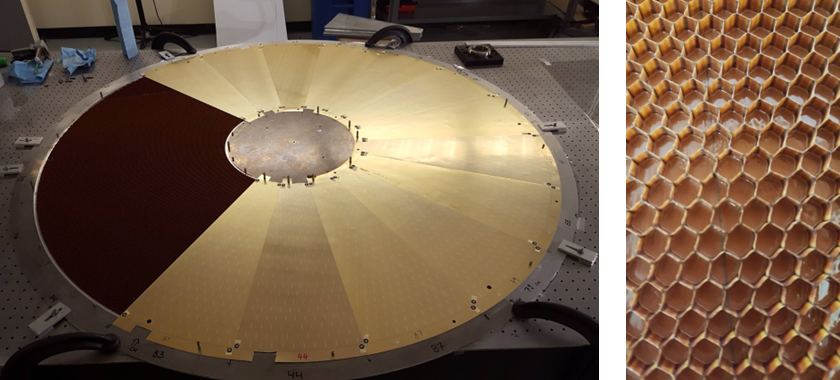}
    \caption{Left: Central cathode being constructed on a flat table at BNL. Right: Honeycomb epoxied to petals.}
    \label{fig:CMConstruction}
\end{figure}
The petals were designed to be small enough to fit into the evaporator system at Stony Brook, where aluminum stripes were evaporated onto them to be utilized by the diffuse laser calibration system. The method of operation of the diffuse laser system is the photoelectric effect. The work function of aluminum is substantially lower than that of gold, allowing the laser to free electrons from the aluminum stripes and not from the rest of the petal surface. The evaporation process involved first evaporating chromium to act as an adhesive surface for the aluminum to stick to, then evaporating the aluminum. Aluminum in particular is fairly robust in its quantum efficiency, i.e. the quantum efficiency is not highly sensitive to parameters such as the thickness, surface roughness, etc. which are challenging to control. The work function of aluminum is $\sim$4.1 eV, corresponding to a photon wavelength of $\sim$300 nm. The work function of both gold and nickel are $\sim$5 eV, corresponding to a photon wavelength of $\sim$250 nm. Thus the 266 nm laser envisioned for use in sPHENIX will produce photons energetic enough to free electrons from the evaporated aluminum stripes, but not from the ENIG coating.\par
\begin{figure}[htbp]
    \centering
    \includegraphics[width=13.9cm]{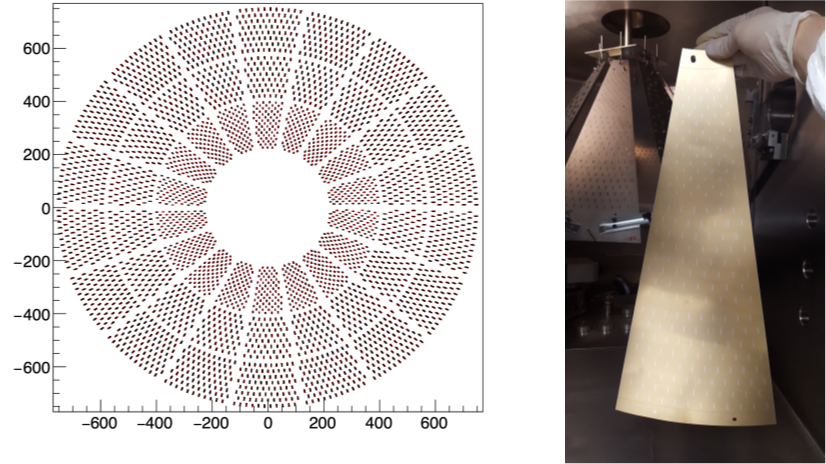}
    \caption{Left: Locations of aluminum stripes on the central cathode in units of mm. Right: Central cathode petal after evaporation of the aluminum stripes.}
    \label{fig:CMPattern}
\end{figure}

The aluminum stripes are evaporated onto the central cathode in a pattern, shown in Fig.~\ref{fig:CMPattern}. This pattern maximizes the ability to determine the direction and magnitude of local field distortions with fine granularity. The evaporated stripes are radially aligned with the pad rows, and roughly one centimeter in length. 
One of the obvious challenges in producing the central cathode out of many individual pieces is ensuring the flatness of all the surfaces with respect to each other. Misaligned or warped petals would obviously be a source of electric field distortions. 
The honeycomb used had a tolerance of $\pm6$ mils on its thickness. A flat optical table\footnote{The typical surface flatness is $\sim$0.005 in. over the size of the central cathode.} at BNL was used as the reference flat surface. A rig was constructed to aid in the positioning of each of the petals, utilizing the precision holes in the optical table.
\begin{figure}[htbp]
    \centering
    \includegraphics[width=13.9cm]{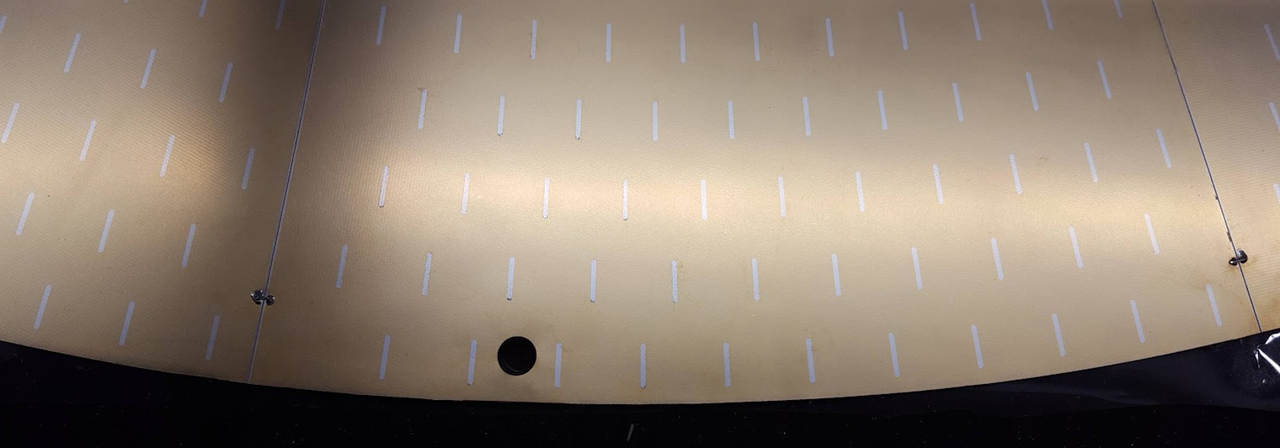}
    \caption{A close up of the edge of the central cathode. The aluminum calibration stripes can be seen in silver. One of the alignment holes is visible, along with the petal segmentation and the jumper solder connection keeping the petals electrically connected to one another.}
    \label{fig:CMStripes}
\end{figure}
A series of holes around the perimeter of the cathode serve as the support points on the field cage chalices. Nylon bolts were used to attach the cathode to the chalices, were were precisely positioned on the field cage.
\begin{figure}[htbp]
    \centering
    \includegraphics[width=13.9cm]{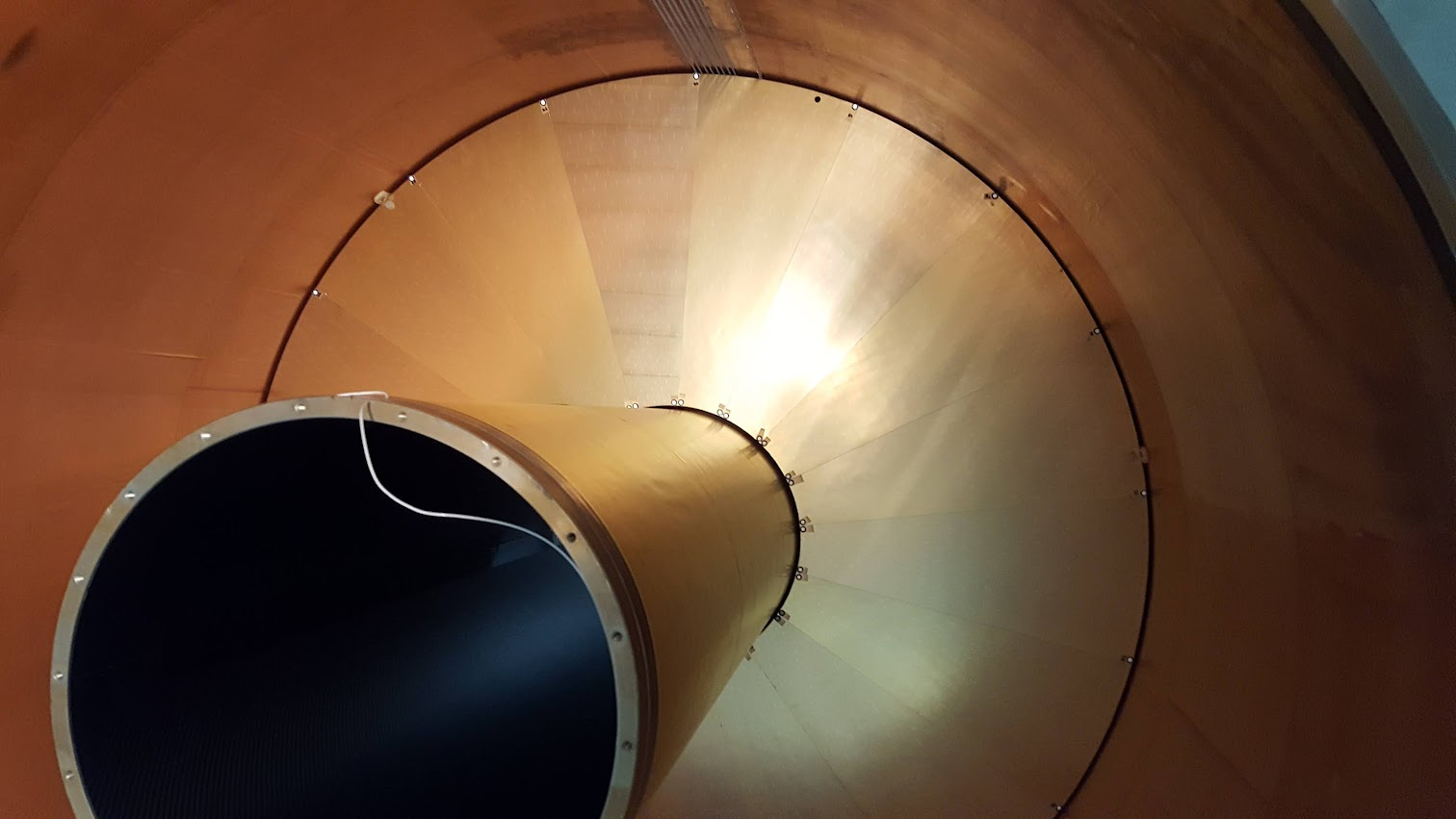}
    \caption{The central cathode installed inside the TPC.}
    \label{fig:CMInstalled}
\end{figure}
After the installation procedure, the central cathode location was precisely surveyed by the BNL survey team before the TPC volume was sealed.
\subsection{Laser Calibration Systems}
\label{lasers}
As mentioned previously, the TPC is equipped with two distinct laser calibration systems. TPCs have historically been equipped with laser calibration systems since the 1980s~\cite{Brand:1989qv,TOPAZ-TPCGroup:1986jsz,ALICETPC:2020ann,Lebedev:2002sp,Blum:1984um}. The uses of these lasers primarily focus on the measurement, and correction of, electron position distortions. The TPC requires known sources of electrons that can accumulate the position distortions, such that they can be subtracted from actual tracks. Additional benefits from the laser calibration systems include the ability to monitor the drift velocity and transverse diffusion of the electrons. The two systems are the diffuse laser which makes use of the photoelectric effect, and the line laser which ionizes impurities in the gas. Both lasers utilize 266 nm laser light for its high capability to extract electrons. Critical for both lasers is that the magnitude of the signals produced by the lasers fall with the dynamic range of the TPC FEE. If the signals are too large, they run the risk of saturating the FEE and losing precision in the charge-sharing.\par
The diffuse laser is an Onda 266 nm laser. This laser offers fairly high power and high repetition rate, allowing it to be operated at up to 50 kHz. The peak of the average pulse power comes at 15 kHz, where the average pulse energy is around 60$~\mu$J. The timing of the laser pulse is around 2 ns, significantly shorter than can be resolved by the TPC electronics. For the purposes of extracting drift velocity, the pulse arrives essentially instantaneously. The laser is expected to degrade  There is an interesting optimization question with regard to firing the laser with high frequency. Firing at high frequency would allow for essentially each event to have its own distortion map. However, since every freed electron will produce $\sim$20 or so ions at the readout, firing the laser more rapidly will increase the amount of space charge in the TPC. The answer to the question of laser firing frequency will likely come after operations begin, and the impact of the space charge produced by the laser can be evaluated against the benefit of running it at high rate. The laser is transported via fiber to the TPC face, where it enters into a sheath with a diffuser situated inside the TPC volume. The diffuser is set at an optimized angle, such that the intensity of the light will be great enough to free sizable numbers of electrons from the central cathode. By discretizing the pattern of stripes on the central cathode, both radial and azimuthal distortions can be monitored, unlike the designs of STAR and ALICE. 
\par
The line laser utilizes a high power 266 nm Q-SPARK laser from Quantum Light Instruments (QLI) capable of ionizing the gas in the TPC. The laser was modified by QLI to contain no magnetic components, to permit stable operation in the sPHENIX magnetic field. The laser itself has a neodymium-doped yttrium aluminum garnet (Nd:YAG) lasing medium, which lases at 1064 nm. The light is frequency doubled twice to produce an output frequency of 266 nm. A Cr:YAG passive element converts the output into 1.5 ns or shorter pulses of around 0.8 mJ that can fire at up to 50 Hz. The ionization energies for CF$_4$ and argon are both too high to be directly ionized by the laser. Instead, impurities at the ppb level can be used as the source of ionization. This technique has been successfully applied in various other collider TPCs, and these impurities generally pose no issue for the drifting electrons due to their low concentration. Unlike ALICE and STAR, where ample radial space was available for ``planes" of laser beams to shine perpendicular to the $z$-axis from various $z$ locations, in sPHENIX the need to utilize all available radial space for active volume, as well as the merging of the gas volume and field cages, makes such an arrangement impossible. Thus, the line laser shines into the TPC via quartz light pipes installed in the wagon wheels. The light pipes are fitted into custom sheaths with NPT threads that allow them to be bolted into the NPT threaded holes in the wagon wheels. Unlike the diffuse laser, the power of the line laser is too high for it to be transported using fibers. A series of mirrors direct the line laser beams through dedicated beam channels flushed with argon gas into the light pipe. To reduce the mechanical complexity and light losses of the beam transport system, the lasers are mounted directly to the wagon wheel, instead of being mounted in a rack and brought to the TPC from a distance as in the diffuse laser.
\begin{figure}[htbp]
    \centering
    \includegraphics[trim={0 1cm 0 0},clip,width=13.9cm]{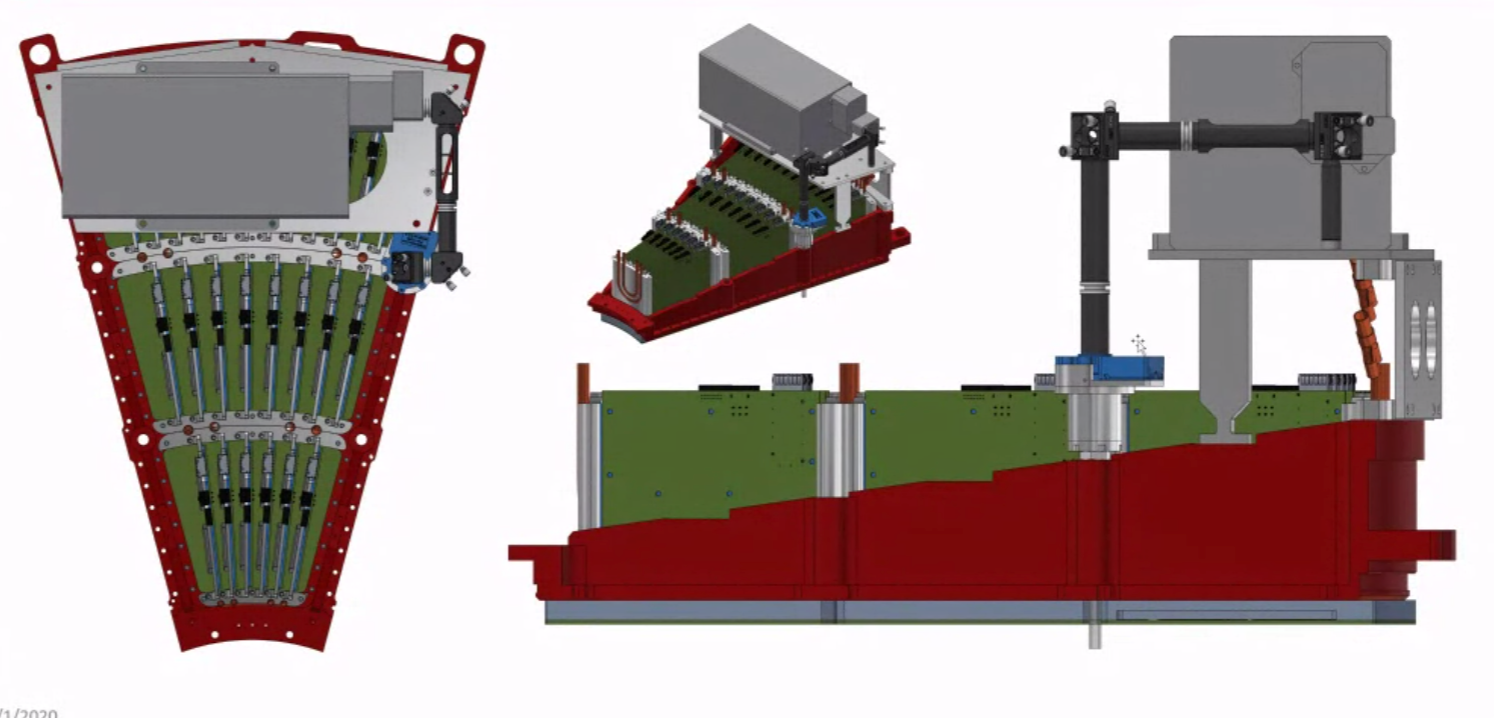}
    \caption{The position and design of the line laser and beam steering system. The wagon wheel is represented in red. The laser itself rests above the R3 module, and the laser is steered into the NPT port situated along the spoke of the wagon wheel between the R2 and R3 radii. In the farthest right view, the extent which the light pipe extends into the TPC volume can be seen.}
    \label{fig:LineLaserAssembly}
\end{figure}
\par
\begin{figure}[htbp]
    \centering
    \includegraphics[width=13.9cm]{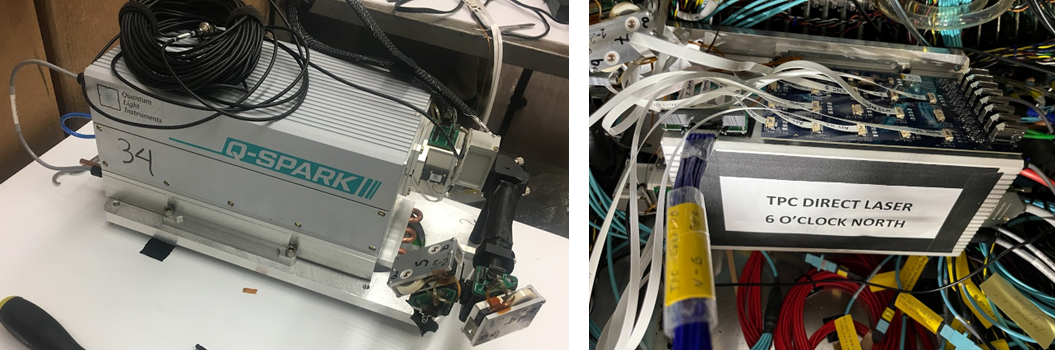}
    \caption{Left: Line laser system pre-installation. The beam transport apparatus can be seen on the right side of the laser. Right: Laser installed in the gravitationally down (6 o'clock) position on the TPC north side. The blue PCB that interfaces the flat flex ribbon cables which are used to operate the steering motors to the HDMI cables that allow communication with them can be seen on top of the laser.}
    \label{fig:LineLaserInstalled}
\end{figure}

The mirrors which direct the line laser are mounted to motors, which can change the angle and location of the laser at the entrance and exit of the light pipe. The ability to articulate the beam inside the TPC by manipulating the entrance angle and location of the beam spot entering the light pipe is essential. The motors must operate reliably and to high precision in the 1.4T magnet, and thus traditional magnetic motors were not an option. Piezoelectric motors were selected for their high precision and magnetic field tolerance. The full system of motors for controlling the line lasers is complex, and a complete description is, unfortunately, beyond the scope of this thesis.
\begin{figure}[htbp]
    \centering
    \includegraphics[trim={0 0 0 2cm},clip,width=13.9cm]{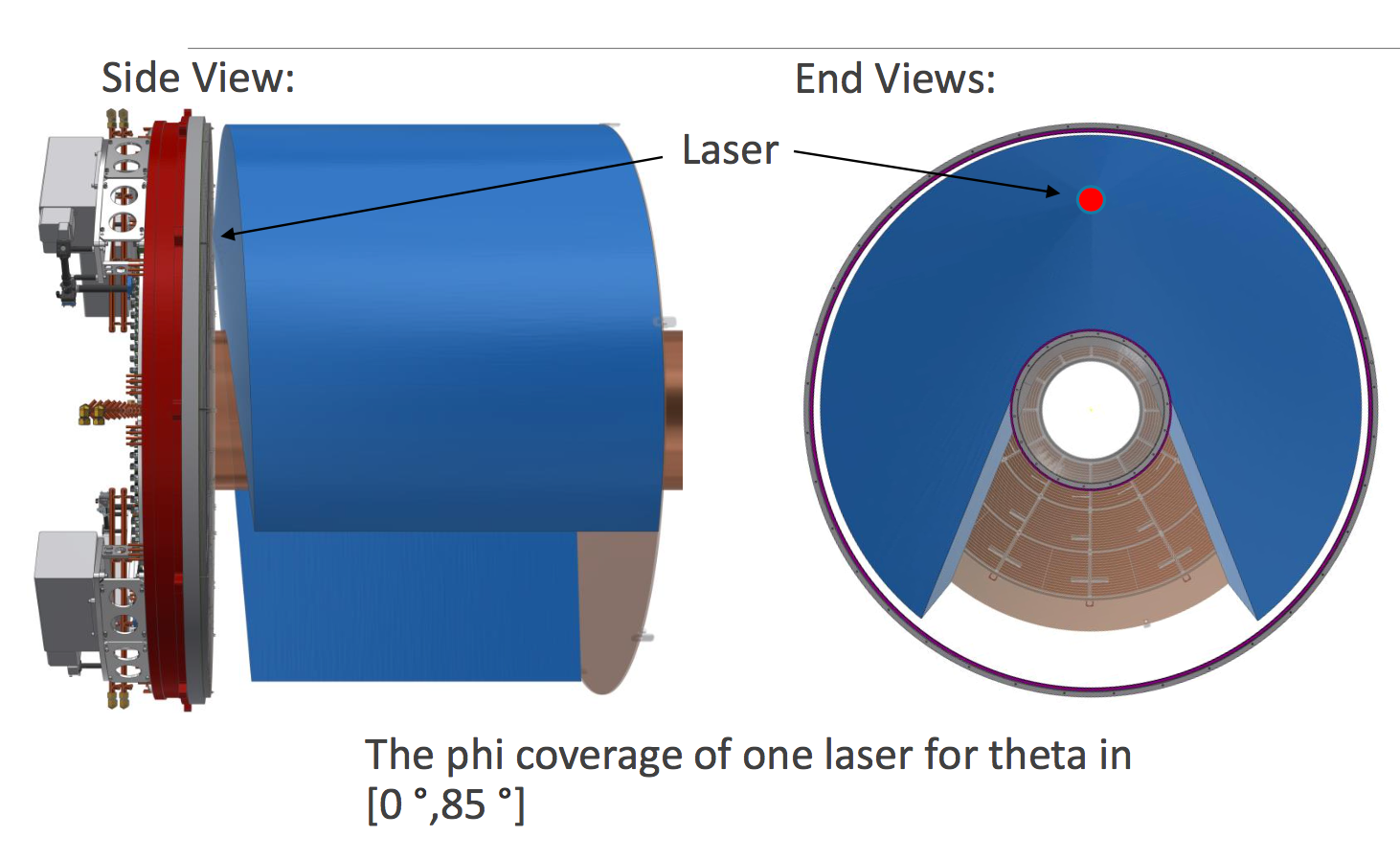}
    \caption{The theoretical range accessible by the line laser. In reality, due to misalignments, losses in the light pipe, etc. the coverage will likely be smaller.}
    \label{fig:LineLaserCoverage}
\end{figure}
The line lasers can be mounted at four positions on both the north and south sides of the TPC. However, due to constraints of time and available parts, only the $\pm x$ and -$y$ locations on the TPC were populated with lasers, i.e. the lasers that would be at the gravitational top of the TPC were not installed. The decision to instrument these laser positions was made to ensure that the region covered by the TPOT, which sits gravitationally below the TPC, would be well covered by the line laser beams. One of the line lasers as built and installed can be see in Fig.~\ref{fig:LineLaserInstalled}
\chapter{Conclusion}
Two projects in pursuit of advancing the understanding of QCD have been presented. The first study of groomed event shapes at HERA and the design and construction of the sPHENIX TPC. \par
A measurement of groomed event shapes was performed using archived H1 data. The data were unfolded and corrected for QED effects. The largest systematic uncertainties on the measurement result from the unfolding procedure, and the imperfect description of the data by the MCs used to unfold. The groomed event shapes as measured appear to deviate from most of the theoretical and numerical predictions. The conception that grooming could reduce or remove the importance of hard fixed-order splittings appears to be invalid in the high $Q^2$ kinematic reach of HERA. This implies that the same will be true for the accessible phase space at the EIC as well. Predictions from SCET should seek to match to fixed-order calculations, even for regions where nominally only resummation effects dominate.\par
The suite of modern MC event generators studied here cannot completely describe the $e+p$ hadronic final state. The high $Q^2$ DIS phase space studied here should provide the closest comparison of $e+p$ collisions to $e^+e^-$ data from LEP and $p+p$ data from the LHC, to which most of the generators have been tuned. The inclusion of higher-order processes is observed to significantly improve the description of the shapes of the groomed hadronic final state in MC event generators. The generators also fail to describe the inclusive DIS cross section at high $Q^2$. The need for improvement of the MCs before the beginning of the EIC is emphasized.\par
A state-of-the-art gaseous TPC was constructed for the sPHENIX experiment at RHIC. The TPC has a unique design driven by the need for very precise tracking at high rate. To achieve these goals, a quadruple GEM readout optimized for IBF suppression, a zig-zag patterned charge-sharing readout pad plane, and a low diffusion gas mixture have been designed and employed. A non-exhaustive description of some aspects of the design and construction were reported in this thesis. At the present time, the sPHENIX experiment is fully installed and eagerly awaiting the first beam from RHIC.
\bibliographystyle{unsrtnat}
\renewcommand{\baselinestretch}{1}
\normalsize

\clearpage
\newpage
\phantomsection%
\addcontentsline{toc}{chapter}{\numberline{}{Bibliography}}%
\bibliography{refs}

\clearpage
\newpage

\appendix
\include{sections/appendixSdkr}

\end{document}